\documentclass[11pt]{article}
\pdfoutput=1
\usepackage{jheppub}
\usepackage[T1]{fontenc}
\usepackage{cancel,tensor,enumitem,fix-cm}
\usepackage{epsfig}
\usepackage{graphicx}
\usepackage[english]{babel}
\usepackage{amsmath,amssymb}
\usepackage{dsfont}
\usepackage{textcomp}
\usepackage{multirow}
\usepackage{booktabs}
\usepackage{tikz}
\usepackage{overpic}
\usepackage{hyperref}
\usepackage{bm}
\usepackage{physics}
\usepackage[lofdepth,lotdepth]{subfig}
\usepackage{float}
\usepackage{xcolor}

\newcommand{\be}{\begin{equation}}
\newcommand{\ee}{\end{equation}}
\newcommand{\bea}{\begin{eqnarray}}
\newcommand{\eea}{\end{eqnarray}}

\newcommand{\beq}{\begin{equation}}
\newcommand{\eeq}{\end{equation}}

\def\({\left(}
\def\){\right)}
\def\[{\left[}
\def\]{\right]}

\subheader{\begin{flushright}
\texttt{BRX-TH-6689}
\end{flushright}}

\title{Sewing spacetime with Lorentzian threads: complexity and the emergence of time in quantum gravity}

\author[a,b]{Juan F. Pedraza,}
\author[a]{Andrea Russo,}
\author[a]{Andrew Svesko}
\author[a]{and Zachary Weller-Davies}
\affiliation[a]{Department of Physics and Astronomy, University College London, Gower st, London WC1E 6BT, UK}
\affiliation[b]{Martin Fisher School of Physics, Brandeis University, 415 South st, Waltham MA 02453, USA}
\emailAdd{j.pedraza@ucl.ac.uk}
\emailAdd{andrea.russo.19@ucl.ac.uk}
\emailAdd{a.svesko@ucl.ac.uk}
\emailAdd{zachary.weller-davies.17@ucl.ac.uk}

\abstract{Holographic entanglement entropy was recently recast in terms of Riemannian flows or `bit threads'. We consider the Lorentzian analog to reformulate the `complexity=volume' conjecture using Lorentzian flows -- timelike vector fields whose minimum flux through a boundary subregion is equal to the volume of the homologous maximal bulk Cauchy slice.  By the nesting of Lorentzian flows, holographic complexity is shown to obey a number of properties. Particularly, the rate of complexity is bounded below by \emph{conditional complexity}, describing a  multi-step optimization with intermediate and final target states. We provide multiple explicit geometric realizations of Lorentzian flows in AdS backgrounds, including their time-dependence and behavior near the singularity in a black hole interior. Conceptually, discretized flows are interpreted as Lorentzian threads or `gatelines'. Upon selecting a reference state, complexity thence counts the minimum number of gatelines needed to prepare a target state described by a tensor network discretizing the maximal volume slice, matching its quantum information theoretic definition. We point out that suboptimal tensor networks are important to fully characterize the state, leading us to propose a refined notion of complexity as an ensemble average. The bulk symplectic potential provides a specific `canonical' thread configuration characterizing perturbations around arbitrary CFT states. Consistency of this solution requires the bulk satisfy the linearized Einstein's equations, which are shown to be equivalent to the holographic first law of complexity, thereby advocating for a principle of `spacetime complexity'.  Lastly, we argue Lorentzian threads
provide a notion of emergent time. This article is an expanded and detailed version of \cite{Pedraza:2021mkh}, including several new results.}

\begin{document}
\maketitle
\flushbottom

\section{Introduction and summary}

\subsection{The big picture}

Spacetime physics and quantum information are fundamentally intertwined. The sharpest realization of the interplay between information and gravity is perhaps best captured by the Ryu-Takayanagi (RT) formula \cite{Ryu:2006bv},\footnote{The RT formula has been generalized in a number of ways, including a covariant formulation \cite{Hubeny:2007xt}; for CFTs dual to higher derivative bulk theories of gravity \cite{Dong:2013qoa,Camps:2013zua}, and when $1/N$ quantum corrections are included \cite{Faulkner:2013ana,Engelhardt:2014gca}.} relating the area of minimal surfaces in a $(d+1)$-dimensional (bulk) curved spacetime to the entanglement entropy of a state of a quantum field theory living on the $d$-dimensional (boundary) spacetime,
\beq
S(A)=\frac{1}{4G_{N}}\,\underset{m\sim A}{\text{min}}\,\text{area}(m(A))\,.\label{eq:RT}
\eeq
This statement is most precisely formulated in the context of the $\text{AdS}_{d+1}/\text{CFT}_{d}$ correspondence, where $S(A)$ is the entanglement entropy of a CFT state confined to a boundary region $A$, and $m(A)$ is the minimal codimension-2 bulk surface that is homologous to $A$. The RT formula can be seen as a generalization of the Bekenstein-Hawking entropy-area relation \cite{Bekenstein:1973ur,Hawking:1974sw}.\footnote{When the minimal area surface is a black hole horizon, the thermodynamic black hole entropy can be understood as the entanglement entropy of a CFT state called the thermofield double state \cite{Maldacena:2001kr}. Conversely, in the CFT vacuum, entanglement entropy can be shown to be equivalent, via conformal mapping, to the thermal entropy of an hyperbolic black hole \cite{Casini:2011kv}. A proof of (\ref{eq:RT}) for general states follows from a generalization of the Gibbons-Hawking derivation of black hole entropy \cite{Lewkowycz:2013nqa,Dong:2016hjy}.} It satisfies all known properties of the von Neumann entropy \cite{Headrick:2013zda}, and can be used to construct other important information theoretic quantities, including  (holographic) mutual information and Reny\'i entropy \cite{Dong:2016fnf}, each of which have dual geometric descriptions akin to (\ref{eq:RT}).

Recently, the entropy-area prescription (\ref{eq:RT}) was reformulated in terms of flows or holographic `bit threads' \cite{Freedman:2016zud}, in which the right hand side is replaced by the maximum flux of a divergenceless, norm-bounded (Riemannian) vector field $v$ through the boundary region $A$,
\beq
S(A)=\underset{v\in\mathcal{F}}{\text{max}}\int_{A}v\,, \qquad\mathcal{F}\equiv\left\{v\,|\,\nabla\cdot v=0\,,\,|v|\leq\tfrac{1}{4G_{N}}\right\}\,.\label{eq:RTasflowint}
\eeq
The equivalence between the two prescriptions follows as a consequence of the continuous version of the max flow-min cut theorem, a well known principle in network theory, where the `min cut' is the minimal surface $m(A)$. This reformulation was proven using convex optimization techniques in \cite{Headrick:2017ucz}, and has been generalized and applied in a number of ways \cite{Chen:2018ywy,Harper:2018sdd,Cui:2018dyq,Hubeny:2018bri,Agon:2018lwq,Du:2019emy,Bao:2019wcf,Harper:2019lff,Agon:2019qgh,Du:2019vwh,Agon:2020mvu,Headrick:toappear,Agon:2021tia,Rolph:2021hgz}. Not only does the flow reformulation (\ref{eq:RTasflowint}) of holographic entanglement entropy have certain technical advantages over the area based picture, it clarifies some conceptual issues surrounding the RT formula (\ref{eq:RT}). In particular, since the flows are defined everywhere in the bulk, the global character of (\ref{eq:RT}) is manifestly captured by the reformulation (\ref{eq:RTasflowint}), such that the bits encoding the microstate of $A$ do not localize on $m(A)$, but rather are carried by threads $v$.\footnote{If the bits encoding $A$ lived on $m(A)$, then the bits would seemingly jump with the minimal area surface whenever the entropy undergoes a phase transition under continuous deformations of the region $A$ (e.g. two disjoint intervals separated by a tunable distance $x$). Bit threads resolve this confusion.} Therefore, a thread emanating from boundary region $A$ which does not return is interpreted as a channel carrying a single independent (qu)bit of information encoding the microstate of $A$ such that the maximum number of such threads gives the entanglement entropy $S(A)$, which may be distilled as Bell pairs. Similarly, an entangled pair of bits between two boundary regions $A$ and $B$ is represented by a thread connecting $A$ and $B$. The bit thread reformulation (\ref{eq:RTasflowint}), moreover, has led to further insights into toy models of holography, in which spacetime is constructed by tensor networks \cite{Vidal:2007hda,Swingle:2009bg,Haegeman:2011uy,Swingle:2012wq,Almheiri:2014lwa,Pastawski:2015qua,Hayden:2016cfa,Harlow:2016vwg,Bhattacharyya:2016hbx}.

Collectively then, statements about gravity can be reinterpreted as relationships between information-theoretic quantities. In particular, gravitational field equations are dual to the first law of entanglement \cite{Lashkari:2013koa,Faulkner:2013ica,Faulkner:2017tkh,Haehl:2017sot} for which the RT formula (\ref{eq:RT}) is a fundamental input. In this way, bulk spacetime dynamics arise from entanglement. Moreover, the bulk spacetime metric itself may be reconstructed solely from boundary entanglement
\cite{Czech:2012bh,Balasubramanian:2013rqa,Balasubramanian:2013lsa,Myers:2014jia,Czech:2014wka,Headrick:2014eia,Czech:2014ppa,Czech:2015qta,Faulkner:2018faa,Roy:2018ehv,Espindola:2017jil,Espindola:2018ozt,Balasubramanian:2018uus,Bao:2019bib,Jokela:2020auu,Bao:2020abm}. Thus, in its deepest form, the RT prescription suggests spacetime connectivity emerges from entanglement \cite{VanRaamsdonk:2010pw,Bianchi:2012ev,Maldacena:2013xja,Balasubramanian:2014sra,Swingle:2014uza}, succinctly summarized by the slogan `entanglement=geometry'.

It has recently been suggested, however, that entanglement entropy alone is insufficient to describe all aspects of bulk gravitational physics \cite{Susskind:2014moa,Susskind:2014rva}. In particular, the late time growth of the Einstein-Rosen bridge inside eternal black holes is not captured by entanglement, but is seemingly instead characterized by complexity. By complexity, one typically means the state complexity, which, in an ordinary quantum mechanical setting, refers to the smallest number of unitary operators (gates) needed to obtain a particular final state from a given initial state within a particular margin of error. While the definition of state complexity in a field theory is still an active area of investigation (c.f. \cite{Chapman:2017rqy,Jefferson:2017sdb,Caputa:2017urj,Caputa:2017yrh,Chapman:2018hou,Hackl:2018ptj,Camargo:2019isp,Flory:2020eot,Flory:2020dja,Chagnet:2021uvi}), it is natural to ask what is the geometric dual of this boundary field theory quantity. Broadly, two proposals have emerged: `complexity=volume' (CV) \cite{Susskind:2014rva,Susskind:2014jwa,Stanford:2014jda,Couch:2016exn} and `complexity=action' (CA) \cite{Brown:2015bva,Brown:2015lvg,Fan:2018wnv}.

More precisely, the CV conjecture says the complexity $\mathcal{C}$ of the boundary CFT state restricted to Cauchy slice $\sigma_{A}$ delimiting a  boundary region $A$ (such that $\partial A=\sigma_{A}$ is dual to the volume of an extremal codimension-1 bulk hypersurface $\Sigma$ homologous to $A$ (for an illustration, see Figure \ref{fig:maxslicenest})
\beq
\mathcal{C}_{V}(\sigma_{A})=\frac{1}{G_{N}\ell}\,\underset{\Sigma\sim A}{\text{max}}\,\text{Vol}(\Sigma(A))\,.\label{eq:CVconj}
\eeq
Here the homology condition $\Sigma\sim A$ is simply that the endpoints of $\Sigma$ are identified with the boundary Cauchy slices $\sigma_{A}=\partial A$, and $\ell$ is some undetermined bulk length scale, \emph{e.g.}, the AdS curvature or the radius of the black hole.\footnote{Here the length parameter $\ell$ is often chosen in an \emph{ad hoc} manner; $\ell=L$  or $\ell=r_{h}$ for black holes large or small compared to $L$, respectively. This \emph{ad hoc} tuning has been a chief reason for preferring the `complexity=action' proposal over CV duality. It turns out, however, as pointed out in \cite{Couch:2018phr}, that for (spherical) black holes in $d+1\geq4$ dimensions, $\ell$ can be replaced by the maximum time $\tau_{f}$ to fall from the horizon to the final maximal cylinder, as $\tau_{f}$ naturally transitions between $L$ or $r_{h}$ when $r_{h}\geq L$ or $r_{h}\leq L$.} Alternatively, the CA conjecture equates the complexity with the gravitational action $I$ evaluated over a specific bulk region known as the Wheeler-De Witt (WDW) patch:
\beq
\mathcal{C}_{I}(\sigma_{A})=\frac{I_{\text{WDW}}}{\pi\hbar}\,.\label{eq:IWDW}
\eeq
Formally, the WDW patch is the domain of dependence of any bulk Cauchy surface that asymptotically approaches the boundary time slice.

While the CV and CA proposals are independent, they share many of the same qualitative features, though not all \cite{Carmi:2016wjl,Chapman:2018lsv,Chapman:2018bqj,Andrews:2019hvq,Bernamonti:2019zyy,Bernamonti:2020bcf}. Notably, in both proposals, the complexity grows linearly in time at a rate characterized by the mass and other thermodynamic potentials of the black hole; the response of either complexity to perturbations follows a `switchback effect', and the complexity of formation of large charged, static and symmetric  black holes is proportional to the black hole entropy.

Given their similar geometric character, it is natural to compare the CV proposal (\ref{eq:CVconj}) to the RT prescription (\ref{eq:RT}). On the one hand, boundary entanglement entropy $S(A)$ is found via a minimization procedure over bulk codimension-2 surfaces, while the boundary complexity $\mathcal{C}(\sigma_{A})$ arises from a maximization procedure over bulk codimension-1 surfaces. In light of the reformulation of the RT relation in the language of bit threads (\ref{eq:RTasflowint}), where the area of the min cut $m(A)$ is replaced with a flow of maximum flux through $A$, one may suspect the CV proposal for holographic complexity (\ref{eq:CVconj}) likewise has a flow based interpretation. Indeed, as first shown in \cite{Headrick:2017ucz} a Lorentzian analog of the max flow-min cut theorem exists, where the Riemannian flows are replaced by Lorentzian flows such that the minimum flux through a boundary region $A$ is equal to the maximum cut of a surface homologous to $A$.

In this article, we use this `min flow-max cut principle' \cite{Headrick:2017ucz} to reformulate the CV conjecture of holographic complexity in terms of Lorentzian flows and explore some of their properties and implications. Precisely, we propose
to compute $\mathcal{C}(\sigma_{A})$ as the minimum flux of a divergenceless, norm-bounded, time-like vector field $v$ through the boundary region $A$
\beq
\mathcal{C}(\sigma_{A})=\underset{v\in\mathcal{F}}{\text{min}}\int_{A}v\,, \qquad\mathcal{F}\equiv\left\{v\,|\,\nabla\cdot v=0\,,\,|v|\geq\tfrac{1}{G_{N}\ell}\right\}\,.\label{eq:cvreformin}
\eeq
which, via the min flow-max cut theorem, should be equal to the maximal volume of a Cauchy slice $\Sigma$ homologous to $A$. In terms of flows, we take advantage of powerful convex optimization techniques to prove a number of general properties holographic complexity is known to satisfy, including a bound on the complexity rate. We are also able to provide a conceptual interpretation of holographic complexity in terms of `Lorentzian threads', which we argue naturally augments the tensor network description of spacetime.

It is worth mentioning that, similar to the bit thread description of the Ryu-Takayangi formula, a crucial difference between CV duality  (\ref{eq:CVconj}) and our reformulation (\ref{eq:cvreformin}) is that while the maximal volume hypersurface is typically unique, the flow $v$ is highly non-unique. In particular, while $v$ takes a specific form on the maximal volume hypersurface, away from this surface $v$ is some arbitrary timelike, divergenceless vector field. Thus, there are infinitely many thread configurations that can achieve CV duality. Since the Lorentzian flows naturally probe the interior of a black hole, they may provide insight into the nature black hole microstates, of which different thread configurations correspond to different classes of microstates. This is motivation enough to study complexity and Lorentzian flows in detail. In this article we take advantage of the non-uniqueness of Lorentzian flows by showing a particular class of configurations -- `canonical' threads defined with respect to the bulk symplectic form --  can be used to show how spacetime dynamics emerges from boundary complexity.

Let us now provide a detailed summary of our central results.

\subsection{Summary of main results}

\textbf{Section \ref{sec:prelims}:} The chief goal of this section is to reformulate CV duality using Lorentzian flows. After briefly reviewing the various prescriptions of boundary complexity and the holographic duals in Section \ref{subsec:prescripts}, we recast CV duality in Lorentzian flow based language, given by (\ref{eq:cvreformin}), in Section \ref{subsec:reformCV}. This is accomplished using a continuum version of the min flow-max cut theorem, first proven in \cite{Headrick:2017ucz}. In Section \ref{subsec:genprops} we use the nesting theorem to derive a number of properties holographic complexity is expected to satisfy, including
\beq \mathcal{C}(\sigma_{A})-\mathcal{C}(\sigma_{AB})\leq \mathcal{C}(\sigma_{A}|\sigma_{AB});,\eeq
and
\beq \Phi(AC)+\Phi(BC)-\Phi(C)-\mathcal{C}(\sigma_{ABC})\leq0\;.\eeq
Here $\mathcal{C}(\sigma_{A}|\sigma_{AB})$ is defined as the minimal flux through region $A$, subject to having a fixed flux through $\Sigma(AB)$, and $\Phi(X)$ corresponds to the minimal flux through a boundary \emph{time-like} region $X$. The first relation can alternatively be understood in terms of a complexity rate, for which we are able to relate the flux through regions not homologous to a single bulk Cauchy slice to the integrated momentum flux, thereby making contact with the `momentum/volume/complexity' correspondence (PVC) \cite{Susskind:2018tei,Susskind:2019ddc,Susskind:2020gnl}.  We also provide a flow reformulation of the superadditivity of CV complexity \cite{Agon:2018zso, Du:2019emy,Caceres:2018blh,Caceres:2019pgf}, and briefly comment on the existence and obstacles in a general proof of `Lorentzian multiflows'.

 In Section \ref{sec:interpretation} we provide a `gateline' interpretation\footnote{The word `gateline' in the context of Lorentzian flows was first coined in \cite{Headrick:2017ucz} to offer a suggestive application of Lorentzian flows. In this article we provide the first explicit realization of Lorentzian flows as gatelines via state preparation.} of the Lorentzian flows in which the complexity $\mathcal{C}(\sigma_{A})$ is understood as the minimum number of threads or gatelines connecting the boundary region $A$ to its complement $\bar{A}$. Equivalently, $\mathcal{C}(\sigma_{A})$ is the minimum number of threads passing through the maximal volume slice $\Sigma(A)$ homologous to $A$. This perspective is conceptually appealing in that it matches to our intuition of circuit complexity being the minimum number of simple gates needed to prepare a target state from a given reference state. More precisely, a CFT state prepared by a Euclidean path integral via the Hartle-Hawking prescription such that its initial data is specified on a maximal volume Cauchy slice in the bulk Euclidean AdS. The threads piercing this bulk slice $\Sigma(A)$ then represent the various unitary gates needed to transform a reference state to a target state. This interpretation allows us to interpret the quantity $\mathcal{C}(\sigma_{A}|\sigma_{AB})$ as a \emph{conditional complexity}: the number of gatelines needed to prepare the state assigned to $A$ given the gatelines used to prepare the state of $AB$.\footnote{Notice that if the state on $AB$ is coincides with the reference state, then $\mathcal{C}(\sigma_{A}|\sigma_{AB})=\mathcal{C}(\sigma_{A})$. Conversely, if the state on $A$ coincides with the state on $AB$, then $\mathcal{C}(\sigma_{A}|\sigma_{AB})=0$.}

\vspace{2mm}

\noindent \textbf{Section \ref{sec:simpleconstructions}:} Lorentzian flows are defined as timelike, divergenceless, future directed vector fields $v$ with a norm bound $|v|\geq\alpha$ for $\alpha$ is some positive constant, where the norm is saturated on the maximal volume slice. As such, Lorentzian flows are highly non-unique. In this section our main goal is to provide explicit geometric constructions of the Lorentzian thread configurations described in Section \ref{sec:prelims}. Two such constructions are considered: starting from a family of integral curves (Section \ref{subsec:geoflows}) or starting with a family of level set surfaces (Section \ref{subsec:levelsetcons}). More precisely, integral line flows are timelike vector fields  which arise from foliating the bulk spacetime with a set of timelike curves, for example, geodesics. Their norm is chosen such that the divergenceless condition is satisfied locally. Level set flows, meanwhile, are timelike vector fields found from a family of nested slices foliating the spacetime, with integral lines orthogonal to these slices. We consider both constructions in empty $\text{AdS}_{n}$ and the BTZ black hole background, where, in the integral curve construction, the interior of the Wheeler De Witt patch is foliated by the geodesics, while in the level set construction the interior of the WDW patch is foliated by surfaces of constant extrinsic curvature. Moreover, to capture the time dependence of complexity and its connection with the second law of complexity, we use the geodesic method to uncover the Lorentzian flows at late times, where the maximal volume slice partially wraps around the singularity. We notice that, even at late times, the threads from this construction foliate the WDW patch, intersecting at its past tip and terminating at the singularity, thus probing Planck scale physics.

\vspace{2mm}

\noindent \textbf{Section \ref{sec:diffformsandeineqs}:} The objectives of this section are twofold. First we show how to recast the Lorentzian flow based formulation of holographic complexity in terms of closed differential forms in Section \ref{subsec:threadsdiffforms}, following a map between divergenceless vector fields $v$ and closed differential forms $u$. Moreover, we find it simplest to work with differential forms when developing the notion of \emph{perturbative} Lorentzian threads,  characterizing linear perturbations to AdS spacetimes, which are dual to excited CFT states, as discussed in Section \ref{subsec:excitedstatespertthreads}. Moreover, for a canonical choice of the perturbative Lorentzian thread form, $\delta u$, we discover the closedness condition $d(\delta u)=0$ encapsulates the linearized Einstein's equations, leading to the second achievement of this section, as we now describe.

It is known the mapping between boundary sources $\tilde{\lambda}$ and bulk initial data holds at the level of the symplectic structure of both boundary and bulk theories \cite{Belin:2018fxe}, which we review in Section \ref{subsec:sympstructrev}. Particularly, the (Euclidean) boundary symplectic form associated with the space of sources is dual to the (Lorentzian) bulk symplectic form on the classical phase space of gravitational configurations,
\beq \Omega_{\text{B}}(\delta\tilde{\lambda}_{1},\delta\tilde{\lambda}_{2})=\int_{\Sigma}\omega^{\text{L}}_{\text{bulk}}(\phi,\delta\phi_{1},\delta\phi_{2})\;,\eeq
where $\Sigma$ is a Lorentzian time slice, $\omega^{\text{L}}_{\text{bulk}}$ is the Lorentzian bulk symplectic 2-form density (the `symplectic current') as a function of arbitrary bulk fields $\phi$, including the metric, and $\Omega_{\text{B}}$ is the boundary symplectic form associated with coherent holographic CFT states prepared by a Euclidean path integral with sources. As shown in \cite{Belin:2018fxe,Belin:2018bpg}, the left hand side of the above relation is equal to the volume of the extremal Cauchy slice when $\Sigma$ is this slice, while $\Omega_{\text{B}}$ with respect to the `new York' deformation, is equal to holographic complexity $\mathcal{C}$. Separately, boundary field complexity defined in terms of the boundary symplectic form is known to satisfy a first law on the space of sources.

In Section  \ref{sec: EEfromComplexity} we show that if one assumes CV duality, and the bulk-boundary symplectic form correspondence, the holographic first law of complexity is equivalent to the linearized Einstein's equations holding in the bulk:
\beq \delta V=\Omega_{\text{B}}\Rightarrow \delta E^{\mu\nu}=0\;.\eeq
Our argument is spiritually and structurally motivated by the derivation of linearized Einstein's equations from the (holographic) first law of entanglement \cite{Lashkari:2013koa,Faulkner:2013ica}, however, physically distinct in that we show bulk gravitational dynamics is encoded in varying boundary complexity. This suggests a notion of `spacetime complexity', and is not only the second central achievement of this section, but, in our view, one of the central results of this manuscript.

We conclude in Section \ref{subsec:bulksymcanthread} by showing the Lorentzian symplectic current $\omega^{\text{L}}_{\text{bulk}}$ provides a canonical choice for the perturbative Lorentzian thread form $\delta u$, providing a solution to the min flow-max cut program, and that the linearized Einstein's equations are captured by the closedness condition $d(\delta u)=0$.

\vspace{2mm}

\noindent \textbf{Section \ref{sec:genprop}:} Here we demonstrate how the gateline interpretation of Lorentzian flows relates to tensor network constructions of spacetime and suggest the need for a more refined notion of holographic complexity. First, in Section \ref{subsec:physint}, we review basics about tensor networks and how holographic entanglement entropy and complexity are computed in this context. We use this to combine the gateline reformulation of CV duality and that complexity can be interpreted as the number of tensors of a tensor network discretization of the maximal volume slice such that the optimal flow solving the min flow-max cut program prepares the tensor network on the maximal volume slice. Motivated by the maximin prescription of the HRT formula \cite{Wall:2012uf}, in Section \ref{subsec:lessonsmaximin} we argue suboptimal flows prepare suboptimal tensor networks, and play a role in a more general definition of holographic complexity. This leads us to propose holographic complexity should be understood as an ensemble average over tensor networks on all Cauchy slices in the Wheeler-De Witt patch, which we explain in Section \ref{subsec:genprop}. We make several comments and compare to the recent holographic dual to the path integral optimization definition of boundary field complexity \cite{Boruch:2020wax,Boruch:2021hqs}.

\vspace{2mm}

\noindent \textbf{Section \ref{sec:disc}:} Here we discuss the main results and, in particular, list how Lorentzian threads suggest a notion of `emergent time'. We conclude with an outline of future research directions we find most promising.

\vspace{2mm}

We also include a number of appendices to keep this article self-contained and pedagogical. Specifically, in Appendix \ref{app:examples} we briefly summarize the bit thread formulation of holographic entanglement entropy and some of their geometric constructions. We include this section as we find it beneficial to compare the features of Riemannian (bit) threads to the features of the Lorentzian threads considered in this paper. Appendix \ref{append:Lorentzmfmc} provides a number of proofs of the various theorems and related corollaries used in Section \ref{sec:prelims} of this article. Notably, we review the proof of the min flow-max cut theorem first given in \cite{Headrick:2017ucz} using convex optimization techniques, and then prove the nesting property and study the possible existence of Lorentzian multiflows. Lastly, we relegate some of the calculational details of deriving the linearized Einstein's equations from the holographic first law of complexity to Appendix \ref{app:verifyEineqs}.  Note this manuscript is an expanded and more detailed version of \cite{Pedraza:2021mkh}.


\section{Holographic complexity and Lorentzian flows} \label{sec:prelims}

\subsection{Prescriptions of holographic complexity: review} \label{subsec:prescripts}

To appreciate our reformulation of holographic complexity using Lorentzian flows and its conceptual interpretation, here we briefly review the various notions of boundary complexity and, when the boundary theory is holographic, their corresponding gravitational duals.

Quantum computational complexity, or circuit complexity, is a key concept in quantum information science described as follows. Given an initial reference state $|\psi_{R}\rangle$, and a finite set of gates (unitary operations) $\{g_{1},...,g_{N}\}$, the complexity $\mathcal{C}(|\psi_{T}\rangle)$ of preparing a target state $|\psi_{T}\rangle$, is the minimum number of such gates needed to construct the unitary operator $U_{TR}$ transforming $|\psi_{R}\rangle$ into $|\psi_{T}\rangle$:
\beq |\psi_{T}\rangle=U_{TR}|\psi_{R}\rangle=g_{i_{n}}...g_{i_{2}}g_{i_{1}}|\psi_{R}\rangle\;.\eeq
In other words, the complexity $\mathcal{C}$ defines the optimal cost required to prepare a specific target state given an initial reference state. The reference and target states, together with the set of gates $\{g_{i}\}$ define the quantum circuit. In this elementary set-up, the growth of complexity as a function of time $t$ obeys an upper bound at late times, first derived by Lloyd \cite{Lloyd:2000}:
\beq \dot{\mathcal{C}}(|\psi_{T}(t)\rangle)\leq\frac{2E}{\pi\hbar}\;,\label{eq:lloydsbound}\eeq
where $E$ is the energy of system and is taken to be constant. This bound places an upper limit on computation speed of a classical computer.

\begin{figure}[t]
\includegraphics[width=8cm]{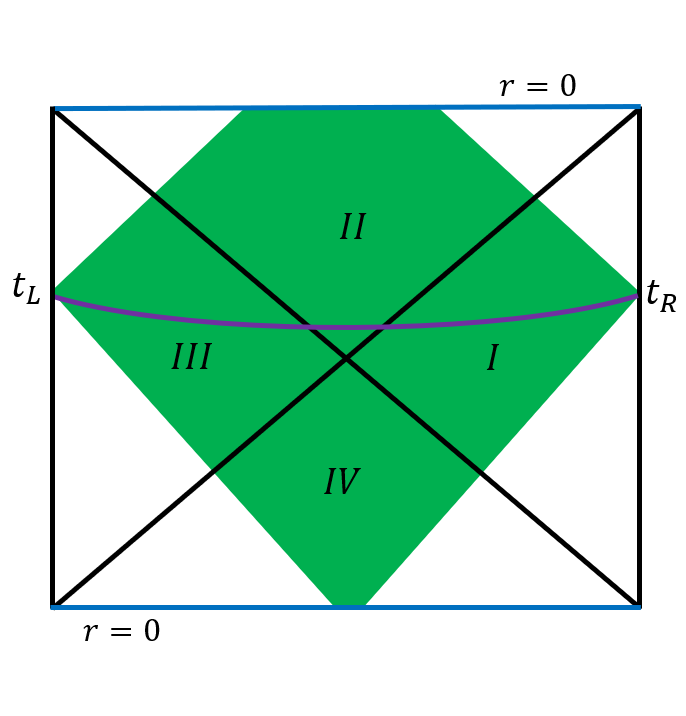}
\centering
\caption{WDW patch in double sided AdS-Schwarzschild black hole. The shaded region (green) is the domain of dependence of the bulk Cauchy slice (purple) asymptoting to the $(r=\infty$) boundary Cauchy surfaces defined at left and right boundary times $t_{L}$ and $t_{R}$. Here we have considered the case when $t_{L}=t_{R}$. More precisely, the left and right corners and upper and lower tips of the causal diamond hit UV regulating surfaces just before reaching the timelike boundaries $r=\infty$ or past and future singularities $r=0$, respectively. CA duality is defined by the gravitational action evaluated on the WDW patch, while CV duality is given by the spatial volume of the maximal hypersurface extending to $t_{L}$ and $t_{R}$.}
\label{fig:wdwads}\end{figure}

The observation that computational complexity grows linearly at late times and obeys Lloyd's bound (\ref{eq:lloydsbound}), aided by the fact entanglement generates wormholes or Einstein-Rosen bridges (`ER=EPR') \cite{Maldacena:2013xja} (see also \cite{Jensen:2013ora,Sonner:2013mba,Chernicoff:2013iga}), act as guiding principles for searching for holographic duals to the boundary complexity. As described earlier, broadly, there are two proposed conjectures or dualities relating complexity of the boundary theory to a bulk geometric quantity. The first of these was `complexity=volume' (CV) \cite{Susskind:2014rva,Susskind:2014jwa,Stanford:2014jda}, where the boundary complexity of a (holographic) CFT state defined on a Cauchy slice $\sigma_{A}$ separating a boundary region $A$ from its complement, such that $\partial A=\sigma_{A}$, is dual to the spatial volume of the maximal hypersurface homologous to $A$,
\beq \mathcal{C}_{V}(\sigma_{A})=\frac{1}{G_{N}\ell}\,\underset{\Sigma\sim A}{\text{Vol}}(\Sigma(A))\;.\label{eq:CVconj2}\eeq
This duality correctly captures the linear growth of complexity at late time, where, the energy $E$ of the system is typically given by the mass of the black hole.\footnote{This is the case for neutral static systems. For more general black holes, the energy $E$ is characterized by the mass and other thermodynamic potentials.} Due to the seeming ambiguity in length scale $\ell$, however, the  `complexity=action' (CA) proposal \cite{Brown:2015bva,Brown:2015lvg} was conjecture. In CA duality the boundary complexity is dual to the on-shell bulk action evaluated on the Wheeler-DeWitt (WDW) patch
\beq \mathcal{C}_{I}=\frac{I_{\text{WDW}}}{\pi\hbar}\;.\label{eq:IWDW2}\eeq
See Figure \ref{fig:wdwads} for a simple illustration.

For charged Reissner-Nordstrom black holes, however, CA duality does not generally obey the Lloyd bound. Recognizing this difficulty, the authors of \cite{Couch:2016exn} proposed a second version of CV duality, dubbed `complexity=volume 2.0' (CV 2.0), where the action of CA duality is replaced with spacetime volume of the WDW patch. Importantly, CV 2.0 duality was shown to satisfy Lloyd's bound for charged RN black holes.\footnote{It is worth pointing out CV 2.0 makes use of extended black hole thermodynamics, such that $\mathcal{C}\sim PV_{\text{spacetime}}$, where $P$ is the thermodynamic pressure. Moreover, it was shown in \cite{Couch:2016exn} the thermodynamic volume of black holes in Einstein-Maxwell gravity is equal to late time growth of the spatial volume of the WDW patch. Recently the thermodynamic volume was also shown to control the complexity of formation of large black holes \cite{Balushi:2020wkt}.} There is also a  `complexity=action 2.0' (CA 2.0) conjecture \cite{Fan:2018wnv}, utilizing similar ideas as CV 2.0 and reduces to CV 2.0 duality for general relativity plus a cosmological constant. There is still debate as to what the correct bulk dual of boundary complexity is, however, the study of holographic complexity tells us the bulk volume and the WDW patch are important new inputs in the holographic dictionary.

Both CV and CA dualities have a conceptual base in computational/circuit complexity, a concept mostly developed in ordinary quantum mechanics. In the context of AdS/CFT, however, the boundary is described by a field theory, and therefore a definition of complexity in field theories is necessary. One approach is to generalize Nielsen's `geometrization' of circuit complexity \cite{Nielsen:2006,Nielsen:2007}, to the case of field theories \cite{Jefferson:2017sdb}. In this method, the unitary gates are given by a continuum representation such that the circuits are represented by geodesics in an auxiliary manifold whose length measures the so-called depth of the associated circuit. The optimal circuit, \emph{i.e.}, the complexity, is then given by the length of the minimal geodesic with endpoints at the reference and target states (see \cite{Chapman:2018hou,Hackl:2018ptj,Bernamonti:2019zyy,Bernamonti:2020bcf} for further developments). This approach to field theory complexity shares many similarities with an alternative approach based on the state dependent Fubini-Study metric \cite{Chapman:2017rqy} (we will have more to say about the Fubini-Study metric in a later section).

In a seemingly independent approach to defining complexity in field theories, the authors of \cite{Caputa:2017urj,Caputa:2017yrh} proposed an optimization procedure for Euclidean path integrals that evaluate CFT wavefunctionals. In particular, the optimization is performed with respect to a particular functional, known as the `path integral complexity', and provides a measure of computational complexity with respect to specific background metrics in the equivalent tensor network description. Thus this alternative approach to field theory complexity provides a concrete realization of the proposed interpretation of AdS/CFT as a tensor network. In the context of two-dimensional CFTs, where the path integral complexity functional is the Liouville action, the authors of \cite{Camargo:2019isp} showed this alternative definition of field theory complexity is in fact equivalent to the circuit complexity advocated by \cite{Jefferson:2017sdb,Chapman:2018hou,Hackl:2018ptj,Bernamonti:2019zyy,Bernamonti:2020bcf}, thereby unifying two seemingly unrelated concepts of field theory complexity (see also \cite{Chandra:2021kdv}).

Let us spend some time describing the path integral optimization description of complexity, as it will help motivate our interpretation of Lorentzian flows later on.  For details, see \cite{Caputa:2017urj,Caputa:2017yrh}. The basic idea is as follows. Begin with a wavefunctional representation of a CFT state defined by Feynman's path integral on a flat Euclidean space with a prescribed set of boundary conditions. Specifically, in the case of a two-dimensional CFT on $\mathbb{R}^{2}$ with coordinates $(\tau,x)$, the ground state wavefunctional of CFT fields $\varphi(\tau,x)$, denoted $\Psi_{\text{CFT}}[\varphi(x)]$, where $\varphi(x)\equiv\varphi(\tau=0,x)$, is formally given by the Euclidean path integral:
\beq \Psi_{\text{CFT}}[\varphi(x)]=\int\prod_{-\infty<\tau\leq0,x}\mathcal{D}\tilde{\varphi}(\tau,x)e^{-I_{\text{CFT}}[\tilde{\varphi}]}\delta(\tilde{\varphi}(0,x)-\varphi(x))\;,\label{eq:cftwavefunc}\eeq
where $I_{\text{CFT}}$ is the Euclidean action of the CFT, and we observe $\Psi_{\text{CFT}}$ is a wavefunctional of the boundary conditions imposed on the CFT fields $\varphi$ at the time $\tau=0$ slice. Formally, this path integral may be evaluated using a lattice discretization, where the flat Euclidean spacetime is characterized by $ds^{2}=\epsilon^{-2}(d\tau^{2}+dx^{2})$, with $\epsilon$ as some UV regularization scale.

The next step is to evaluate this Euclidean path integral utilizing conformal symmetry by performing a conformal transformation of the Euclidean metric whilst all boundary conditions are kept fixed. Explicitly, one considers the conformally flat Euclidean metric $ds^{2}=e^{2\phi(\tau,x)}(d\tau^{2}+dx^{2})$. The measure of the CFT fields $\mathcal{D}\varphi$ in (\ref{eq:cftwavefunc}) undergo an anomalous change under this conformal transformation, such that wavefunctional (\ref{eq:cftwavefunc}) becomes
\beq \Psi_{\text{CFT}}[\varphi(x)]\to e^{I_{L}[\phi]-I_{L}[\phi_{0}]}\Psi_{\text{CFT}}^{(\phi_{0})}[\varphi(x)]\;,\eeq
where $\phi_{0}$ is defined by the boundary condition $e^{2\phi(0,x)}=\epsilon^{-2}\equiv e^{2\phi_{0}}$, and $I_{L}[\phi]$ is the Liouville action on a flat Euclidean space.

Intuitively, the path integrals on different conformally deformed Euclidean geometries can be interpreted as continuous tensor networks preparing the same quantum state. The size of a tensor network -- the number of tensors needed to prepare a specific quantum state -- provides a notion of complexity. As such, one seeks to optimize over different conformal factors so as to compute the path integral which generates the tensor network of the smallest size. This procedure is called `path integral optimization' and corresponds to minimizing the `path integral complexity' functional, in this case given by the Liouville action, over the background metric. Minimizing the Liouville action amounts to solving the Liouville equations of motion. As shown in \cite{Caputa:2017urj,Caputa:2017yrh}, solving the Liouville equation is equivalent to the condition that the two-dimensional Euclidean geometry be a hyperbolic space, and, in particular, the optimal metric to prepare the vacuum state of a CFT confined on a line is a time slice of empty $\text{AdS}_{3}$; to prepare the TFD state, the optimal metric is a time slice of the BTZ black hole. Therefore, the path integral optimization presents a continuum version of the AdS/tensor network correspondence, in which AdS geometry emerges from a tensor network.

While the method of path integral optimization is most precise for two-dimensional CFTs, the authors of \cite{Caputa:2017urj,Caputa:2017yrh} provided general arguments to write down a higher dimensional generalization. Even in the two-dimensional case, however, there were still open questions, particularly concerning the preparation of excited states which seemingly required one to minimize the quantum Liouville theory, understood as the path integral, rather than the classical Liouville theory. At the semiclassical level, this was shown to give the expected gravity metric dual to a primary state with $1/c$ corrections (a time slice of $\text{AdS}_{3}$ with a deficit angle due to a bulk point particle backreaction). Very recently these open questions were resolved by showing path integral optimization is equivalent to the maximization of the Hartle-Hawking wavefunctional on a Euclidean AdS background \cite{Boruch:2020wax,Boruch:2021hqs}. We will return to this new interpretation in Section \ref{sec:disc}.

Lastly, let us briefly summarize a recent holographic complexity relation, known as the  `complexity/momentum' correspondence \cite{Susskind:2018tei} (see also  \cite{Susskind:2019ddc,Susskind:2020gnl}).  In this set-up, increase in bulk radial momentum is dual to the growth of boundary operators, which, via complexity's relation to operator size, links the momentum generator to growth in complexity. The intuition behind the complexity/momentum correspondence is that bulk gravitational clumping is holographically dual to an increase in complexity of the corresponding boundary quantum state. This is indeed the case for black holes, but has been argued to be true for any gravitationally infalling matter. In particular, in the case of collapsing thin shells (c.f. \cite{Magan:2018nmu,Lin:2019qwu,Mousatov:2019xmc,Chapman:2018dem,Chapman:2018lsv,Barbon:2019tuq}) one finds the complexity rate of the collapsing shells $\dot{\mathcal{C}}$ is related to the energy-momentum tensor $T_{\mu\nu}$ of the background spacetime \cite{Barbon:2019tuq},
\beq \dot{\mathcal{C}}_{\text{shell}}=-\int_{\Sigma}n^{\mu}T_{\mu\nu}\zeta^{\nu}\;,\label{eq:pvcshell}\eeq
where $n^{\mu}$ is the unit timelike vector to the maximal volume hypersurface $\Sigma$, and $\zeta^{\mu}$ is an inward pointing radial field tangent to $\Sigma$.\footnote{More precisely, $\zeta$ is a vector tangent to $\Sigma$ that is asymptotically equal to a radial, inward-pointing vector field whose modulus is given by the angular sphere at infinity.} The right hand side is understood to be the (infalling) momentum flux through $\Sigma$. Since $\Sigma$ plays a prominent role in relating the momentum and complexity, Eq. (\ref{eq:pvcshell}) is referred to as the momentum/volume/complexity (PVC) relation. More generally, it was proven (\ref{eq:pvcshell}) is essentially the momentum constraint of general relativity, such that one has the `generalized PVC' relation \cite{Barbon:2020olv,Barbon:2020uux}
\beq \dot{\mathcal{C}}_{V}=P_{\zeta}[\Sigma]+R_{\zeta}[\Sigma]\;,\quad \label{eq:PVCgen}\eeq
with integrated momentum flux $P_{\zeta}=\int_{\Sigma}\mathcal{P}_{\zeta}$, where $\mathcal{P}_{\zeta}=\mathcal{P}_{a}\zeta^{a}$ with $\mathcal{P}_{a}=-n^{\mu}T_{\mu\nu}e^{\nu}_{a}$ is the pulled-back momentum flux through $\Sigma$, and $R_{\zeta}[\Sigma]$ is a `remainder' arising from an integration of the momentum constraint of general relativity.\footnote{The momentum flux is related to the (contracted) Codazzi equation $\nabla^{a}K_{ab}-\nabla_{b}K=-8\pi G\mathcal{P}_{b}$, where $K_{ab}$ is the extrinsic curvature of the bulk codimension-1 Cauchy slice, and the remainder term is $R_{\Sigma}=-\frac{1}{8\pi G}\int_{\Sigma}K^{ab}\nabla_{a}\zeta_{b}$.} In the event the field $\zeta$ is an exact conformal Killing vector field, \emph{e.g.}, for smooth, spherically symmetric backgrounds, or the background is a solution to Einstein's equations in $2+1$ dimensions, $R_{\zeta}=0$ such that the complexity rate is solely attributed to the momentum flow through $\Sigma$.

As we will explore below, our reformulation of holographic complexity using Lorentzian flows provides a new interpretation of holographic complexity, allows us to make contact with the PVC relation, and deepens the connection between spacetime and tensor networks.

\subsection{Reformulating `complexity=volume' using min flow-max cut} \label{subsec:reformCV}

The Ryu-Takayanagi (RT) formula relating holographic entanglement entropy $S(A)$ across a boundary region $A$ to the area of a bulk minimal surface  homologous to $A$, denoted as $m(A)$, was reformulated in terms of holographic `bit threads' using Riemannian flows $v$ \cite{Freedman:2016zud}:
\beq S(A)=\underset{v}{\text{max}}\int_{A}v\;.\label{eq:RTbtv1}\eeq
Here we are using the notation that $\int_{A}v$ represents the flux of flow $v$ through the bulk surface $m(A)$ homologous to $A$. (For full details see \cite{Freedman:2016zud}, or a condensed discussion in Appendix \ref{app:examples}). The reformulation of the RT equation (\ref{eq:RTbtv1}) is a consequence of the max flow-min cut theorem for \emph{Riemannian} flows,
\beq \underset{v}{\text{max}}\int_{A}v=\alpha\underset{m\sim A}{\text{min}}\text{area}(m)\;,\label{eq:Riemfmcv1}\eeq
where $\alpha$ is a constant bounding the normalization of $v$, $|v|\leq\alpha$, and $m\sim A$ denotes the homology constraint. The left hand side describes a (non-unique) flow $v$ that maximizes the flux through $A$, while the right hand side gives the minimal area of the bulk surface $m$ homologous to $A$. When $\alpha\equiv\frac{1}{4G_{N}}$, by the RT formula (\ref{eq:RT}), one arrives to the reformulation (\ref{eq:RTbtv1}). The continuous version of the max flow-min cut theorem (\ref{eq:Riemfmcv1}) was proven  using convex optimization techniques in \cite{Headrick:2017ucz} after restating the question as a convex optimization problem.

Analogously, the authors of \cite{Headrick:2017ucz} proved a Lorentzian analog of the max flow-min cut theorem, appropriately called the \emph{min flow-max cut} theorem, again using convex optimization techniques. In this subsection we briefly provide a statement of the theorem as described in \cite{Headrick:2017ucz}, with minor changes in notation, and demonstrate how it is related to holographic complexity, before moving onto the properties and interpretational questions in later subsections.

\begin{figure}[t]
\includegraphics[width=8cm]{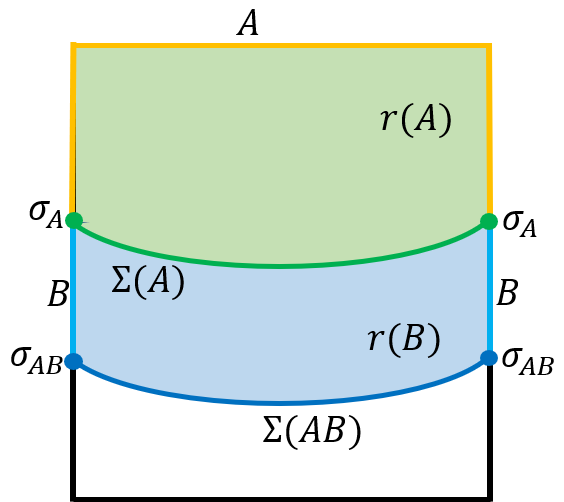}
\centering
\caption{Asymptotically AdS $M$  in Lorentzian signature foliated by maximal bulk slices $\Sigma$. Here the boundary subregion $A$ (gold)  has boundary $\partial A=\sigma_{A}$ on both timelike boundaries of $M$, anchoring the maximal volume slice $\Sigma(A)$. Disconnected boundary subregion $B$ (light blue) is disjoint from $A$ and has boundary $\sigma_{AB}$ anchor $\Sigma(AB)$. The surfaces $A$ and $AB\supset A$ are nested boundary regions. The upper shaded region (light green) is the bulk region $r(A)$, while the middle shaded region (blue) is the bulk region $r(B)$. Bulk regions $r(A)$ and $r(B)$ are nested inside the bulk region of $AB$, $r(AB)=r(A)\cup r(B)$. }
\label{fig:maxslicenest}\end{figure}

Let $M$ be a $d+1$-dimensional compact, oriented and time-oriented Lorentzian manifold with boundary $\partial M$. Here we will mostly consider when $M$ is an asymptotically AdS background (with a timelike $d$-dimensional conformal boundary and Euclidean past and future boundaries), however, the statement of the theorem holds for other Lorentzian spacetimes as well, including flat and de Sitter space. A bulk region, denoted by $r$, is an embedded compact codimension-0 submanifold (with boundary) of $M$, while a boundary region, denoted by $R$, is defined similarly with respect to $\partial M$. A slice $\Sigma$ is an embedded compact oriented codimension-1 submanifold-with-boundary of $M$ whose interior is contained inside the interior of $M$. The orientation of the hypersurface $\Sigma$ is identified using a normal covector $n_{\mu}$, for which we take to be future directed ($n_{0}>0$). We will take $\Sigma$ to be a Cauchy slice of $M$ providing a Cauchy development of $M$. For an illustration, see Figure \ref{fig:maxslicenest}.

For our purposes we will assume the timelike boundary of $M$ is foliated by (boundary) Cauchy slices $\sigma$ (in Figure \ref{fig:maxslicenest} these are denoted as points on the left and right timelike boundaries). We will take the bulk slice $\Sigma$ to be anchored at these boundary slices $\sigma$. Sometimes we will denote $\sigma$ by a boundary time parameter $t$ and typically assume that the left and right boundary times are equivalent. We emphasize, however, the left and right boundary times need  not be equivalent for the statement or proof of the min flow-max cut theorem. Alternatively, the boundary of a boundary region $A$ ends at a point $\sigma_{A}$, and so we sometimes denote $\partial A=\sigma_{A}$.

The homology condition relating a slice $\Sigma$ to a boundary region $A$ is analogous to the equivalent statement in the Riemannian case. More precisely, $\Sigma$ is homologous to $A$ (relative to $R$), $\Sigma\sim A$, when there exists a bulk region $r(A)$ satisfying
\beq \partial r\setminus \partial M=-(\Sigma\setminus\partial M)\;,\label{eq:relhomo}\eeq
such that the intersection between the bulk region $r(A)$ and the complement $R^{c}$ is equal to boundary region $A$, \emph{i.e.}, $r\cap R^{c}=A$. An essential difference from the Riemannian set-up is that we require an additional necessary and sufficient condition such that the homology constraint (\ref{eq:relhomo}) holds: The causal future of $A$, denoted $J^{+}(A)$, intersected with the complement of boundary region $R^{c}$ is equal to $A$, $J^{+}(A)\cap R^{c}=A$. Consequently, for asymptotically AdS spacetimes cutoff by spacelike and timelike boundaries, the boundary region $R$ may be the entire or a subset of the timelike boundary. Here we will often assume $R=\emptyset$ such that $A=J^{+}(\sigma)\cap\partial M$, and $\partial A=\sigma_{A}$, in which case $\Sigma$ is a Cauchy slice anchored by the left and right $\sigma_{A}$ Cauchy slices.

With this set-up out of the way, let us now define Lorentzian flows and state the min flow-max cut theorem and how we can use it to reformulate the complexity=volume conjecture (\ref{eq:CVconj2}), analogous to what was accomplished for holographic entanglement entropy \cite{Freedman:2016zud}. A Lorentzian flow\footnote{In the event $R\neq\emptyset$, then we also demand $\sqrt{h}n_{\mu}v^{\mu}|_{R}=0$, such that $v$ has vanishing flux through $R$. This is a consequence of the relative homology constraint (\ref{eq:relhomo}).} $v$ is  a divergenceless, future oriented vector field on $M$ with norm $|v|\geq\alpha$, \emph{i.e.},
 \beq \nabla_{\mu}v^{\mu}=0\;,\;\; v^{0}>0\;,\;\;-v_{\mu}v^{\mu}\geq \alpha\;,\;. \label{eq:Lorflowdef}\eeq
Here, $\alpha$ is some real positive constant which we will set momentarily, though we can choose the normalization such that $-v^{2}\geq1$.

Together, the homology constraint (\ref{eq:relhomo}), and the definition of the flow (\ref{eq:Lorflowdef}) yields
\beq \int_{A}v\equiv\int_{A}\sqrt{h}n_{\mu}v^{\mu}=\int_{\Sigma}\sqrt{h}n_{\mu}v^{\mu}\geq\alpha \mathcal{V}(\Sigma)\;,\eeq
where $n_{\mu}$ is the unit normal covector\footnote{Since boundary region $A$ can have both timelike and spacelike sections, covector $n_{\mu}$ could likewise be timelike or spacelike. On the maximal volume slice $\Sigma(A)$, however, $v^{\mu}|_{\Sigma}=\alpha n^{\mu}$, such that $n^{\mu}$ is purely timelike on $\Sigma$.} to $A$ and $\sqrt{h}$ the induced volume element. The logic is as follows. The first equality on the left hand side is the notational definition of the flux through $A$, which is precisely equal to the flux across $\Sigma(A)$ since $\Sigma(A)\sim A$, where we have made use of the divergenceless condition. The inequality arises since $n_{\mu}v^{\mu}\geq\alpha$, and we have used $\int_{\Sigma(A)}\sqrt{h}\equiv\mathcal{V}(\Sigma)$, with $\mathcal{V}$ being the volume of $\Sigma$.

The min flow-max cut theorem is the non-trivial statement that the inequality is saturated for a (non-unique) minimizing flow
\beq \underset{v}{\text{min}}\int_{A}v=\alpha\,\underset{\Sigma\sim A}{\text{max}}(\mathcal{V}(\Sigma))\;.\label{eq:mxflmnct1}\eeq
In words, a flow $v$ which minimizes the flux through boundary region $A$, has its flux equal to the maximal volume of the bulk slice $\Sigma$ homologous to $A$. Therefore, the inequalities $\int_{A}v\geq\alpha\mathcal{V}(\Sigma)$  for all of the different members $\Sigma$ of the homology class are the only obstructions to decreasing the flux, where the strongest is the volume maximizing representative $\Sigma(A)$, an \emph{inverse bottleneck}. The proof of this theorem can be found in \cite{Headrick:2017ucz}, which follows the same steps as the Riemannian case, using convex optimization techniques.

We now have all of the necessary ingredients to reformulate the complexity=volume conjecture in terms of Lorentzian flows. With Lorentzian manifold $M$ being a static asymptotically AdS bulk spacetime, $A$ a region of its conformal boundary, and $\Sigma(A)$ the bulk Cauchy slice homologous to $A$ with maximal volume $\mathcal{V}(\Sigma)$, setting $\alpha=\frac{1}{G_{N}\ell}$ we uncover
\beq \mathcal{C}(\sigma_{A})=\int_{A}v(A)=\frac{1}{G_{N}\ell}\mathcal{V}(\Sigma)\;.\label{eq:CVflowv1}\eeq
Here it is understood $v(A)$ is a minimizing flow obeying the min flow-max cut theorem (\ref{eq:mxflmnct1}). In this way, the maximal bulk slice serves as the (inverse) bottleneck for the Lorentzian flow. We emphasize that here we are advocating to replace the complexity=maximal volume picture with a Lorentzian flow picture. This leads to some interesting differences. First, recall the maximal volume will in general be divergent at the timelike and asymptotic boundaries. That is, it is necessary to introduce regulating cutoff surfaces at the boundaries to define the globally maximal slice. In the flow picture, however, one can give a definition of a minimal flow that applies even when its flux is infinite. This is accomplished by diminishing the flow $v$ by a vector field $\Delta v$ with negative flux through $A$ such that $v+\Delta v$ is also a flow. A minimal flow is then one which cannot be diminished \cite{Headrick:2017ucz}.

\subsection{Properties of Lorentzian flows} \label{subsec:genprops}

Before discussing the conceptual interpretation of Lorentzian flows, let us now uncover interesting properties we expect holographic complexity to satisfy, inherited from conditions on Cauchy slices of maximal volumes.

\subsubsection{Nesting and complexity rates}

The first important property is that the bulk regions bounded by maximal bulk Cauchy slices are `nested', analogous to the nesting properties of bulk regions bounded by minimal RT surfaces in the context of holographic entanglement entropy. This follows as a lemma from the min flow-max cut theorem, whose proof follows straightforwardly from the Lorentzian analog of the nesting lemma in the Riemannian case.

\vspace{2mm}

\noindent \textbf{Two nested regions}

\vspace{2mm}

For concreteness, consider two nested regions $AB$, and $AB\supset A$ of the boundary, which, without loss of generality obey, $A\cap B=\emptyset$. Let $\Sigma(A)$ and $\Sigma(AB)$ be unique maximal cuts anchored at boundary Cauchy slices $\sigma_{A}$ and $\sigma_{AB}$, respectively, where $\sigma_{A}$ lies entirely to the future of $\sigma_{AB}$, \emph{i.e.}, the boundary is foliated by Cauchy slices. When the strong energy condition and the Einstein's equations hold in the bulk, this boundary foliation will induce a bulk foliation by non-intersecting maximal volume slices $\Sigma(A)$ and $\Sigma(AB)$, with $\Sigma(A)\sim\sigma_{A}$ and $\Sigma(AB)\sim\sigma_{AB}$, such that slice $\Sigma(A)$ lies entirely to the future of $\Sigma(AB)$ \cite{Couch:2018phr}. This is the picture we have in mind, however, the nesting theorem need not make this assumption about the bulk. Either way, the bulk region $r(A)$ is nested inside $r(AB)$, $r(A)\subset r(AB)$. See Figure \ref{fig:maxslicenest} for an illustration.  Notice region $B$ is not homologous to a single bulk slice $\Sigma(B)$. From the Lorentzian flow perspective, the nesting of maximal cuts tells us there exists a flow $v(A,AB)$ that simultaneously minimizes the flux through $A$ and $AB$, but not through $A$ and $B$ simultaneously.\footnote{A brief comment on notation. In \cite{Freedman:2016zud} $v(A,B)$ denotes a Riemannian flow which maximizes through $A$ and $AB$, and minimizes through $B$. Instead, we denote the Lorentzian analog by $v(A,AB)$ which we find illustrative.} Equivalently, $v(A,AB)$ \emph{maximizes} the flux through $B$, given the condition it also minimizes through $AB$.

\begin{figure}[t]
\includegraphics[width=15cm]{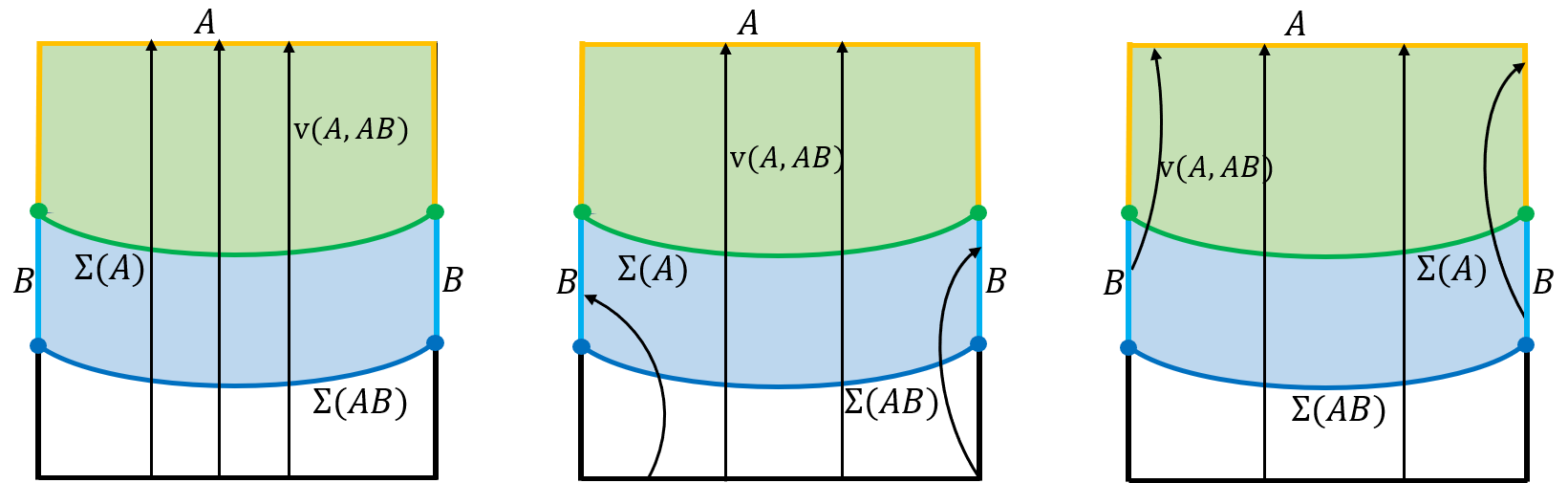}
\centering
\caption{Examples of flows $v(A,AB)$ for two nested regions. Left: $\mathcal{C}(AB)=\mathcal{C}(A)$, occurs when there is no flux through $B$; all flux passing through $\Sigma(AB)$ also passes $\Sigma(A)$. Middle: $\mathcal{C}(AB)>\mathcal{C}(A)$, where flux crosses $\Sigma(AB)$ but exits through $B$. Right: $\mathcal{C}(AB)<\mathcal{C}(A)$, when flux enters through $B$ and exits through $A$.}
\label{fig:threadsnested2re}\end{figure}

It is worth emphasizing that Lorentzian flows through a boundary region $A$ homologous to any Cauchy slice $\Sigma$ will \emph{always} have positive flux. This follows from the norm bound and the reverse Cauchy-Schwarz inequality for timelike vectors $v^{\mu}$ and $n_{\mu}$, where $n_{\mu}v^{\mu}\geq|n_{\mu}||v^{\mu}|\geq\alpha$ (when evaluated on $\Sigma$). Note, however, not all Lorentzian flows have positive flux. In particular, the flux of $v(A,AB)$ through the timelike region $B$ may be positive or negative. This is because that $B$ obeys $J^{+}(B)\cap\partial M\neq B$, \emph{i.e.}, it is not homologous to a bulk slice $\Sigma(B)$. We will see the consequences of this momentarily.

Via CV duality (\ref{eq:CVflowv1}), we can use the nesting property to uncover a number of interesting behaviors holographic complexity must satisfy. For starters, by the nesting property there exists a unique flow $v(A,AB)$ that simultaneously minimizes the flux through boundary regions $A$ and $AB$, such that
\beq \mathcal{C}(\sigma_{AB})-\mathcal{C}(\sigma_{A})=\int_{AB}v(A,AB)-\int_{A}v(A,AB)=\int_{B}v(A,AB)\;,\label{eq:condcomp2}\eeq
where the final equality simply follows from $\int_{A}v+\int_{B}v=\int_{AB}v$. Moreover, given that $v(A,AB)$ is the maximum flux on $B$, we have the manifestly non-positive quantity:
\beq \int_{B}v(AB)-\int_{B}v(A,AB)\leq0\;,\label{eq:CMI1}\eeq
where $v(AB)$ denotes the minimizing flow through $AB$, but not necessarily through $A$ or $B$.\footnote{We point out we are restricting to flow $v(AB)$ rather than the flux minimizing through $B$, $v(B,AB)$, which also satisfies inequality (\ref{eq:condcomp2}). This is because the flux of $v(AB)$ through $B$ is bounded from below, whereas $v(B,AB)$ is generally unbounded from below.} Incidentally, combining inequalities (\ref{eq:condcomp2}) and (\ref{eq:CMI1}), we arrive at the following
\beq \mathcal{C}(\sigma_{A})-\mathcal{C}(\sigma_{AB})\leq-\int_{B}v(AB)\equiv\mathcal{C}(\sigma_{A}|\sigma_{AB})\;,\label{eq:inequality1}\eeq
where we have suggestively defined a new quantity $\mathcal{C}(\sigma_{A}|\sigma_{AB})$ as the flux of minimizing flow through $AB$. We will ascribe meaning to this quantity as well as (\ref{eq:inequality1}) momentarily.

A few comments are in order. From (\ref{eq:condcomp2}), we see (net) flux through region $B$ determines how complexity changes in time. We have three cases: zero flux, flux leaving, and flux entering. When there is net zero flux entering or leaving through $B$, $\mathcal{C}(\sigma_{AB})=\mathcal{C}(\sigma_{A})$, \emph{i.e.}, there is no change in complexity between a CFT state at $\sigma_{A}$ and $\sigma_{AB}$, and is characterized by a flow configuration which enters through region $R$ and exits through $A$. Alternatively, a flow which has integral lines entering through $R$ and exiting through $A$, in addition to integral lines entering through $R$ and exiting through region $B$ describes $\mathcal{C}(\sigma_{AB})>\mathcal{C}(\sigma_{A})$, \emph{i.e.}, complexity decreases to the future. Lastly, when a flow has integral lines entering through $R$ and ending on $A$, but also has lines entering through $B$, avoiding $\Sigma(AB)$, and ending on $A$,  then $\mathcal{C}(\sigma_{A})>\mathcal{C}(\sigma_{AB})$, such that complexity increases to the future.\footnote{For general spacetimes, there is nothing to prevent complexity from increasing or decreasing in time. In black hole spacetimes, however, complexity is known to obey a second law \cite{Brown:2017jil}, and is easily justified in the language of flows. The change in volume between Cauchy slices is entirely accounted for by the flux of the flow $v$ through the boundaries of the slice because $v$ is divergenceless. As pointed out in \cite{Couch:2018phr}, for slices restricted inside the horizon of a black hole, the change in volume is quantified by the flux through the horizon, a null surface. Since flows are taken to be future directed, the flux through the future event horizon is positive such that the interior volume can only increase. Provided the apparent horizon has only spacelike sections, such that flux through it is positive, by the CV relation one recovers the second law. Though here we considered $B$ a timelike region, one can consider a null region $B$ such that flux through $B$ is positive, leading to a second law. Proving this carefully requires extending the definition of flows to the case of degenerate metrics \cite{Headrick:2017ucz}.} For an illustration distinguishing each of these scenarios, see Figure \ref{fig:threadsnested2re}. Collectively, we emphasize $\mathcal{C}(\sigma_{A}|\sigma_{AB})$ describes a two-step optimization procedure: a flow first passes through $\Sigma(AB)$, such that additional flux through $B$ then passes through $A$ at a later time. We will more carefully interpret this procedure in a moment.

As the left hand side of (\ref{eq:condcomp2}) is computing a change in complexity, in the limit we shrink region $B\to0$, we actually uncover a complexity rate. More precisely, let $\sigma_{AB}$ be the boundary Cauchy slice at time $t_{AB}$, and $t_{A}$ the time denoting $\sigma_{A}$, with $t_{A}>t_{AB}$, such that $t_{A}=t_{AB}+\delta t$ with $\delta t>0$. As $B\to0$, then $t_{A}-t_{AB}=\delta t\to0$, and we have
\beq -\frac{d\mathcal{C}}{dt}=\lim_{B,\delta t\to0}\frac{1}{\delta t}\int_{B}v(A,AB)\;.\eeq
Thus, the rate of complexity is characterized by the maximizing flux through region $B$, and, by way of inequality (\ref{eq:inequality1}), is bounded below by the flow minimizing through $AB$, $\mathcal{C}(\sigma_{A}|\sigma_{AB})$.

The above rate allows us to make contact with the PVC correspondence. Specifically, since in the limit $B\to0$ the bulk Cauchy slices $\Sigma(A)$ and $\Sigma(AB)$ tend to the same maximal volume surface $\Sigma$, we may relate the rate of maximal flux through $B$ to the momentum flux through $\Sigma$ upon invoking the generalized PVC relation (\ref{eq:PVCgen}) \cite{Barbon:2020olv,Barbon:2020uux},
\beq \lim_{B,\delta t\to0}\frac{1}{\delta t}\int_{B}v(A,AB)=\int_{\Sigma}T_{\mu\nu}n^{\mu}\zeta^{\mu}-R_{\Sigma}\;,\eeq
The first term on the right hand side is the integrated momentum flux $P_{\zeta}$, where $\zeta$ is an `infalling' vector field tangent to $\Sigma$, asymptotically equal to a radial, inward pointing vector with modulus given by the radius of the angular sphere at long distances. The remainder term $R_{\Sigma}$ arises from integrating the momentum constraint in general relativity and vanishes whenever $\zeta$ is a conformal killing vector field. Hence, for spherically symmetric configurations, or when the background Lorentzian manifold $M$ is a solution to Einstein's equations in $2+1$-dimensions, $R_{\Sigma}=0$ and the right hand side collapses to $-P_{\zeta}$. Notice the background energy-momentum tensor has been left unspecified, which suggests the flux through $B$ may be controlled by various energy conditions.

\begin{figure}[t]
\includegraphics[width=8cm]{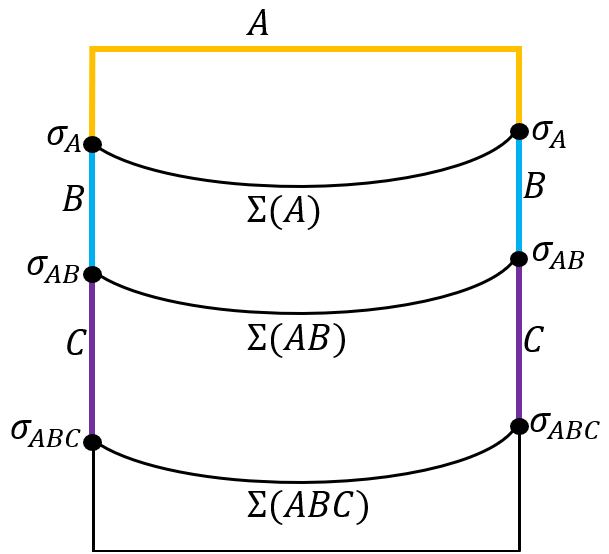}
\centering
\caption{Nested cuts associated with the three disjoint boundary regions $A,B,C$.}
\label{fig:nestedABCv1}\end{figure}

\vspace{2mm}

\noindent \textbf{Three nested regions}

\vspace{2mm}

It is straightforward to generalize to more nested regions, where we uncover additional properties holographic complexity obeys. Consider the case we have three boundary regions $A,B$ and $C$, such that $\Sigma(A)$ lies to the future of $\Sigma(AB)$ which lies to the future of $\Sigma(ABC)$ (see Figure \ref{fig:nestedABCv1}). Let $\Phi(X)$ denote the minimal flux through a boundary region $X$, and consider the following flows:  $v(C,AC)$ which minimizes the flux through $C$ and $AC$ and $v(BC,ABC)$ which minimizes the flux through $BC$ and $ABC$. We have
\beq
\begin{split}
\Phi(AC)+\Phi(BC)-\Phi(C)-\mathcal{C}(\sigma_{ABC})&=\int_{AC}\hspace{-3mm}v(C,AC)+\int_{BC}\hspace{-3mm}v(BC,ABC)\\
&-\int_{C}v(C,AC)-\int_{ABC}\hspace{-5mm}v(BC,ABC)\\
&=\int_{A}v(C,A)-\int_{A}v(BC,A)\;,
\end{split}
\label{eq:condmutcomp1}\eeq
where we used the fact $v(C,AC)$ and $v(BC,ABC)$ are both maximizing flows through $A$.  By the nesting property, $v(C,AC)$ will also minimize the flux through $ABC$, and $v(BC,ABC)$ minimizes on $C$, allowing us to write (\ref{eq:condmutcomp1}) as
\beq \Phi(AC)+\Phi(BC)-\Phi(C)-\mathcal{C}(\sigma_{ABC})=\int_{A}v(C,AC,ABC)-\int_{A}v(BC,ABC,C)\;,\eeq
where $v(C,AC,ABC)$ minimizes the flux through $A$ and $v(BC,ABC,C)$ maximizes the flux on $A$. Since the minimum is less than the maximum, we have
\beq \Phi(AC)+\Phi(BC)-\Phi(C)\leq\mathcal{C}(\sigma_{ABC})\;.\label{eq:weaksupadd}\eeq
Thus the complexity of all boundary regions $\mathcal{C}(\sigma_{ABC})$ is bounded below by the net flux through regions $AC$, $BC$, and $C$.\footnote{The above calculation is the Lorentzian analog of the strong subadditivity of holographic entanglement entropy using Riemannian flows \cite{Freedman:2016zud}.} Unfortunately, since flux entering through regions $AC$, $BC$, and $C$ are generally unbounded from below, not much can be gleaned from (\ref{eq:weaksupadd}) without imposing further conditions on the background spacetime.

The case of three nested regions does lead to an interesting relation on the second derivative of the complexity. Shrink regions $B$ and $C$ such that the boundary Cauchy slices $\sigma_{ABC}$, $\sigma_{AB}$ and $\sigma_{A}$, denoting boundary time slices $t_{ABC}, t_{AB}$ and $t_{A}$, respectively, are infinitesimally separated, \emph{i.e.}, $t_{A}=t_{AB}+\delta t$ and $t_{AB}=t_{ABC}+\delta t$ (assuming the time differences between nested regions $A$ and $AB$, and $AB$ and $ABC$ are equal to $\delta t$). Then, Taylor expanding $\mathcal{C}(\sigma_{A})$ and $\mathcal{C}(\sigma_{ABC})$ about $t_{AB}$, the second derivative of $\mathcal{C}$ is given by
\beq \ddot{\mathcal{C}}|_{t=t_{AB}}=\frac{1}{(\delta t)^{2}}\left[\mathcal{C}(\sigma_{A})-\mathcal{C}(\sigma_{AB})-(\mathcal{C}(\sigma_{AB})-\mathcal{C}(\sigma_{ABC}))\right]\to\frac{1}{\delta t}[P^{B}_{\zeta}-P_{\zeta}^{C}+R_{B}-R_{C}]\;.\eeq
To get to the final expression we used the generalized PVC relation (\ref{eq:PVCgen}) to express the change in complexity in terms of the integrated momentum flux through regions $B$ and $C$, $P_{\zeta}^{B}$ and $P_{\zeta}^{C}$. Taking $\delta t\to0$ and assuming the infalling vector field $\zeta$ is a CKV we recover the known result that the second derivative of complexity is given by the time derivative of the momentum $P_{\zeta}$ \cite{Susskind:2018tei,Susskind:2019ddc}:
\beq \ddot{\mathcal{C}}=\dot{P}_{\zeta}\;.\eeq
As argued in \cite{Susskind:2019ddc} (see also \cite{Barbon:2020olv}), working locally in the Lorentzian spacetime such that we may assume the Newtonian limit of a collection of point particles, one recovers Newton's second law in a gravitational background such that the force of attraction, clumping, is equal to the second time derivative of the complexity, suggesting gravitational dynamics may be holographically equivalent to laws obeyed by complexity. We will verify this in Section \ref{sec:diffformsandeineqs}.

\subsubsection{A comment on Lorentzian multiflows} \label{sec:multiflowscomp}

With our Lorentzian flow reformulation of the CV proposal, from the nesting property alone we uncovered holographic complexity satisfies two bounds (\ref{eq:inequality1}) and (\ref{eq:weaksupadd}) depending on the number of nested regions. The second of these follows from the Lorentzian analog of the derivation of strong subadditivity of entanglement entropy using Riemannian flows \cite{Freedman:2016zud}. It is natural to ask how far we may extend the metaphor between holographic complexity and holographic entanglement. Specifically, one may wonder if there is a type of monogamy relationship for Lorentzian flows analogous to the monogamy of holographic mutual information (MMI) \cite{Hayden:2011ag}.

To prove MMI in (Riemannian) flow based language, it was shown in \cite{Freedman:2016zud} that the nesting theorem is not enough. Rather, the MMI concerns a collection of flows, and thus one is required to introduce Riemannian \emph{multiflows} \cite{Cui:2018dyq}. Multiflows are a natural generalization of single Riemannian flows and is a set of vector fields simultaneously occupying the same geometry which satisfy the same conditions as a single flow (reviewed in Appendix \ref{app:examples}). In proving the existence of a \emph{max multiflow} -- a multiflow which maximizes the flux on each boundary subregion -- the authors of \cite{Cui:2018dyq} verified the MMI.

In the Lorentzian context, the goal then is to search for a collection of flows $v_{ij}$ such that their linear combination, a \emph{min multiflow}, minimizes the flux out of the non-overlapping timelike boundary regions. If such a flow exists it contains correlated components, and we might hope to find new, generalized inequalities, such as a monogamy-like relationship. A natural, but ultimately naive definition is the following. Given a Lorentzian manifold $M$ with boundary $\partial M$ covered by a finite collection of $n$ non-overlapping boundary regions $\{A_{i}\}$ for $i=1,...n$  ($\cup_{i}A_{i}=\partial M$). A \emph{Lorentzian multiflow} is a set of divergenceless, timelike and causal vector fields $v_{ij}$ on $M$ satisfying the following
\beq
\begin{split}
\nabla\cdot v_{ij}=0\;,\quad \sum_{i<j}|v_{ij}|\geq\alpha\;,\quad & \hat{n}\cdot v_{ij}=0 \;\;(\text{on}\;\; A_{k}\;\;\text{for}\;k\neq i,j)\;,\quad v_{ij}=-v_{ji}\;.
\end{split}
\label{eq:lorentzmultiflowv1}\eeq
As we detail in Appendix \ref{append:Lorentzmfmc}, working with this definition has a number of technical and conceptual challenges. For example, the antisymmetry $v_{ij}=-v_{ji}$, telling us a flow directed to the future, connecting regions $A_{j}$ to $A_{i}$ for $A_{j}<A_{i}$, is the negative of the past directed flow from $A_{i}$ to $A_{j}$, seems counter to the demand Lorentzian flows be future directed and causal.\footnote{In the Riemannian case there is no future directed condition and so $v_{ij}$ and $-v_{ji}$ equally satisfy the definition of a flow.} Relaxing this conceptual hurdle, the antisymmetry to tells us there exist $n(n-1)/2$ independent vector fields. Then, up to the future directed condition, for constant coefficients $\xi_{ij}\in[1,\infty]$, the linear combination
\beq v=\sum_{i<j}^{n}\xi_{ij}v_{ij}\;,\eeq
is a timelike flow obeying the norm bound.\footnote{This readily follows from the \emph{reverse} triangle inequality for future directed timelike vectors; $|v|+|w|\leq|v+w|$. Specifically, for $\xi_{ij}\in[1,\infty]$, we have $|v|=|\sum_{i<j}^{n}\xi_{ij}v_{ij}|\geq\sum_{i<j}^{n}|v_{ij}|\geq\alpha$.} Moreover, given a multiflow $\{v_{ij}\}$, we can define $n$ vector fields $v_{i}$
\beq v_{i}\equiv\sum_{j=1}^{n}v_{ij}\;.\label{eq:multiflowsmin}\eeq
Note, however, unlike the Riemannian case, the linear combination of $v\equiv \sum_{i=1}v_{i}=\sum_{i,j}v_{ij}=0$, is no longer a flow as it does not obey the norm bound. While we don't work with $v$ directly, the fact the linear combination $v$ is not a flow is inconsistent, meaning there is something awry with the antisymmetry condition $v_{ij}=-v_{ji}$.

Collectively then, the definition (\ref{eq:lorentzmultiflowv1}) is unsatisfactory and one must seek an alternative definition and see whether a min multiflow can be proven to exist. For the interested reader, in Appendix \ref{append:Lorentzmfmc} we review such an alternative, and other obstructions in defining Lorentzian multiflows, as well as a proof for an uncorrelated multiflow.

We should point out while (\ref{eq:lorentzmultiflowv1}) is insufficient in describing a Lorentzian multiflow, it is already known min multiflows exist for certain backgrounds. For example, in \cite{Couch:2018phr} it was shown when the strong energy condition holds, a foliation of the boundary induces a foliation of the bulk by maximal volume slices. Consequently, there exists a minimally packed flow, $n^{\mu}$, where $n^{\mu}$ is the unit normal to the foliation, and is easily seen to be divergenceless due to the fact that $\nabla_{\mu} n^{\mu} = K =0$. In such spacetimes, \emph{e.g.}, AdS, the flux out of nested regions $A_i$ is simultaneously minimized by $n^{\mu}$ and such that $v_i^{\mu} = n^{\mu}$ for all $i$. Moreover, $v_i$ are maximally correlated with each other and this was exploited to prove a second law of complexity for black hole horizons.

\subsubsection{Subregion complexity and superadditivity}

CV complexity is known to satisfy a number of other properties, including superadditivity and weak superadditivity \cite{Agon:2018zso, Du:2019emy,Caceres:2018blh,Caceres:2019pgf}. The superadditivity relations concerns itself with \emph{subregion complexity}: given that states on boundary subregions are dual to respective entanglement wedges \cite{Czech:2012bh,Almheiri:2014lwa,Dong:2016eik},  the volume of a maximal volume slice within the entanglement wedge is dual to the `subregion complexity' of the reduced state on the associated boundary subregion \cite{Alishahiha:2015rta,Carmi:2016wjl}.\footnote{For static bulk geometries, CV duality considers the volume of the portion of the constant-$t$ slice bounded by the boundary subregion and the corresponding Ryu-Takayangi surface \cite{Alishahiha:2015rta}. For time dependent geometries, this proposal is naturally extended using the HRT prescription where one considers the (maximal) volume of a bulk slice bounded by the boundary region and the HRT surface \cite{Carmi:2016wjl}.} Subregion complexity thus implies a notion of mixed state complexity. One possible definition of mixed state complexity is `purification complexity' \cite{Agon:2018zso}: the minimum number of gates required to prepare an arbitrary purification of a given mixed state.\footnote{A second definition of mixed state complexity is the `basis complexity', which is roughly the minimum number of gates required to prepare a specific mixed state sharing the same eigenvalue spectrum as the initial minimal complexity state \cite{Agon:2018zso}.} Such a definition was explored in the context of the complexity=action conjecture in \cite{Agon:2018zso} and the complexity=volume conjecture in \cite{Caceres:2018blh,Caceres:2019pgf} (see also \cite{Ghodrati:2019hnn}).

\begin{figure}[t]
\centering
\includegraphics[scale=0.7]{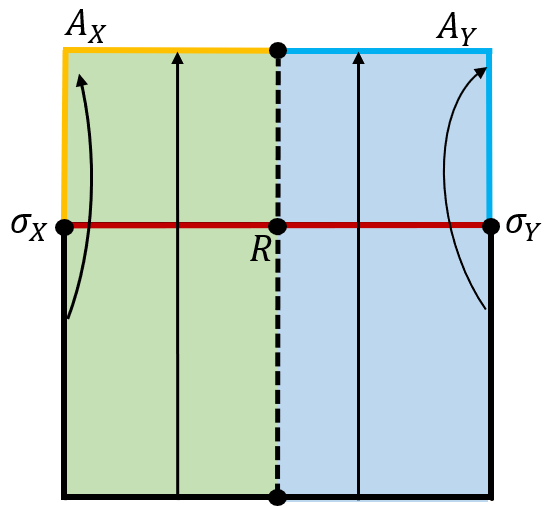}
\caption{A partition of a boundary region $A\subset \partial M$ into $A=A_{X}\cup A_{Y}$, where $A_{X}\cap A_{Y}=\emptyset$, and with associated non-intersecting boundary Cauchy slices $\sigma_{A_{X}}\cup\sigma_{A_{Y}}=\sigma_{A}$. The red line connecting $\sigma_{X}$ and $\sigma_{Y}$ denotes the maximal Cauchy slice $\Sigma$ containing the HRT surface $R$. The minimal flux of a Lorentzian flow $v$ through the boundary $A^{X,Y}$ computes the subregion complexity of the boundary regions $\sigma_{X,Y}$, given by the maximum volumes of $V_{A_{X}}$ and $V_{A_{Y}}$, respectively. The flow which computes the volume of the entire boundary slice will always have more flux through $A^{X}$ and $A^{Y}$ than the minimal flux, leading to superadditivity \eqref{eq: superAdditivityProof}.}
\label{fig:partition}\end{figure}

 In the latter case, by comparing the volumes of different Cauchy slices in the bulk, it was found subregion complexity via CV duality obeys a superadditivity relation. To be precise, define subregion complexity $C_S$ according to the prescription of \cite{Carmi:2016wjl} as follows. Let $\sigma_{A}$ be a boundary Cauchy slice of a boundary region $A$ containing spatial subregions $\sigma_{X}$ and $\sigma_{Y}$, such that $\sigma_{X}\cup\sigma_{Y}=\sigma_{A}$ and $\sigma_{X}\cap\sigma_{Y}=\emptyset$. Denote the HRT surface dividing $\sigma_{A}$ by $R$.\footnote{For simplicity we will assume the state on $\sigma_{A}$ is pure such that $\sigma_{X}$ and $\sigma_{Y}$ share the same HRT surface $R$. If the state is mixed, as is the case in black hole backgrounds, we first purify the state such that $\sigma_{X}$ and $\sigma_{Y}$ have the same HRT surface.} Then, the subregion complexity $C_{S}(\sigma_{X})$ is given by finding the maximal volume slice anchored at $\sigma_{X}\cup R$, and similarly for $C_{S}(\sigma_{Y})$.  The superadditivity of subregion complexity then states
\begin{equation}\label{eq: subregionSuperadditivity}
C_{S}(\sigma_{X}\cup\sigma_{Y}) \geq C_{S}(\sigma_{X}) + C_S(\sigma_{Y})\;.
\end{equation}

We leave the study of a flow formulation of subregion complexity more generally to future work, however, it is easy to prove \eqref{eq: subregionSuperadditivity} using flows. To see this, first note that since $\sigma_{X},\sigma_{Y}$ are non-overlapping boundary slices, then their corresponding entanglement wedges $w_A,w_B$ are non-overlapping \cite{Czech:2012bh}.  Extending this into the bulk means we can consider the bipartition of a boundary region $A\subset \partial M$ into two non-overlapping boundary regions $A=A_{X}\cup A_{Y}$ (with $A_{X}\cap A_{Y}=\emptyset$ and $A_{X,Y}$ are not nested regions; see Figure \ref{fig:partition}). Let $\sigma_{A_{X}}$ and $\sigma_{A_{Y}}$ be the associated non-intersecting spatial slices of $A_{X}$ and $A_{Y}$, respectively, obeying $\sigma_{A_{X}}\cup\sigma_{A_{Y}}=\sigma_{A}$. A solution to the min-flow problem for subregion complexity induces a flow $v$ on $M$ which simultaneously computes the minimal fluxes through the spatial slices anchored at $\sigma_{A_{X}}\cup R$ and $\sigma_{A_{Y}}\cup R$. The minimum flux through each slice yields the maximum volumes $V_{A_{X}}$ and $V_{A_{Y}}$, and is less than or equal to the volume of the maximal slice $\Sigma(A)$. Therefore, the subregion complexity of a state reduced to the Cauchy slice $\sigma_{X}\cup \sigma_{Y}$ is given
\begin{equation}\label{eq: superAdditivityProof}
C_{S}(\sigma_{X}\cup\sigma_{Y}) = \int_{\sigma_{A}} v= \int_{A_{X}} v + \int_{A_{Y}} v \geq C_{S}(\sigma_{X}) + C_{S}(\sigma_{Y}),
\end{equation}
where the inequality follows from the fact the flux of the flow through $A_{X}$ will always be greater than the minimum flux through  $A_{X}$, which computes $C_{S}(\sigma_{X})$ and similarly for $A_{Y}$.

 Equivalently, superadditivity (\ref{eq: superAdditivityProof}) can be rephrased in terms of the mutual complexity $\Delta C_{S}$ \cite{Caceres:2019pgf},
\beq \Delta C_{S}=C_{S}(\sigma_{X})+C_{S}(\sigma_{Y})-C_{S}(\sigma_{XY})\;,\eeq
where $\Delta C_{S}<0$ implies (\ref{eq: superAdditivityProof}). It would be interesting to extend subregion complexity to multiple boundary regions, in which flows may be technically advantageous.


\subsection{Interpretation: Lorentzian threads as `gatelines'\label{sec:interpretation}}

Above we used Lorentzian flows $v$ to reformulate the complexity=volume conjecture (\ref{eq:CVflowv1}), and showed, using the nesting property, holographic complexity satisfies the inequality (\ref{eq:inequality1})  such that the (negative) change in complexity evaluated at two boundary times is bounded below by the minimal flux through the combined region, $\mathcal{C}(\sigma_{A}|\sigma_{AB})$. In this section we focus on developing a physical interpretation of each of these statements.

Similar to the `bit thread' interpretation\footnote{Namely, the set of (unoriented) integral curves of the flow $v$ with density $|v|$ everywhere and are only allowed to end on the boundary, which can be interpreted as a continuous path in a network. As reviewed in Appendix \ref{app:examples}, the maximum number of bit threads for a given configuration $N_{A\bar{A}}$ quantifies the entanglement between $A$ and $\bar{A}$, $S(A)=\text{max}\;N_{A\bar{A}}$.} of Riemannian flows \cite{Freedman:2016zud}, there is a unique mapping (up to unimportant Planck scale effects) between Lorentzian flows and what we will call `Lorentzian threads' or `gatelines'. Specifically, a thread configuration is defined as a set of future oriented  timelike curves on the Lorentzian manifold $M$ satisfying: (i) they only end on the boundary $\partial M$, and (ii) everywhere the thread density obeys $|v|\geq\frac{1}{G_{N}\ell}$. We emphasize the norm bound condition tells us the threads cannot be packed more loosely than $1/G_{N}\ell$. Thus, on macroscopic AdS scales, where $\ell=L_{\text{AdS}}$, one has $G_{N}\sim N^{-2}$ and $L\sim N$ (in the usual gauge/gravity parlance), such that the threads are bounded below by $\sim N$, and saturate this density on the maximal bulk surface.

Let us denote the number of threads connecting a timelike boundary region $A$ to its complement $\bar{A}$ in a specific configuration by $N_{A\bar{A}}$. Then it is straightforward to show, given our reformulation (\ref{eq:CVflowv1}), that holographic complexity is the \emph{minimum} number of threads connecting $A$ to $\bar{A}$. Equivalently, $N_{A\bar{A}}=N_{A}$, where $N_{A}$ are the threads piercing homologous bulk Cauchy slice $\Sigma(A)$ to $A$:
\beq
\mathcal{C}(\sigma_{A})=\text{min}\; N_{A}\,.\label{eq:minnumthreads}
\eeq
The proof of (\ref{eq:minnumthreads}) is as follows. First,  given any flow $v$ we can construct a thread configuration by choosing a set of integral curves with density $|v|$ everywhere, such that the $N_{A\bar{A}}$ -- the number of threads passing from $\bar{A}$ to $A$ (and vice versa) -- is bounded above by the flux of $v$ on $A$. When this flow is a minimizing flow $v(A)$, this is equal to the complexity $\mathcal{C}(\sigma_{A})$, $N_{A\bar{A}}\leq\int_{A}v(A)=\mathcal{C}(\sigma_{A})$. Alternatively, $N_{A\bar{A}}$ is bounded below by the volume of any slice $\Sigma$ homologous to $A$, multiplied by the density $|v|=\alpha$, \emph{i.e.}, $N_{A\bar{A}}\geq\alpha\,\text{vol}(\Sigma)$.\footnote{This can be shown by considering a bulk region surrounding $\Sigma$ of some thickness $T$, such that the (spacetime) volume $T\text{vol}(\Sigma(A))$ multiplied by $\alpha$ is a lower bound to the length of the threads within the bulk region. Moreover, any thread connecting $A$ to $\bar{A}$, and  must pass through $\Sigma$, and such that the total length of threads passing through this region is $TN_{A\bar{A}}$. } When $\Sigma$ is the maximum volume slice, we have $N_{A\bar{A}}\geq\alpha\Sigma(A)=\mathcal{C}(\sigma_{A})$. Combining this with the previous bound, we find the saturation (\ref{eq:minnumthreads}).

 Thus, the holographic complexity $\mathcal{C}$ associated with a boundary subregion $A$ can be understood as the minimum number of threads connecting $\bar{A}$ to $A$, for $\bar{A}$ to the past of $A$. Equivalently, $\mathcal{C}(\sigma_{A})$ is the minimum number of threads passing through the maximal volume slice $\Sigma(A)$ homologous to $A$. This latter observation naturally encourages the idea these threads should really be thought of as `gatelines' preparing the state on the homologous maximal volume slice from a specific reference state defined on the infinite past of the manifold.\footnote{The word `gateline' in the context of Lorentzian flows was introduced in \cite{Headrick:2017ucz} to offer a suggestive application of Lorentzian flows. Here we provide the first explicit realization of Lorentzian flows as gatelines via state preparation.}

To make this picture more precise, recall that bulk Lorentzian AdS spacetimes really describe the time evolution of a specific CFT state, in which the CFT state is prepared by a Euclidean path integral \cite{Skenderis:2008dh,Skenderis:2008dg}. That is, the initial CFT state is described using a wavefunctional approach following the Hartle-Hawking prescription where the functional is computed by summing over bulk field configurations on a half of Euclidean AdS$_{d+1}$, a southern hemisphere, whose boundary is the half cylinder $S^{d-1}\times\mathbb{R}^{-}\approx B^{d}$, where the origin of the sphere is mapped to negative infinite Euclidean time. The cap of the southern hemisphere is given by a spatial Cauchy slice, which here we label as $\Sigma_{-}$. On the gravity side then, all that is required to describe a CFT state is initial data, \emph{i.e.}, data on $\Sigma_{-}$ given by the value of the bulk fields and their normal derivatives on $\Sigma_{-}$. The former values are interpreted as sources that are turned on, while the latter specify the choice of `reference state' on $\Sigma_{-}$. With this initial data, the time evolution of the CFT state is determined by solving the bulk equations of motion using the boundary conditions imposed at the timelike AdS boundary. Furthermore, a Lorentzian AdS cylinder is glued along $\Sigma$, whose length describes the duration of the time evolution in real time.  Note that this prescription of real-time gauge/gravity duality was generalized to characterize excited \cite{Botta-Cantcheff:2015sav} and coherent CFT states \cite{Marolf:2017kvq}, where coherent states are prepared by Euclidean path integrals with sources turned on.\footnote{In fact, we can be a bit more precise here. Following the observations made in \cite{Belin:2020zjb}, turning on Euclidean sources for single-trace operators corresponds to looking for regular solutions to the bulk equations of motion satisfying suitable boundary conditions such that the bulk fields match the sources, \emph{i.e.}, given some sources there exists a map determining the corresponding Lorentzian initial data preparing a state. Turning this map around, it turns out the problem of using arbitrary initial data to determine the associated boundary sources is ill-posed, \emph{e.g.}, localized initial data leads to divergent CFT sources \cite{Marolf:2017kvq}, and the initial data cannot be completely generic but rather analytic. Moreover, to obtain arbitrary initial data, one must include delta function sources in the bulk, located at singularities in the bulk equations of motion. To have a smooth Euclidean section, the initial data must obey a non-local condition specified by a particular integral equation. Thus, the bulk equations of motion reveal which sources are needed to prepare a state, and we assume we work with analytic initial data such that the Euclidean section is singularity free.} We will make extensive use of these generalizations in Section \ref{sec:diffformsandeineqs}.

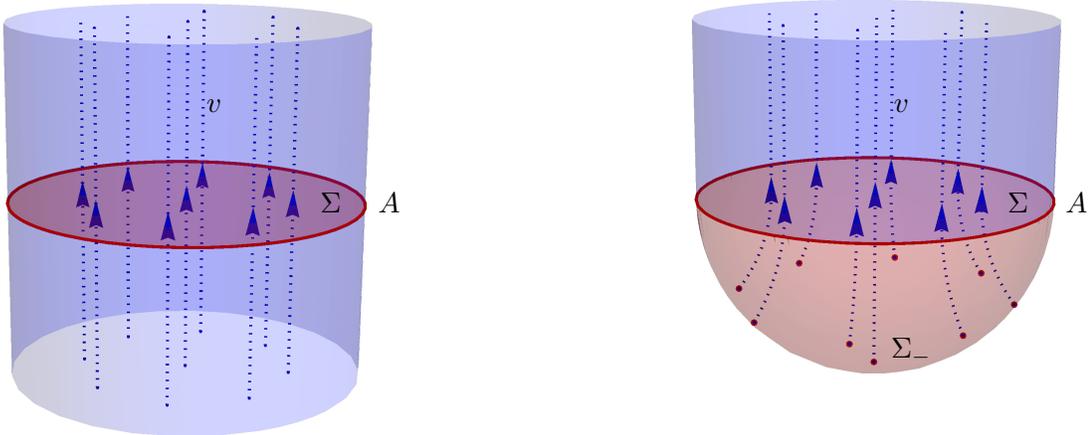
\begin{figure}[t]
\centering
 \includegraphics[width=3in]{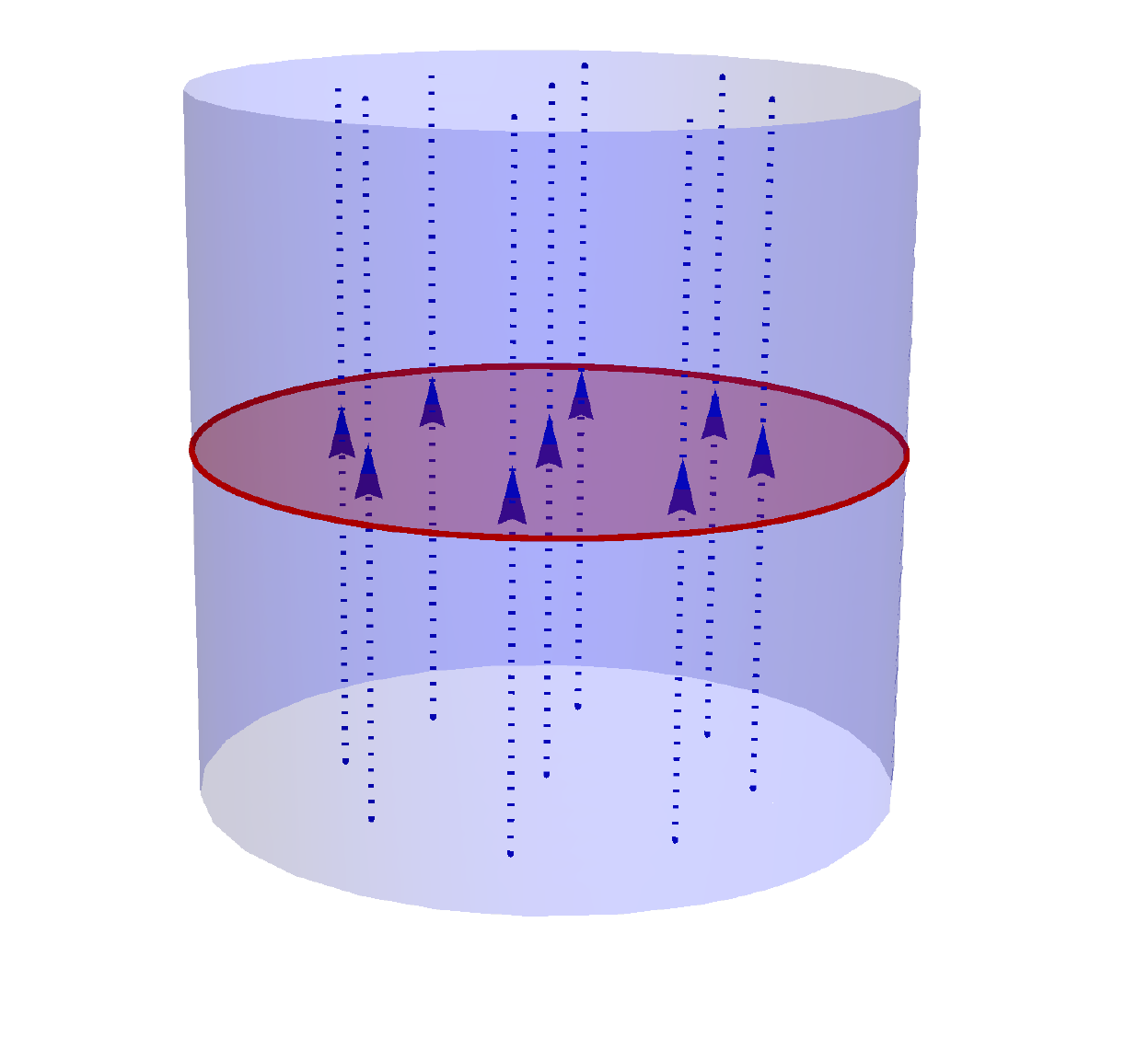}$\qquad\qquad$\includegraphics[width=3in]{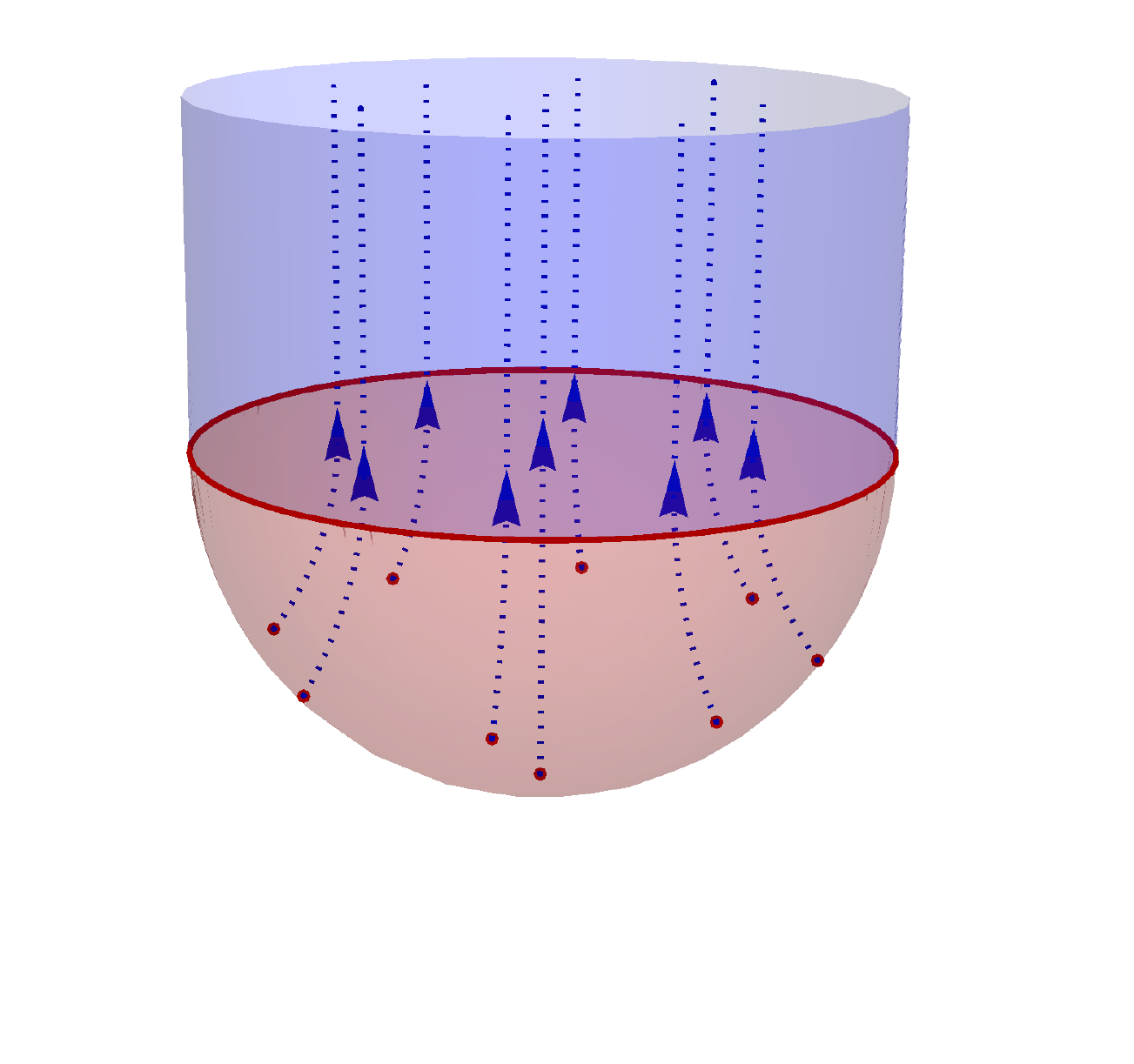}
 \begin{picture}(0,0)
\put(-65,127){$\Sigma$}
\put(-43,127){$A$}
\put(195,127){$\Sigma$}
\put(217,127){$A$}
\put(151,72){$\Sigma_{-}$}
\put(-108,165){$v$}
\put(152,165){$v$}
\end{picture}
\vspace{-1cm}
\caption{Left: Visualization of Lorentzian threads moving in the Lorentzian cylinder. Right: State preparation by a Euclidean path integral, where the optimal thread configuration prepares the CFT state on the maximal volume hypersurface $\Sigma$, the cap of the southern hemisphere of Euclidean AdS, which is glued along the Lorentzian cylinder. Complexity is then understood to be the minimum number of Lorentzian threads (\emph{i.e.} `gatelines') passing through this bulk slice or, equivalently, the minimum number of gates required to prepare the state on $\Sigma$ from a reference state on $\Sigma_{-}$.
\label{fig:gatelines}}
\end{figure}

We can now consider Lorentzian threads associated to a boundary region $A$, flowing in the AdS background. These threads pass through Cauchy slices foliating the spacetime, including the maximal hypersurface $\Sigma$, whose volume is given by the minimum number of threads passing through. More generally, due to the divergenceless condition, we can consider any surface $\Sigma'$ homologous to $A$. The norm bound then enforces $\Sigma'=\Sigma$. When the bulk region in the past of $\Sigma$ is replaced with the southern hemisphere as in the Hartle-Hawking description, then the Lorentzian flow $v$ with minimal flux through $\Sigma$ prepares the CFT state on this maximal volume slice.  A visualization of this is given in Figure \ref{fig:gatelines}. From this perspective, it is natural to interpret Lorentzian threads as gatelines: trajectories in spacetime that represent the various unitary gates needed to transform a reference state to a target state. Complexity, then, is simply the minimum number of gatelines through the hypersurface preparing the reference state.

With this interpretation we can provide conceptual insight to the inequality (\ref{eq:inequality1}) and the quantity $\mathcal{C}(\sigma_{A}|\sigma_{AB})$. First, $\mathcal{C}(\sigma_{AB})$ is the determined by the minimum number of gatelines needed to prepare the state on Cauchy slice $\Sigma$. The inequality (\ref{eq:inequality1}) simply tells us the minimizing flux through region $B$ is the amount of gatelines used to prepare the state on slice $\Sigma(A)$ given the state on $\Sigma(AB)$ (which is now taken as the reference state). In other words, the minimum number of gatelines needed to prepare state $A$ given the minimum number of gatelines preparing $AB$. It is thus natural to interpret $\mathcal{C}(\sigma_{A}|\sigma_{AB})$ as a `conditional complexity': the complexity of state $A$ conditioned by the state on $AB$.  In the event the flux through $B$ is zero, then the same minimum number of gatelines prepare the state on the timeslices $\Sigma_{AB}$ and $\Sigma_{A}$, such that the states have the same complexity. From the boundary perspective, this is not surprising as the same number of boundary sources are used to the prepare the state at two different times. Alternatively, for gatelines exiting $B$ there are more gatelines needed to prepare the state on $\Sigma_{AB}$ than on $\Sigma_{A}$, \emph{i.e.}, $\mathcal{C}(\sigma_{AB})>\mathcal{C}(\sigma_{A})$, while threads entering through $B$ indicate additional gatelines are needed prepare the state on $\Sigma_{A}$ such that it has a higher complexity than the state on $AB$ (more boundary sources are needed to prepare the state on $\Sigma_{A}$). Nonzero flux $\Phi(B)$ thus can either reduce or increase the complexity of the state prepared on $\Sigma_{A}$ which we can think of as characterizing the rate of complexity or `uncomplexity'. Lastly, $\mathcal{C}(\sigma_{A}|\sigma_{AB})$ exemplifies the aforementioned two step optimization procedure: an intermediate state is prepared on $\Sigma(AB)$ via some number of gatelines before the state on $\Sigma(A)$ is prepared by a potentially different number of gatelines.

There are several more comments in order. First, for the Lorentzian setting to work it is important the full manifold $M$ defining the state preparation be compact, accomplished by gluing Euclidean portions to the past $M_{-}$  to the future $M_{+}$ along the Lorentzian cylinder. The  boundary region $A$ then covers at least the boundary of $M_{+}$. Second, while the full Lorentzian manifold specifies the time evolution of the boundary state, the portion of the flows to the future of slices where a state is prepared do not affect the preparation of the target state. That is, the complexity of the state prepared on $\Sigma_{A}$ is characterized by the number of threads that enter $A$ from the past, which pass through the boundary part of the manifold to the past of $\Sigma_{A}$ -- the part of the manifold to the future of $\Sigma$ does not affect this number of threads, and hence complexity.

Next, in our interpretation, specifying the reference state is particularly important, unlike the traditional CV or CA dualities, where one usually does not define the reference state. Our prescription, which is tightly connected to the problem of state preparation in AdS/CFT, the reference state is found by specifying (normalizable) boundary conditions on $\Sigma_{-}$. The target state is then specified as a boundary condition on $\Sigma$, where one asks what sources (non-normalizable modes) are needed to be turned on at $\Sigma_{-}$ to reach the target state. Thus, if one were to ask what is the complexity of the vacuum state, in our prescription we would first need to know what the reference state is, \emph{i.e.}, the complexity of the vacuum is dependent on the reference state. For example, if the reference state is the vacuum -- where all sources are turned off -- then complexity is zero. In this case, the part of the manifold to the past of $\Sigma$ (including the Euclidean southern hemisphere $M_{-}$) would collapse to a point and the volume (or, equivalently, number of threads) goes to zero. If, however, the reference state is other than the vacuum, the complexity is non-zero as some sources must be turned on, and the past to $\Sigma_{-}$ is not collapsed to a point.

We would like to also briefly highlight our interpretation here offers a notion of `emergent time evolution'. More precisely, sitting along the bulk surface $\Sigma$ where the CFT state has been prepared, the forward time evolution of this state arises by following the trajectory of the Lorentzian flow passing through $\Sigma$. Earlier, moreover, we noted time evolution of the boundary state is determined by solving bulk (gravitational) equations of motion subject to specific boundary conditions. This suggests Lorentzian threads, or at least a particular thread configuration, encodes information about bulk field equations. We detail this  in Section \ref{sec:diffformsandeineqs}.

Lastly, our intuition above focused on thread configurations which solve the min flux problem, namely, optimal flows. However, non-optimal flows are expected to play a role and naturally fit withina tensor network description of AdS. This suggests an alternative notion of complexity. We offer a proposal for such an alternative and describe how gatelines prepare tensor networks in Section \ref{sec:genprop}.

\section{Simple geometric constructions} \label{sec:simpleconstructions}

In Section \ref{sec:prelims} we proposed holographic complexity, as understood in the `complexity=volume' conjecture, is given by the minimum flux of divergenceless bulk timelike vector fields $v$ whose norm satisfies $|v|\geq1$.\footnote{Here we normalize our flows such that $\alpha=1$.} The definition (\ref{eq:Lorflowdef}) of Lorentzian flows is highly non-unique, as infinitely many timelike vector fields can be made to satisfy its properties. As such, so far  we have described Lorentzian flows and their interpretation rather generically.

 In this section we provide explicit realizations of these Lorentzian threads. Broadly we consider two types of constructions:

\vspace{2mm}

\noindent (i) \emph{integral lines}: timelike vector fields found by foliating the bulk Lorentzian spacetime with timelike curves, such as geodesics, where the norm is specifically chosen such that the divergenceless condition is locally satisfied.

\vspace{2mm}

\noindent (ii)  \emph{level set flows}: timelike vector fields obtained whose integral lines are orthogonal to a family of (nested) slices foliating the spacetime. By construction, the norm of the vector saturates the bound $|v|=1$. In the case where these are taken to be maximal volume slices, they lead to the so-called minimally packed flows which we have already discussed above in relation to the property of superadditivity.

\vspace{2mm}

\noindent The Riemannian analogs of these constructions were developed in \cite{Agon:2018lwq} (also briefly summarized in Appendix \ref{app:examples}), however, we will see some key differences in this context.  For example, when the bulk spacetime under consideration is asymptotically AdS, the geodesic flows are constructed to foliate inside and outside the Wheeler-DeWitt (WDW) patch, though we build flow configurations via level sets \emph{only} in the interior of the WDW patch, augmenting the exterior by geodesic flows. Because of this, let us first describe integral line constructions in detail.

\subsection{Method of integral lines} \label{subsec:geoflows}

Here we consider a construction using a set of integral lines obeying certain properties such that the Lorentzian flow is described by a vector field tangent to the integral curves. A specific subclass of such lines, as we will show, are future directed timelike geodesics which foliate the bulk spacetime.

Flows based on integral lines are built following a rather generic algorithm. For this construction, we assume the following: a Lorentzian bulk spacetime endowed with metric $g_{\mu\nu}$ with a (compact) boundary $\partial M$, and any connected boundary region $A$ which has an associated bulk maximal volume slice denoted by $\Sigma(A)$, such that $\Sigma(A)|_{\partial M}=\partial A$. There are essentially two steps to this construction: (i) consider a family of integral curves satisfying the properties described below, from which we are able to identify the vector $\hat{\tau}$ tangent to the flows, and (ii) the magnitude $|v|$ is found by demanding the divergenceless condition be satisfied. These two steps will allow us to construct the desired vector field, $v=|v|\hat{\tau}$, characterizing the holographic complexity.

More precisely, the family of integral curves we propose are taken to satisfy:

\vspace{2mm}

\noindent (1) The tangent vector $\hat{\tau}$ is taken to be unique and well defined everywhere. Consequently, the flows themselves are continuous and non self-intersecting, except possibly at a set of bulk points of measure zero. If these points exist, the amount of integral lines coming in and out should exactly cancel out, such that total flux across a surface enclosing the point is zero.

\vspace{2mm}

\noindent (2) The tangent vector $\hat{\tau}$ must be equal to the unit normal $\hat{n}_{m}$ at the maximal volume slice $\Sigma(A)$. This must hold since the bound on the flow $|v|=1$ on $\Sigma(A)$, as demanded by the min flow-max cut theorem.

\vspace{2mm}

\noindent (3) The vector field $v$ must be divergenceless, such that, as we will see, the integral lines should begin and end at the boundary of the manifold. 

\vspace{2mm}

Once we have picked a family of integral curves satisfying properties (1)-(3), the final task is to find an appropriate norm $|v|$ such that $|v|\geq1$ holds everywhere. A brute force method is to simply scale the tangent vector $\hat{\tau}$ by some function of the coordinates and demand $\nabla\cdot v=0$. This can result in a partial differential equation which may be difficult to solve, particularly for spacetimes which don't exhibit much symmetry. A more natural way to obtain the appropriate norm is by enforcing the integral version of Gauss's theorem, so that the flux through an infinitesimal volume element is constant through the Lorentzian threads. This is accomplished by first parametrizing the curves $X(x_{m},\lambda)$ by the point $x_{m}$ at which the flows intersect the maximal volume slice, and where $\lambda$ is an affine parameter that runs along the curve. We then follow the set of integral curves that leave from an infinitesimal region $\delta A(x_{m})$ surrounding $x_{m}$ at $\Sigma(A)$. The volume of the bulk slice will propagate along the flow, defining a co-dimension zero region in the bulk. Since this bulk region is made out of the proposed integral curves, flux does not enter or leave the region such that the divergenceless condition is imposed by choosing norm $|v|$ so the flux through any transverse section of the bulk region is constant. More precisely, we choose $|v|$ such that
\beq \int_{\delta A(x_{m})}\sqrt{h_{\lambda}}d^{d-1}x_{m}|v|=\text{constant}\;,\eeq
where $h_{\lambda}$ is the induced metric on the plane orthogonal to the flow line at point $x_{m}$. Since $|v(x_{m},\lambda_{m})|=1$ (the location of the maximal slice), where $\lambda_{m}$ is the value of $\lambda$ at which the flow intersects $\Sigma(A)$, one arrives to the following expression for $|v|$ along the integral curves:
\beq |v(x_{m},\lambda)|=\frac{\sqrt{h(x_{m},\lambda_{m})}}{\sqrt{h(x_{m},\lambda)}}\;.\label{eq:conteq}\eeq
From this continuity equation, we have $|v|\geq1$ if and only if the volume satisfies $\sqrt{h(x_{m},\lambda_{m})}<\sqrt{h(x_{m},\lambda)}$, \emph{i.e.}, the volume decreases everywhere away from the maximal volume slice. This condition, however, must be explicitly verified after the fact: if the norm bound $|v|\geq1$ is not satisfied, we must modify our choice of curves until the bound is achieved.

A natural subclass of flows which follow from the above algorithm are so-called geodesic flows. In particular, as we will show, timelike geodesics foliating the interior of the WDW patch can be used to construct particular flows using our algorithm. By the choice of boundary conditions in the interior of the WDW patch the integral lines in this case will not foliate the region outside the WDW patch, however, we will fill this region by translating and twisting the interior geodesic curves.

We will apply this general algorithm specifically when the background geometry is empty AdS and the BTZ black hole, where our proposed thread configurations are the integral curves to timelike geodesics foliating the interior of the WDW patch. As we will see, there is an important difference between the constructions in empty AdS and the BTZ black hole. Due to the higher degree of symmetry in empty AdS, without loss of generality, the WDW patch is built by picking a point at the origin in radial coordinates -- regardless of the value of the time coordinate -- from which the null rays emanating from the origin form the boundary of the WDW patch. Consequently, the maximal volume slice may be determined independently from imposing orthogonality to the hypersurface on the solutions to the timelike geodesic equations. Rather, the WDW patch and the maximal volume slice appear automatically from the emission of all possible radially outgoing geodesics from the origin. Alternatively, the WDW patch in the BTZ black hole background is heavily influenced by the presence of a double timelike boundary as well as the past and future singularities. We therefore impose orthogonality relations on the timelike geodesics such that the associated left and right corners of the WDW patch intersecting the timelike boundaries are anchored at the same boundary time (we take left and right boundary times to be equal, $t_{L}=t_{R}$). This is in fact standard in the context of studying the growth of the volume of the wormhole behind the black hole horizon in the complexity=volume scenario (see, for example, \cite{Carmi:2017jqz}). We will also explore the change in Lorentzian threads due to this time dependence by explicitly construcing the threads configuration for the late time surface, which we then use to explain how the second law of complexity can be interpreted in the light of these gatelines.

 Lastly, note that radial timelike geodesics are not the only class of flows we could have considered, \emph{e.g.}, integral curves of timelike Killing vectors, however, timelike geodesics carry a physical interpretation: they represent timelike trajectories of backreactionless observers probing the geometry of the entire spacetime.  It is also worth pointing out the foliation of the WDW patch is obtained by a congruence of observers which probe the entirety of spacetime at the same instant in coordinate time $t$. By this we mean the maximal volume slice lies exactly where every point in the spacetime slice is reached by a geodesic foliating the WDW patch. This always happens in vacuum AdS and happens in the BTZ spacetime when the boundary times at which the WDW patch is anchored are the same time.


\subsubsection{Flows in vacuum AdS}

\subsection*{Geodesic foliation of WDW patch}

\begin{figure}[t]	
\begin{center}
\includegraphics[width=8cm]{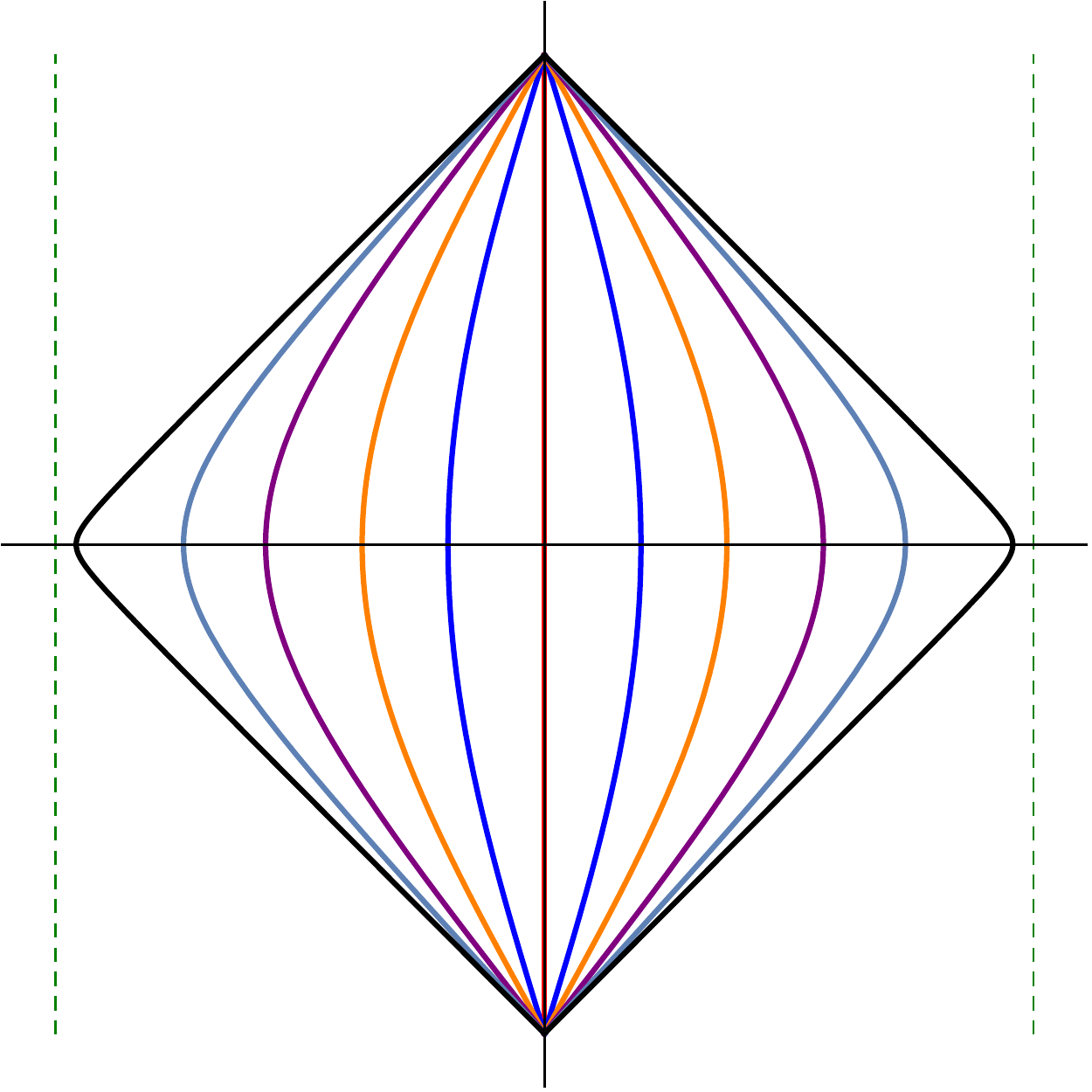}
\end{center}
 \begin{picture}(0,0)
\put(219,258){$t$}
\put(227,36){$t_{past}$}
\put(227,243){$t_{fut}$}
\put(337,141){$\rho$}
\end{picture}
\caption{Foliation of the Wheeler-De Witt patch by timelike geodesics in compactified coordinates, $r=\tan(\rho)$. The boundary of the WDW patch is generated by null curves originating at the $t_{past}=-\frac{\pi L}{2}$, hitting the conformal boundary at $t=0$ and ending at $t_{fut}=\frac{\pi L}{2}$. Higher initial energies $E$ of the timelike geodesics correspond to higher initial radial velocities, asymptotically approaching, but never reaching the conformal boundary. All the geodesics cross at the past and future tip of the WDW patch. Here we have set $L=1$. Curves of different colors correspond to different values of $E$: $E=1$ (red), $E=1.05$ (blue), $E=1.2$ (orange), $E=1.6$ (violet), $E=2.5$ (cyan), and $E=15$ (black).  As the energy increases, the timelike geodesics tend to the null geodesics forming the boundary of the WDW patch. Geodesics at the right side of the $\rho$ axis are at an angle $\theta_0$, while those on the left side are at an angle $\theta_0 + \pi$, thus correspond to their reflected versions.}
\label{fig:adsgeodesics}
\end{figure}
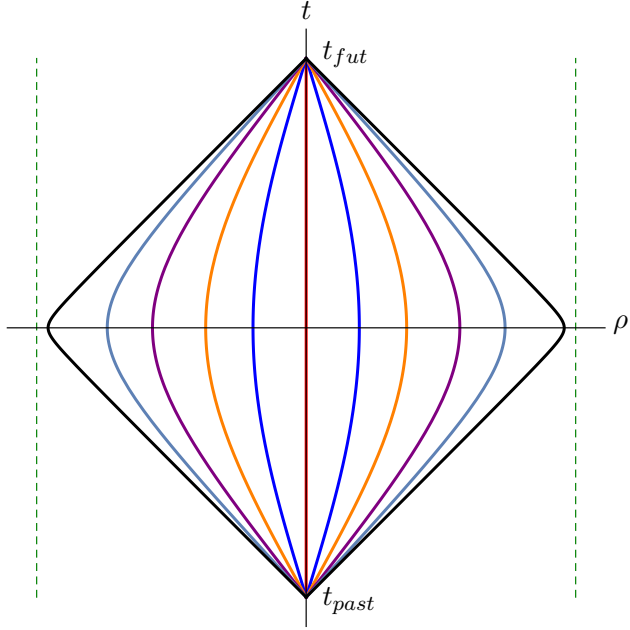

For simplicity and pedagogy, let us first work in $\text{AdS}_{3}$. We will return to its higher dimensional generalization momentarily. In global coordinates the line element takes the form
\beq ds^{2}=-f(r)dt^{2}+f^{-1}(r)dr^{2}+r^{2}d\theta^{2}\;,\quad f(r)=1+\frac{r^{2}}{L^{2}}\;,\label{eq:ads3metglob}\eeq
where $L$ is the AdS length. As mentioned above, we have the freedom to place the past tip of the WDW patch at $r=0$, for any coordinate $t$ value. Future null rays emanating from the origin $r=0$ then hit the timelike boundary before bouncing back and ending at the future tip $r=0$ (see Figure \ref{fig:adsgeodesics} for an illustration).

A natural candidate for the integral curves satisfying the desired properties (1) -- (3) of our aforementioned algorithm are non-intersecting timelike geodesics of $\text{AdS}_{3}$. In particular, we will show timelike geodesics foliate the interior of the WDW patch, such that the tangent vector $\hat{\tau}$ to the integral lines of the geodesic flows is equal to the timelike unit normal vector $n^{\mu}$ to the maximal volume slice, and, using the continuity equation (\ref{eq:conteq}), the associated vector field will indeed satisfy $v=|v|\hat{\tau}$ everywhere inside the WDW patch.

Thus, our first task is to work out the timelike and null geodesics (since the boundary of the WDW patch is generated by the null geodesics) for global $\text{AdS}_{3}$, and verify they foliate the interior of the WDW patch. The geodesic equations are simply:
\beq
\begin{split}
 &\ddot{t}+\frac{2r}{L^{2}f}\dot{t}\dot{r}=0\;,\quad \ddot{r}+\frac{rf}{L^{2}}\dot{t}^{2}-\frac{r}{L^{2}f}\dot{r}^{2}-rf\dot{\theta}^{2}=0\;,\quad\ddot{\theta}+\frac{2}{r}\dot{r}\dot{\theta}=0\;,
\end{split}
\eeq
from which we uncover two conserved quantities $E$, and $\ell$,
\beq E\equiv\dot{t}f\;,\quad \ell\equiv\dot{\theta}r^{2}\;,\eeq
where we take the `energy' $E$ to be positive. Due to conservation of angular momentum, we need only to consider a constant angle $\theta=\theta_0$.

To solve the geodesic equations, we consider future directed geodesics emanating from the past tip of the WDW patch, which we fix to be the point $p_{past}=(-\frac{\pi L}{2},r=0)$. Consequently, the maximal volume slice is the $t=0$ slice, where the null rays intersect the (right) conformal boundary, and the geodesics all end at the future tip of the WDW patch located at $p_{fut}=(\frac{\pi L}{2},r=0)$. The integral curves $x^{\mu}$ to the \emph{null} geodesics satisfying the initial condition may be parametrized by either an affine parameter $\lambda$ or coordinate time $t$:
\beq
\begin{split}
&x^{\mu}_{\lambda}(\lambda)=[t(\lambda)\;,\; r(\lambda)\;,\;\theta(\lambda)]=\left[L\arctan\left(\frac{E\lambda}{L}\right)-\theta_0\;,\; E\lambda\;,\;\theta_0\right]\;,\\
&x^{\mu}_{t}(t)=[t\;,\;r(t)\;,\;\theta(t)]=\left[t\;,\;L\biggr|\text{cot}\left(\frac{t}{L}\right)\biggr|\;,\;\theta_0\right]\;.
\end{split}
\label{eq:nullgeoads3}\eeq
It is easy to verify $x_{t}^{\mu}(t)=(-\frac{\pi L}{2})=x_{\lambda}^{\mu}(\lambda=0)$.  Similarly, we may parametrize the \emph{timelike} geodesics in terms of proper time $\tau$ or coordinate time $t$:
\beq
\begin{split}
&x^{\mu}_{\tau}(\tau)=[t(\tau)\;,\; r(\tau)\;,\;\theta(\tau)]=\left[\frac{EL}{\sqrt{1+\omega^{2}}}\arctan\left(\sqrt{1+\omega^{2}}\tan\left(\frac{\tau}{L}\right)\right)-\frac{\pi L}{2}\;,\;\omega L\sin\left(\frac{\tau}{L}\right)\;,\;\theta_0\right]\;,\\
& x^{\mu}_{t}(t)=[t\;,\; r(t)\;,\;\theta(t)]=\left[t,\frac{L\sqrt{E^{2}-1}}{\sqrt{1+\frac{E^{2}}{\tan^{2}\left(\frac{t}{L}+\frac{\pi}{2}\right)}}}\;,\;\theta_0\right]\;.
\end{split}
\label{eq:timelikegeoads3}\eeq
Here $\omega\equiv\frac{dr(\tau=0)}{d\tau}$, and we see $r(\tau=0)=0$.

A few comments are in order with respect to integral lines (\ref{eq:nullgeoads3}) and (\ref{eq:timelikegeoads3}). First, notice the coordinate functions $x^{\mu}$ are real if and only if $E\geq1$. This tells us timelike geodesics are bounded above by the null rays emanting from $p_{past}$. Indeed, for timelike geodesics (\ref{eq:timelikegeoads3}) the timelike condition $g_{\mu\nu}\dot{x^{\mu}}_{\tau}\dot{x}^{\nu}_{\tau}<0$, where $\dot{x}^{\mu}_{\tau}\equiv\frac{dx^{\mu}_{\tau}}{d\tau}$, yields
\beq g_{\mu\nu}\dot{x^{\mu}}_{\tau}\dot{x}^{\nu}_{\tau}<0\Rightarrow 0\leq\omega<E\;.\eeq
Consequently, since $E$ controls the velocity of the null geodesics and $\omega$ controls the velocity of the timelike geodesics, we have the null rays bound the timelike geodesics emanating from $p_{past}$.  This observation, moreover, tells us that while the null rays reach infinity $(r=\infty,t=0)$ after a time $t=\frac{\pi L}{2}$, bounce back, and meet again at $r=0$ after total coordinate time $\Delta t=\pi L$; the timelike geodesics never reach infinity.\footnote{This is easily seen by looking at $r(t)$ in (\ref{eq:timelikegeoads3}) for which we see $E$ must take imaginary values for $r(t=0)=\infty$, a contradiction.} An illustration of the WDW patch is presented in Figure \ref{fig:adsgeodesics}, where we use radial compactified coordinates, $r=L\tan\rho$, with $0<\rho<\frac{\pi}{2}$, for which the line element (\ref{eq:ads3metglob}) becomes
\beq ds^{2}=-\sec^{2}(\rho)dt^{2}+\sec^{2}(\rho)d\rho^{2}+L\tan^{2}(\rho)d\theta^{2}\;.\label{eq:compactAdS3coord}\eeq

Second, the timelike geodesics completely foliate the interior of the WDW patch, \emph{i.e.}, for any point inside of the patch that does not belong to the null boundary, there exists a unique timelike geodesic which passes through $q$. To see this, let $q=(t_{q},r_{q})$ be a point inside the WDW patch and does not lie on the boundary. For the range $-\frac{\pi L}{2}<t_{q}<\frac{\pi L}{2}$, we can always invert the integral curves (\ref{eq:timelikegeoads3}) to find $E$ as a function of coordinates $r$ and $t$  and identify the timelike geodesics emanating from $p_{past}$ which passes through $q$. The geodesic which passes through $q$ will have a unique energy $E=E_{q}$ given by
\beq E_{q}^{2}=f(r_{q})\left[1-\frac{r^{2}_{q}}{L^{2}\tan^{2}\left(\frac{t_{q}}{L}+\frac{\pi}{2}\right)}\right]^{-1}\;.\eeq
Thus, for every point $q$ inside the WDW patch there exists a unique timelike (radial) geodesic passing through $q$. Due to the angular symmetry of the background spacetime, this argument need not require we fix $\theta$.

Lastly, we see the tangent vector to the curves $\hat{\tau}$ is well-defined everywhere and is equal to the unit normal $\hat{n}$ to the maximal volume slice at $t=0$, \emph{e.g.},
\beq \hat{\tau}^{\mu}=\frac{dx^{\mu}_{t}}{dt}=\left(1\;,\;-\frac{E^{2}\sqrt{E^{2}-1}}{1+E^{2}\tan^{2}\left(\frac{t}{L}\right)}\sec^{2}\left(\frac{t}{L}\right)\tan\left(\frac{t}{L}\right)\;,\;0\right)\;,\eeq
with $\hat{\tau}|_{t=0}=(1,0,0)=\hat{n}$. The unit vector, however, is neither divergenceless, nor has norm satisfying $|\tau|\geq1$ everywhere.

\subsection*{Timelike geodesics as Lorentzian threads}

Thus far we have shown the integral lines associated to the timelike geodesics foliating the interior of the WDW patch meet the requirements outlined by the three criteria in our general algorithm. Indeed, by construction, the tangent vector $\hat{\tau}$ is unique and well defined everywhere,  is equal to the timelike unit normal on the maximal volume slice $\Sigma$, and the flows begin and end at the past and future tips of the WDW patch. It remains to be seen whether the properly normalized vector field is divergenceless and obeys the norm bound $|v|\geq1$.

To find the norm we parametrize the curves by the point $x_{m}$ at which the flows intersect the maximal volume slice $\Sigma$ at $t=0$. The integral curves to the timelike geodesics  parameterized by coordinate time $t$ in (\ref{eq:timelikegeoads3}) evaluated at $t=0$ obey
\beq
x^{\mu}_{t}(0)=\left[0\;,\;L\sqrt{E^{2}-1}\;,\;\theta_0\right]\;,
\eeq
from which we see the energy $E$ defines the intersection point $r_{m}\equiv L\sqrt{E^{2}-1}$. Thus, for any point $(t,r)$ in the interior of the WDW patch, the integral lines $x^{\mu}_{t}$ may be expressed in terms of $x_{m}$,
\beq x^{\mu}_{m}(t)=\left[t\;,\;\frac{r_{m}}{\sqrt{1+\left(1+\frac{r_{m}^{2}}{L^{2}}\right)\tan^{2}\left(\frac{t}{L}\right)}}\;,\;\theta_0\right]\;.\label{eq:intlinesxm}\eeq
Moving forward, it behooves us to work in radial compactified coordinates (\ref{eq:compactAdS3coord}), for which the integral lines (\ref{eq:intlinesxm}) take the form
\beq
x^{\mu}_{m}(t)=\left[t\;,\;\text{arctan}\left(\frac{\tan(\rho_{m})}{\sqrt{1+\sec^{2}(\rho_{m})\tan^{2}t}}\right)\;,\theta_0\right]\;,\label{eq:intlinesxm2}
\eeq
where we have set $L=1$ for convenience, and $\rho_{m}$ denotes the intersection point of the integral curves at the $t=0$ maximal volume hypersurface.

Our task now is to find the norm bound following our algorithm and using the continuity equation (\ref{eq:conteq}). This requires we determine the induced metric in the adapted coordinates $h_{\mu\nu}(\rho_{m},t)$. With the following coordinate transformation
\begin{equation}
\tan (\rho(t,\rho_m))  = \frac{\tan(\rho_m)}{\sqrt{1+\sec^2(\rho_m)\tan^2(t)}}\;,
\end{equation}
we may write the $\text{AdS}_{3}$ metric (\ref{eq:compactAdS3coord}) in adapted coordinates $(t,\rho_{m})$ as
\begin{equation}
g_{\mu \nu}(t,\rho_m) = \left(
\begin{array}{ccc}
 -\frac{\sec ^2(\rho_m) \sec ^4(t)}{\left(1+\sec ^2(\rho_m) \tan ^2(t)\right)^2} & -\frac{\cos(\rho_m)\cot(t)\csc^2(t)\sin(\rho_m)}{(1+\cos(\rho_m)^2\cot^2(t)} & 0\\
-\frac{\cos(\rho_m)\cot(t)\csc^2(t)\sin(\rho_m)}{(1+\cos(\rho_m)^2\cot^2(t)}  & \frac{\sec ^2(a_m) \sec ^2(t)}{\left(1+\sec ^2(\rho_m) \tan ^2(t)\right)^2} & 0 \\
0 & 0 & -\frac{1}{1+\csc(\rho_m)^2\sec(t)^2}
\end{array}
\right).
\end{equation}

With the above metric we identify the timelike unit normal vector field $\hat{\tau}$ everywhere tangent to the integral curves to be
\begin{equation}
\tau^{\mu} = \frac{1}{\sqrt{- g_{tt}}} (1,0,0) = \left(\frac{1+ \sec^2(\rho_m)\tan^2(t)}{\sec^2(t) \sec(\rho_m)}\;,\;0\;,\;0\right)\;.
\end{equation}
The induced metric $h_{\mu\nu}=g_{\mu\nu}+\tau_{\mu}\tau_{\nu}$ characterizing surfaces of constant $t$ in adapted coordinates $(t,\rho_{m})$ is then given by
\begin{equation}
h_{\mu\nu}(t,\rho_m) = \left(
	\begin{array}{cc}
\frac{\cot^2(t)(\cos^2(\rho_m)\csc^2(t)+\sin^2(\rho_m))}{(1+\cos^2(\rho_m)\cot^2(t))^2} & 0\\
0 & -\frac{1}{1+\csc(\rho_m)^2\sec(t)^2} \\
	\end{array}
\right).
\end{equation}

We now use the continuity equation (\ref{eq:conteq}) to construct the Lorentzian thread vector field:
\beq
\begin{split}
v^{\mu} &=\sqrt{\frac{h(0,\rho_m)}{h(t,\rho_m )}} n^{\mu} =\frac{(1+\sec^2(\rho_m)\tan^{2}(t))}{4\cos(\rho_{m})}|\cos(2t)-3-2\cos^2(t)\cos^2(2\rho_m)|\partial^{\mu}_{t}\;.
\end{split}
\label{eq:vecfieldads3com}\eeq
By construction, it is straightforward to verify, $v^{\mu}$ is divergenceless, and obeys $|v|\geq1$, saturating the bound at $t=0$ on the maximal volume slice. Moreover, we can transform $v(t,\rho_{m})$ back to the radial compactified AdS$_3$ coordinates $(t,\rho)$,
\begin{equation}
v^{\mu}(t,\rho)  = \left(2\sqrt{2}\sqrt{\frac{\cos^2(t)\cos^8(\rho)}{(\cos(2t)+\cos(2\rho))^3}}\;,\;-\sqrt{2}\sin(2t)\sqrt{\frac{\cos^6(\rho)\sec^2(t)\sin^2(\rho)}{(\cos(2t)+\cos(2\rho))^3}}\;,\;0   \right)\;.
\end{equation}
The vector field remains divergenceless and satisfies $|v|\geq1$ everywhere, and are tangent to the integral curves foliating the interior of the WDW patch. For an illustration, see Figure \ref{fig:vecfieldads3OUT}.

\subsection*{Generalizing to $\text{AdS}_{n}$}

Before moving on to discuss the thread configuration outside the WDW patch, we point out the above construction generalizes to higher dimensional empty AdS in a straightforward way. In $n$ spacetime dimensions, empty AdS in global coordinates is
\begin{equation}
ds^2=-f(r)dt^2+f^{-1}(r)dr^2+r^2d\Omega^2_{n-1}\;, \quad d\Omega^2_{n-1}=d\theta+sin^2(\theta)d\Omega^2_{n-2},
\end{equation}
where $d\Omega^{2}_{n-2}$ is the $(n-2)$-dimensional spherical boundary line element. In compactified radial coordinates $r=\tan\rho$ (where we maintain $L=1$) this becomes,
\begin{equation}
ds^2_{AdS_n}=-\sec^2(\rho)dt^2+\sec^2(\rho)d\rho^2+ \tan^2(\rho)\left[ d\theta^2+\sum_{i=1}^{n-3}\left(\prod_{j=1}^{i}\sin^2(\phi_{j-1}) \right)d\phi_{i}^2\right], \label{eqn: AdSNmetric}
\end{equation}
where $\phi_0=\theta\in[0,\pi /2]$ and $\phi_{i\geq 1}\in[0,2\pi]$ are angular coordinates.

A nearly identical analysis of the geodesic structure follows, from which we find the normal $\hat{\tau}$ to surfaces of constant $t$ in coordinates adapted to the foliation is given by
\begin{equation}
\tau^\mu=\frac{1+\sec^2(\rho_m)\tan^2(t)}{\sec^2(t)\sec(\rho_m)}(1\;,\;0\;,\;...,\;0)\;.
\end{equation}
The induced  metric $h_{\mu\nu}(t,\rho_{m})$ characterizing the constant $t$-surfaces follows from $h_{\mu\nu}=g_{\mu\nu}+\tau_{\mu}\tau_{\nu}$, such that now
\begin{equation}
\frac{h(0,\rho_m)}{h(t,\rho_m)}=\left[\frac{\sec^2(t)\sec^2(\rho_m)}{4}(3-\cos(2t)+2\cos^2(t)\cos(2\rho_m))\right]^{n-1}\;,
\end{equation}
which leads to the norm
\begin{equation}\label{eq: normBoundMonotonicMain}
|v| = (1+ \frac{1}{4} \tan^2(t) \sec^2 (\rho_m))^{(n-1)/2},
\end{equation}
and is easily seen to satisfy the norm bound inside the WdW patch. In fact, one can see that the norm is monotonically increasing away from the maximal volume slice at $t=0$. Consequently, the Lorentzian thread vector field (\ref{eq:vecfieldads3com}) in radial compactified coordinates becomes
\begin{equation}
v^\mu=\left(\frac{2}{\cos(2t)+\cos(2\rho)}\right)^\frac{n}{2}\left(|\cos(t)|\cos^{n+1}(\rho) ,-2^{-1}\sin(2t)\sec(t)\sin(\rho)\cos^{n}(\rho),\;0\;,...,\;0 \right).
\end{equation}
When $n=3$, after some simplifications, we recover the vector field in (\ref{eq:vecfieldads3com}).

\subsection*{Thread configuration outside the WDW patch}

Above we constructed a Lorentzian thread configuration inside the WDW patch of empty AdS for any number of dimensions using radial timelike geodesics. To build these configurations we specified boundary conditions such that the flows foliate the interior of the WDW patch. Due to these boundary conditions, the timelike geodesics inside the WDW patch do not foliate the entire spacetime, leaving the outside of the patch untouched by the flow. Nevertheless, in order for the min flow-max cut theorem to apply under the conditions stated in Section \ref{sec:prelims}, we would like to foliate the entire manifold. Therefore, we proceed with the construction of a Lorentzian vector field which will be able to fill the gaps left by the previously constructed flow.  The vector field outside the WDW patch is built by twisting and translating the timelike geodesics that have been used for the inside. In this way, we are confident the curves will neither intersect with the curves inside the patch nor with each other.

\bigskip

For simplicity we work in $\text{AdS}_{3}$. Shifting the integral curves  (\ref{eq:intlinesxm2}) by $\frac{\pi}{2}$ in $t$ and $\rho$ such that we are outside of the WDW patch, upon rescaling the parameter $t$, we have
\begin{equation}
x^\mu=\left[t, \frac{\pi}{2}-\arctan\left(\frac{\cot(\rho_m)}{\sqrt{1+\csc^2(\rho_m)\cot^2(t)}} \right),\theta \right]\;,
\end{equation}
where we use $t=\frac{\pi}{2}$ as the intersection surface. The tangent to the curves are easily obtained by taking the time derivative, which we write in compactified radial coordinates by making the following substitution.
\begin{equation}
\rho_m=\arccot\left(\sqrt{\frac{-1+\cos(2\rho)}{\cos(2t)+\cos(2\rho)}} \cot(\rho)\right)\;.
\end{equation}
The resulting vector field takes the simple form of
\begin{equation}
V^\mu=\left(1,-\cot(t)\cot(\rho),0 \right)\;.
\end{equation}

Note that vector field $V^{\mu}$ is not divergenceless and therefore does not meet the criteria for a valid Lorentzian flow. Nonetheless, let us normalize $V^{\mu}$, resulting in
\begin{equation}
\tau^\mu=\frac{1}{\sqrt{\sec^2(\rho)-\cot^2(t)\csc^2(\rho)}}\left(1,-\cot(t)\cot(\rho),0 \right)\;.
\end{equation}
We can construct a divergenceless vector field by brute force by multiplying the normalised vector field by a scalar function and solve a differential equation to set the divergence to zero, and then verify \emph{ex post facto} the norm bound is satisfied. Multiplying $\tau^{\mu}$ by the function $f(t,\rho)$ and demanding $\nabla\cdot (f\tau)=0$, the solution is non-unique and is solved when $f$ is given by the following class of functions:
\begin{equation}\label{eq: functionOutsideWdW}
f(t,\rho)=\frac{\sin(t)}{2\sqrt{1-\cos^2(\rho)}}C(-2\cos(\rho)\csc(t))\;,
\end{equation}
where $C$ can be any function of $(-2\cos(\rho)\csc(t))$. We will take advantage of this arbitrariness momentarily.

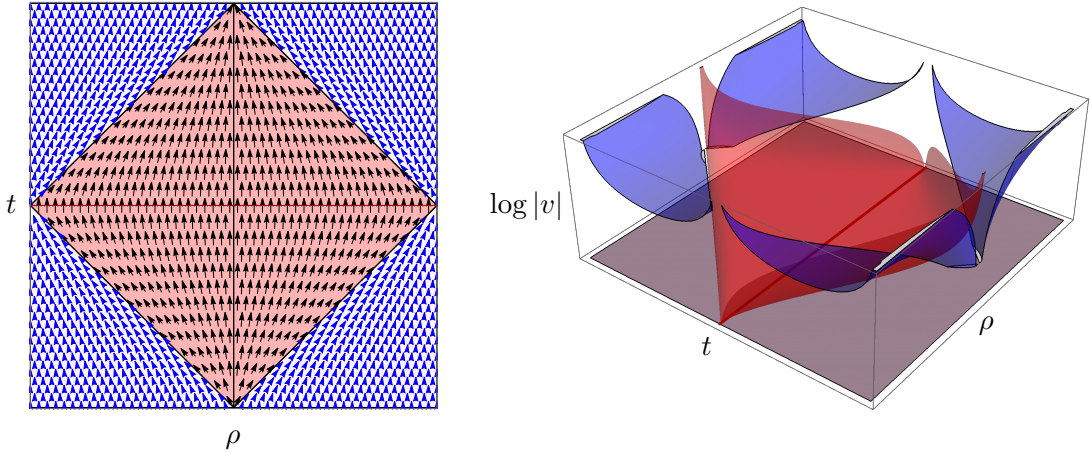
\begin{figure}[t]
\begin{center}
\includegraphics[width=0.35\textwidth]{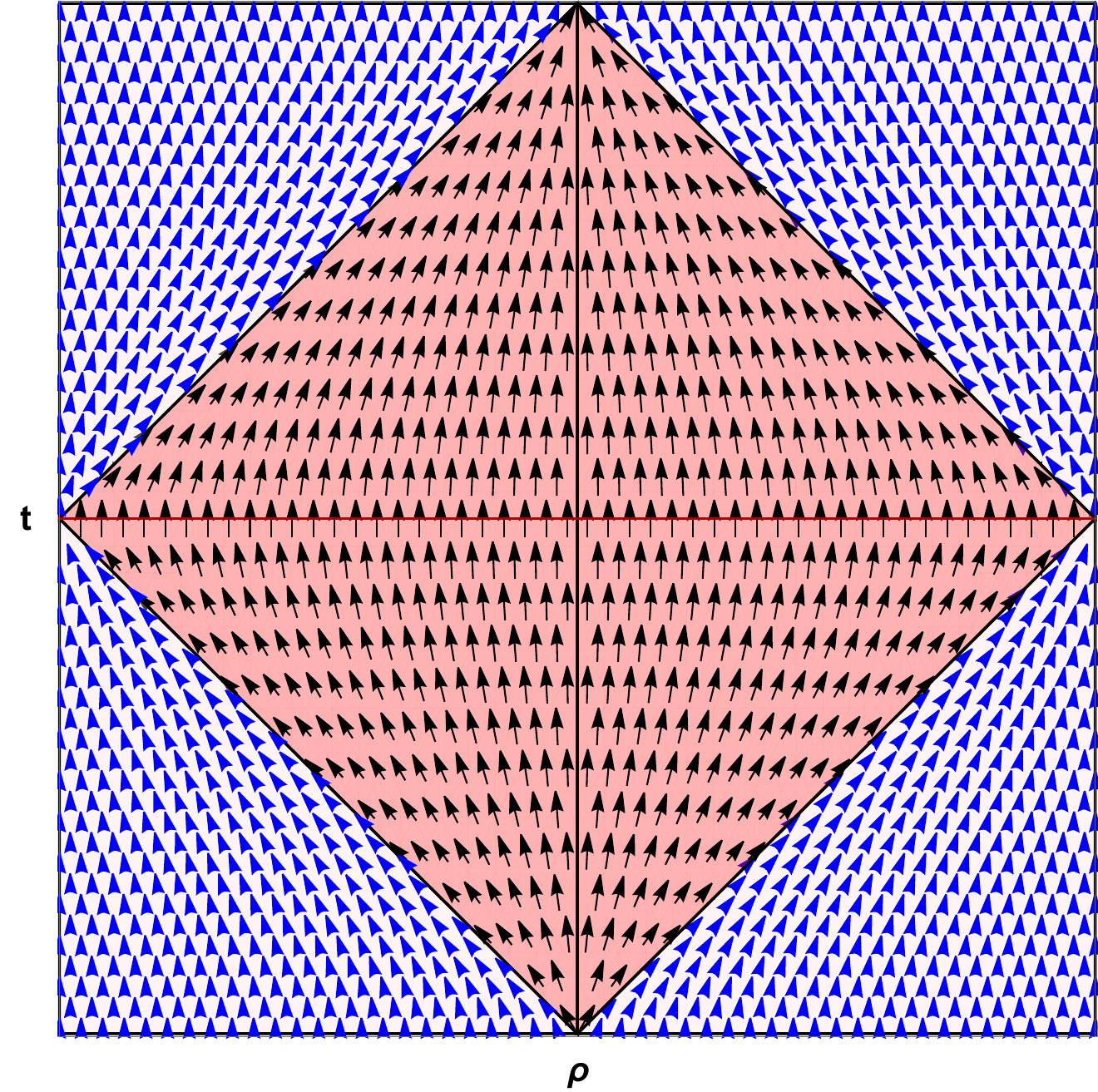}$\qquad\qquad$
\includegraphics[width=0.45\textwidth]{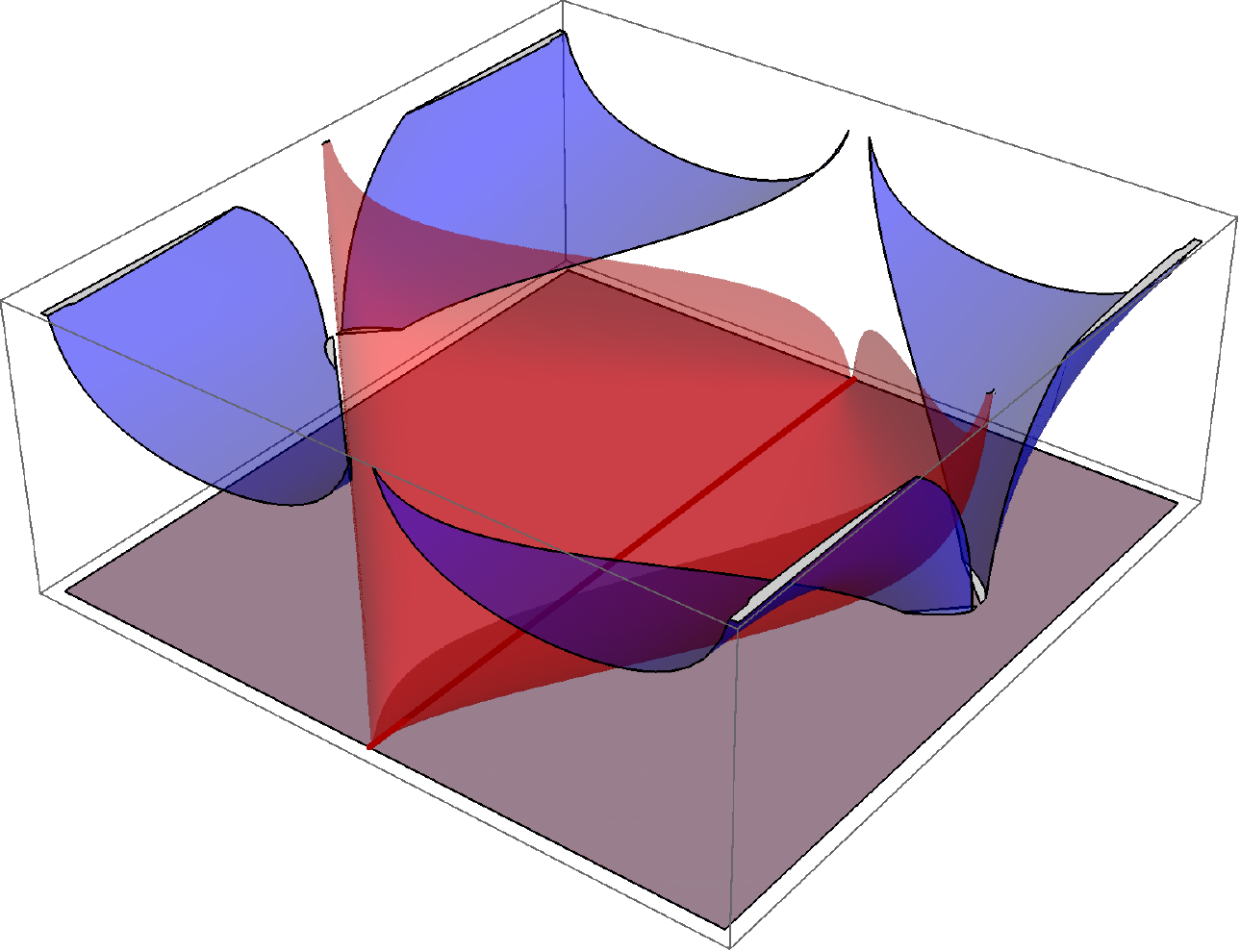}
\end{center}
 \begin{picture}(0,0)
\put(275,48){$t$}
\put(380,58){$\rho$}
\put(195,102){$\log|v|$}
\put(12,102){$t$}
\put(95,15){$\rho$}
\end{picture}
\caption{Left: Plot of the complete Lorentzian thread field in AdS spacetimes. The distinction between the inside and the outside of the WDW patch can be easily seen. The inside field has been constructed from timelike geodesics, while the outside field had to be constructed from translated geodesics. Right: Logarithmic plot of the full norm of the vector field inside and outside the WdW patch in AdS$_3$. In both plots the time coordinate $t$ goes from $-\frac{\pi}{2}$ to $\frac{\pi}{2}$, while the radial coordinate $\rho$ goes from $0$ to $\frac{\pi}{2}$. The norm satisfies $|v|\geq 1$, saturating the norm bound $|v|= 1$ on the maximal volume slice at $t=0$, which is indicated by a dark red line on the norm plot. The angular coordinate has been suppressed for all values except 0 and $\pi$, which are represented by the right and left side of the $\rho$ axes.}
\label{fig:vecfieldads3OUT}
\end{figure}

Now that the divergenceless condition has been satisfied, it remains to be checked that the norm bound, $|v|\geq1$, is satisfied. In particular, we want our choice of $f(t,\rho)$ to be at its minimum on the maximal volume slice. Given the choices made so far in the construction of the WDW patch, the Lorentzian thread will cross the maximal volume surface at ($t=0$, $\rho=\frac{\pi}{2}$). Technically, this point is on the boundary of the patch and not outside of it. Thus a sufficient condition for the total Lorentzian thread field, constructed by summing the inside and the outside vector field contributions, outlining the maximal volume surface, is for the norm outside the patch to never saturate the lower bound. This can be achieved by choosing $f(t,\rho)$ such that the Lorentzian thread field $v^\mu=f(t,\rho)\tau^\mu$ is given by:
\begin{equation}
v^\mu=\left(\frac{\sec(\rho)\sin^2(t)}{2\epsilon\sin(\rho)\sqrt{\sec^2(\rho)-\cot^2(t)\csc^2(\rho)}},-\frac{\cos(t)\csc(\rho)\sin(t)}{2\epsilon\sin(\rho)\sqrt{\sec^2(\rho)-\cot^2(t)\csc^2(\rho)}},0\right)\;.
\end{equation}
Here we have introduced a scalar number $\epsilon$ in the expression for the thread vector field using the arbitrariness of $f$. See Figure \ref{fig:vecfieldads3OUT} for an illustration.

Note that generically the norm $|v^{\mu}|$ will violate the bound as we approach the anchoring point of the WDW patch at $(t=0,\rho=\frac{\pi}{2})$. This violation of the norm bound is an artifact of competing divergences as we near this point. The role of $\epsilon$ is to constrain this violation exactly to the anchoring point such that as $\epsilon\ll1$ but finite, the violation is tuned away. In fact, the norm bound violation at the anchoring point can be ignored by either taking a cutoff at infinity, as is usually done when studying holographic complexity, or by recognizing the point lies on the boundary of the WDW patch, which lies outside the domain of the exterior Lorentzian thread configuration. Excluding this point, the norm bound in the manifold outside the patch does then behave as it should, expressed by
    \begin{equation}
        |v^\mu|=\Big |\frac{\csc^2(2\rho)\sin^4(t)}{\epsilon^2}\Big |
    \end{equation}
and is depicted in Figure \ref{fig:vecfieldads3OUT}.  Consequently,  we have a valid Lorentzian thread configuration outside of the WDW patch which complements the interior construction.


\subsection*{Generalisation to AdS$_n$}

It is straightforward to generalize the construction for the threads outside the WDW patch to $n$ dimensions. The vacuum AdS$_n$ metric is given by (\ref{eqn: AdSNmetric}). Given that the time and radial part remain invariant, the approach used to construct the field outside the WDW patch remains the same. Even the function $C(-2\cos(\rho)\csc(t))$ solves the divergenceless condition. The only change that occurs in higher dimensions is, given that $f(t,\rho)$ presents higher powers of the trigonometric functions, the choice of $C(t,\rho)$ has to be modified accordingly. To be precise, in $n$-dimensional AdS:
    \begin{equation}
        f_{(n)}(t,\rho)=\frac{\sin^{n-2}(t)}{2^{n-1}(1-\cos^2(\rho))^{\frac{n-2}{2}}}C(-2\cos(\rho)\csc(t)).
    \end{equation}
To obtain the same result as in AdS$_3$, we pick $C(t,\rho)_{(n)}=(-2\cos(\rho)\csc(t)))^{-n+2}$ and multiply by $-\frac{2^{n-2}}{\epsilon^{n-2}}$ such that:
    \begin{equation}
    v^\mu_{(n)}=\left(\frac{\sec^{(n-2)}(\rho)\sin^{(n-2)}(t)}{(2\epsilon)^{(n-2)}\sqrt{\sec^2(\rho)-\cot^2(t)\csc^2(\rho)}},
    -\frac{\cos(t)\sin^{(2n-5)}(t)\csc^{(n-1)}(\rho)\sec^{(n-3)}(\rho)}{(2\epsilon)^{(n-2)}\sqrt{\sec^2(\rho)-\cot^2(t)\csc^2(\rho)}} ,0,...,0 \right).
    \end{equation}
Consequently, the norm for $n$-dimensional thread fields can be expressed as:
    \begin{equation}
|v^\mu|_{(n)}=\Big | \frac{\csc^{2}(2\rho)\sin^{4}(t)}{\epsilon}    \Big |^{(n-2)}\;,
    \end{equation}
obeying the norm bound.


\subsubsection{Flows in the BTZ black hole}

\subsection*{Geodesic foliation of WDW patch}

Let us now explicitly construct the Lorentzian thread configuration in an eternal (double sided) $2+1$-dimensional BTZ black hole background using our general algorithm. Motivated by our construction of Lorentzian threads in empty AdS, it is natural to guess the integral lines to the timelike geodesics foliating the WDW patch are good candidate vector fields satisfying the properties of our general algorithm.  Unlike the empty AdS case, any spacelike slice, including the maximal volume slice, is anchored to the left and right boundaries at, in principle, two different times $t_{L}$ and $t_{R}$, respectively.\footnote{We will make the simplification $t_{R}=t_{L}$.} Consequently, the WDW patch is double sided and we will look for a geodesic foliation of the interior of the patch before using a level construction to describe the thread configuration outside of the patch. We also emphasize studying Lorentzian threads in a  BTZ black hole background is of interest as the BTZ black hole has been an illuminating case study in holographic complexity. This is because, in the context of AdS/CFT, the BTZ black hole is dual to a thermofield double (TFD) state \cite{Maldacena:2001kr}, and the TFD state is one of the few examples where the boundary complexity has been analyzed in detail \cite{Chapman:2018hou}. Holographically, the growth of the boundary complexity due to the time evolution of the TFD state maps to the time dependent growth of the spatial volume of the black hole interior \cite{Stanford:2014jda,Carmi:2017jqz}.

The metric of the static, neutral BTZ black hole \cite{Banados:1992wn,Banados:1992gq} in Schwarzschild coordinates takes the form
\begin{equation}
ds^2=-\left(\frac{r^2}{L^2}-M\right)dt^2+\left(\frac{r^2}{L^2}-M\right)^{-1}dr^2+r^2d\phi^2\;,
\label{eq:staticBTZ}\end{equation}
where $L$ is the AdS length, $M$ is its ADM mass, and $r=L\sqrt{M}$ is the location of the horizon. It is prudent for us to work in the maximal Kruskal extension of the BTZ solution since we want the integral curves to the Lorentzian threads to be globally defined, especially when crossing the event horizon. Working in Kruskal coordinates is contrary to what is typically done in the literature, where one typically uses Eddington-Finkelstein coordinates to model shockwaves \cite{Stanford:2014jda}.  In null Kruskal coordinates $(u,v)$, which are related to
 Schwartzchild $(t,r)$ coordinates by,
\begin{equation}
u=-\sqrt{\frac{r-L\sqrt{M}}{r+L\sqrt{M}}}e^{-t}\;,\quad v=\sqrt{\frac{r-L\sqrt{M}}{r+L\sqrt{M}}}e^{t}\;,
\label{eqn: kruscaltrans}
\end{equation}
the Schwarzschild line element becomes
\begin{equation}
ds^2=-\frac{4dudv}{(1+uv)^2}+\frac{(1-uv)^2}{(1+uv)^2}d\phi^2\;.
\label{eq:BTZkruskal}
\end{equation}
The horizon is now positioned at $u=v=0$, the $\text{AdS}_{3}$ boundary is at $uv=-1$, and past and future singularities are located at $uv=1$.

\begin{figure}[t]	
	\begin{center}
\includegraphics[width=4cm, angle=45,trim=1.5in 1.5in 1.5in 1.5in]{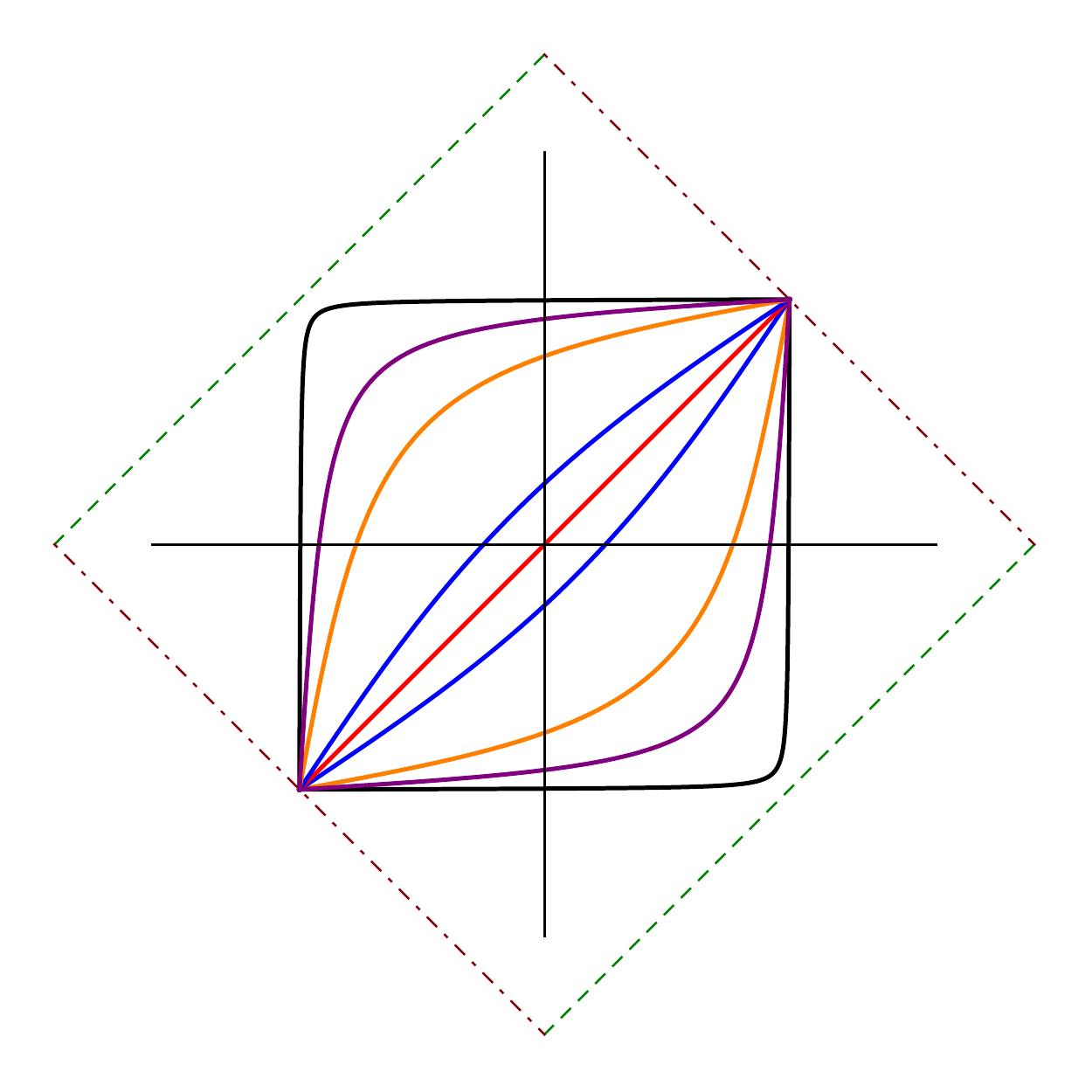}
	\end{center}
 \begin{picture}(0,0)
\put(138,184){$\rho$}
\put(298,184){$\sigma$}
\put(185,105){$\textbf{III}$}
\put(216,120){$\textbf{II}$}
\put(245,105){$\textbf{I}$}
\put(215,85){$\textbf{IV}$}
\end{picture}
\caption{Plot of the compactified BTZ timelike geodesics inside the WDW patch for different crossing points at the maximal volume $t=0$ slice. As the crossing point approaches $L=1$, the timelike geodescs tend to the null geodesics which form the boundary of the causal diamond. The crossing points values are color coordinated: $|s_{m}|=0$ (red), $|s_{m}|=0.1$ (blue), $|s_{m}|=0.4$ (orange), $|s_{m}|=0.6$ (violet), and $|s_{m}|=0.9$ (black).}
\label{fig:compactbtz}
\end{figure}
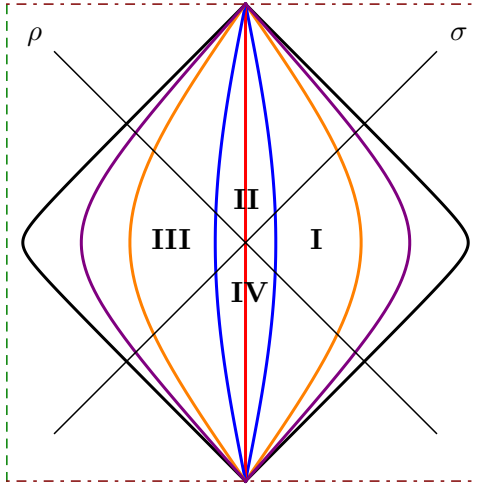

 Since the geodesic foliation of the WDW patch in the BTZ black hole is more involved than in empty AdS, let us describe the geodesic analysis in some detail. Ingoing null radial geodesics are described by
\begin{equation}
\frac{d^2v}{d\lambda^2}-\frac{2u}{1+uv}\left(\frac{dv}{d\lambda}\right)^2=0\;,
\end{equation}
where $\lambda$ is an affine parameter. For geodesics crossing the  horizon in quadrant \textbf{I}, the null condition imposes $u$ to take a constant value, $u=u_0\geq 0$. Consequently, null geodesics trace straight lines to the future singularity. Similarly, infalling radial null geodesics crossing the horizon from quadrant \textbf{III}  tend towards the future singularity and have constant $v=v_0$. Solving the geodesics equation and imposing the conditions $v(\lambda\rightarrow -\infty)=-\frac{1}{u_0}$ and $v(0)=\frac{1}{u_0}$, the two null rays identify two boundaries of the WDW patch anchored at $t_L=t_R=0$. Analogously, we identify the remaining sides of the WDW patch by sending null geodesics toward the past singularity. See Figure \ref{fig:compactbtz} for an illustration.

 We now construct the interior of the WDW patch by demonstrating every point in the interior volume belongs to a timelike geodesic orthogonal to the (Schwarzschild) time $t=0$ slice. The timelike geodesic equation and its timelike condition form the following system of ordinary differential equations
\begin{equation}
\frac{d^2v}{d\tau^2}-\frac{2u}{1+uv}\left(\frac{dv}{d\tau}\right)^2=0\;,\quad -\frac{4}{(1+uv)^2}\dot{u}\dot{v}=-1\;,
\end{equation}
where $\dot{u}=\frac{du}{d\tau}$ and $\dot{v}=\frac{dv}{d\tau}$, with $\tau$ being the proper time parametrizing the geodesic. We find an analytical solution to the set of equation by first extracting from the geodesic equation an expression for $u(\tau)$ in terms of $v(\tau)$ and its derivatives, and then substituting it into the timelike condition such that we arrive at the following third order ODE
\begin{equation}
\frac{d^3v}{d\tau^3}-\frac{1}{2}v'(\tau)-\frac{3}{2}\frac{(v''(\tau))^2}{v'(\tau)}=0\;.
\label{eq:thirdorderodebtz}\end{equation}
Using the initial conditions $v(0)=-v_0$, $u(0)=u_0$, which sets the initial point to lie in quadrant \textbf{I}, and $\frac{v'(\tau)}{u'(\tau)}\Big|_{\tau=0}=\frac{dv}{du}\Big|_{\tau=0}=1$, such that the curves are normal to the $t=0$ slice, the solution to (\ref{eq:thirdorderodebtz}) is
\beq
\begin{split}
&v(\tau)=\frac{(1-s_{m}^2)\sec\left( \frac{\tau}{2} +\arctan(s_{m})\right)\sin\left( \frac{\tau}{2} \right)}{\sqrt{1+s_{m}^2}}-s_{m}\;,\quad u(\tau)=\frac{s_{m}\cos\left( \frac{\tau}{2} \right)+\sin\left( \frac{\tau}{2} \right)}{\cos\left( \frac{\tau}{2} \right)+s_{m}\sin\left( \frac{\tau}{2} \right)}\;.
\end{split}
\label{eq:uvtaubtz}\eeq
Here, moreover, we fixed  $u_0=v_0=s_{m} \geq 0$ such that the geodesics at $\tau=0$ lie on the $t=0$ slice and have coordinates $(u,v)|_{\tau=0}=(s_{m},-s_{m})$. The point $s_{m}$ is analogous to the intersection point $r_{m}$ in empty AdS, which characterize geodesics of different `energy' $E$. As we will see momentarily, we will treat $(\tau, s_{m})$ as adapted coordinates describing the geodesics.

Inverting the relations (\ref{eq:uvtaubtz}) and using $uv=1$, we find the proper time $\tau_{sing}$ at which the timelike geodesics reach the past and future singularities is $\tau_{sing}=\pm\frac{\pi}{2}$.\footnote{More explicitly, $uv=1$ yields $\frac{2-2s_{m}^2}{1-s_{m}^2+\cos(\tau)+s_{m}^2\cos(\tau)}-1=1$, which implies $\tau_{sing}=\pm\frac{\pi}{2}$.} Thus, all geodesics meet at the past and future singularities at the same proper time at which they cross with each other, independent of their starting position along the $t=0$ surface.

It is straightforward to repeat the above analysis with timelike geodesics in quadrant \textbf{III}, which cross the horizon from the other side of the wormhole and meet at the black hole singularities. The geodesic equations are of the same form, except now $s_{m}\leq 0$. Considering all geodesics coming from both quadrants \textbf{I} and \textbf{III}, it follows we can construct the interior of the WDW patch, provided the timelike geodesics foliate the spacetime inside the boundary of the patch identified by the null geodesics, as we will now verify.

Proving the timelike geodesics foliate the interior of the WDW patch is most easily visualized in compactified coordinates
\begin{equation}
u=\tan(\sigma), \quad v=\tan(\rho),
\end{equation}
with $-\frac{\pi}{2}\leq \sigma,\rho\leq \frac{\pi}{2}$, such that the Kruskal metric (\ref{eq:BTZkruskal}) transforms to
\begin{equation}
ds^2=-4\sec^2(\rho-\sigma)d\sigma d\rho+\cos^2(\rho+\sigma)\sec^2(\rho-\sigma)d\phi^2.
\label{eq:compactbtz}\end{equation}
where we have set $L=M=1$ for convenience.  In $(\sigma,\rho)$ coordinates the BTZ spacetime boundaries at the singularities and spatial infinity limit the range of the coordinates to $|\sigma|+|\rho|\leq\frac{\pi}{2}$. Therefore, the WDW patch anchored at $t_L=t_R=0$ is now identified with $-\frac{\pi}{4}\leq \sigma,\rho\leq \frac{\pi}{4}$.

Let us now show the timelike geodesics foliate the interior of the WDW patch. The argument is nearly identical to the empty AdS case.  Let $p$ be any point inside the patch that does not lie on one of the null rays, and has (compactified) Kruskal coordinates $(\sigma_p,\rho_p)$. The WDW patch is considered foliated by the geodesics if, for every point inside the patch, there exists a unique geodesic which passes through the point. Given that timelike geodesics in the patch are identified by their crossing point $s_{m}$, we can invert the geodesic equation to find
	\begin{equation}
s_{m}=\sqrt{\cos(2\rho_p)\cos(2\sigma_p)}-\cos(\rho_p+\sigma_p)\csc(\rho_p-\sigma_p)\;.
	\end{equation}
Since the coordinates inside the WDW patch are bounded from above and below ($-\frac{\pi}{4} < \sigma,\rho < \frac{\pi}{4}$), this identifies a unique geodesic to which the point belongs. Moreover, we can discern the proper time $\tau_{p}$ the curve intersects the crossing point by explicitly inverting the equation to express $\tau$ as a function of the point coordinates.\footnote{Specifically,
$$
\tau_{p}=\frac{ \sqrt{\cos(2\rho_p)}\cos(\sigma_p)-\cos(\rho_p)\sqrt{\cos(2\sigma_p)} }{ \sqrt{\cos(2\rho_p)}\sin(\sigma_p)-\sin(\rho_p)\sqrt{\cos(2\sigma_p)}}\;.
$$}
Therefore, each point inside the WDW patch belongs to a unique timelike geodesic, proving that the patch is foliated by this congruence of integral curves.


\subsubsection*{Timelike geodesics as Lorentzian threads}

Let us move forward and use the integral lines to the timelike geodesics as our proposed Lorentzian thread configuration.  Indeed, the curves $x^{\mu}(\tau)=\{v(\tau),u(\tau)\}$ are continuous and non-intersecting (except at the singularities), such that the tangent to the curves $\tau^{\mu}=\frac{dx^{\mu}}{d\tau}$ is well-defined everywhere and is equal to the unit normal $\hat{n}$ to the maximal volume slice at $t=0$.  We will use the continuity equation (\ref{eq:conteq}) to further construct the vector field $v^{\mu}$ which is divergenceless and has norm bounded from below.

As in empty AdS,  first we express the geodesics in adapted coordinates $(\tau,s_{m})$,
\begin{align}
&\sigma(\tau,s_{m})=\arctan\left[ \frac{s_{m}\cos\left( \frac{\tau}{2} \right)+\sin\left( \frac{\tau}{2} \right)}{\cos\left( \frac{\tau}{2} \right)+s_{m}\sin\left( \frac{\tau}{2} \right)}  \right]\;,   \\
&\rho(\tau,s_{m})=\arctan\left[ \frac{(1-s_{m}^2)\sec\left( \frac{\tau}{2} +\arctan(s_{m})\right)\sin\left( \frac{\tau}{2} \right)}{\sqrt{1+s_{m}^2}}-s_{m}\right]\;,
\end{align}
where the metric in compactified coordinates (\ref{eq:compactbtz}) becomes
\begin{equation}
ds^2=-d\tau^2 + \frac{4\cos^2(\tau)}{(1-s_{m}^2)^2}ds_{m}^2+ \frac{(1+s_{m}^2)^2\cos^2(\tau)}{(1-s_{m}^2)^2}d\phi^2\;.
\end{equation}

In these adapted coordinates, the vector tangent to the integral curves is
\begin{equation}
\tau^\mu=\left(\frac{1}{\sqrt{-g_{\tau\tau}}},0,0\right)=(1,0,0)\;,
\end{equation}
such that the induced metric on surfaces of constant $\tau$, $h_{\mu\nu}=g_{\mu\nu}+\tau_{\mu}\tau_{\nu}$, is
\begin{equation}
ds^2=\frac{4\cos^2(\tau)}{(1-s_{m}^2)^2}ds_{m}^2+ \frac{(1+s_{m}^2)^2\cos^2(\tau)}{(1-s_{m}^2)^2}d\phi^2.
\end{equation}
Knowing that the point of intersection is at $\tau=0$, we can construct the Lorentzian threads following (\ref{eq:conteq})
\begin{equation}
v^\mu=\sqrt{\frac{h(0,s_m)}{h(\tau,s_m)}}\tau^\mu=(\sec^2(\tau),0,0)\;.
\end{equation}
It is straightforward to verify that $v^{\mu}$ is divergenceless and has norm bound satisfying $|v|\geq1$. To better visualise the Lorentzian threads, we transform from adapted coordinates back into compactified Kruskal coordinates, leading to
\begin{equation}
v^\mu=\frac{\cos(\rho-\sigma)}{2}\sec^2\left[2\arctan\left( \frac{ \sqrt{\cos(2\rho)}\cos(\sigma)-\cos(\rho)\sqrt{\cos(2\sigma)} }{ \sqrt{\cos(2\rho)}\sin(\sigma)-\sin(\rho)\sqrt{\cos(2\sigma)}}\right) \right]\left(\sqrt{\frac{\cos(2\sigma)}{\cos(2\rho)}},\sqrt{\frac{\cos(2\rho)}{\cos(2\sigma)}},0  \right).
\end{equation}
The norm is simply
	\begin{equation}
|v^\mu|=\sec^4\left[2\arctan\left( \frac{ \sqrt{\cos(2\rho)}\cos(\sigma)-\cos(\rho)\sqrt{\cos(2\sigma)} }{ \sqrt{\cos(2\rho)}\sin(\sigma)-\sin(\rho)\sqrt{\cos(2\sigma)}}\right) \right],
	\end{equation}
where we see the minimum coincides at the maximal volume slice. An illustration of the Lorentzian threads and the norm is given in Figure \ref{fig:BTZCOMPLETE}.

\subsection*{Thread configuration outside the WDW patch}

Now that we have constructed the Lorentzian threads inside the WDW patch of BTZ, let us construct the field outside. As we explained when constructing the threads outside the WDW patch of vacuum AdS, there is no naturally occurring set of integral curves that complements the future directed timelike geodesics -- we simply translated and twisted the flows foliating the interior. Therefore, as before, we manually construct the integral curves by translating the interior geodesics appropriately. Given the symmetry of the BTZ Penrose diagram, we need to fill four corners which form the complement of the WDW patch. Luckily, it is enough for us to translate the geodesics in compactified Kruskal coordinates by $\pm \frac{\pi}{2}$, namely, $\sigma \pm \frac{\pi}{2}$ and $\rho \pm \frac{\pi}{2}$. The resulting Lorentzian vector field will then naturally fill the remaining corners appropriately.

We start by translating the $\rho$ coordinate to $\rho-\frac{\pi}{2}$, keeping $\sigma$ fixed,\footnote{Note that given the symmetry of the tangent function $v=\tan(\rho)$ from which $\rho$ is defined, we realize the Lorentzian threads obtained from the translation $\rho-\frac{\pi}{2}$ are exactly the same as those obtained from $\rho+\frac{\pi}{2}$. Therefore, we will fill two corners outside the WDW patch in one go.} such that the compactified null coordinates in terms of $\tau$ and $s_m$ become
    \begin{align}
&\sigma(\tau,s_m)=\arctan\left[\frac{s_m \cos\left(\frac{\tau}{2}\right) + \sin\left(\frac{\tau}{2}\right)}{\cos\left(\frac{\tau}{2}\right) + s_m \sin\left(\frac{\tau}{2}\right)} \right],\\
&\rho(\tau,s_m)=\arctan\left[\frac{(1 - s_m^2) \sec\left(\frac{\tau}{2}+ \arctan(s_m)\right) \sin\left(\frac{\tau}{2}\right)}{\sqrt{1 + s_m^2}} \right]-\frac{\pi}{2}.
    \end{align}
We then pullback the metric into adapted coordinates
    \begin{equation}
ds^2=-\left(\frac{e^{i\tau}(1-s_m^2)}{s_m(1+e^{2i\tau})}\right)^2 d\tau^2+\frac{ds_m^2}{s_m^2} + \frac{(1-s_m^2)^2\tan^2(\tau)}{4s_m^2}d\theta^2.
    \end{equation}
The unit vector along the curves is then easily read off from the metric,
    \begin{equation}
\tau^\mu=\left(\frac{1}{\sqrt{-g_{\tau\tau}}},0,0\right),
    \end{equation}
and used to induce it on surfaces of constant $\tau$;
    \begin{equation}
ds^2=\frac{ds_m^2}{s_m^2} + \frac{(1-s_m^2)^2\tan^2(\tau)}{4s_m^2}d\theta^2.
    \end{equation}

Following the construction inside the WDW patch, we would like to construct the Lorentzian threads
    \begin{equation}
        v^\mu=\sqrt{\frac{h(\tau_0,s_m)}{h(\tau,s_m)}}\tau^\mu=\left(\frac{2s_m\cos(\tau)\cot(\tau)\tan(\tau_0)}{1-s_m^2},0,0 \right),
    \end{equation}
where we have left $\tau_0$ unspecified for the moment. Even with arbitrary $\tau_{0}$, $v$ is divergenceless. One can also easily  transform back into compactified coodinates $(\rho,\sigma)$, though the resulting expressions are not terribly enlightening. While $v$ remains divergenceless,  the norm bound  is not identically satisfied and depends on the choice of $\tau_0$. Ideally, we would like to set $\tau_0=-\frac{\pi}{2}$. Unfortunately, this is not possible since this is where the singularity lies, and the vector field diverges at that point. Nevertheless, we are allowed to get arbitrarily close by setting $\tau_0=-\frac{\pi}{2}+\epsilon$. For arbitrary $\epsilon$ the norm bound will be violated, however as  $\epsilon\rightarrow 0$, the violation gets pushed to the edge of the WDW patch at spacelike infinity. Then, as in vacuum AdS, we again use the cutoff argument to eliminate the norm violation.

\begin{figure}[t]
\begin{center}
\includegraphics[width=0.35\textwidth]{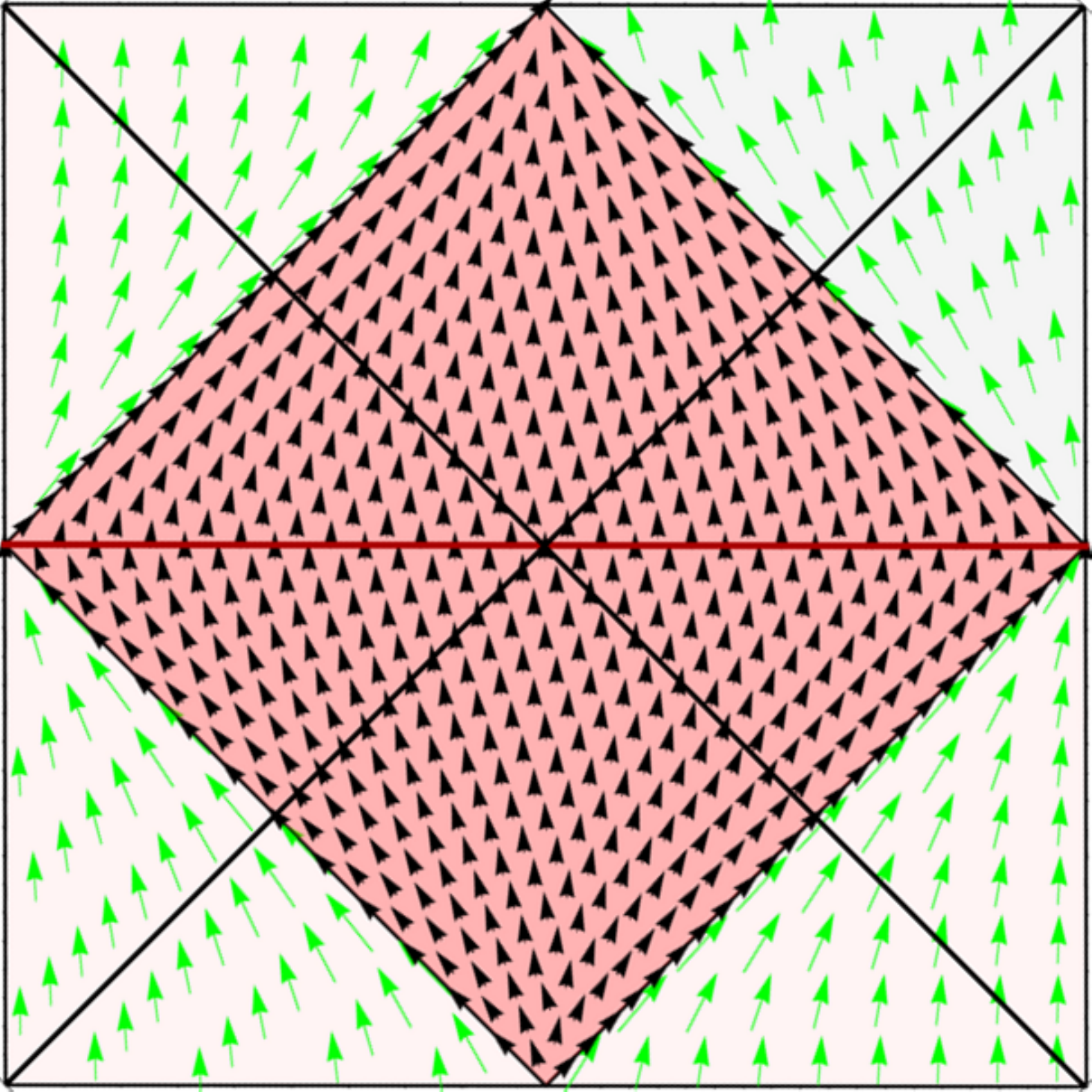}$\qquad\qquad$
\includegraphics[width=0.45\textwidth]{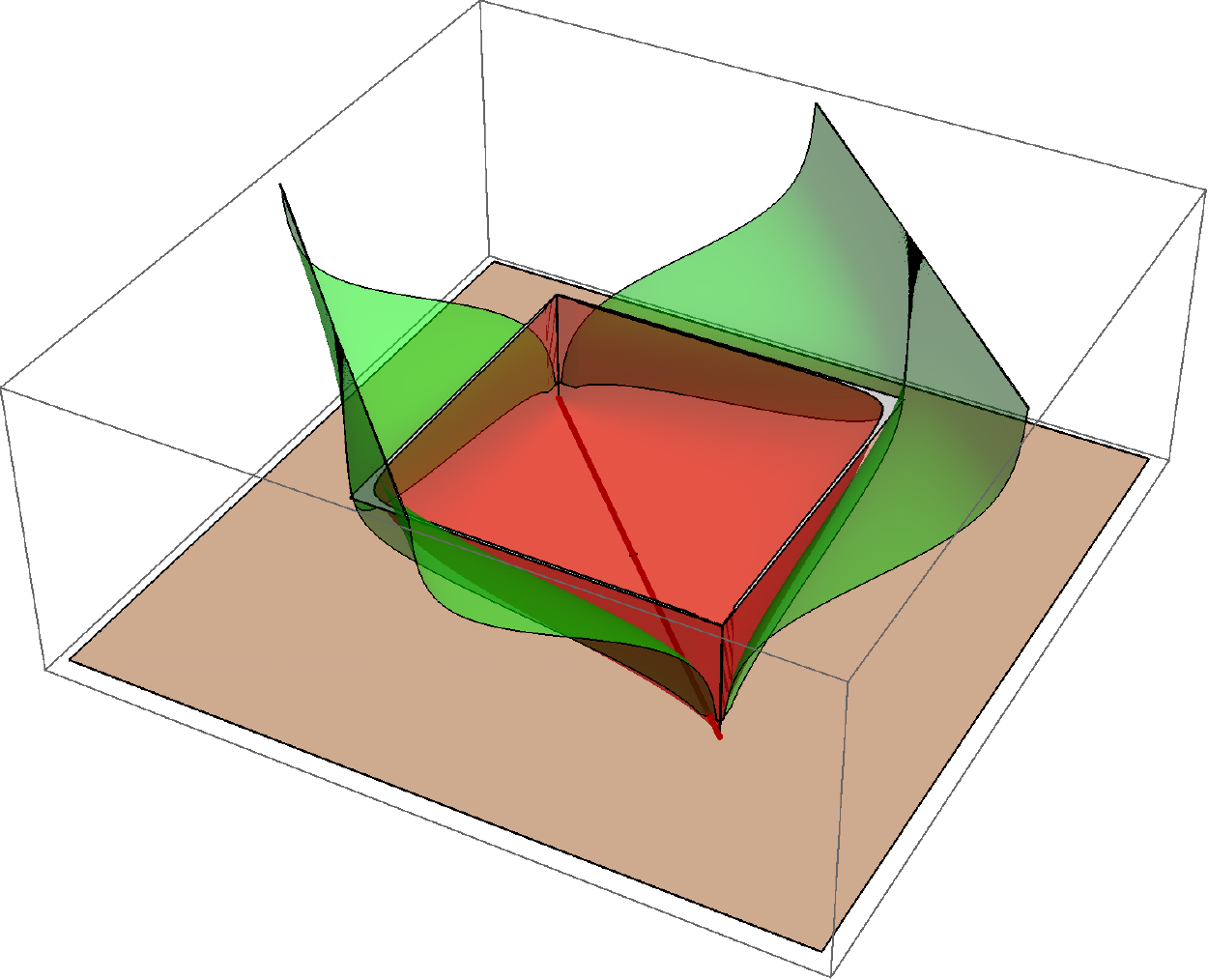}
\end{center}
 \begin{picture}(0,0)
\put(285,46){$\sigma$}
\put(390,67){$\rho$}
\put(195,102){$\log|v|$}
\put(33,173){$\rho$}
\put(159,173){$\sigma$}
\end{picture}
\caption{Left: Plot of the Lorentzian flows for the full BTZ spacetime in compactified Kruskal coordinates. The coordinates are bounded by $\pm \frac{\pi}{4}$. The maximal volume slice runs from the left to the right corner. The plot is divided in the 4 quadrants representing two asymptotically AdS spacetimes connected by a wormhole. For this choice of WDW patch, the past and future tips are located precisely on the past and future singularities. Right: Norm of the vector field for the full BTZ in compactified Kruskal coordinates. The maximal volume slice runs from the left to the right corner and coincides with the minimum value of the norm, indicated by a darker red line.}
\label{fig:BTZCOMPLETE}
\end{figure}
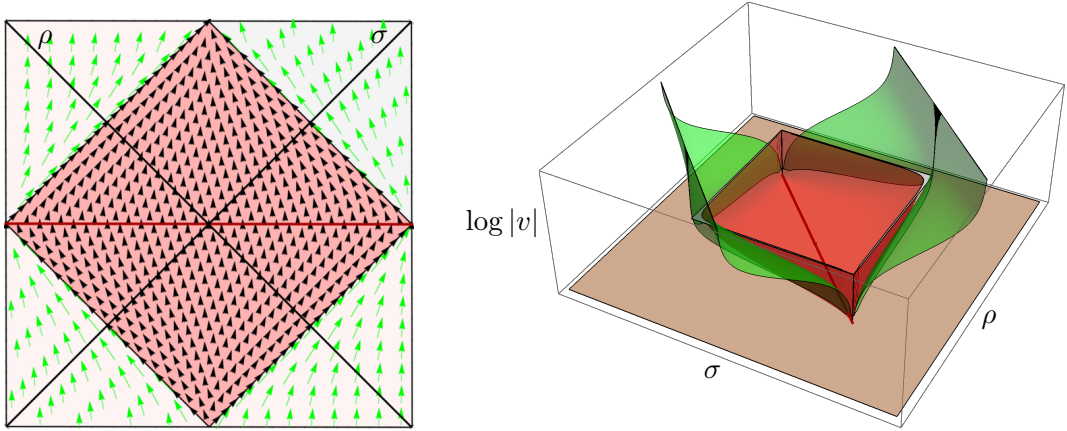

The remaining two corners outside of the WDW patch are built in exactly the same way, where now we keep $\rho$ fixed but translate $\sigma\to\sigma-\frac{\pi}{2}$. The adapted and induced metric are the same, and the resulting thread vector field is
    \begin{equation}
v^\mu=\left(-\frac{1}{2}\cos(\rho-\sigma)\cos(2\sigma)\cot(\epsilon)\sec(\rho+\sigma),\frac{1}{2}\cos(\rho-\sigma)\cos(2\rho)\cot(\epsilon)\sec(\rho+\sigma),0\right)
    \end{equation}
with norm
    \begin{equation}
|v^\mu|=\cos(2\sigma)\cos(2\rho)\cot^2(\epsilon)\sec^2(\rho+\sigma)
    \end{equation}
A complete plot of the threads and their norm can be seen in Figure \ref{fig:BTZCOMPLETE}.

\subsubsection*{Late time behavior and second law of complexity}

In the previous section, we showed how Lorentzian threads explicitly identify the maximal volume slice passing through the bifurcation surface,  such that the WDW patch is anchored at the boundary time $t_L=t_R=0$. As the boundary time moves forward, the portion of the maximal volume surface trapped in the interior of the black hole grows. This represents the growth of the volume of the wormhole connecting the asymptotically AdS spacetimes and it is holographically dual to the growth in complexity of the boundary TFD state \cite{Stanford:2014jda}. As time continues forward, the maximal volume surface tends to a surface of constant radius,  entirely confined in the interior of the black hole, and begins to wrap around the future singularity. This slice is referred to as the \emph{late time surface} and describes the complexity of the boundary state as $t\rightarrow\infty$. Eventually the complexity saturates and plateaus at some value until the recurrence time, which is doubly exponential in the entropy \cite{Brown:2017jil}.

Here we wish to study the late time behavior of the geodesic construction of the Lorentzian thread configurations. The construction which we will present holds for any of the surfaces at $t>0$, but we focus on the late time surface both for simplicity, as we know an explicit expression for the slice, and due to the greater importance that this slice has. The strategy is as follows. First we consider geodesics emanating from the maximal volume surface at late time, where we see the late time surface corresponds to the WDW patch anchored at infinity. We then adapt the algorithm used above to construct Lorentzian flows, such that even at late times the norm bound $|v|\geq1$ is saturated at the late time surface. Moreover, as we will comment on briefly, the Lorentzian flows nicely illustrate the second law of complexity. The late time surface for a BTZ black hole is given by the constant $r$-surface
    \begin{align}
        &r=\frac{1-\tan(\sigma)\tan(\rho)}{1+\tan(\sigma)\tan(\rho)}=\frac{1}{\sqrt{2}}\;,
    \end{align}
where we expressed the radius in terms of Kruskal coordinates using (\ref{eqn: kruscaltrans}). The radius value can be obtained through a method analogous to \cite{Estabrook:1973ue}, where we treat the maximal volume surface as a lower dimensional geodesic problem, extremizing the functional in Eddington-Finkelstein coordinates
    \begin{equation}
        \mathcal{L}=\sqrt{r^2(-f(r)\dot{v}^2+2\dot{v}\dot{r})}
    \end{equation}
where $v$ is the ingoing null coordinate.
The result has been computed numerically in \cite{Zhang:2019pzd}. Imposing the timelike condition and orthogonality of the radial geodesics to the surface, it is straightforward to show the geodesics $x^{\mu}=\{x^{\sigma},x^{\rho},x^{\theta}\}$  emanating from the late time surface in compactified coordinates are
    \begin{align}
        &x^\sigma=\arccot\left[\cot(s_m)\csc\left(\frac{\pi}{8}+\frac{\tau}{2}\right)\left(\cos\left(\frac{\tau}{2}\right)\sin\left(\frac{\pi}{8}\right)+(2\sqrt{2}-3)\sin\left(\frac{\tau}{2}\right)\cos\left(\frac{\pi}{8}\right) \right) \right],\\
        &x^\rho= -\arctan\left[\cot(s_m)\left(-1+\sqrt{2}+(-2+\sqrt{2})\cos\left(\frac{\pi}{8}-\frac{\tau}{2}\right)\sec\left(\frac{\pi}{8}+\frac{\tau}{2}\right)\right)\right],\\
        &x^\theta=\theta\;,
    \end{align}
where $\tau$ is the proper time along the geodesics, with $\tau=0$ indicating the time at which geodesics intersect the maximal volume surface, and $s_m$ being the parameter indicating the point of intersection with the maximal volume surface.

     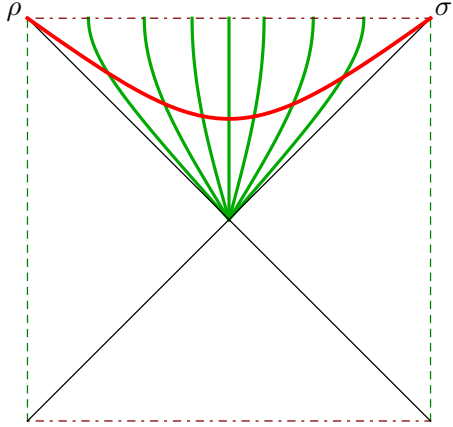
\begin{figure}[t]	
	\begin{center}
\includegraphics[width=4cm, angle=45, trim=2cm 2cm 2cm 2cm]{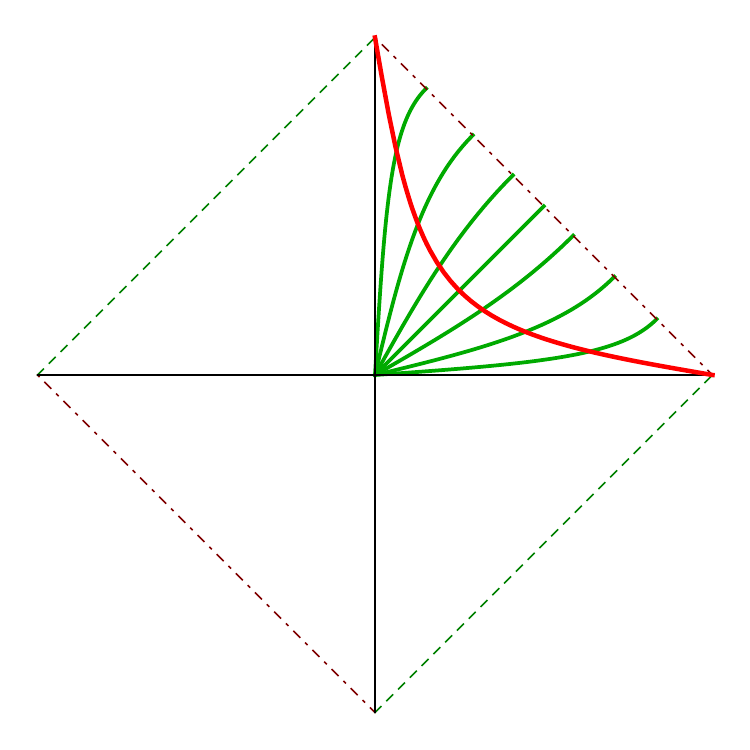}
	\end{center}
 \begin{picture}(0,0)
\put(137,185){$\rho$}
\put(299,185){$\sigma$}
\end{picture}
\vspace{-1cm}
\caption{The late time maximal volume slice (in red), and the timelike geodesics passing through. The timelike geodesics meet at the past tip of the late time WDW patch and end on the singularity.}
\label{fig:latetimegeod}
\end{figure}

A quick computation shows all of the geodesics meet the singularity at $\tau_{sing}=\frac{\pi}{4}$ and meet at the bifurcation surface at $\tau_{b}=-\frac{\pi}{4}$ as shown in Figure \ref{fig:latetimegeod}. Moreover, as seen from the figure, geodesics meet at the bifurcation surface. Incidentally, this intersection point is where the past tip of the WDW patch lies, the closest point to the past singularity.

This allows for the null rays emitted from the tip to reach the boundaries without ending on the singularity. Indeed, the construction of Lorentzian threads which we are about to present goes through smoothly if one starts from the past tip of the WDW patch and follows all possible future time directed geodesics. If the past tip is then positioned at the furthest possible future, the geodesics coincide with those showed here. This is how we originally individuated the maximal volume surface to lie at $r=\frac{1}{\sqrt{2}}$.

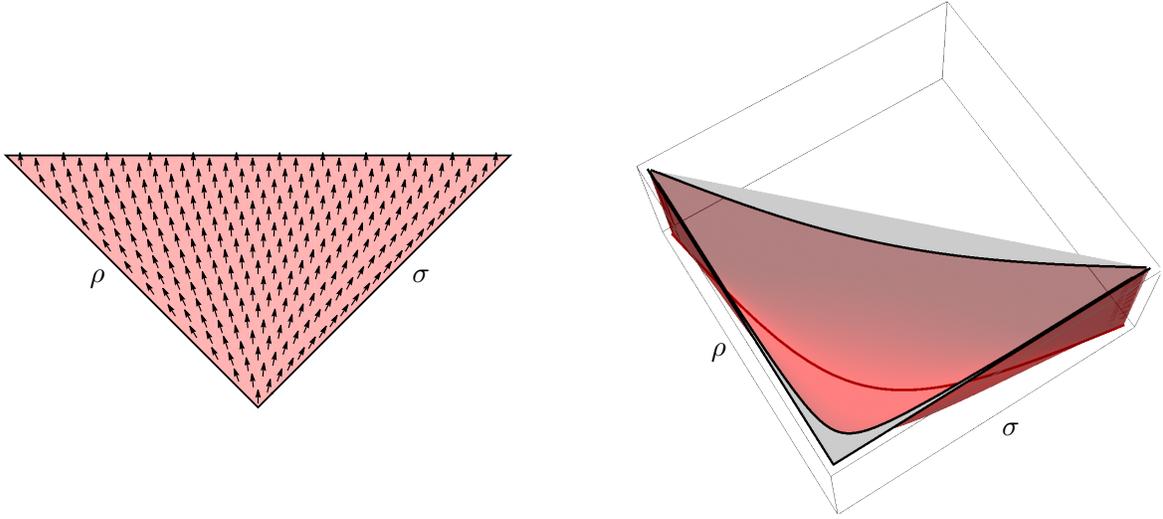
\begin{figure}[t]
\begin{center}
	   $\qquad\qquad\qquad$\includegraphics[angle=45,width=0.45\textwidth,trim=-0.5in -0.5in 0.5in 0.5in]{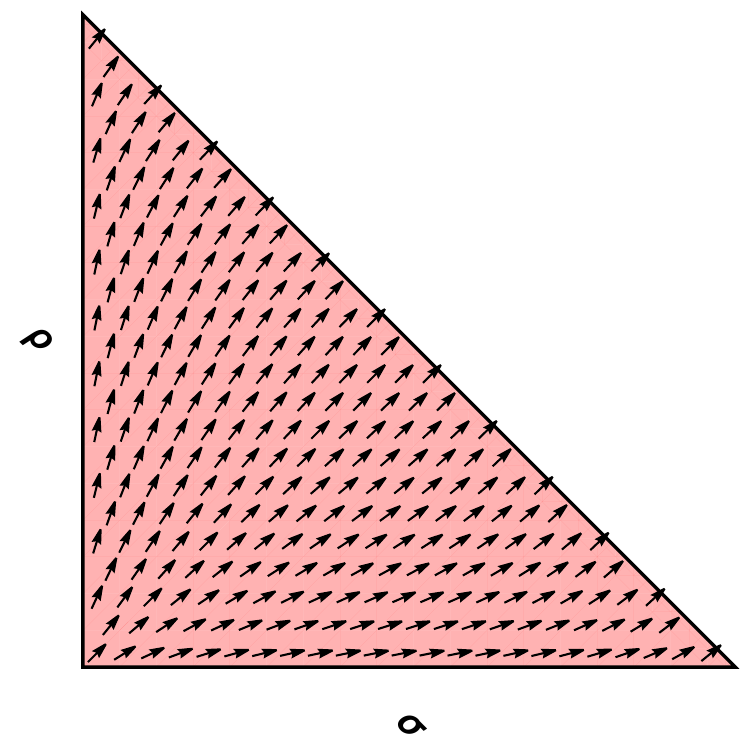}$\qquad\qquad$\includegraphics[width=0.45\textwidth]{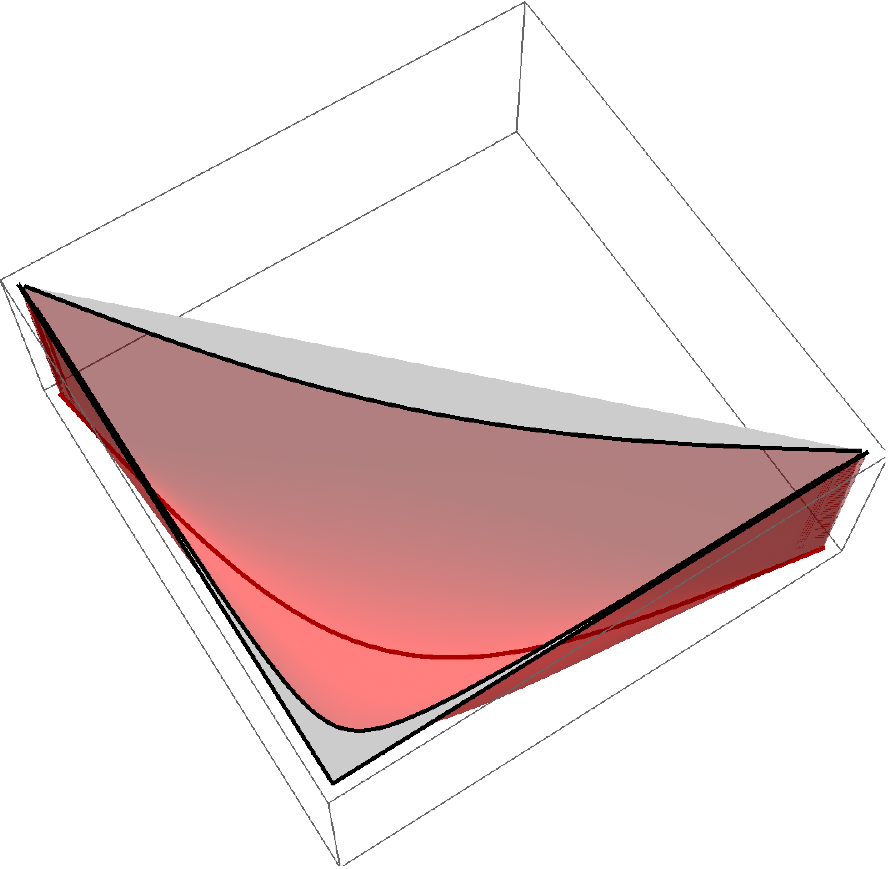}
 \begin{picture}(0,0)
\put(-185,103){$\rho$}
\put(-63,103){$\sigma$}
\put(50,75){$\rho$}
\put(160,45){$\sigma$}
\end{picture}
\caption{Lorentzian thread vector field at late times and its norm. The norm satisfies $|v|\geq1$, saturating the lower bound on the maximal slice (thick line in red).}\label{fig:latetime}	
\end{center}
\end{figure}

Following the general algorithm, we now find the vector tangent to the integral curves. Since the tangent vector field to these curves is not divergenceless, we opt for the Gaussian method to construct the thread configurations. Pulling back the metric to coordinates adapted to the geodesics, we obtain a diagonal metric where $\tau$ and $s_m$ are the coordinates parallel and perpendicular to the geodesics. While the metric itself is quite cumbersome to write, the unit vector field along the curves dramatically simplifies to a unit vector orthogonal to surfaces of constant $\tau$
    \begin{equation}
        \tau^\mu=\left(\frac{1}{\sqrt{-g_{\tau\tau}}},0,0\right)=(1,0,0).
    \end{equation}
Rather nicely,  the induced metric on surfaces of constant $\tau$ leads to a remarkably simple expression for the metric determinant:
    \begin{equation}
        h(\tau,s_m)=\frac{1}{4}\cos^2(2\tau)\csc^2(s_m)\sec^2(s_m),
    \end{equation}
from which we construct the vector field  $v$,
    \begin{equation}
        v^\mu=\sqrt{\frac{h(0,s_m)}{h(\tau,s_m)}}=\big(\sec(2\tau),0,0\big).
    \end{equation}
Transforming back to compactified Kruskal coordinates, we obtain,
    \begin{align}
        &v^\mu=\xi(\sigma,\rho)\Big(\cos(\rho-\sigma)\csc(\rho)\sin(\sigma),\cos(\rho)(\cos(\rho)+\sin(\rho)\tan(\sigma)),0\Big),\\
        &\xi(\sigma,\rho)=\cos^2(\rho-\sigma)\sec(\rho+\sigma)\sqrt{\csc(\rho)\csc(\sigma)\sec(\rho)\sec(\sigma)\cot(\sigma)\;,\tan(\rho)}
    \end{align}
from which it is straightforward to verify $v^{\mu}$ is divergenceless, $\nabla_\mu v^\mu=0$. \\
Meanwhile, the associated norm is
    \begin{equation}
        |v^\mu|=-\frac{1}{16}\cos^4(\rho-\sigma)\sec^2(\rho+\sigma)\csc(\rho)\csc(\sigma)\sec(\rho)\sec(\sigma)\;.
    \end{equation}
We plot the vector field and its norm in Figure \ref{fig:latetime}, where we observe $|v|$ reaches its minimum on the maximal volume surface, as expected.

Before moving to the level set construction, it is interesting to consider what the Lorentzian thread flow can tell us about the second law of quantum complexity. In the case of a double sided black hole, it has been shown that complexity grows monotonically with the boundary time until it reaches a plateau value, \emph{i.e.}, black holes obey a second law of complexity. The maximum plateau value can be interpreted in different ways. For example, from a circuit complexity point of view, it can be seen as the volume of the space of operators $SU(2^K)$ \cite{Brown:2017jil}, while from a graph theory perspective the max value can be instead interpreted as the size of the graph at which the no-collision assumptions breaks down \cite{Susskind:2018pmk}.

In a manner similar to \cite{Couch:2018phr}, we can think of the growing volume inside the double sided black hole as the increase in the number of Lorentzian threads which cross the portion of the maximal volume slice inside the horizon (thus, reaching the singularity) as we move the associated WDW patch forward in time. Alternatively, by the same reasoning, a decrease in the complexity of the state as it evolves forward from past infinity  is associated to the decrease in the number of threads crossing the portion of the maximal volume slice inside the white hole. Thus, we can connect the thread picture to the second law of quantum complexity: as the WDW patch moves forward in time, additional  threads are forced to cross the event horizon before meeting the maximal volume surface and ultimately ending at the singularity.  This leads to a monotonic increase in the associated complexity of the boundary state.

With this point of view in mind, it is possible to interpret the maximal complexity as the complexity associated to the `final state' in which \emph{all} the threads cross the maximal volume surface inside the black hole and reach the singularity. Therefore, the complexity has the same plateau value at past and future infinity, and the semiclassical boundary states associated to the time in between the two correspond to a dip in complexity, occurring once every recurrence time \cite{Brown:2017jil}, corresponding to the threads switching their crossing points with the maximal volume surface from inside the white hole to inside the black hole, with the minimum of the dip corresponding to the transition time at which the crossing points lie completely outside the black hole except for a set of measure zero on the bifurcation surface.

We conclude the section with a thought on the CV duality. It has been argued that one of the reasons why the CV duality is preferable over the CA duality is that, even at late times, the maximal volume slice does not intersect the singularity, thus is robust against quantum corrections. Among other things, this allows one to circumvent the difficulties in handling the boundary terms which appear when computing the complexity through the action of the WDW patch.
Given the above observation the Lorentzian thread reformulation of CV duality shows that the threads do, in fact, probe the singularity, unlike Riemannian Bit-threads. It would be interesting to explore the consequences of this. Indeed, since the region near the singularity is expected to encode Planck scale physics it is likely that the late time surface and the corresponding threads will need quantum corrections to correctly characterize the CFT state. In particular, without quantum corrections, the thread gatelines commute, however, when quantum effects are included, this commutativity property may be lost.


\subsection{Method of level sets} \label{subsec:levelsetcons}

The second method of constructing Lorentzian thread configurations has us specify a family of level set hypersurfaces, satisfying the below properties, where the vector field is orthogonal to the level sets and is divergenceless. Our construction is motivated by the Riemannian analog for holographic bit threads introduced in \cite{Agon:2018lwq,Agon:2020mvu} and reviewed in Appendix \ref{app:examples}.

We propose  a family of level sets satisfying:

\vspace{2mm}

\noindent (1) One of the members is the maximal bulk slice $\Sigma_{A}$ homologous to boundary region $A$.

\vspace{2mm}

\noindent (2) All surfaces are continuous and do not self-intersect.

\vspace{2mm}

\noindent (3) Closed bulk surfaces are not included.

\vspace{2mm}

\noindent (4) All hypersurfaces are homologous to $A$; not just the maximal bulk slice.

\vspace{2mm}

From the first three properties it is possible to construct a divergenceless vector field with a desired boundary condition. This is seen as follows. Given a family of level sets satisfying at least (1)--(3), one generates the corresponding integral lines orthogonal to each hypersurface in the family. Once the integral lines have been found, then the problem reduces to the one described in Section \ref{subsec:geoflows}.

All that remains is checking the norm bound is satisfied. Condition 4, while not necessary, is useful to assume since it will ensure $|v|\geq1$ is satisfied everywhere. This follows from the min flow-max cut theorem, where, since $|v|_{\Sigma_{A}}=1$, then $|v|\geq1$  at any other member of the family of level sets. Note that a simple vector field configuration generated by level sets which are \emph{not} all homologous to the same boundary region $A$ are the minimally packed flows described in Section \ref{sec:prelims}. In that context, recall, the level set surfaces constituted a family of nested maximal volume slices, containing $\Sigma_{A}$ as one of its members, and the flux of the thread $v$ was minimal through multiple regions, satisfying $|v|=1$ in a given bulk region. An explicit geometric realization of such flows were also described in \cite{Couch:2018phr}, where one introduces the notion of a `volume current' $V$. The volume current was defined given a foliation of the bulk spacetime by maximal volume hypersurfaces induced by a Cauchy foliation of the boundary by slices orthogonal to an asymptotic Killing flow affiliated with time translations. Specifically, the volume current $V$ is the unit timelike vector field orthogonal to the bulk foliation with a divergence proportional to the extrinsic curvature $K$ of each bulk Cauchy slice. Since all slices are assumed to be maximal, $K=0$, the divergenceless conditions trivially holds. Our explicit construction below will not be of this minimally packed type (we assume condition 4 above), however, we will foliate the bulk spacetime by constant $K$ slices (a `constant mean curvature' (CMC) slicing).

Let us now describe in some detail the general construction from level sets, and then apply this to a specific context. We specify our level hypersurfaces as $\varphi=\text{constant}$ surfaces, for some appropriate scalar function $\varphi(x^{i})$. We then consider the gradient flow
\beq v=\Upsilon(x^{i})\nabla\varphi\;,\eeq
where $\Upsilon$ is some unspecified scalar function. The unit normal vector $\hat{\tau}=v/|v|$ is given by
\beq \hat{\tau}=\frac{\nabla\varphi}{|\nabla\varphi|}\;.\eeq

Note further that the covector $v_{\mu}=\Upsilon(\varphi,g)\partial_{\mu}\varphi$, such that the boundary condition at the maximal bulk slice implies
\beq |\Upsilon^{2}g^{\mu\nu}\partial_{\mu}\varphi\partial_{\nu}\varphi|_{\Sigma_{A}}=1\;.\eeq
This condition gives the value of $\Upsilon$ at $\Sigma_{A}$:
\beq \Upsilon(\varphi,g)|_{\Sigma_{A}}=|\partial\varphi|_{g}^{-1}\biggr|_{\Sigma_{A}}\;,\quad |\partial\varphi|_{g}\equiv\sqrt{g^{\mu\nu}\partial_{\mu}\varphi\partial_{\nu}\varphi}\;.\label{eq:bclevelcurvegen}\eeq
It remains to find $\Upsilon$ away from $\Sigma$. This is accomplished by imposing the divergenceless condition $\nabla\cdot v=0$, leading to a first order differential equation:
\beq (\nabla\varphi)\cdot (\nabla\Upsilon)+(\Box\varphi)\Upsilon=0\;. \label{eq:diffeqlevelsetgen}\eeq
Solving this differential equation subject to the boundary condition (\ref{eq:bclevelcurvegen}) at $\Sigma_{A}$ yields a unique solution for Lorentzian thread configuration $v$.

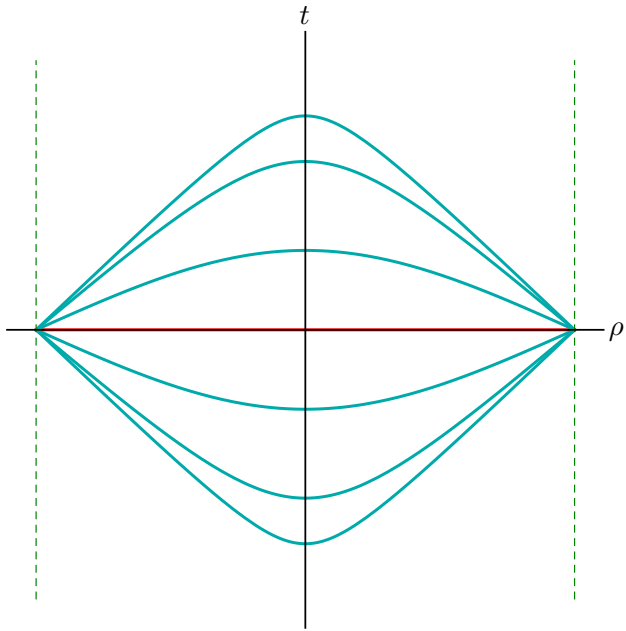
\begin{figure}[t]
    \centering
    \includegraphics[width=8cm]{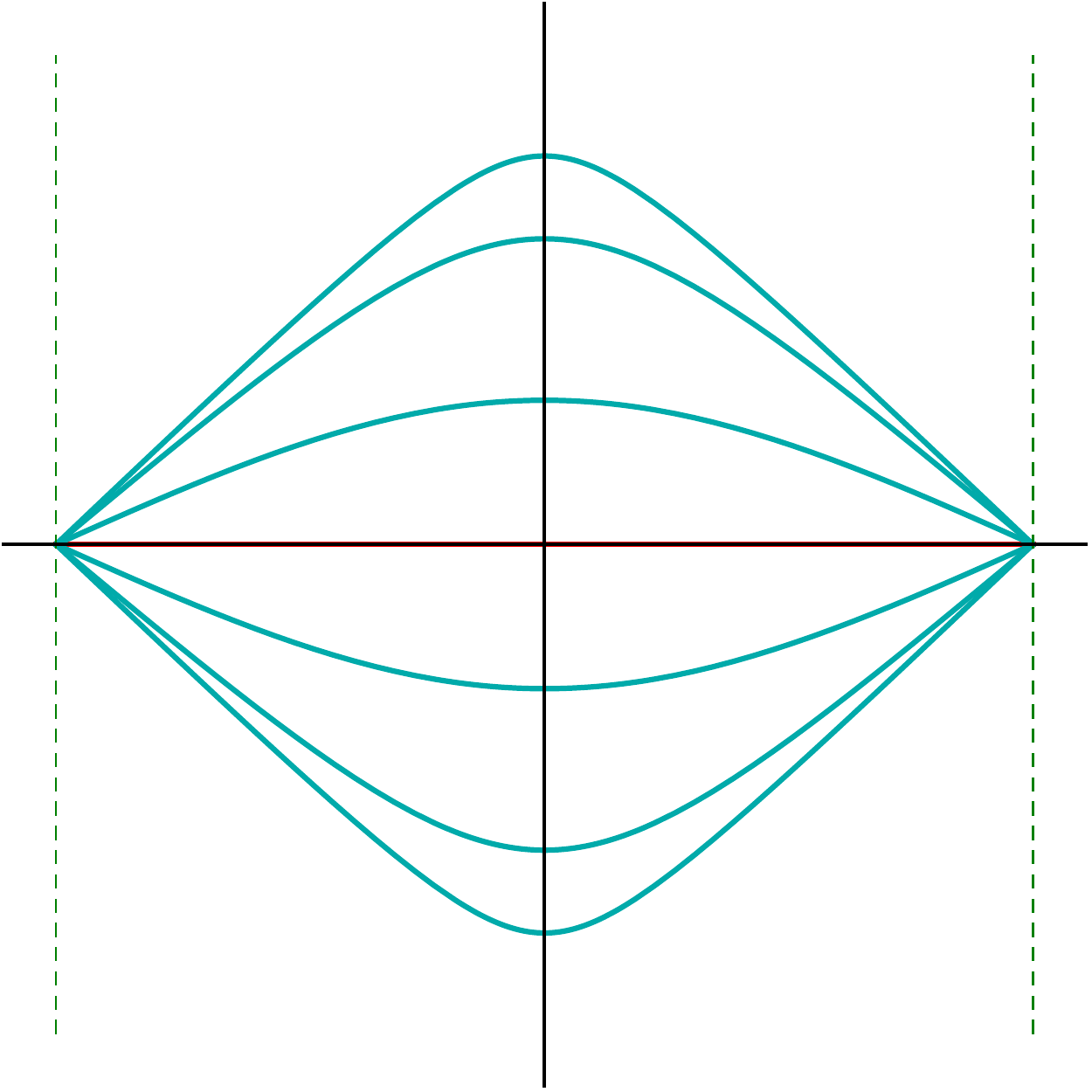}
 \begin{picture}(0,0)
\put(-120,230){$t$}
\put(-3,112){$\rho$}
\end{picture}
    \caption{Plot of the level sets for AdS$_3$. The CMC slicing of the WDW patch are curves of constant $K$. For $K= 0$, the corresponding slice is the maximal volume slice, represented by the red line along the $\rho$ axis. We suppressed the angular coordinate apart from  $\theta=0 $ and $\theta=\pi$, which are represented on the right and left side of the time axis.}
    \label{fig:AdSlevelsets}
\end{figure}

Below we carry out this procedure in the case of empty AdS where our level sets are specified as constant mean curvature slices $K=\text{constant}$, where the maximal hypersurface $K=0$ is a member of the family of hypersurfaces.

\subsubsection{Flows in vacuum AdS}

Here we build Lorentzian thread configurations using a family of constant mean curvature $K$ hypersurfaces in empty  AdS$_{n+1}$. Such a  CMC slicing for empty $\text{AdS}_{n+1}$ is given in terms of Wheeler- De Witt (WDW) coordinates
 \begin{equation}
   ds^2=-d\tau^2+\cos^2(\tau)d\Sigma^2_{n},
    \label{eq:wdwcoord}\end{equation}
where $d\Sigma^2_{n}=h_{ab}dx^a dx^b$ is a $\tau$ independent Einstein metric satisfying $(R_{n})_{ab}=-(n-1)h_{ab}$. In these coordinates AdS is foliated by surfaces of constant $\tau$, or, equivalently, surfaces of constant $K$ \cite{Belin:2018bpg}:
\begin{equation}
    K=-n\tan(\tau)\;.
   \label{eq:constKcmc}\end{equation}

  \begin{figure}[t]
	   \centering
	   \includegraphics[width=0.40\textwidth]{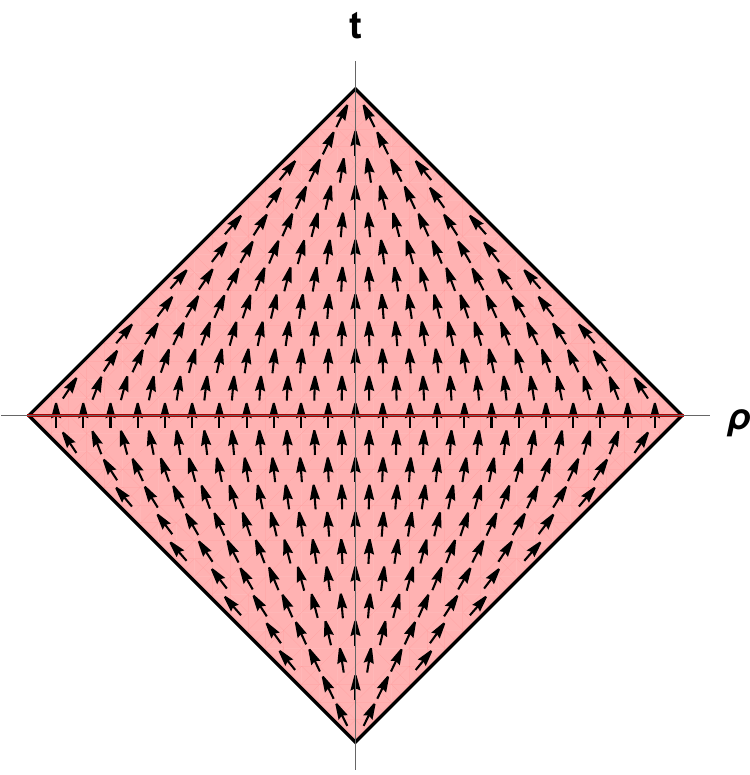}$\qquad\qquad$
	   \includegraphics[width=0.45\textwidth]{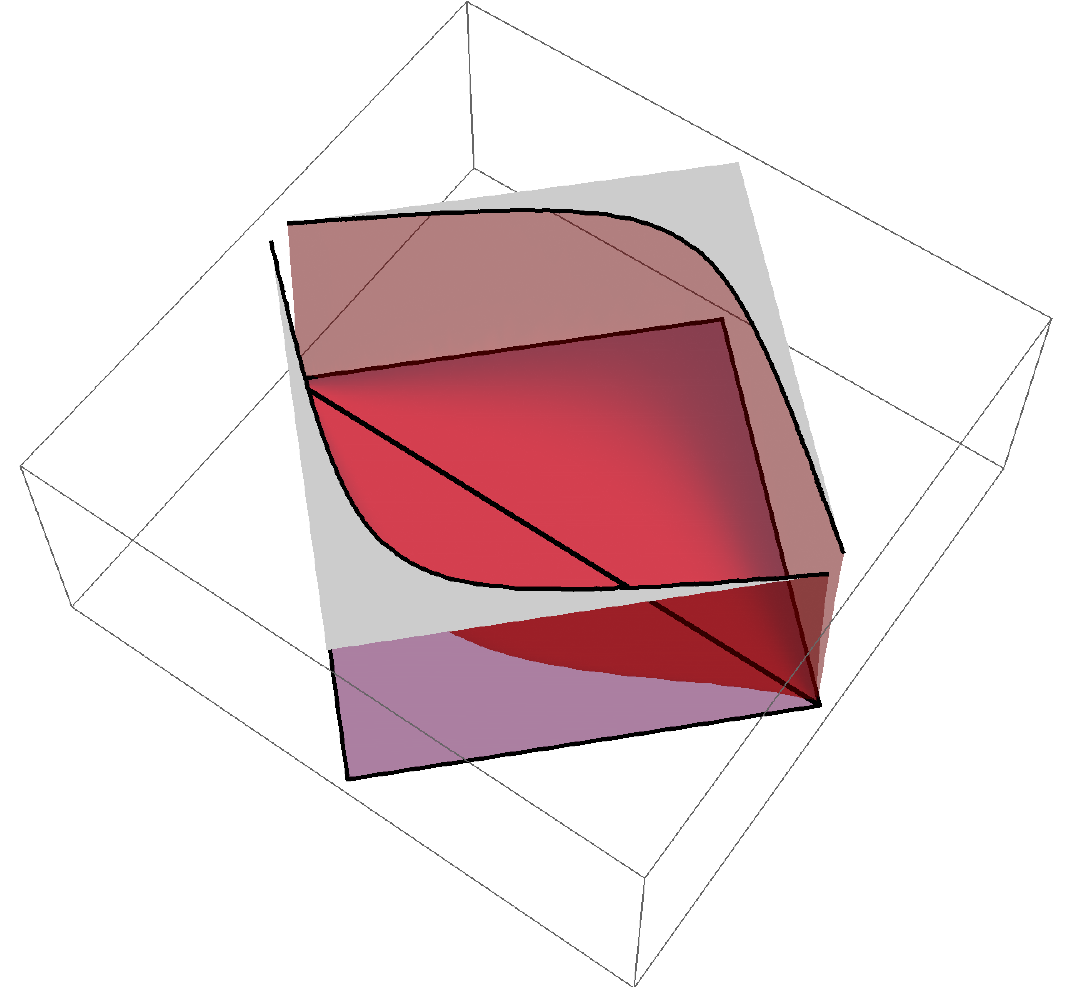}
 \begin{picture}(0,0)
\put(-342,180){$t$}
\put(-250,87){$\rho$}
\put(-165,146){$t$}
\put(-151,29){$\rho$}
\end{picture}
   \caption{Plot of the Lorentzian thread vector field in AdS$_3$ and  of its norm. The maximum volume surface is indicated by the darker horizontal line. As for the geodesic construction, the angular coordinate is suppressed for all values except $\theta=0$ and $\theta=\pi$.}
    \label{fig:AdSThreadlevelsets}
\end{figure}
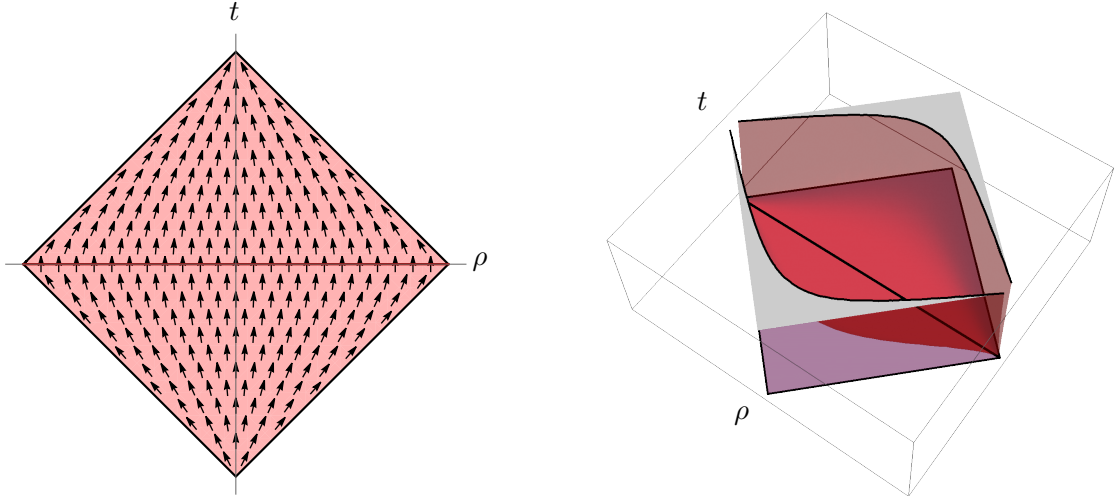

We seek a function $\varphi(x^{i})$ characterizing these level sets, where $x^{i}$ are global compactified coordinates $(t,\rho,\theta_{i})$. This is accomplished by first translating WDW coordinates (\ref{eq:wdwcoord}) into Poincar\'e coordinates $(T,Z)$ and then into global coordinates
\beq \tau=\arcsin\left(\frac{T}{Z} \right)\;,\quad T=\frac{\sin(t)}{\cos(t)+\sin(\rho)\Omega_{n-1}}\;,\quad Z=\frac{\cos(\rho)}{\cos(t)+\sin(\rho)\Omega_{n-1}}\;,\eeq
where $\Omega_{n-1}$ is the solid angle and we have set $L=1$. Using $K(\tau)$ above, we may express global coordinate $t$ in terms of $K$ and $\rho$ such that our level sets are defined by
    \begin{equation}
        t+\arcsin\left(\frac{K\cos(\rho)}{\sqrt{n^2+K^2}} \right)=0\;.
    \end{equation}
Inverting this relation to express $K=K(t,\rho)$, the function $\varphi$ characterizing our level sets is defined to be:
    \begin{equation}
\varphi(t,\rho)\equiv K(t,\rho)=-\frac{n\sin(t)}{\sqrt{\cos^2(\rho)-\sin^2(t)}}\;.
    \end{equation}
We present illustration of our level set construction in Figure \ref{fig:AdSlevelsets}. As depicted, we see our level sets satisfy properties (1)--(4) described above.

The gradient $(\nabla\varphi)$ vector, denoted by $V^{\mu}_{(n)}$ is
    \begin{equation}
        V^\mu_{(n)}=\left(\frac{(n-1)\cos(t)\cos^4(\rho)}{\sqrt{(\cos^3(\rho)-\sin^2(t))^3}},-\frac{(n-1)\cos^3(\rho)\sin(t)\sin(\rho)}{\sqrt{(\cos^3(\rho)-\sin^2(t))^3}},0,...,0 \right)\;.
    \end{equation}
It remains to be seen whether this vector field satisfies the norm bound and is divergenceless. It turns out in $n=3$, $\nabla\cdot V_{(3)}=0$ automatically.\footnote{While the divergenceless condition is automatically satisfied in $n=3$, $V^{\mu}_{(3)}$ does not satisfy the norm bound, requiring we multiply by an overall scalar function.} In higher dimensions $V^{\mu}_{(n)}$ is not divergenceless, thus we multiply by an overall scalar function $\Upsilon(t,\rho)$ and determine $\Upsilon$ by imposing the divergenceless condition, where we demand the norm bound be saturated at the $K=0$ surface. It is straightforward to show the function solving the differential equation (\ref{eq:diffeqlevelsetgen}) is
    \begin{equation}
        \Upsilon(t,\rho)_{(n)}=\sec^{(n-1)}(t)\sqrt{\left(\frac{1-\sin^2(\rho)}{1-\sec^2(t)\sin^2(\rho)}\right)^{n-1}}.
    \end{equation}
The resulting vector field is
\beq v^{\mu}\equiv \Upsilon(t,\rho)V^{\mu}_{(n)},\eeq
depicted in Figure \ref{fig:AdSThreadlevelsets}, from which we see  $v$ is a valid Lorentzian thread configuration.


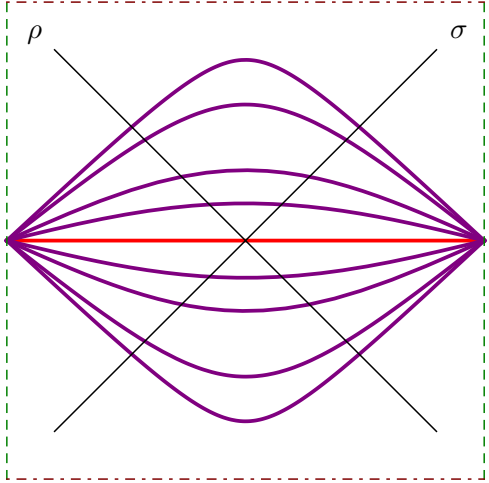
\begin{figure}[t]
	\begin{center}
\includegraphics[width=4cm, angle=45,trim=1.5in 1.5in 1.5in 1.5in]{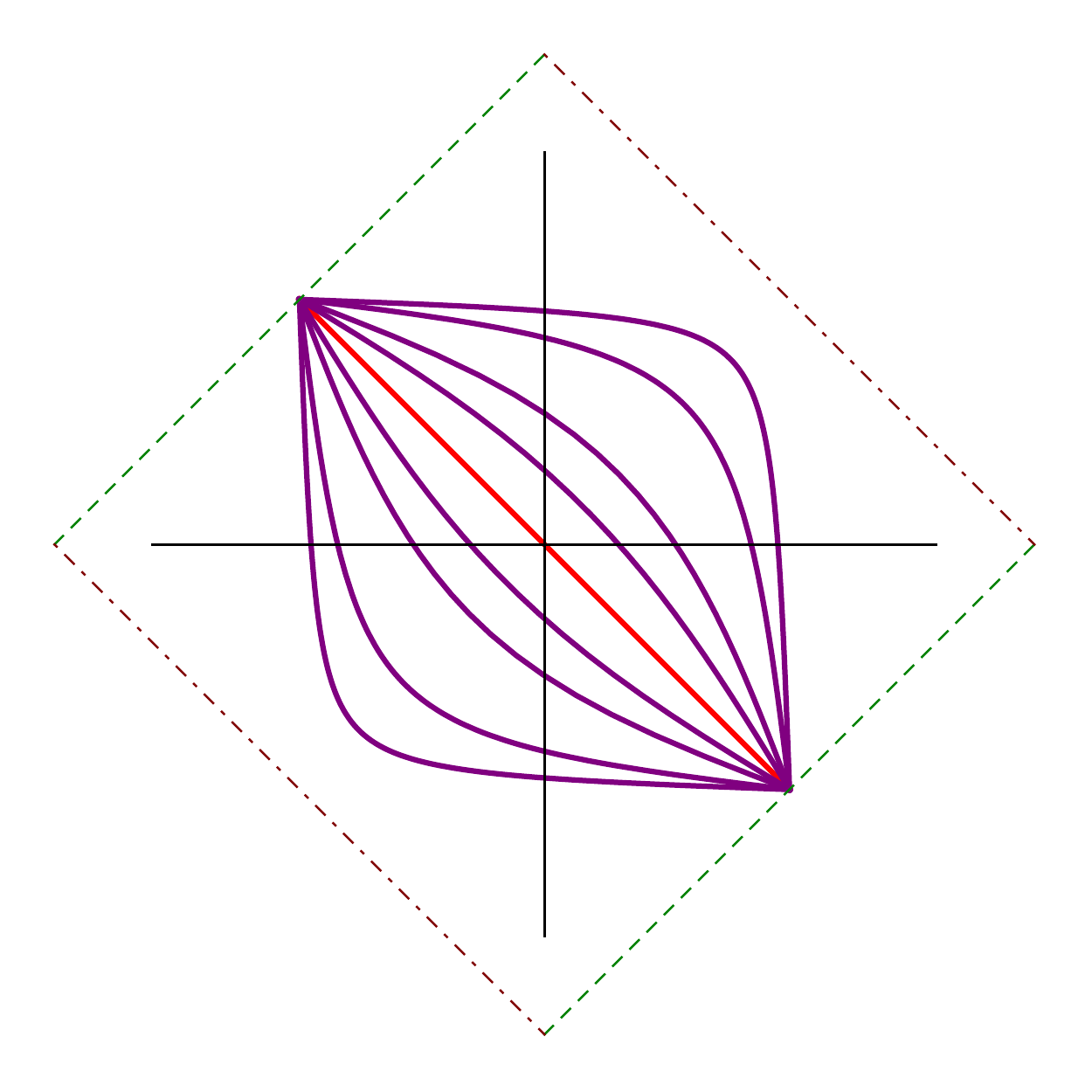}
	\end{center}
 \begin{picture}(0,0)
\put(138,184){$\rho$}
\put(298,184){$\sigma$}
\end{picture}
    \caption{Plot of the level sets for the BTZ black hole. The CMC slicing of the WDW patch are curves of constant $K$. $K=0$ is the maximal volume slice and it is represented by the red line.}
    \label{fig:BTZlevelsets}
\end{figure}

\subsubsection{Flows in the BTZ black hole}

Let us now find a Lorentzian flow using a similar level set construction in a BTZ background. This is rather straightforward given our construction above and the fact the BTZ black hole and $\text{AdS}_{3}$ have similar group structures, such that both solutions can be put into Poincar\'e form. In particular, the transformation of the static BTZ black hole (\ref{eq:staticBTZ}) in Schwarzschild coordinates $(t,r,\phi)$ to Poincar\'e coordinates $(T,Z,X)$ is (see, \emph{e.g.}, Eq. (2.9) in \cite{Carlip:1995qv} in the static limit):
\beq T=\sqrt{\frac{r^2-1}{r^2}}\sinh(t)e^{\phi}\;,\quad Z=\sqrt{\frac{1-r^2}{r^2}}e^{\phi}\;,\quad X=\sqrt{\frac{r^{2}-1}{r^{2}}}\cosh(t)e^{\phi}\;,
 \label{eqn: BTZpoincare}\eeq
where for convenience we have set $L=M=1$ such that $r_{s}=1$.

 In terms of Poincar\'e coordinates, we may now use the CMC slicing (\ref{eq:wdwcoord}) with extrinsic curvature $K$ (\ref{eq:constKcmc}) for $n=2$.\footnote{We are using the fact the static BTZ black hole in CMC coordinates is a quotient of $\text{AdS}_{3}$ by a discrete subgroup of the $SO(2,2)$ isometry group, with topology $\Sigma\times\mathbb{R}$, where $\Sigma$ is an annulus \cite{Krasnov:2000zq}.} Then, combining the transformation (\ref{eqn: BTZpoincare}) with Kruskal coordinates (\ref{eqn: kruscaltrans}), we may write an equation for the level sets in global compactified Kruskal coordinates $(\sigma,\rho)$ as follows
    \begin{equation}
        \sigma+\arctan\left(\frac{K-\sqrt{4+K^2}\tan(\rho)}{K\tan(\rho)-\sqrt{4+K^2}} \right)=0\;.
    \end{equation}
Inverting this relation to express $K=K(\sigma,\rho)$, the function $\varphi$ characterizing the BTZ level sets is given by
    \begin{equation}
\varphi(t,\rho)\equiv K(\sigma,\rho)=-\frac{2(\tan(\sigma)+\tan(\rho))}{\sqrt{(1-\tan^2(\sigma))(1-\tan^2(\rho))}}\;.
    \end{equation}
These level sets are presented in Figure \ref{fig:BTZlevelsets}, from which we see the BTZ level sets satisfy general properties (1)--(4) above.

   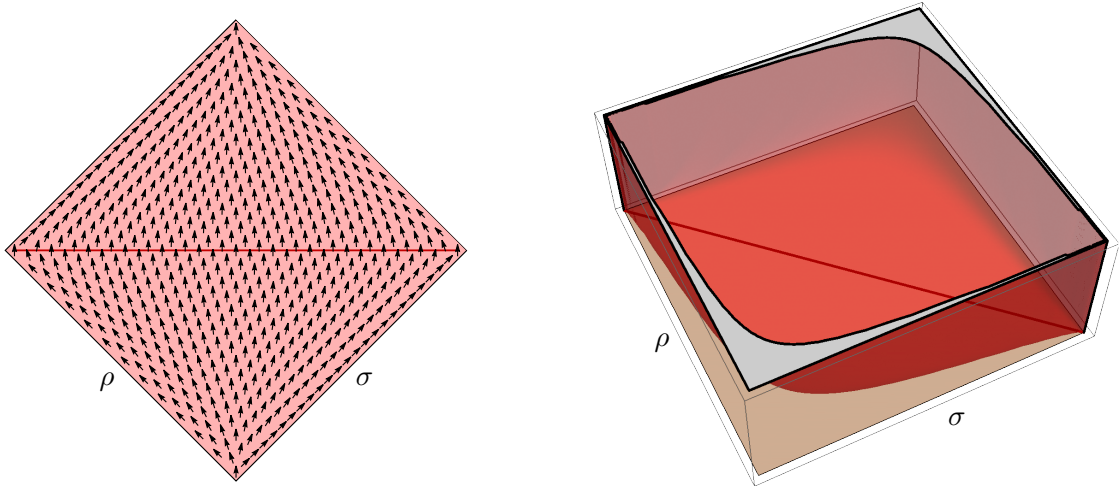
\begin{figure}[t]
	   \centering
	   \includegraphics[angle=45,width=0.40\textwidth]{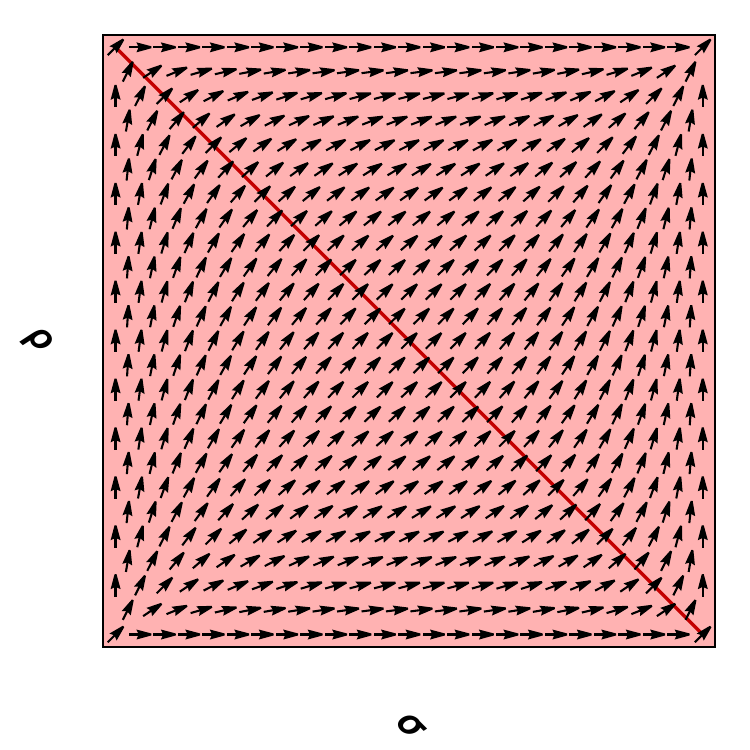}$\qquad\qquad$
	   \includegraphics[width=0.45\textwidth]{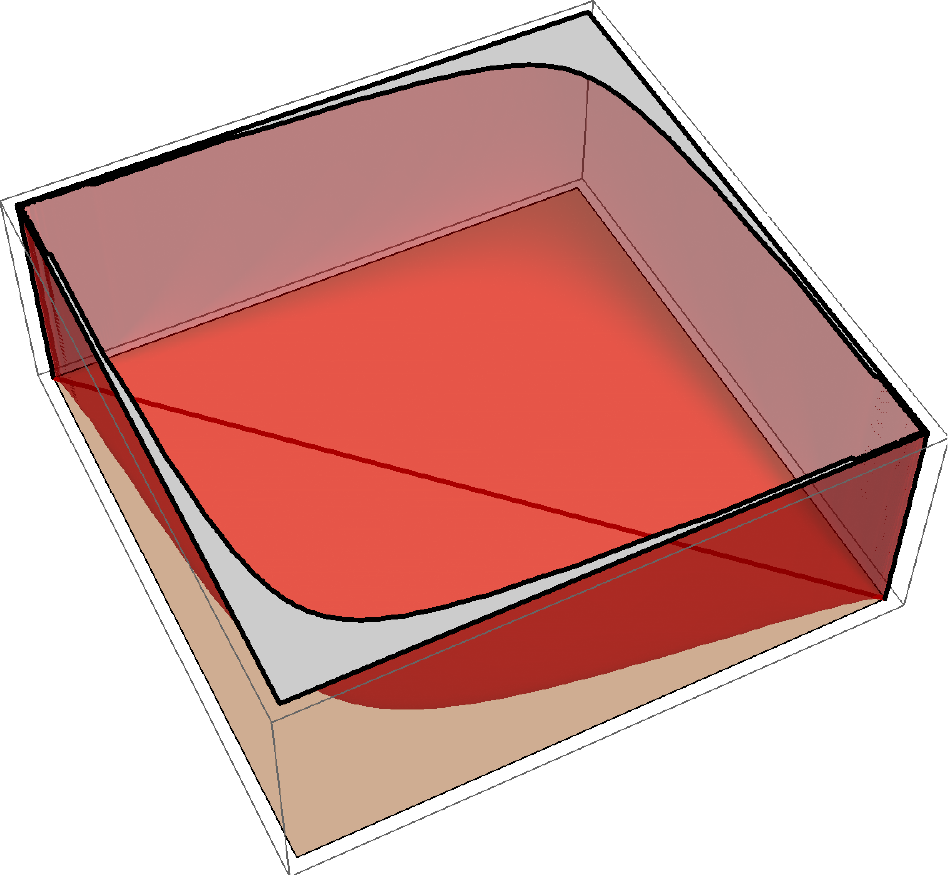}
 \begin{picture}(0,0)
\put(-390,38){$\rho$}
\put(-293,38){$\sigma$}
\put(-180,53){$\rho$}
\put(-69,23){$\sigma$}
\end{picture}
   \caption{Plot of the Lorentzian thread vector field in BTZ and of its norm. The maximum volume surface is indicated by a darker line.}
    \label{fig:BTZThreadlevelsets}
\end{figure}

The gradient of this function, which we will denote by $\nabla^{\mu}\varphi \equiv V^\mu$, is
    \begin{equation}
        V^\mu=\left(\frac{(1-\tan^2(\sigma))(\cos(\sigma)+\sin(\sigma)\tan(\rho))^3 \sec(\sigma)}{\sqrt{[(1-\tan^2(\sigma))(1-\tan^2(\rho))]^3}},
        \frac{(1-\tan^2(\rho))\sec(\rho)(\cos(\rho)+\sin(\rho)\tan(\sigma))^3}{\sqrt{[(1-\tan^2(\sigma))(1-\tan^2(\rho))]^3}},0 \right)\;.
    \end{equation}
It is easy to verify $V^{\mu}$ is divergenceless. The norm is minimum at the maximal volume hypersurface, however, the bound is not saturated. Happily, unlike the empty AdS case, since the norm on the surface is constant we do not need introduce a function $\Upsilon$ and solve the differential equation (\ref{eq:diffeqlevelsetgen}) to fix $\Upsilon$. Rather, it is enough to multiply $V^{\mu}$ by the scalar function $\Upsilon(\sigma,\rho)=\frac{1}{4}$ such that the final Lorentzian thread field is then given by
    \begin{equation}
        v^\mu\equiv\frac{1}{4}V^\mu\;.
    \end{equation}
The vector field along with its norm is plotted in Figure \ref{fig:BTZThreadlevelsets}.

\subsection*{Outside the WDW patch}

Note we only considered level set constructions using CMC slices which foliate the interior of the WDW patch in both empty AdS and the BTZ black hole. In principle one can consider a similar level set construction outside the patch, however, the differential equation needed to solve for the form $\Upsilon$ is more difficult to evaluate due to a difference in a change in boundary conditions. Alternatively, one can always invoke the geodesic construction outside of the patch explored before, in which case the thread configuration of the whole spacetime is an example of a \emph{mixed flow}: threads orthogonal to CMC slices inside the WDW patch combined with integral lines tangent to radial timelike geodesics outside of the patch.


\section{Perturbative threads and Einstein's equations} \label{sec:diffformsandeineqs}

Thus far we have developed a Lorentzian thread reformulation of complexity=volume duality and provided multiple explicit geometric constructions of these threads. In particular, we considered geodesic flows obtained via timelike geodesic foliations of the WDW patch of the bulk geometry, where the geodesics are identified with the integral lines of the corresponding divergenceless Lorentzian thread vector field $v$. We also considered a level set configuration, in which we foliated the interior of the WDW patch with a set of codimension-1 hypersurfaces of constant mean curvature (CMC) $K$.

This CMC slicing has played a prominent role in understanding CV duality before \cite{Belin:2018bpg}. As we will describe in more detail below, in the phase space formulation of general relativity the Hamiltonian constraint can be understood as a differential equation in the spatial volume density such that, in a CMC slicing, the extrinsic curvature $K$ can be interpreted as a time parameter, the so-called `York time' \cite{York:1972sj}. In this slicing, the spatial volume $V$ of the constant $K$ curvature slice is interpreted as a Hamiltonian whose variation coincides with the covariant bulk symplectic form when $V$ is the volume of an extremal slice. Furthermore, using that the bulk symplectic form is dual to the boundary symplectic form \cite{Belin:2018fxe}, it was shown \cite{Belin:2018bpg} the variation $\delta V$ is proportional to the variation of boundary complexity $\delta\mathcal{C}$.

The bulk symplectic form is useful when studying linear perturbations around the bulk spacetime since one need not make explicit reference to the background metric, and it encodes information about gravitational field equations. This suggests a deep relationship between boundary complexity and bulk spacetime dynamics. Here we will expand on this insight and show a first law of holographic complexity leads to the linearized Einstein's equations. Furthermore, we will demonstrate an equivalence between Einstein's equations and a condition on the Lorentzian threads. Specifically, using a map between the divergenceless timelike vector fields $v$ and closed forms $u$, we show \emph{perturbative Lorentzian threads} -- Lorentzian threads corresponding to linear perturbations around AdS -- are naturally identified with the bulk symplectic current such that the linearized Einstein's equations are encoded in the closedness condition of the perturbative Lorentzian thread form.

Since our argument is rather detailed, let us provide a very brief summary of this section. First in Section \ref{subsec:excitedstatespertthreads}, we develop the notion of perturbative Lorentzian threads, and how they relate to linear perturbations to vacuum AdS, which are dual to excited CFTs. We then cast Lorentzian threads in the language of differential forms including their perturbative counterparts in Section \ref{subsec:threadsdiffforms}. This set-up will allow us to directly connect the closedness condition of the perturbative thread form to the linearized Einstein's equations. To illustrate this, we first review the relationship between the boundary and bulk symplectic structure in Section \ref{subsec:sympstructrev}, including the definition of the York deformation and how it relates to CV duality \cite{Belin:2018fxe,Belin:2018bpg}. We then provide the first detailed proof showing an equivalence between a first law of holographic complexity and the linearized Einstein's equations in Section \ref{sec: EEfromComplexity}. We emphasize our argument is spiritually and structurally very similar to the derivation of Einstein's equations from the holographic first law of entanglement \cite{Lashkari:2013koa,Faulkner:2013ica}. In Section \ref{subsec:bulksymcanthread} we conclude by showing a canonical choice of the perturbative thread is the bulk symplectic current, such that the closedness condition of the thread form is equivalent to linearized Einstein's equations being satisfied.


\subsection{Excited states and perturbative Lorentzian threads} \label{subsec:excitedstatespertthreads}

By the holographic dictionary, bulk vacuum AdS is dual to the vacuum state of the boundary CFT, while linear perturbations to vacuum AdS correspond to perturbative excited CFT states. This fact was utilized in \cite{Lashkari:2013koa,Faulkner:2013ica} to show the first law of entanglement is equivalent to linearized gravitational field equations being satisfied in the bulk. To encode metric perturbations in the bit thread reformulation of holographic entanglement, one needs to generalize the bit thread construction to time dependent backgrounds, where, moreover, the threads themselves are perturbed. Such `perturbative bit threads' characterize entanglement in excited boundary states and were explicitly constructed in \cite{Agon:2020mvu}. It will benefit us to review this argument as we will consider an analogous scenario for the Lorentzian threads.

 While a fully covariant bit thread formulation of HRT is currently underway \cite{Headrick:toappear}, one may instead use the maximin principle of covariant holographic entanglement entropy \cite{Wall:2012uf} to describe bit threads living in a dynamical bulk spacetime. That is, one first picks a specific Cauchy slice $\Sigma_{t}$ extending to the timelike boundary characterized by time coordinate $t$, perform the area minimization over this slice to find the minimal RT surface, and then maximizes over all such possible Cauchy slices $\Sigma_{t}$. The  entanglement entropy is then given by the area of a maximin surface $m(A)$. This leads to a  `maximax' version of Riemannian flows \cite{Agon:2019qgh,Agon:2020mvu}, where one first finds the Riemannian flow with maximum flux through the boundary region $\Sigma_{t}\cap D[A]$, a Cauchy slice intersecting the boundary domain of dependence of (boundary) region $A$, and then maximizes over all such Cauchy slices $\Sigma_{t}$. It turns out Cauchy slices $\Sigma_{t}$ away from the maximin surface $m(A)$ are highly degenerate \cite{Wall:2012uf}.

This observation about degeneracy was used to argue that to first order in a general time dependent perturbation to an otherwise static metric, $\Sigma_{t}$ may always be chosen to be the Cauchy slice $\Sigma$ associated with the unperturbed metric, or any spacelike surface $\Sigma_{\eta}$ \cite{Agon:2020mvu}. Here $\Sigma$ corresponds to the Cauchy slice at $t=0$, while $\Sigma_{\eta}$ is the spacelike hypersurface perturbatively close to $\Sigma$, where $\eta$ is a small parameter characterizing the bulk metric perturbation, $g_{\mu\nu}^{\eta}=g_{\mu\nu}+\eta\delta g_{\mu\nu}+\mathcal{O}(\eta^{2})$ (which is assumed to satisfy Einstein's equations). Cauchy slices  far away from $\Sigma_{0}$ correspond to highly excited CFT states and are neglected since the unique maximin surface is not necessarily minimal on such slices. Correspondingly, the perturbative bit thread $v_{\eta}$ associated to a perturbatively excited CFT state is given by
\beq v_{\eta}=v+\eta\delta v+\mathcal{O}(\eta^{2})\;.\label{eq:pertvf}\eeq
As a flow, $v_{\eta}$ is expected to satisfy the usual criteria, thereby constraining $\delta v$.

Since either holographic complexity conjecture naturally incorporates time dependent backgrounds, naively we need not worry about the maximin principle. This is not quite accurate, however, as,  in vacuum, the maximal volume slice $\Sigma_{0}$ is `foliated' by all RT surfaces of all subregions. Thus, by the above discussion, only for perturbative excited states, where $\Sigma_{\eta}$ is equivalent to $\Sigma$ at leading order, will the maximin surface be contained in the maximal volume slice. The point we are making here deserves emphasis: the CV proposal, and hence our Lorentzian flow reformulation, holds perturbatively. For highly excited states, where the minimax surface may not be contained in the maximal volume slice at $t=0$, it may be more suitable to define an alternative notion of holographic complexity. We will come back to this point in Section \ref{sec:disc}.

Let us now briefly describe perturbative Lorentzian threads in some generality. For a perturbed metric of the form  $g_{\mu\nu}^{\eta}=g_{\mu\nu}+\eta\delta g_{\mu\nu}$, for the vector field (\ref{eq:pertvf}) to be a Lorentzian flow, we must have
\beq \nabla_{\eta}\cdot v_{\eta}=0\;,\quad v_{\eta}^{0}>0\;,\quad |v_{\eta}|\geq1\;,\quad v^{\mu}_{\eta}|_{\Sigma}=n^{\mu}\;,\quad n_{\mu}v^{\mu}_{\eta}|_{R}=0\;,\eeq
where $\nabla_{\eta}$ is the covariant derivative with respect to perturbed metric $g^{\eta}_{\mu\nu}$. Here $v=v_{\eta=0}$ is assumed to already satisfy each of these properties. The future directed  and flux vanishing on boundary region $R$ conditions tell us, respectively,
\beq v^{0}>\eta\delta v^{0}\;,\quad \eta n_{\mu}(\delta v_{\eta}^{\mu})|_{R}=0\;.\eeq
Using $g^{\mu\nu}_{\eta}=g^{\mu\nu}+\eta \delta g^{\mu\nu}$, and working to linear order in parameter $\eta$, we find the norm bound leads to
\beq -g_{\eta}^{\mu\nu}v_{\mu}^{\eta}v_{\nu}^{\eta}=-g^{\mu\nu}v_{\mu}v_{\nu}+\eta(2g^{\mu\nu}v_{\mu}\delta_{\nu}+v_{\mu}v_{\nu}\delta g^{\mu\nu})|_{\Sigma}\geq1\;.\eeq
When evaluated on the maximal volume slice $\Sigma$, we have that the second term vanishes on $\Sigma$. Lastly, the divergenceless condition, gives
\beq 0=\nabla_{\eta}\cdot v_{\eta}=\frac{1}{\sqrt{g_{\eta}}}\partial_{\mu}(\sqrt{g_{\eta}}v_{\eta}^{\mu})=\frac{\sqrt{g}}{\sqrt{g_{\eta}}}(\nabla\cdot v)+\frac{\eta}{\sqrt{g_{\eta}}}\partial_{\mu}[v^{\mu}\delta(\sqrt{g_{\eta}})+\sqrt{g}\delta v^{\mu}]\;,\label{eq:annoydivcond}\eeq
such that, upon using $\nabla\cdot v=0$, the last term in brackets is equal to a spacetime constant.

Altogether, given a background metric $g_{\mu\nu}$ and a solution to the min flow problem $v^{\mu}$, we can always solve the problem of minimizing the flux where the metric $g^{\eta}_{\mu\nu}$ is perturbatively close to the original one, \emph{i.e.}, for bulk geometries dual to perturbative excited states. Due to the presence of the metric and its variations, working with perturbative vector fields becomes a bit cumbersome. As we will show below, it will be more convenient to work in the language of differential forms.


\subsection{Threads as differential forms and linear perturbations of complexity} \label{subsec:threadsdiffforms}

Our reformulation of CV duality (\ref{eq:cvreformin}) relied on an equivalence between maximal volumes and minimal flows $v$, where $v$ is a divergenceless timelike vector field with norm satisfying $|v|\geq1$. Thus, our reformulation (\ref{eq:cvreformin}) makes explicit reference to the background metric. Here, however, we will be interested in studying bulk locality and how Einstein's equations, which are manifestly background independent, are encoded in the Lorentzian threads. To make the background independence explicit, we opt to work with differential forms.\footnote{Replacing (Riemannian) flows with closed $(d-1)$-forms was recently accomplished in \cite{Agon:2020mvu} to express the bit thread reformulation of the Ryu-Takayanagi relation in the language of differential forms. For a brief review see Appendix \ref{app:examples}. The equivalence of flows and closed forms was also previously remarked in \cite{Freedman:2016zud,Headrick:2017ucz}.} This is rather straightforward, because, in the presence of a $D$-dimensional spacetime metric $g_{\mu\nu}$, divergenceless vector fields $v$ map directly to closed $(D-1)$-forms $u$:
\begin{equation}\label{eq: flowFormDef}
\begin{aligned}
v^{\mu}& =g^{\mu\nu}(\star u)_{\nu}\;,  \\
 (\star u)_{\nu} & \equiv \frac{1}{(D-1) !} \sqrt{g} u^{\mu_{1} \ldots _{\mu_{D-1}}} \varepsilon_{\mu_{1} \ldots \mu_{D-1} \nu}\;,
\end{aligned}
\end{equation}
where $\star$ is the Hodge star, and $\varepsilon_{ \mu_1 \dots \mu_{D-1}\nu}$ represents the Levi-Civita symbol such that $\epsilon = \sqrt{g} \varepsilon$ is the volume form on manifold with metric $g_{\mu\nu}$. In terms of the natural volume form $\epsilon$,
\beq \epsilon=\frac{1}{D!}\epsilon_{\mu_{1}...\mu_{D}}dx^{\mu_{1}}\wedge...\wedge dx^{\mu_{D}}\;,\eeq
we formally express $u$ as
\beq u=\frac{1}{(D-1)!}\epsilon_{\mu_{1}...\mu_{D-1}\nu}v^{\nu}dx^{\mu_{1}}\wedge...\wedge dx^{\mu_{D-1}}\;.\label{eq:uformgen}\eeq
In components we have $u_{\mu_{1}...\mu_{D-1}}=\epsilon_{\mu_{1}...\mu_{D-1}\nu}v^{\nu}$.

It is now clear how to connect divergenceless vector fields $v$ to closed forms $u$. Taking the exterior derivative of (\ref{eq:uformgen}), one has
\beq du=(\nabla_{\mu}v^{\mu})\epsilon\;,\quad \nabla_{\mu}v^{\mu}=0\;\;\Leftrightarrow\;\; du=0\;.\eeq
The norm bound $|v|\geq 1$, meanwhile, now becomes $\langle u, u \rangle  \geq 1$, where $\langle u, \tilde{u} \rangle$ is a norm on the space of forms defined via
\begin{equation}
\langle u, \tilde{u}\rangle_{g}=\frac{1}{(D-1)!} g^{\mu_{1} \nu_{1}} \cdots g^{\mu_{D-1} \nu_{D-1}}u_{\mu_{1} \ldots \mu_{D-1}} \tilde{u}_{\nu_{1} \ldots \nu_{D-1}} \ .
\label{eq: normDefinitionForms}
\end{equation}
The future oriented condition $v^{0}>0$ is also easy to state as $g^{0\nu}(\star u_{\nu})>0$.\footnote{For spacetimes where $g^{0i}=0$ and $g^{00}<0$, it follows $\ast u_{0}<0$.}

We can now express our reformulation of CV duality (\ref{eq:cvreformin}) (equivalently, the min flow-max cut optimization program) in terms of $u$. To do this we write down the value of $u$ pulled back to a hypersurface $\Sigma$ of $M$ with normal covector $n_{\mu}$,
\begin{equation}
u|_{ \Sigma} = (n_\mu v^\mu ) \tilde{ \epsilon}\;,
\end{equation}
where $\tilde{ \epsilon}$ is the pull back of the volume form to $\Sigma$, $ \epsilon_{ \mu_1 \dots \mu_{d-1}\nu} = d \tilde{\epsilon}_{\left[\mu_{1} \ldots \mu_{D-1}\right.} n_{\nu]}$. In particular, when $\Sigma$ is a maximal volume slice, the norm bound is saturated on $\Sigma$, resulting in\footnote{Here we have already made use of Stokes' theorem in terms of differential forms, namely, $0=\int_{r}du=\int_{A}u-\int_{\Sigma}u$, where bulk region $r$ has boundary $\partial r=-(\Sigma/\partial M)$.}
\begin{equation}
\int_{\Sigma} u = \int_{ \Sigma} ( n_\mu v^\mu ) \tilde{ \epsilon} = \int_{ \Sigma} \tilde{ \epsilon} = \operatorname{ vol} ( \Sigma)\;,
\end{equation}
where we used $n_{\mu}v^{\mu}=1$; otherwise $n_{\mu}v^{\mu}\geq1$. Thus, in terms of forms, the full min-flow max-cut program reads
\beq \underset{u}{\text{min}}\int_{A}u=\underset{\Sigma\sim A}{\text{max}}\;\text{vol}(\Sigma)\;.\eeq
Consequently, the flow version of  CV duality (\ref{eq:cvreformin}) becomes
\beq \mathcal{C}(\sigma_{A})=\underset{u}{\text{min}}\int_{A}u\;,\eeq
where it is understood $u$ is a closed $(D-1)$-form.

We can also use differential forms to describe the perturbative Lorentzian threads introduced above. Particularly, for spacetime metrics perturbatively close to $g_{\mu\nu}$, \emph{i.e.}, $g_{\mu\nu}^{\eta} = g_{\mu\nu} + \eta\delta g_{\mu\nu}$, we denote the perturbed $(D-1)$ form by  $u^{\eta} = u + \eta\delta u$. The closedness condition of the solution $u^{\eta}$ implies
\begin{equation} \label{eq: perturbedClosed}
d ( u + \eta\delta u ) = 0 \implies d( \delta u ) =0\;,
\end{equation}
which is far simpler over the divergenceless condition (\ref{eq:annoydivcond}). Moreover, by direct calculation the norm bound $- \langle u^{\eta} , u^{\eta} \rangle$ is easily seen to be
\begin{equation}
\label{eq: perturbedNormBound}
- \langle u, u \rangle_g + \eta \left [ 2 \langle u, \delta u \rangle_g + \langle u, u \rangle_{ \delta g} \right]  \geq 1
\;.
\end{equation}
Lastly, using that to first order in $\eta$ the maximal volume slice $\Sigma$ is unaltered, we know
\begin{equation}\label{eq: peturbedVolume}
( u + \eta \delta u ) |_{\Sigma} = \tilde{\epsilon} + \eta\delta \tilde{ \epsilon} \implies \delta u|_{\Sigma} = \delta \tilde{\epsilon}.
\end{equation}
Hence, when studying linear perturbations of complexity around a background metric we seek to find a closed $(D-1)$ form $\delta u $ satisfying $\delta u|_{ \Sigma} = \delta \tilde{ \epsilon}$ and the norm bound in \eqref{eq: perturbedNormBound}.

\subsection{Symplectic structure  and York time: review} \label{subsec:sympstructrev}

Here we review  the central results of \cite{Belin:2018fxe,Belin:2018bpg}: the boundary dual of the bulk symplectic form and effect of `new York' time translations on the boundary. This review will introduce necessary tools to show an equivalence between the (Euclidean) bulk Einstein's equations and a boundary first law of complexity. Later we will show for linear perturbations of the metric, the `new York' transformation of the bulk symplectic form provides a canonical choice of $\delta u$ and hence a solution to the peturbed min-flow max-cut program and which also characterizes Einstein's equations.


\subsubsection{Boundary dual of the bulk symplectic form}\label{sec: boundaryDualOfBulkSymplecticForm}

Building off of the `piece-wise holography' developed in \cite{Skenderis:2008dh,Skenderis:2008dg} and its generalization to excited states \cite{Botta-Cantcheff:2015sav}, the authors of \cite{Marolf:2017kvq} demonstrated coherent states of the boundary CFT are dual to bulk coherent states, \emph{i.e.}, semi-classical geometries. More precisely, coherent CFT states $|\lambda\rangle$  with wavefunctionals prepared via a Euclidean path integral
\begin{equation}\label{eq: coherentstates}
|\lambda\rangle=T e^{-\int_{\tau<0} d \tau d^{d-1} \vec{x} \lambda_{\alpha}\left(\tau, \vec{x}\right) O_{\alpha}\left(\tau, \vec{x}\right)}|0\rangle\;,\quad  \langle\lambda|=\langle 0| T e^{-\int_{\tau>0} d t_{E} d^{d-1} \vec{x} \lambda^{\ast}_{\alpha}\left(-\tau, \vec{x}\right) O_{\alpha}^{\dagger}\left(\tau, \vec{x}\right)}
\end{equation}
are understood as bulk coherent states in that $|\lambda\rangle\sim e^{\lambda_{i}a^{\dagger}_{i}}|0_{\text{bulk}}\rangle$ where $|0_{\text{bulk}}\rangle$ is the bulk vacuum state, and $\{a_{i}^{\dagger}\}$ are a collection of commuting bulk creation operators, dual to derivative operators acting on the conformal descendants of CFT primaries $O(x)$. Thus, $\lambda$ acts as a source for $O(x)$ and the corresponding semi-classical geometries are understood to be states prepared by Euclidean path integrals with sources turned on.

\begin{figure}[t]
\centering
\includegraphics[scale=0.25]{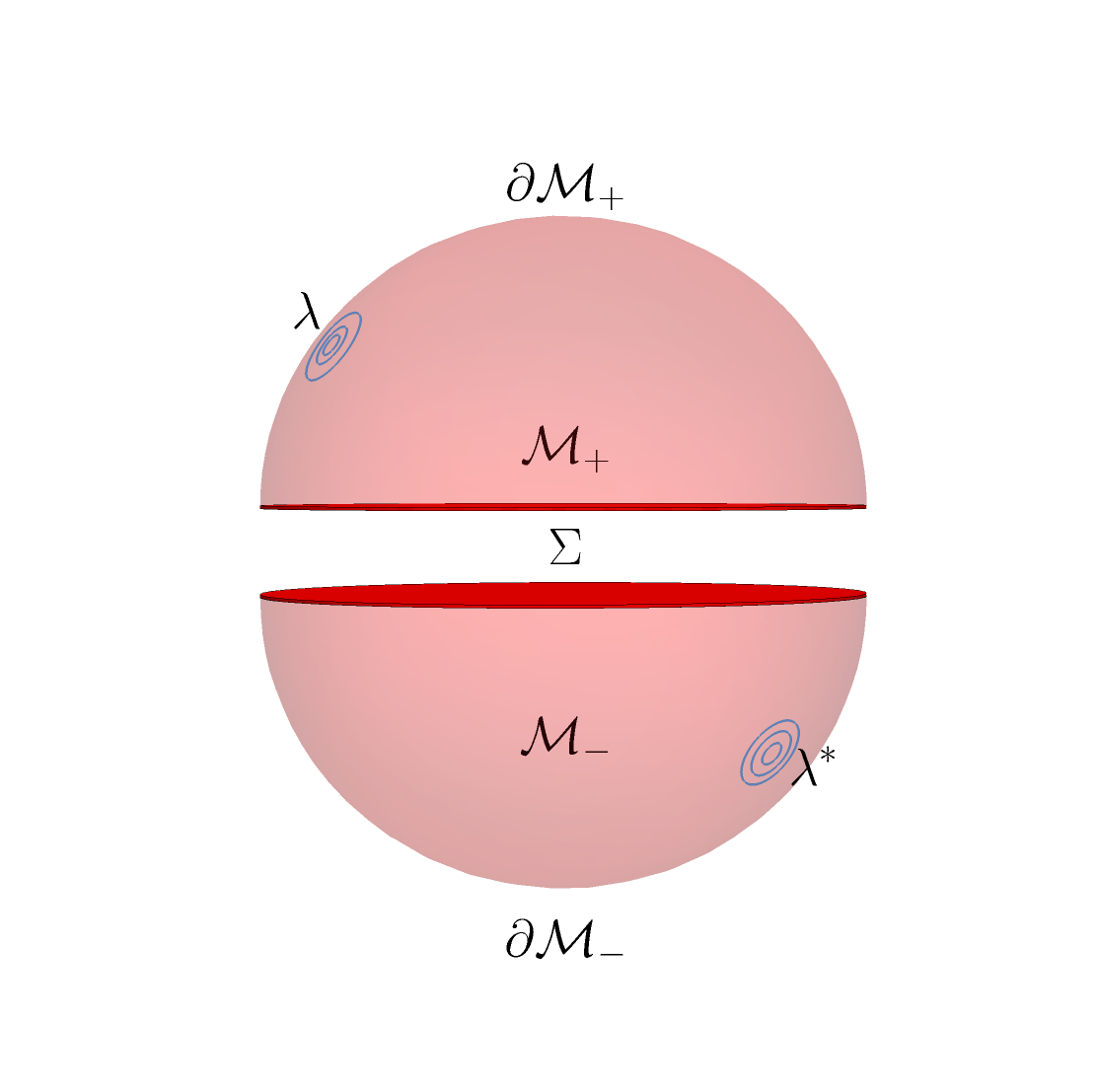}
\caption{Coherent state preparation. The CFT state $|\lambda \rangle$ represents a state prepared by a Euclidean path integral over the $\tau<0$ section of the boundary of Euclidean AdS $\partial \mathcal{M}_{-}$ whilst inserting sources $\lambda_{\alpha}$ on the boundary. The dual state $\langle \lambda|$ is prepared by integrating over the $\tau>0$ section of the boundary inserting the sources $\lambda^{\ast}(-\tau, x)$ and the inner product $\langle \lambda | \lambda \rangle$ is computed by gluing the two manifolds together.}
\label{fig: sourcePrep}
\end{figure}

To clarify, $|\lambda\rangle$ is understood to be a path integral over the boundary of Euclidean $\text{AdS}_{d+1}$ and is really the state prepared by initial data on the boundary of the southern hemisphere $S^{d-1}\times\mathbb{R}^{-}\sim B^{d}$, denoted $\partial\mathcal{M}_{-}$. The conjugate $\langle \lambda|$ corresponds to sources $\lambda^{\ast}$ inserted in the northern hemisphere $\mathcal{M}_{+}$. Since the two submanifolds $\partial\mathcal{M}_{\pm}$ are diffeomorphic, functions and operators in either section  can be extended to $\partial \mathcal{M}$ via the dualization map $\lambda(\tau,x) \to \lambda^{\ast}( -\tau, x)$, which may be combined into a single joint source profile $\tilde{\lambda}(x)$, which equals $\lambda(x)$ for $\tau<0$ and $\lambda^{\ast}(x^{T})$ for Euclidean time $\tau>0$. Lastly, the vacuum CFT wavefunction is represented by a Euclidean path integral over $\partial\mathcal{M}_{-}$ in terms of the Euclidean action $S^{\text{CFT}}_{E}$,  $|0\rangle\equiv\int_{\tau<0}[D\phi(\tau)]e^{-S^{\text{CFT}}_{E}}$. For an illustration, see Figure \ref{fig: sourcePrep}.

The key insight of  \cite{Belin:2018fxe} is that the mapping between boundary sources and initial data continues to persist at the level of the symplectic structure. That is, for holographic CFTs, the symplectic form on the classical phase space of gravitational configurations is dual to the CFT symplectic form obtained from the Fubini-Study metric pulled back to the space of Euclidean path integral states. To see this, note the coherent states (\ref{eq: coherentstates}) satisfy the holomorphicity condition
\beq \partial_{\lambda^{\ast}_{\alpha}}|\lambda\rangle=0\;,\quad \partial_{\lambda_{\alpha}}\langle\lambda|=0\;,\label{eq:holomorphiccond}\eeq
such that the space of coherent CFT states $\{|\lambda\rangle,\langle\lambda|\}$ is described by a K{\"a}hler manifold with K{\"a}hler potential and a closed K{\"a}hler 2-form, respectively,
\beq \mathcal{K}=\log\langle\lambda|\lambda\rangle\;,\quad  \Omega_{\text{B}}=i\partial_{\lambda_{\alpha}}\partial_{\lambda^{\ast}_{\alpha'}}\log\langle\lambda|\lambda\rangle d\lambda_{\alpha}\wedge d\lambda_{\alpha'}^{\ast}\;.\eeq
The  form $\Omega_{\text{B}}$ is a symplectic form,\footnote{In fact the symplectic form can be interpreted as the Berry curvature 2-form associated to the Berry connection on the complex manifold, $\mathcal{A}=i\langle\Psi_{\lambda}|d|\Psi_{\lambda}\rangle$ with $|\Psi_{\lambda}\rangle=\frac{|\lambda\rangle}{\sqrt{\lambda|\lambda\rangle}}$.} and thus the complex manifold is symplectic.  Denoting global coordinates on $\mathcal{M}$ by $\tilde{\lambda}=(\lambda,\lambda^{\ast})$, one has the symplectic form in terms of variations of $\tilde{\lambda}$ given by
\beq \Omega_{\text{B}}(\delta_{1}\tilde{\lambda},\delta_{2}\tilde{\lambda})=i\partial_{\lambda_{\alpha}}\partial_{\lambda_{\alpha'}^{\ast}}\log\langle\lambda|\lambda\rangle[\delta_{1}\lambda_{\alpha}\delta_{2}\lambda_{\alpha'}^{\ast}-\delta_{2}\lambda_{\alpha}\delta_{1}\lambda_{\alpha'}^{\ast}]\;.\eeq
Physically, the K{\"a}hler potential is interpreted  in terms of the partition function $Z_{\text{CFT}}[\tilde{\lambda}]\equiv\langle \lambda|\lambda\rangle$ of the CFT with sources, $\mathcal{K}=\log Z_{CFT}[\tilde{\lambda}]$, where now the K{\"a}hler potential is understood as a functional of half-sided sources $(\lambda,\lambda^{\ast})$, and its K{\"a}hler form becomes
\beq \Omega_{\text{B}}(\delta_{1}\tilde{\lambda},\delta_{2}\tilde{\lambda})=i(\delta_{1}^{\ast}\delta_{2}-\delta_{2}^{\ast}\delta_{1})\log Z_{CFT}[\tilde{\lambda}]= i \int_{\tau>0} d x\left(\delta \lambda_{1}^{*} \delta_{2}\langle O\rangle-\delta \lambda_{2}^{*} \delta_{1}\langle O\rangle\right)
\label{eq: boundarysymplecticform}\;.\eeq
The expression in \eqref{eq: boundarysymplecticform} can seen to localize on $\tau>0$ due to the $Z_2 + C$ symmetry of the sources, but we could write an equivalent expression in terms of quantities on $\tau<0$.

When the boundary CFT is holographic, the standard AdS/CFT dictionary yields
\beq \langle\lambda|\lambda\rangle=e^{-S_{E,\text{grav}}^{\text{on-shell}}[\tilde{\lambda}]}\;,\eeq
with $\mathcal{K}=-S_{E,\text{grav}}^{\text{on-shell}}$, and $\tilde{\lambda}$ set the boundary conditions for the bulk fields following the piece-wise holography prescription \cite{Skenderis:2008dh,Skenderis:2008dg}. In this holographic context, the authors of \cite{Belin:2018fxe} showed the K{\"a}hler form can be cast in terms of the bulk symplectic form,
\begin{equation}
\label{eq: bulktoboundarysymplecticform}
\Omega_{\text{B}}( \delta_1 \tilde{\lambda}, \delta_2 \tilde{\lambda} ) = i ( \delta_2^{\ast} \delta_1 - \delta^{\ast}_1 \delta_2 ) S^{\text{on-shell}}_{\text{grav}}[ \lambda, \lambda^{\ast}] = i \int_{\partial \mathcal{M}_-} \omega^{\text{E}}_{\text{bulk}}(\phi, \delta_1 \phi, \delta_2 \phi)\;.
\end{equation}

To see the above statement, recall that on a bulk $D$-dimensional spacetime $\mathcal{M}$ with volume form $\epsilon$, the Lagrangian $D$-form is $\mathbf{L}=L\epsilon$, where $L$ is the scalar Lagrangian, and is local in arbitrary (bulk) fields $\phi$. The variation of the (Euclidean) Lagrangian form is
\beq \delta \mathbf{L}=-E_{\phi}\delta\phi+d\theta(\phi,\delta\phi)\;,\label{eq: actionvariation}\eeq
where $E_{\phi}$ is a $D$-form characterizing the equations of motion of fields $\phi$, and $\theta$ is the symplectic potential, a $(D-1)$-form, or, equivalently on phase space, the symplectic one-form density.  For on-shell field configurations $E_{\phi}=0$, such that upon integrating (\ref{eq: actionvariation}) over the Euclidean AdS bulk $\mathcal{M}$, one has a boundary term
\beq \delta S_{E,\text{grav}}^{\text{on-shell}}[\tilde{\lambda}]=\int_{\partial \mathcal{M}=S^{d}}\theta(\tilde{\lambda},\delta\tilde{\lambda})\;.\eeq

From the joint source $\tilde{\lambda}$ profile, we have
\beq \delta \delta_{1}S^{\text{on-shell}}_{\text{grav}}=\int_{\tau<0}\theta(\tilde{\lambda},\delta\tilde{\lambda})\;,\quad \delta \delta^{\ast}_{1}S^{\text{on-shell}}_{\text{grav}}=\int_{\tau<0}\theta(\tilde{\lambda},\delta\tilde{\lambda})\;.\eeq
Then, using the bulk symplectic 2-form density, \emph{i.e.}, the symplectic current $(D-1)$-form,
 \beq \omega_{\text{bulk}}(\phi;\delta\phi_{1},\delta\phi_{2})=\delta_{1}\theta(\phi,\delta_{2}\phi)-\delta_{2}\theta(\phi,\delta_{1}\phi)\;,\label{eq:sympcurrent}\eeq
the holographic boundary symplectic form $\Omega_{\text{B}}$ (\ref{eq: bulktoboundarysymplecticform}) follows, where the extrapolate dictionary has been used to relate the sources $\tilde{\lambda}$ to the boundary values of the dual bulk fields $\phi$. We emphasize that in arriving to (\ref{eq: bulktoboundarysymplecticform}) we explicitly used that the bulk background fields $\phi$ are solutions to the equations of motion, $E_{\phi}=0$.

\begin{figure}[t]
\centering
\includegraphics[scale=0.7]{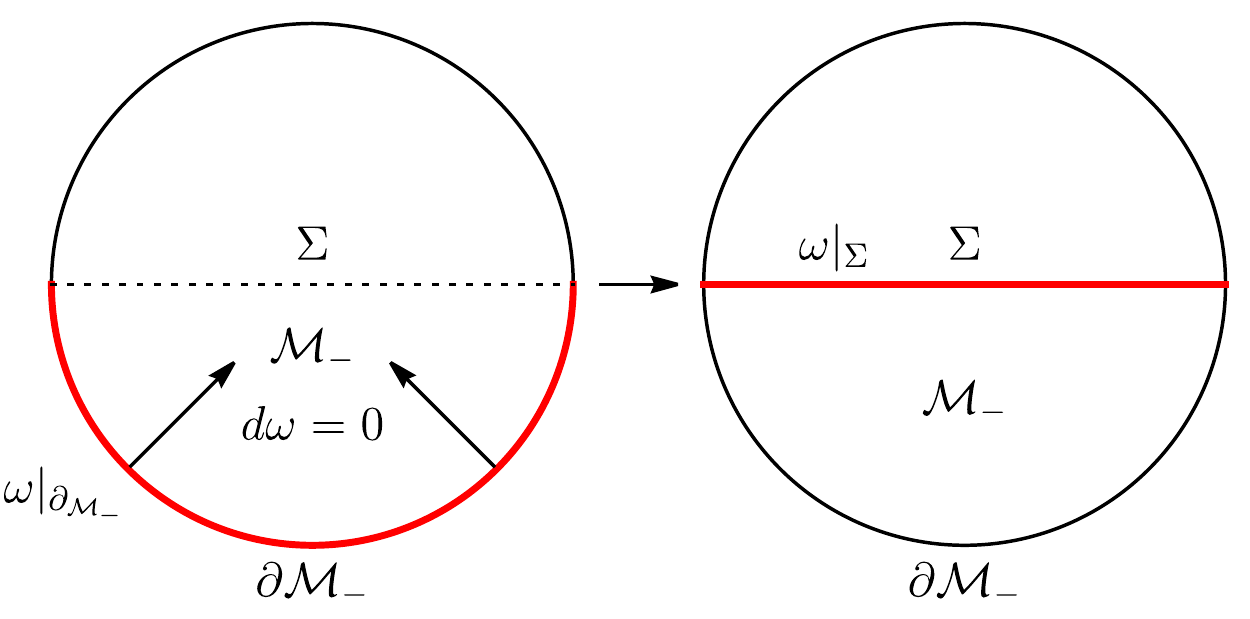}
\caption{Using the on-shell conservation of the bulk symplectic form $d \omega =0$, one can use the boundary-to-bulk Green's functions to push the boundary symplectic form to any surface $\Sigma$ anchored at $\tau=0$ on the boundary. If the fields continue smoothly to Lorentzian initial data on $\Sigma$, this gives an equality between the boundary symplectic form and the symplectic form on initial data, described by Equation \eqref{eq: boundaryEqualsInitial}. }
\label{fig: pushToBulk}
\end{figure}

Further, note that while the fields $\phi$ satisfy the bulk equations of motion, $E_{\phi}=0$, the generally independent variations $\delta_{1}\phi$ and $\delta_{2}\phi$ need not satisfy the equations of motion, even to linear order. However,  when the field variations $\delta_{1}\phi,\delta_{2}\phi$ are solutions to the \emph{linearized} equations of motion, the bulk symplectic 2-form $\omega_{\text{bulk}}$ in (\ref{eq:sympcurrent}) will be closed, $d\omega_{\text{bulk}}=0$. Indeed, since variations $\delta_{1}$ and $\delta_{2}$ commute and $d\delta\theta=\delta d\theta$, it is quick to see
\beq
\begin{split}
 d\omega_{\text{bulk}}&= (\delta_{1}E_{\phi})\delta_{2}\phi-(\delta_{2}E_{\phi})\delta_{1}\phi\;,
\end{split}
\eeq
which vanishes when $\delta_{1,2}E_{\phi}=0$ at linear order in $\delta_{1}\phi,\delta_{2}\phi$. When this occurs, $\omega_{\text{bulk}}$ can be `pushed' to other codimension-1 surfaces. Specifically, using the boundary-to-bulk Green's function, one can push the southern hemisphere $\partial\mathcal{M}_{-}$ to a surface $\Sigma$ anchored at $\tau=0$ in the Euclidean bulk $\mathcal{M}$ (see Figure \ref{fig: pushToBulk} for an illustration).  To relate this push to the bulk symplectic structure, the symplectic flux is integrated on a \emph{Lorentzian} initial value surface $\Sigma$ (a Lorentzian $t=0$ time slice). Consequently, the holographic dual of the boundary symplectic form is the bulk symplectic form of the initial data on the Lorentzian time slice:
\beq \Omega_{\text{B}}(\delta\tilde{\lambda}_{1},\delta\tilde{\lambda}_{2})=\int_{\Sigma}\omega^{\text{L}}_{\text{bulk}}(\phi,\delta\phi_{1},\delta\phi_{2})=\Omega^{\text{L}}_{\text{bulk}}(\phi,\delta_{1}\phi,\delta_{2}\phi)\;.
\label{eq: boundaryEqualsInitial}\eeq
Here $\omega^{\text{L}}_{\text{bulk}}$ denotes the Lorentzian  symplectic 2-form density.

\subsubsection{The `new York' time transformation}

The authors of \cite{Belin:2018fxe,Belin:2018bpg} further showed there is a special deformation of boundary sources, and hence bulk symplectic form, which gives rise to the change in volume on extremal surfaces. This deformation is dubbed the `new York' transformation due to the similarities with York time \cite{York:1972sj}. We briefly introduce this deformation and later we will show how it gives rise to a canonical thread construction.

In Einstein gravity, the bulk symplectic form can be directly read off from the variation of the action, \eqref{eq: actionvariation}. Working in Lorentzian signature, consider a $d+1$ split of the gravitational action, where the Lorentzian manifold $M$ is foliated by surfaces  $\Sigma_t$ of constant $t$, with normal $n_{\mu}$, where $\mu$ runs over the full spacetime index. The  induced metric on $\Sigma_{t}$ is $h_{\mu\nu}= g_{\mu\nu} + n_{\mu} n_{\nu} $, and the canonical momentum conjugate to $h_{\mu\nu}$ is
\beq \pi^{\mu \nu} = -\frac{\sqrt{h}}{16 \pi G_{N}}( K^{\mu \nu} - h^{\mu \nu} K )\;,\eeq
where $K_{\mu \nu} = h_{\mu}^{\lambda} \nabla_{\lambda} n_{\nu}$ is the extrinsic curvature to $\Sigma_t$ in $M$. Let $\xi^{\mu} = \delta^{\mu}_t$ be a timelike vector field responsible for $t$ translations. generated by the Hamiltonian $H_{\xi}$. The vector field $\xi$ in the ADM decomposition is represented as $\xi^{\mu} = -Nn^{\mu} + N^{ \mu}$, where the $t$-derivative is given by $\dot{h}_{\mu \nu} := h_{\mu}^{\alpha} h_{\nu}^{\beta} \mathcal{L}_{\xi} h_{\alpha \beta}$, with $\mathcal{L}_{\xi}$ being the Lie derivative along $\xi$.

Assuming bulk matter is minimally coupled, schematically write $\phi_m, \pi_m$ for the collection of matter fields and conjugate momenta, such that the gravitational action takes the form
\begin{equation}
S_{\text{grav}} = \int dt  d^d x \left[ \pi^{\mu \nu} \dot{h}_{\mu \nu} + \pi_m \dot{\phi}_m - H_{ \xi}\right]\;,
\end{equation}
where $H$ is the ADM Hamiltonian. The gravitational pre-symplectic form $\theta_{g}$ is read off from the metric variation of $S_{\text{grav}}$, $\theta_g = \pi^{\mu \nu} \delta h_{ \mu \nu}$, from which the bulk symplectic 2-form density (a $d$-form on $M$) is defined by
\begin{equation}\label{eq: gravitationalSymplecticDensity}
\omega^{\text{L}}(\delta_1 \phi, \delta_2 \phi) = \delta_1 \pi^{\mu \nu} \delta_2 h_{\mu \nu} - \delta_2 \pi^{\mu \nu} \delta_1 h_{\mu \nu}\;.
\end{equation}
The pullback of $\omega^{\text{L}}_{\text{bulk}}$ to the hypersurface $\Sigma_t$ one has the bulk symplectic form
\begin{equation}
\Omega^{\text{L}}_{\text{bulk}}( \delta_1 \phi, \delta_2 \phi) = \int_{\Sigma_t} ( \delta_1 \pi^{ij} \delta_2 h_{ij} - \delta_2 \pi^{ij} \delta_1 h_{ij})
\end{equation}
where we use the $i,j$ indices to emphasize that $ h_{ij}, \pi^{ij}$ are the (proper) induced metric and its conjugate momentum, which live as tensors on $\Sigma_{t}$ (as opposed to being tensors in $M$ lying tangent to hypersurfaces $\Sigma_t$) and we write $L$ to remind us that we are now in the Lorentzian bulk. Conservation of the bulk symplectic form follows from the important fact that
\begin{equation}
d \omega^{\text{L}}_{\text{bulk}} = \delta_1 E_{\phi}\delta_2 \phi - \delta_2 E_{\phi}\delta_1 \phi
\end{equation}
vanishes, so long as the perturbations are on-shell. To be on-shell, the phase space coordinates must satisfy the Hamiltonian constraint $\mathcal{H}  = \frac{\delta H }{\delta N }=0$ and momentum constraint $\mathcal{H}_{i} = \frac{\delta H}{\delta N^{i} } =0$, or, respectively, working with variables pulled back to $\Sigma_{t}$,
\begin{equation} \label{eq: covarianthamiltonianMomentum Constraints}
\begin{aligned}
\mathcal{ H} & = \frac{1}{16 \pi G_{N}} \left[ -R_d +2 \Lambda + (16 \pi G_{N})^{2} h^{-1} \left( \pi_{ij} \pi^{ij} -\frac{1}{d-2} \pi^{2} \right) \right] + \mathcal{H}_m =0\;,\\
\mathcal{H}^i& = -2 D_{j} (  h^{-1/2} \pi^{ij}) + \mathcal{H}^{i}_m=0\;.
\end{aligned}
\end{equation}
Here $\mathcal{H}_m, \mathcal{H}_m^{i}$ are the matter contributions of the lapse and shift variations. The momentum constraint is associated with diffeomorphisms inside the surface $\Sigma_{t}$, while the Hamiltonian constraint is associated with diffeomorphisms which change the initial value surface.

The tensors $h_{ij}$ and $\pi^{ij}$ may not be `good' phase space variables
; a `good' set of phase variables are those which satisfy the momentum and Hamiltonian constraints. It turns out the momentum constraint may always be satisfied by fixing a gauge, while the Hamiltonian constraint is not always solvable. However, when one provides initial data on a surface with constant mean curvature, there is a general method for solving the constraint developed by York \cite{York:1972sj}. The basic idea is to separate the induced metric into a scale captured by the volume element $\sqrt{h}$ on $\Sigma_{t}$, and a conformal metric $\bar{h}_{ij}=|h|^{-1/d}h_{ij}$. In these new variables, one has $\pi^{ij}\delta h_{ij}=\pi_{V}\delta\sqrt{h}+\bar{\pi}^{ij}\delta \bar{h}_{ij}$, where
\beq \pi_{V}=\frac{2(1-d)}{d}K\;,\quad \bar{\pi}^{ij}=|h|^{\frac{1}{d}+\frac{1}{2}}(K_{ij}-\frac{1}{d}Kh^{ij})\;.\eeq
When $\pi_{V}$ is constant, the Hamiltonian constraint may be interpreted as a differential equation in the volume density $\sqrt{h}$, known as the Lichnerowicz equation. This equation is solvable and allows us to interpret $\sqrt{h[\bar{h},\pi_{V},\bar{\pi}]}$ as a functional of the remaining phase space variables. Working in a constant mean curvature (CMC) slicing, where each slice has constant $K$, $\pi_{V}$ is a number parametrizing each of the slices and can be interpreted as time -- the York time. Meanwhile, the volume $V=\int\sqrt{h}$ can be understood as a Hamiltonian for the remaining variables.

The authors of \cite{Belin:2018fxe,Belin:2018bpg} utilized this decomposition due to York and provided a boundary interpretation of $V$. In particular, they showed a special deformation of the Euclidean boundary data, and therefore the bulk symplectic form, gave rise to the change of volume of maximal slices $\Sigma$. This transformation was dubbed the `new York' transformation, denoted by $\delta_Y$.  Explicitly, in terms of the new variables,
\beq \delta_{Y}\pi_{V}=2(d-1)\alpha\;,\quad \delta_{Y}\bar{\pi}^{ij}=\delta_{Y}\sqrt{h}=\delta_{Y}\bar{h}_{ij}=0\;,\eeq
where $\alpha$ is some constant unrelated to the normalization of our Lorentzian flows. In terms of the extrinsic curvature, the above reads
\begin{equation} \label{eq: yorkTransformation}
\delta_{Y}h_{ij}=0\;,\quad \delta_{Y} K_{ij}=\alpha h_{ij}\;.
\end{equation}
Since the new York transformation is generically not a diffeomorphism, it does not evolve the gauge invariant initial data in (York) time. Rather $\delta_{Y}$ copies the initial data to a neighboring slice. We will see, however, for deformations about empty AdS, the new York transformation (\ref{eq: yorkTransformation}) is indeed a diffeomorphism.

One can directly verify this leads to the variation of the bulk symplectic form \cite{Belin:2018fxe,Belin:2018bpg}
\begin{equation}
\Omega_{\text{B}}(\delta_{Y}\tilde{\lambda},\delta\tilde{\lambda})=\Omega_{\text{bulk}}^{\text{L}}( \delta_Y \phi, \delta \phi) = \int_{\Sigma} \frac{(d-1) \alpha}{8 \pi G_{N}} \delta(\sqrt{h}) = \frac{(d-1) \alpha}{8 \pi G_{N}}  \delta V\;,
\label{eq:bulksymdeltav}\end{equation}
where $V=\int\sqrt{h}$ is the spatial volume of the slice $\Sigma$.\footnote{In a less generic context, the relation (\ref{eq:bulksymdeltav}) was also derived in detail \cite{Jacobson:2018ahi} (see their Appendix B). Nicely, the parameter $\alpha$ appearing in the York deformation is shown to be equal to the proper velocity of the conformal factor defining a conformal Killing vector generating a causal diamond in a maximally symmetric background evaluated on a $s=0$ slice of conformal Killing time $s$. That is, the conformal Killing transformation and new York transformation coincide on the maximal volume slice.}

In general the deformation (\ref{eq: yorkTransformation}) will not be on-shell, \emph{i.e.}, the Hamltonian and momentum constraints are not generally preserved under $\delta_Y$. In particular, while the momentum constraint is automatically preserved, the  Hamiltonian constraint reads
\begin{equation}
\delta_Y \mathcal{H} = 2 ( d-2) K\;.
\end{equation}
Thus, $\delta_{Y}$ is an on-shell perturbation of initial data provided the deformation occurs on a maximal slice $\Sigma$, where $K=0$. In other words, $\delta_{Y}$ is on-shell when $V$ is the volume of the maximal hypersurface $\Sigma$. In choosing $\alpha$ such that the coefficient becomes $\alpha\equiv (8\pi/\ell(d-1))$, we see by CV duality $\Omega_{\text{bulk}}^{\text{L}}(\delta_{Y}\phi,\delta\phi)$ encodes a notion of varying complexity, $\delta\mathcal{C}$. More carefully, by the equivalence of boundary and bulk symplectic forms, $\delta V$ is equal to the boundary symplectic form (on $\partial\mathcal{M}$) and may be interpreted as  type of field theory complexity. We will return to this point later on.



\subsubsection*{The new York deformation of sources in vacuum AdS} \label{sec: NYVaccumAdS}

Since it will be useful momentarily, let us consider the  new York deformation of sources on the Euclidean boundary which will give rise to the new York deformation on $\Sigma$. It is easiest to work with a CMC slicing in Wheeler De Witt coordinates \cite{Belin:2018bpg}, but it shall prove useful for us to study the transformation in Euclidean Poincar\'e coordinates, with Euclidean time $\tau$,
\begin{equation}
ds^2 = \frac{1}{z^2}( d\tau^2 + dz^2 + dx^i dx^i )\;.
\label{eq:adspoinc}\end{equation}
In vacuum AdS, the new York deformation is a diffeomorphism \cite{Belin:2018bpg} (and in fact equivalent to translation in York time) generated by the vector field
\begin{equation}
 \xi^{\mu} \partial_{\mu} = \frac{\alpha  z^2}{\sqrt{\tau^2+z^2}} \partial_\tau - \frac{\alpha  \tau z}{\sqrt{\tau^2+z^2}}\partial_z\;,
\label{eq:vecfchiyork}\end{equation}
under which the metric changes according to
\begin{equation}
\label{eq: NYBulk}
\delta_Y g_{\mu \nu}  = \frac{2 \alpha  \tau^3}{z^2 \left(\tau^2+z^2\right)^{3/2}}d\tau^2  +\frac{4 \alpha  \tau^2}{z \left(\tau^2+z^2\right)^{3/2}}d\tau dz + \frac{2 \alpha  \tau}{\left(\tau^2+z^2\right)^{3/2}}dz^2  + \frac{2 \alpha  \tau}{z^2 \sqrt{\tau^2+z^2}} dx^i dx^i .
\end{equation}
The fact $\delta_{Y}$ acts as a diffeomorphism in vacuum AdS will prove crucial for our derivation of Einstein's equations.

Near the boundary, the new York transformation acts like
\begin{equation}
\label{eq: NYBoundary}
\delta_{Y}|_{z\to0}\to-\alpha z \,\text{sign}(\tau)\, \partial_z\;,
\end{equation}
such that $\delta_{Y}$ on the boundary metric $\gamma_{ \mu \nu}$ retains the $Z_2 +C$ symmetry under the new York deformation. Thus, near the boundary, the transformation \eqref{eq: NYBoundary} acts as a Weyl-like transformation. Imposing a cut-off at $z= \epsilon$, one can read off the sources from the change in boundary metric due to the new York transformation explicitly
\begin{equation}
\delta_Y \gamma =  \frac{2 \alpha  \tau^3}{\left(\tau^2+\epsilon^2\right)^{3/2}}d\tau^2 + \frac{2 \alpha  \tau}{ \sqrt{\tau^2+\epsilon^2}} dx^i dx^i \;.
\end{equation}


\subsection{Einstein's equations from varying complexity} \label{sec: EEfromComplexity}

We now have all of the ingredients to explicitly derive the linearized Einstein's equations from varying complexity. Before we demonstrate this, however, it is worthwhile to compare our argument with previous derivations of gravitational equations of motion from physical principles other than varying the action.

\subsubsection{Comparison to other derivations of Einstein's equations}

Most famously, perhaps, is Jacobson's derivation of the full non-linear Einstein's equations from the Clausius relation \cite{Jacobson:1995ab}. In this formulation, one assumes: (i) local holography, such that local patches of spacetime carry a thermodynamic entropy proportional to the area of local Rindler horizons and a local Unruh-Davies temperature, and (ii) the Clausius relation $Q=T\delta S$, where $T$ is the Unruh-Davies temperature associated with locally accelerating observers, holds. Here $Q$ corresponds to the heat seen by accelerating observers and depends on the matter stress tensor. These two inputs are enough to show
\beq Q=T\delta S\Rightarrow G_{\mu\nu}+\Lambda g_{\mu\nu}=8\pi G_{N}T_{\mu\nu}\;,\eeq
where the right hand side is the full non-linear Einstein's equations. This argument holds for other theories of gravity, where the entropy $S$ is replaced with the Wald entropy functional (see, \emph{e.g.}, \cite{Padmanabhan:2007en,Guedens:2011dy,Dey:2016zka,Parikh:2017aas}). Thus, gravity emerges from spacetime thermodynamics. A similar idea appeared in \cite{Verlinde:2010hp}, where gravity was supplanted with an `entropic force'.

The spirit of Jacobson's argument has also been utilized in the context of spacetime entanglement \cite{Lashkari:2013koa,Faulkner:2013ica,Faulkner:2017tkh,Haehl:2017sot}. The basic idea is that the first law of entanglement, $\delta S_{A}=\delta\langle H_{A}\rangle$,  is dual to the linearized Einstein's equations due to linearized perturbations on an AdS background.  More precisely, for small perturbations over a reference state $\rho_{A}=\rho_{A}^{(0)}+\lambda\delta\rho$, where one takes $\rho_{A}^{(0)}$ to be the vacuum CFT state confined to a ball $A$ of radius $R$, the modular Hamiltonian $H_{A}\equiv e^{-\rho_{A}}$ may be cast in terms of the CFT stress-energy tensor \cite{Hislop:1981uh,Casini:2011kv}, such that, for holographic CFTs dual to Einstein gravity, the leading order variation $\delta \langle H_{A}\rangle$ is given in terms of bulk metric fluctuations evaluated on the boundary. The leading order variation of the entanglement entropy $\delta S_{A}$, meanwhile, is given by the Ryu-Takayangi prescription, cast in terms of the volume form $\epsilon$ on the RT surface $\gamma_{A}$.

Rather elegantly, the Iyer-Wald formalism \cite{Iyer:1994ys,Wald:2005nz} can be used to define a $(d-1)$-form $\chi$ in terms of a timelike Killing vector $\xi$, the volume $(d-1)$-form $\epsilon$, and the metric fluctuation such that
\beq \delta S^{\text{grav}}_{A}=\int_{\gamma_{A}}\chi\;,\quad \delta E^{\text{grav}}_{A}=\int_{A}\chi\;\label{eq:entmodchi}\eeq
where the nomenclature is as above: $S_{A}^{\text{grav}}$ is the holographic dual of the boundary entanglement entropy, and $E_{A}^{\text{grav}}$ is the holographic dual of the expectation value of the boundary modular Hamiltonian.  Evaluating $\chi$ on a constant $t$ Cauchy slice containing $\gamma_{A}$ and $A$, one finds
\beq d\chi=-2\xi^{t}\delta E^{g}_{tt}\epsilon^{t}\;,\eeq
where $\delta E^{g}_{tt}$ is the $tt$ component of the linearized Einstein's equations. The other components to Einstein's equations may be obtained by considering other Cauchy slices. Thus, when $\chi$ is closed, the linearized Einstein's equations are satisfied. By Stokes' theorem and (\ref{eq:entmodchi}), when $\chi$ is closed one has
\beq \delta S^{\text{grav}}_{A}=\delta E^{\text{grav}}_{A}\Rightarrow \delta E^{g}_{\mu\nu}=0\;.\eeq
In other words, for holographic field theories where the RT prescription holds, the first law of entanglement is dual to the bulk linearized Einstein's equations being satisfied in perturbations about empty AdS. In this way, gravity emerges from `spacetime entanglement'.\footnote{A distinct but related equivalence between Einstein's equations and the first law of entanglement is given in terms of Jacobson's entanglement equilibrium proposal \cite{Jacobson:2015hqa}, where the first law of entanglement gives the non-linear Einstein's equations, linearized higher curvature field equations \cite{Bueno:2016gnv}, and is intimately tied to the principles of spacetime thermodynamics \cite{Svesko:2018qim}.}

Here we propose a new derivation of the linearized Einstein's equations, involving varying complexity. In lieu of CA duality, it is natural to expect Einstein's equations will arise from varying complexity, as the complexity is dual to the bulk gravitational action. This was argued for in \cite{Czech:2017ryf} by showing vacuum solutions of pure Einstein gravity in $\text{AdS}_{3}$ arise from optimizing path integral complexity. Much was left to be desired, however, as it was not clear how the covariant nature of Einstein's equations could be made manifest. We thus seek an argument similar to \cite{Lashkari:2013koa,Faulkner:2013ica} which shows the leading variation of the complexity is equivalent to Einstein's equations linearized about AdS. What is needed is a quantity that relates the form $\chi$ to the variation of complexity. As already eluded to, such a quantity is given by the bulk symplectic form, evaluated with respect to the new York time translation. Then, using the fact the bulk symplectic form is dual to the boundary symplectic form \cite{Belin:2018fxe}, which is equal to the variation of complexity \cite{Belin:2018bpg}, we will show the first law of holographic complexity is equal to the linearized Einstein's equations. Thus, Einstein's equations arise from `spacetime complexity'.\footnote{In using `complexity=momentum' proposal \cite{Susskind:2018tei,Susskind:2020gnl}, it was shown the attractive nature of gravity, as understood by Newton's laws, are a consequence of the inevitable increase in complexity \cite{Susskind:2019ddc}. This argument has some similarities with the entropic force viewpoint advocated in \cite{Verlinde:2010hp}.} It is important to note our argument relies on CV duality and is motivated by the derivation invoking the first law of entanglement  \cite{Lashkari:2013koa,Faulkner:2013ica}.


\subsubsection{Varying complexity and Einstein's equations}

Above we reviewed the equivalence between the boundary and bulk symplectic forms $\Omega_{\text{B}}$ and $\Omega^{\text{L}}_{\text{bulk}}$ at the conformal boundary. The new York transformation of the bulk symplectic form (\ref{eq:bulksymdeltav}) was shown to be proportional to the variation of the spatial volume of the Cauchy slice $\Sigma$, which for on-shell perturbations, is the maximal volume hypersurface in Lorentzian AdS. Assuming the complexity=volume conjecture, the new York transformation therefore provides a first law relating $\delta V$ to the boundary symplectic form $\Omega_{\text{B}}$, where $\Omega_{\text{B}}$ is interpreted as the variation of the boundary complexity, $\Omega_{\text{B}}( \delta_Y\tilde{\lambda}, \delta\tilde{\lambda}) = \delta \mathcal{C}$. Precisely, the duality between boundary and bulk symplectic forms in addition to the fact the boundary symplectic form is pushed to the maximal bulk slice when the metric perturbations satisfy the linearized  bulk gravitational equations of motion, tells us
\beq \delta E_{\mu\nu}=0\Rightarrow \delta V=\Omega_{\text{B}}\;,\label{eq: 1stlaw}\eeq
where one further identifies $\delta\mathcal{C}=\frac{1}{8G_{N}\ell}\delta V$ upon assuming CV duality.

Here we show the converse statement: assuming CV duality, the holographic first law (\ref{eq: 1stlaw}) implies the (linearized) Einstein's equations must be satisfied. Our argument can be viewed in a similar vein as gravity from entanglement \cite{Lashkari:2013koa,Faulkner:2013ica}, but rather, gravity from complexity. Unlike the derivation given by  \cite{Lashkari:2013koa,Faulkner:2013ica}, or the thermodynamic reasoning of Jacobson \cite{Jacobson:1995ab}, a weakness of our argument is that the first law of complexity is not well established. That is, while  the first law of entanglement and the Clausius relation are firmly established principles which, when applied in a particular context, have a gravitational character, the first law of complexity lacks the same robustness. In part this is due to a lack of a precise definition of field theory complexity.

There have been recent advances in establishing a first law of complexity, however. One adapts Nielsen's geometrization program for circuit complexity to field theories \cite{Bernamonti:2019zyy,Bernamonti:2020bcf}. Under a change in target state, the first order variation of complexity is given by a `cost' function $F(x^{a},\dot{x}^{a})$, which depends on paths $x^{a}(s)$ and tangent vectors $\dot{x}^{a}$ to these paths in the space of unitaries defining the circuit, and satisfies a number of niceness conditions \cite{Jefferson:2017sdb,Hackl:2018ptj}.

In the set-up under consideration here, namely, the space of boundary souces $\lambda$, the boundary symplectic form has a natural interpretation as `complexity', without invoking CV duality. As argued in \cite{Belin:2018bpg}, distances in the space of sources are given in terms of the K{\"a}hler metric $g_{ab}=(\delta^{+}_{a}\delta^{-}_{b}+\delta_{b}^{+}\delta_{a}^{-})\log Z[\tilde{\lambda}]$, where the minimal path in this space is found by minimizing the `cost' function $F[g_{ab}\dot{\lambda}^{a}\dot{\lambda}^{b}]$. Assuming the cost function is the kinetic energy,\footnote{The kinetic energy is $F[y]=y$ and is additive for tensor product states. A natural alternative is the geodesic distance $F[y]=\sqrt{y}$, however, it is not generally additive \cite{Belin:2018bpg,Brown:2017jil}.} the boundary complexity associated with the Euclidean path integrals is given by
\beq \mathcal{C}(s_{i},s_{f})=\int_{s_{i}}^{s_{f}}ds g_{ab}\dot{\lambda}^{a}\dot{\lambda}^{b}\;,\eeq
where $s$ parametrizes trajectories in the space of sources. Thus, finding the complexity is analogous to identifying a `particle' trajectory minimizing its kinetic energy. Moreover, using the definition of the boundary symplectic form, varying the complexity with respect to the endpoint coordinate $\lambda(s_{f})\equiv\lambda_{f}$ results in the following first law of complexity \cite{Belin:2018bpg}:
\beq \delta_{\lambda_{f}}\mathcal{C}=(\dot{\lambda}^{a}|_{\lambda_{f}})g_{ab}\delta\lambda^{b}_{f}\;.\eeq
Provided the reference state $\lambda_{i}$ does not equal the target state $\lambda_{f}$ (for which $\delta\mathcal{C}=0$), the variation in complexity is expected to be non-zero.  In what follows, it will be this boundary first law we have in mind. In fact, when the tangent vector to the minimal trajectory in source space $\dot{\lambda}^{a}|_{\lambda_{f}}$ is identified with the complex structure of the K{\"a}hler manifold $J[\delta_{Y}\lambda]$, then
\beq \delta_{\lambda_{f}}\mathcal{C}=\Omega_{\text{B}}(\delta_{Y}\tilde{\lambda},\delta\tilde{\lambda})\;.\label{eq:bdryfirstlaw}\eeq
Note that this is entirely a boundary statement.  For holographic CFTs, the equivalence between boundary and bulk symplectic forms leads to $\delta \mathcal{C}\sim\delta V$ on the maximal volume slice $\Sigma$.\footnote{As pointed out in \cite{Belin:2018bpg}, around the AdS vacuum the variation $\delta V$ is purely divergent. This is remedied by choosing the reference state to be the vacuum state, \emph{i.e.}, $\lambda_{i}=0$.}

Our argument is now easy to state. We claim that when the holographic version of the boundary first law (\ref{eq:bdryfirstlaw}) holds, where in lieu of CV duality, $\delta \mathcal{C}\sim \delta V$, then the linearized Einstein's equations are required to hold in the bulk. That is, assuming (i) the first law of complexity, and (ii)  CV duality, then
\beq \delta V=\Omega_{\text{B}}\Rightarrow \delta E_{\mu\nu}=0\;,\eeq
where $\delta E_{\mu\nu}=0$ is the linearized Einstein's equations for perturbations about vacuum AdS.

\begin{figure}[t]
\centering
\includegraphics[scale=0.4]{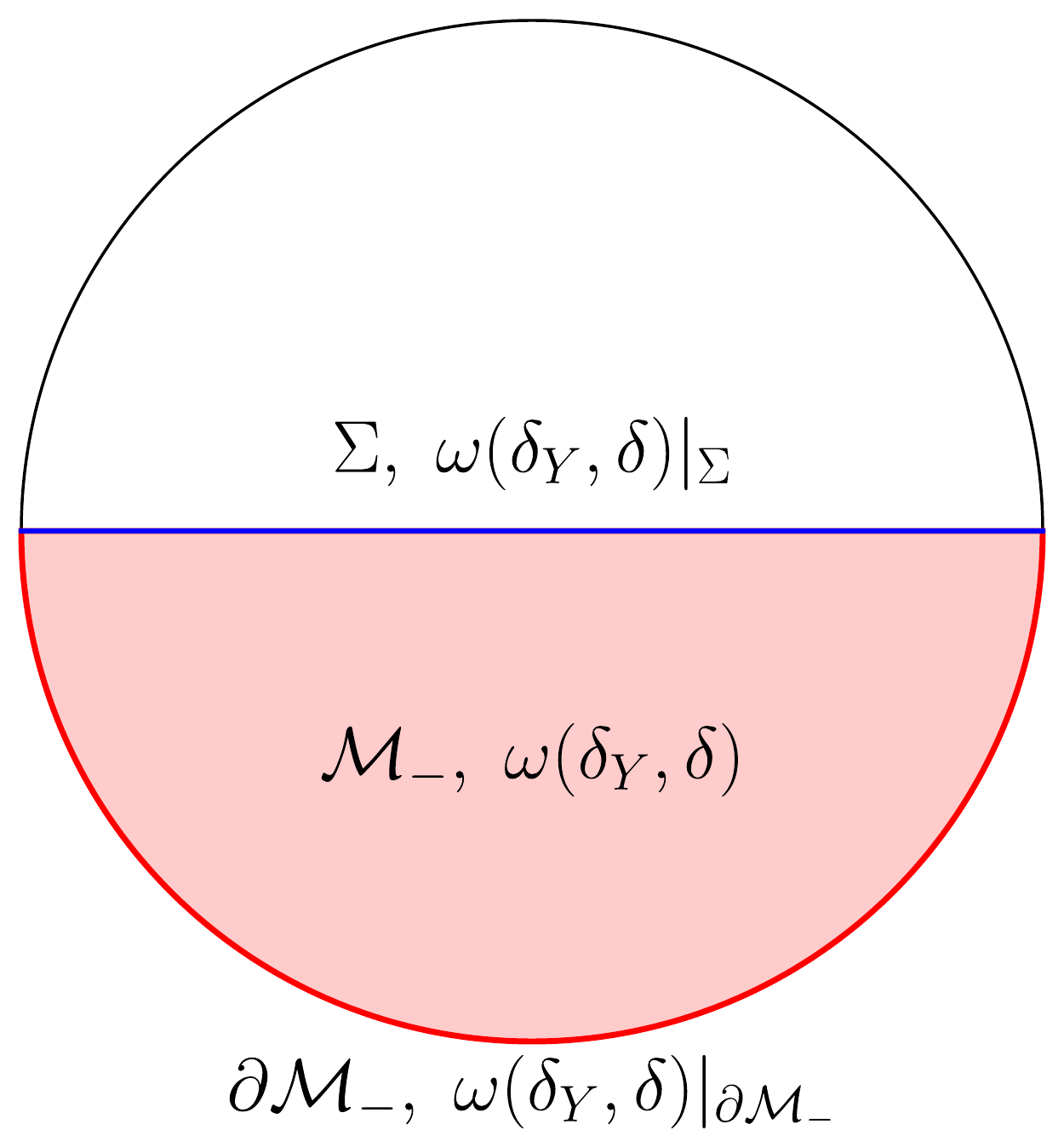}
\caption{An illustration showing how we relate the boundary symplectic form to the symplectic form on $\Sigma$ through $d\omega$ in the bulk.}
\end{figure}

Let us now begin our explicit derivation. We start with the following identity
\begin{equation}\label{eq: integralofdomega0}
i \int_{\mathcal{M}_{-}} d \omega^{\text{E}}_{\text{bulk}} = i \int_{\partial \mathcal{M}_{-}} \omega^{\text{E}}_{\text{bulk}} - i   \int_{\Sigma} \omega^{\text{E}}_{\text{bulk}}\;,
\end{equation}
which must  hold for all variations that yield real Lorentzian initial data on $\Sigma$. To derive the Einstein's equations from (\ref{eq: integralofdomega0}), it is sufficient to consider a subset of variations\footnote{More generally we could consider arbitrary variations $\delta$ in the bulk which preserve the $Z_2 + C$ symmetry at the boundary, but that is not necessary for our purposes.} in the bulk of $\mathcal{M}$, namely, $\delta = \delta^+ + \delta^-$, where $\delta^{\pm}$ are the variations which  localize to $\mathcal{M}_{\pm}$, respectively, and $\delta^{+}(\tau,x) = (\delta^{-})^{\ast}(-\tau,x)$. For the variations to agree on $\Sigma$ means $\delta^{+} \phi(t,x)|_{ \Sigma}  =\delta^{-} \phi(t,x)|_{ \Sigma}$ and $ \partial_t (\delta \phi^{+} )^{\ast}|_{\Sigma}  = - \partial_t ( \delta \phi^{-})|_{ \Sigma} $ from which it follows that `good' Lorentzian initial data $(\varphi_{L},\pi_{L})$ can be defined by
 \begin{equation}
 \delta \varphi_L = \text{Re} [ (\delta^+ \phi)|_{ \Sigma} ]\;,\quad  \delta \pi_L = \text{Im} [ (\partial_t \delta^+ \phi)|_ \Sigma]\;.
 \end{equation}
It follows
\begin{equation}\label{eq: integralofdomega}
i \int_{ \mathcal{M}_-} d \omega^{\text{E}}_{\text{bulk}} = i \int_{\partial\mathcal{M}_-} \omega^{\text{E}}_{\text{bulk}} - i   \int_{\Sigma} \omega^{\text{E}}_{\text{bulk}}   =  \Omega_{\text{B}} ( \delta_Y\tilde{\lambda}, \delta\tilde{\lambda}) - \int_{ \Sigma} \omega^{\text{L}}_{\text{bulk}} ( \delta_Y \phi, \delta\phi)\;,
\end{equation}
where the last equality appears since we wish to perturb around Lorentzian initial data cancelling the factor of $i$, and where we used (\ref{eq: bulktoboundarysymplecticform}) to exchange the integral over $\omega^{\text{E}}_{\text{bulk}}$ for $\Omega_{\text{B}}$.\footnote{We remind the reader that in this step we made use of the fact the background is a solution to Einstein's equations, $E_{\phi}=0$. Here we are aiming to show $\delta E_{\phi}=0$, which does not trivially follow from $E_{\phi}=0$ generally.}

Now, in the case of Einstein gravity,  we know from the new York transformation the Lorentzian symplectic 2-form density $\omega^{\text{L}}_{\text{bulk}}$ is equal the variation of the spatial volume on $\Sigma$,
\begin{equation}
\int_{ \Sigma} \omega^{\text{L}}_{\text{bulk}} ( \delta_Y \phi, \delta\phi) = \delta V\;.
\label{eq:omegaLanddeltav}\end{equation}
Collectively then, we have
\begin{equation}
i\int_{ \mathcal{M}_-} d \omega_{\text{bulk}}^{\text{E}}( \delta_Y \phi, \delta\phi) =\Omega_{\text{B}} ( \delta_Y \tilde{\lambda}, \delta\tilde{\lambda}) - \delta V\;.
\end{equation}
When the right hand side vanishes, $\Omega_{\text{B}}=\delta V$, it follows $d \omega_{\text{bulk}}^{\text{E}}( \delta_Y, \delta)$ must vanish in all of $\mathcal{M}_-$. Similarly, since the background field is real and we are considering the subset of transformations for which $\delta^{+}(t,x) = \delta^{-}(-t,x)$, $d \omega( \delta_Y, \delta)$ must also vanish in all of $\mathcal{M}_+$, \emph{i.e.}, everywhere inside the manifold where the path integral is prepared.\footnote{As we will show in the next section, $\delta_Y g_{ \mu \nu} |_{ \Sigma}$ vanishes due to the new York transformation and so $d \omega_{\text{bulk}}^{\text{E}}( \delta_Y, \delta)$ must vanish in all of $\mathcal{M}$.}

Let us now explore the consequences of $d \omega_{\text{bulk}}^{\text{E}}( \delta_{Y}\phi, \delta\phi)=0$. Taking the exterior derivative of the symplectic form density \eqref{eq: gravitationalSymplecticDensity} and substituting in the variation \eqref{eq: actionvariation} we arrive at
\begin{equation}
d \omega_{\text{bulk}}^{\text{E}}( \delta_Y \phi, \delta\phi) = \delta_Y E_{\phi} \delta \phi - \delta E_{\phi} \delta_Y \phi .
\label{eq: bulkintegral}
\end{equation}
As discussed in Section \ref{sec: NYVaccumAdS}, the new York deformation $\delta_{Y}$ is a diffeomorphism for perturbations around vacuum AdS. Consequently, $\delta_Y E_{\phi} =0$ and the derivative \eqref{eq: bulkintegral} reduces to
\begin{equation}
d \omega_{\text{bulk}}^{\text{E}}( \delta_Y \phi, \delta\phi) = - \delta E_{\phi} \delta_Y \phi = - \epsilon \delta E^{\mu \nu} \delta_Y g_{\mu \nu}\;.
\end{equation}
where in the last equality we are considering only perturbations to the bulk AdS metric, with $\delta E^{\mu \nu} = \frac{1}{\sqrt{g}}\frac{\delta S_{\text{grav}}}{\delta g_{ \mu \nu}}$, such that $E_{\phi}$ is no longer a $d$ form on $\mathcal{M}$. More generally, $\delta_{Y}$ is not a diffeomorphism on general backgrounds, however, $\delta_{Y}$ is on-shell, such that $\delta_{Y}E_{\mu\nu}=0$ for perturbations around any on-shell background. We will return to this point momentarily.

Demanding that $\int_{\mathcal{M}} d\omega_{\text{bulk}}^{\text{E}}( \delta_Y, \delta)$ vanishes for all variations therefore implies
\begin{equation}\label{eq: yorkdefconstraint}
\delta E^{\mu \nu} \delta_Y g_{\mu \nu} =0\;.
\end{equation}
We shall  now argue that demanding this holds for all Lorentzian initial data is equivalent to the linearized Einstein's equations $\delta E^{\mu \nu}$ being satisfied in the Euclidean bulk $\mathcal{M}$.


\subsubsection{First law of complexity implies the linearized Einstein equations}

Here we explicitly study the consequences of \eqref{eq: yorkdefconstraint} for variations around vacuum AdS. We first study the case when the Euclidean bulk is glued to the Lorentzian bulk along a \emph{maxima}l hypersurface $\Sigma$ in a fixed Lorentz frame described by coordinates $t=0$. Working in Euclidean  Poincar\'e cooordinates (\ref{eq:adspoinc}), the new York transformation is generated by the vector field (\ref{eq:vecfchiyork}), under which the metric changes according to (\ref{eq: NYBulk}). Substituting (\ref{eq: NYBulk}) into the constraint \eqref{eq: yorkdefconstraint} yields
\begin{equation}
\label{eq: variationPoincareCoords}
\delta E^{\tau\tau} \frac{2 \alpha  \tau^3}{z^2 \left(\tau^2+z^2\right)^{3/2}}  + \delta E^{\tau z} \frac{4 \alpha  \tau^2}{z \left(\tau^2+z^2\right)^{3/2}}+ \delta E^{zz}\frac{2 \alpha  \tau}{\left(\tau^2+z^2\right)^{3/2}} +  \sum_{i} \delta E^{ii}\frac{2 \alpha  \tau}{z^2 \sqrt{\tau^2+z^2}}  =0 .
\end{equation}
Multiplying through by $( \tau^2 + z^2)^{3/2}$ this can be slightly simplified to arrive at
\begin{equation}
\label{eq: numeratorPoincare1}
2 \tau^3 {\delta E}^{\tau\tau}+4 \tau^2 z {\delta E}^{\tau z}+2 \tau \left(\tau^2+z^2\right) {\delta E}^{xx}+2 \tau \left(\tau^2+z^2\right) {\delta E}^{yy}+2 \tau z^2 {\delta E}^{zz} =0\;.
\end{equation}
We now demand this constraint hold for all Lorentzian maximal slices $\Sigma$ which provide data for different Lorentz observers in the Lorentzian space-time, a fact we shall use to conclude the linearized Einstein's equations must be satisfied in the Euclidean bulk.

\begin{figure}[t]
\centering
\includegraphics[scale=0.4,trim=0 1in 0 2.5in]{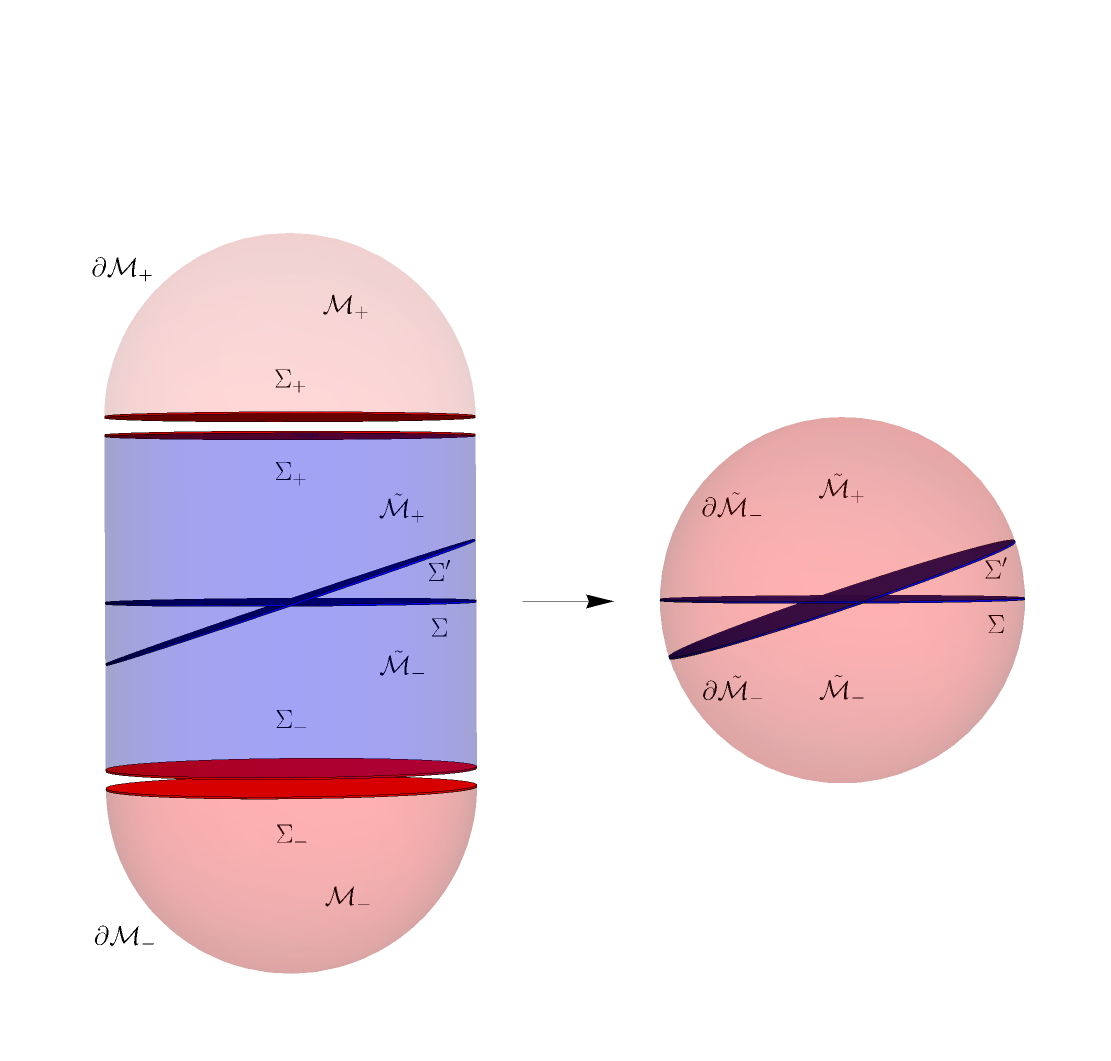}
\caption{An illustration from \cite{Botta-Cantcheff:2015sav}, where the Lorentzian space-time is cut-into two sections along a surface $\Sigma$. By Wick rotating each section, then we can interpret $\tilde{\mathcal{M}}_- \cup \tilde{\mathcal{M}}_-$ and $\tilde{\mathcal{M}}_+ \cup \mathcal{M}_+$ as describing preparation of an initial (and final) state on $\Sigma$, respectively. The slice $\Sigma'$ is an equally valid bulk Lorentzian slice which also partitions the two regions. In particular, we could pick $\Sigma'$ to be the constant time surface of a Lorentz boosted observer.}
\label{fig: prepforlorentzian}
\end{figure}

So far we have been considering initial states $\Sigma$ which can be viewed as initial data for a $t=0$ slice in the Lorentzian bulk. As mentioned before, more generally, according to the Skenderis van-Rees prescription \cite{Skenderis:2008dh,Skenderis:2008dg}, we can compute real-time correlation functions by performing a Euclidean path integral over $\mathcal{M}_-$, gluing this to a Lorentzian space-time along $\Sigma_-$ and evolving in Lorentzian time to the surface $\Sigma_+$, where the path integral is closed over $\mathcal{M}_+$. Furthermore, the authors of \cite{Botta-Cantcheff:2015sav} considered when the Lorentzian space-time $\tilde{\mathcal{M}}$ is split in two sections $\tilde{\mathcal{M}}_-, \tilde{\mathcal{M}}_+$ along a surface $\Sigma$ (see Figure \ref{fig: prepforlorentzian}).

If we perform a Wick rotation in each of the Lorentzian pieces $\tilde{\mathcal{M}}_{\pm}$ then we can interpret $\tilde{\mathcal{M}}_- \cup \mathcal{M}_-$ and $\tilde{\mathcal{M}}_+ \cup \mathcal{M}_+$ as describing preperation of an initial (and final) state on $\Sigma$. Indeed $\tilde{\mathcal{M}}_{ \pm} \cup \mathcal{M}_{ \pm}$ is diffeomorphic to half of Euclidean AdS and the Euclidean path integral is performed by joining the two hemispheres. However, there is nothing special about the slice $\Sigma$ and we could have instead chosen a different bulk slice $\Sigma'$, for example, the constant time surface of a Lorentz boosted observer in the bulk. In this case, after Wick rotating the two regions $\tilde{\mathcal{M}}_{ \pm}^{\prime}$ we again end up with a Euclidean path integral over the sphere, but which now prepares Lorentzian initial data on the slice $\Sigma'$. In particular, in boosted coordinates, the surface provides Lorentizian initial data along the $t'=0$ surface (see Figure \ref{fig: prepforlorentzian}).

\begin{figure} [t]
\includegraphics[scale=0.6]{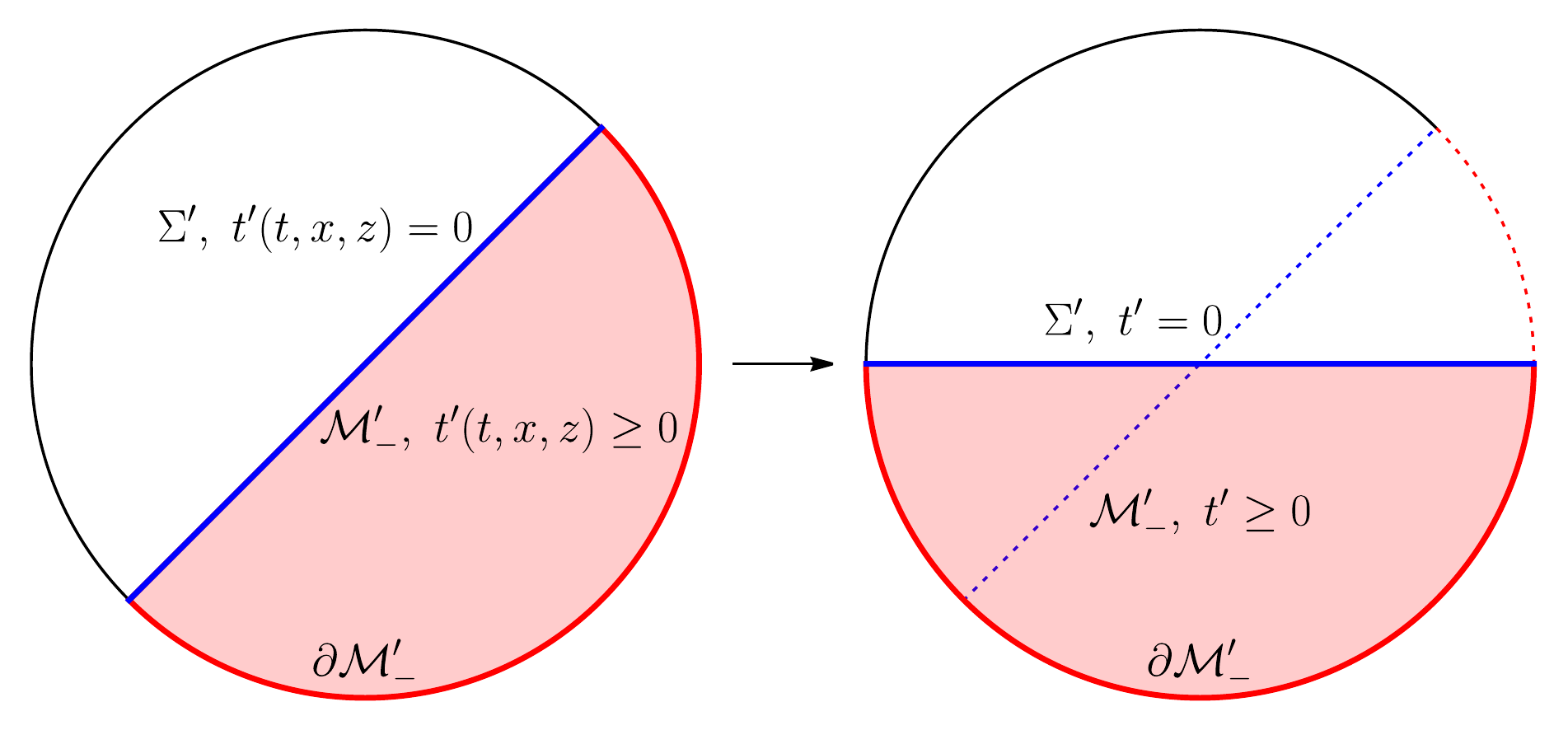}
\centering
\caption{In the Euclidean bulk,the background is real and we can imagine picking any other maximal surface $\Sigma'$ in the Euclidean manifold, which we view as preparing initial states on the different possible Lorentzian slices $\Sigma$ illustrated in Figure \ref{fig: prepforlorentzian}. On the left hand side is the boosted surface $\Sigma'$, $t'(t,x,z)=0$ as drawn in $(t,x,z)$ co-ordinates, whilst on the right hand side the surface is drawn in $(t',x',z')$ co-ordinates, and the boosted surface lies along the $t'=0$ plane. }
\label{fig: pickADifferentSlice}
\end{figure}

More generally, since the background is real in Euclidean AdS, we can imagine picking any other maximal surface $\Sigma'$ in the Euclidean manifold, described by coordinates $t'(t,z,x)=0$. We view this as preparing initial states on the different possible Lorentzian slices $\Sigma$ as illustrated in Figures \ref{fig: prepforlorentzian} and \ref{fig: pickADifferentSlice}. To draw a consistent picture, we find it useful to change coordinates so that the surface $\Sigma'$ lies along the $t'=0$ surface.

In particular, if $\Sigma'$ is related to $\Sigma$ by an isometry of AdS, then in the new coordinates the Euclidean metric will take the same form, as will the initial data on $\Sigma'$. As a consequence the York deformation, which is defined with respect to deformations of initial data also remains unchanged and so we know that under an isometry $(\delta_Y g)'(x'(x)) = (\delta_Y g)(x)$. However, the equations of motion will transform according to
\begin{equation}
\delta E^{\prime \mu\nu}(x'(x)) = \Lambda^{\mu}_\gamma \Lambda^\nu_\delta \delta E^{\gamma\delta}(x)\;,
\end{equation}
where $\Lambda^\mu_\nu =\frac{ \partial x^{\prime \mu}}{\partial x^{\nu}} $. Since $d \omega_{\text{bulk}}( \delta_{Y}\phi, \delta\phi)$ must also vanish in the primed coordinates, we deduce
\begin{equation}
\delta E^{\prime \mu \nu} (\delta_Y g)^{\prime}_{ \mu \nu}(x'(x))  = \Lambda^{ \mu}_{ \rho} \Lambda^{ \nu}_{ \sigma} \delta E^{ \rho \sigma} \delta_Y g_{ \mu \nu}(x) =0\;.
\end{equation}
Thus, demanding the constraint \eqref{eq: yorkdefconstraint} holds for all maximal slices $\Sigma$ corresponding to different Lorentz observers in the bulk means
\begin{equation} \label{eq: moreGeneral}
\Lambda^{ \mu}_{ \rho} \Lambda^{ \nu}_{ \sigma} \delta E^{ \rho \sigma} \delta_Y g_{ \mu \nu} =0
\end{equation}
 for any rotation of Euclidean AdS (including the identity).

It turns out this statement, in addition to the Bianchi identities, $\nabla_\mu \delta E^{\mu\nu} =0$, which follows directly from the diffeomorphism invariance of the Einstein action, is enough to conclude that the linearized Einstein's equations must hold in the bulk $\mathcal{M}$. In Appendix \ref{app:verifyEineqs}, we provide an explicit demonstration of this point for spacetimes in $d=3+1$ dimensions.

To summarize, we see that when we assume (i) complexity=volume, (ii) the first law of complexity holds, and (iii) the boundary symplectic form is dual to the bulk symplectic form, we find the first law of (holographic) complexity implies the linearized Einstein's equations must be satisfied around vacuum AdS:
\beq \delta V=\Omega_{\text{B}}\;\;\Rightarrow\;\; \delta E^{\mu\nu}=0\;.\eeq
Bulk spacetime dynamics emerges from boundary complexity.

Before we move onto describe how Lorentzian threads encode the bulk equations of motion, a few comments are in order. First, we point out our derivation of linearized Einstein's equations can be adapted to derive equations of motion for higher derivative gravity theories as well. That is, provided one suitably modifies the definition of `volume' in CV duality, namely, the generalized volume in \cite{Bueno:2016gnv}, the first law for holographic complexity is equivalent to linearized higher curvature equations of motion being satisfied. We say more about this in Section \ref{sec:disc}. Second, our derivation above holds for perturbations over general states, not just around vacuum AdS.

 Let us clarify this second point a bit. As noted in \cite{Belin:2018fxe}, the bulk symplectic form is tightly related to the problem of state preparation using the Euclidean path integral, where one replaces the part of the manifold in the past of the slice  $\Sigma$ of interest by a Euclidean hemisphere. One then imposes `initial conditions' in $\Sigma_{-}$ (both normalizable and non-normalizable modes) and let the bulk equations determine the state on $\Sigma$. This is done generally, either in empty AdS, or even on a two sided BH background following the Hartle-Hawking prescription. In practice, what the new York deformation $\delta_Y$ implements is an analogous problem, where one instead specifies `boundary conditions' (normalizable modes) both on $\Sigma_{-}$ and $\Sigma$, such that bulk equations of motion determine which sources on $\Sigma_{-}$ are needed (see  \emph{e.g.} \cite{Belin:2020zjb}). Provided the states on $\Sigma_{-}$ and $\Sigma$ are reasonable, one can solve the problem, and the resulting geometry should be smooth, solving the Einstein equations.\footnote{There is a caveat, however, as sometimes this problem is not so well-defined and one ends up with singularities in the sources. Intuitively, we believe this could be interpreted to the impossibility of going from the reference state to the target state (with a given precision) using the allowed gates that we have at disposal in a perturbative setting. } This combined with the factor $\delta_{Y}$ is on-shell tells us our derivation of the linearized equations holds for more perturbations about more general backgrounds

This second point, moreover, potentially leads to something rather profound. Since the bulk-boundary symplectic form equivalence (\ref{eq: bulktoboundarysymplecticform}) holds for perturbations over general states,  then, as suggested in \cite{Lewkowycz:2018sgn}, \emph{any} asymptotically AdS spacetime obeying CV duality and the first law for first order state/metric deformations around the background should satisfy the full non-linear Einstein equations. Though not shown rigorously, this would constitute a derivation of the full non-linear Einstein equations from holographic complexity.


\subsection{Einstein's equations encoded in Lorentzian threads} \label{subsec:bulksymcanthread}

Now that we have demonstrated how the linearized Einstein's equations emerge from the first law of complexity, let us show how Einstein's equations are encoded in a specific Lorentzian thread configuration. To do this, we will verify the symplectic density $\omega_{\text{bulk}}(\delta_{Y}\phi, \delta\phi)$ defines a canonical pertubative Lorentzian thread $d$-form $\delta u$, explicitly in the case of perturbations about vacuum AdS.\footnote{Our derivation of Einstein's equations was accomplished explicitly for perturbations about vacuum AdS, however, as noted above, we expect the derivation to extend to other spacetimes. Since here we focus on vacuum perturbations about pure AdS, $\phi$ represents the metric $g_{\mu\nu}$ of perturbed AdS only.} Indeed,  it is clear that if we perturb \textit{initial data} according to the new York transformation \eqref{eq: yorkTransformation} (with the suitable $\alpha$), then, when restricted to the maximal slice $\Sigma$,
\beq \omega_{\text{bulk}}^{\text{L}}( \delta_{Y}\phi, \delta\phi)|_{\Sigma} = \delta \tilde{\epsilon}\;,\eeq
is the same condition the perturbative thread $d$-form $\delta u$ satisfies,  requirement \eqref{eq: peturbedVolume}. This suggests then $\delta u=\omega_{\text{bulk}}^{\text{L}}(\delta_{Y}\phi,\delta\phi)$. To make this identification, we must show $\omega_{\text{bulk}}^{\text{L}}(\delta_{Y}\phi,\delta\phi)$ meets the other criterion the perturbative thread $u_{\eta}$ satisfies, namely, the closedness condition (\ref{eq: perturbedClosed}) and the norm bound (\ref{eq: perturbedNormBound}). If so, even if only to linear order in $\eta$ for background $g^{\eta}_{\mu\nu}=g_{\mu\nu}+\eta\delta g_{\mu\nu}$, then we may safely conclude $\omega_{\text{bulk}}^{\text{L}}(\delta_{Y}\phi,\delta\phi)$ represents a perturbative thread form configuration which solves the min flow-max cut problem.

\begin{figure}[t]
\centering
\includegraphics[scale=0.78]{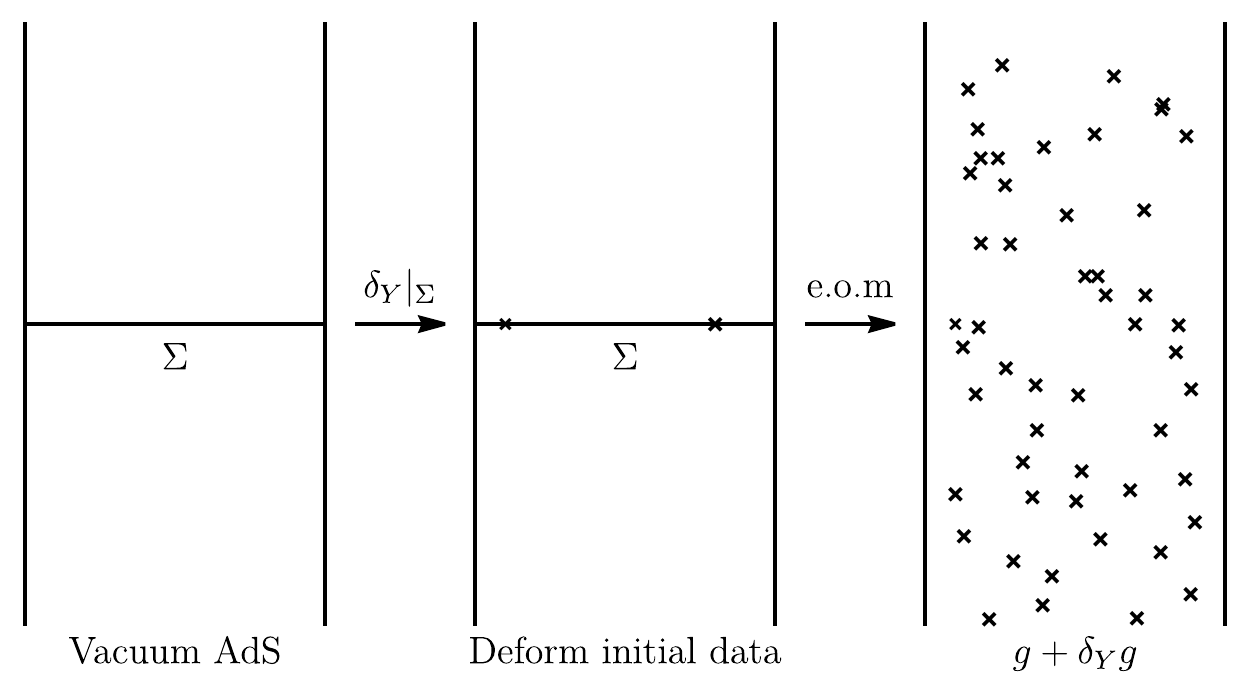}
\caption{The new-York transformation acts to deform the initial data on the Lorentzian slice $\Sigma$. Given a foliation of AdS into space-like slices $\Sigma_t$, with initial data on $\Sigma_{t=0}$ we use the equations of motion in the ADM formalism to build up the deformed four metric $\delta_Y g$, from which we can define a symplectic form on the full Lorentzian space-time.}
 \label{fig:buildingUpSymplecticFormCartoon}
\end{figure}

 In order to satisfy these remaining requirements, we first need to extend $\omega$ to a form on the Lorentzian space-time $\mathcal{M}$ using the ADM equations of motion. With the new initial data on $\Sigma$, $( h_{ij}, \pi^{ ij} + \delta_Y \pi^{ij}, \phi_m, \pi^m)$, we solve Einstein's equations in the ADM formalism to build a varied  metric on $\mathcal{M}$. See Figure \ref{fig:buildingUpSymplecticFormCartoon} for an illustration of this point.

Denote the York deformation of the background metric as $\delta_{Y} g_{\mu \nu}$, from which we can calculate the variation of the (extended) induced metric and conjugate momenta $\delta_Y h^{\mu \nu}, \delta_Y \pi^{\mu \nu}$. We can input this into the definition of the gravitational symplectic density on $\mathcal{M}$ from \eqref{eq: gravitationalSymplecticDensity} to define the (phase space) 1-form
\begin{equation}
\delta u = \omega_{\text{bulk}}^{\text{L}}(\delta_Y g, \delta g) = \delta_Y \pi^{\mu \nu} \delta h_{\mu \nu} - \delta \pi^{\mu \nu} \delta_Y h_{\mu \nu}
\end{equation}
which we emphasize now defines a $d$-form on $\mathcal{M}$ whose pullback to the maximal volume slice takes the form
\begin{equation}\label{eq: changeInVolumeOnMaximalSlice}
\delta u |_{ \Sigma} = \omega_{\text{bulk}}^{\text{L}}(\delta_Y g, \delta g)|_{ \Sigma} = \frac{1}{2} \sqrt{h} h^{ij} \delta h_{ ij} = \delta \sqrt{h} = \delta \tilde{\epsilon}\;.
\end{equation}
Furthermore, since the York transformation is on-shell when the initial slice is maximal, the form $\delta u$ is closed since
\begin{equation}
d (\delta u)= d \omega_{\text{bulk}}^{\text{L}}( \delta_Y \phi, \delta\phi) =-\delta E_{\phi}\delta_Y \phi=0\;,
\end{equation}
 provided we consider on-shell variations, $\delta E_{\phi}=0$. Thus far then, $\omega_{\text{bulk}}^{\text{L}}( \delta_Y \phi, \delta\phi)$ is a closed, $d$-form whose integral over $\Sigma$ gives rise to the change in volume on $\Sigma$.

It remains to show the norm bound of (\ref{eq: perturbedNormBound}) is satisfied to leading order in $\eta$. To discuss the norm bound, we could proceed with the definition given in \eqref{eq: perturbedNormBound}, however, we find it slightly easier to work with the corresponding perturbed vector field $v^{\mu} + \eta \delta v ^{\nu} = g_\eta^{\mu \nu} \star ( u + \eta \delta u)_{ \nu}$. The reason for this is purely technical: the norm on the space of forms, as defined in  \eqref{eq: normDefinitionForms}, contains products of the metric, which can be more cumbersome. Thus we map the variation of the perturbed form $\delta u = \omega( \delta_Y \phi, \delta\phi)$ to the corresponding vector field. In components, the Hodge star operation relating forms and vector fields reads
 \begin{equation}\label{eq: hodgeDualComponents}
v^\mu = g^{\mu \nu} \frac{1}{(D-1)!} \sqrt{g} u^{\mu_1 \dots \mu_{D-1}}\varepsilon_{ \mu_1 \dots \mu_{D-1} \nu}\;.
\end{equation}
A quick calculation shows the variation of the vector field can be written as
\begin{equation}\label{eq: changeInVectorField}
\delta v^{\mu} = \delta v_{0}^\mu  - \frac{1}{2}v^\mu g^{ \rho \sigma} \delta g_{ \rho \sigma}\;,
\end{equation}
where $ \delta v_0^{\mu}$ is defined as the Hodge dual of the form $\delta u$ in the unperturbed background $\delta v_0^{\mu}$, $\delta v_0^\mu = g^{\mu \nu} \star( \delta u)_{ \nu}$. More explicitly, using
\begin{equation}
\delta u_{ \mu_1 \dots \mu_{ D-1}} = \omega_{\text{bulk}}^{\text{L}}\left( \delta_{Y} g, \delta g\right)_{ \mu_1 \dots \mu_{ D-1}} =\frac{1}{16 \pi G_{N}} \epsilon_{a \mu_1 \dots \mu_{D-1}} P^{\alpha \beta \gamma \delta \epsilon \zeta}\left(\delta g_{\beta \gamma} \nabla_{\delta} \delta_{Y} g_{\epsilon \zeta}-\delta_Y g_{\beta \gamma} \nabla_{\delta} \delta g_{\epsilon \zeta}\right)\;,
\end{equation}
with
\begin{equation}
P^{\alpha \beta \gamma \delta \epsilon \zeta }=g^{\alpha \epsilon} g^{\beta \zeta} g^{\gamma \delta}-\frac{1}{2} g^{\alpha \delta} g^{\beta \epsilon} g^{\gamma \zeta}-\frac{1}{2} g^{\alpha \beta} g^{\gamma \delta} g^{\epsilon \zeta}-\frac{1}{2} g^{\beta \gamma} g^{\alpha \epsilon} g^{f d}+\frac{1}{2} g^{\beta \gamma} g^{\alpha \delta} g^{\epsilon \zeta}\;,
\end{equation}
 we can write $\delta v_0^{\mu}$ in terms of variations of the metric as
\begin{equation}
\delta v_0^{\mu} = \frac{1}{16 \pi G_{N} } P^{\mu \beta \gamma \delta \epsilon \zeta}\left(\delta g_{\beta \gamma} \nabla_{\delta} \delta_{Y} g_{\epsilon \zeta}-\delta_Y g_{\beta \gamma} \nabla_{\delta} \delta g_{\epsilon \zeta}\right).
\end{equation}
It is useful to note, since the form $u$ is closed, $\delta v_0^{\mu}$ is divergenceless with respect to the background metric $g_{\mu \nu}$, $(\nabla_{\eta} \delta v_0)|_{\eta=0}  = 0$.

From \eqref{eq: changeInVectorField} we see $\delta v^\mu$ depends not only on the variation of the background form, but also on the background flow $v^\mu$. As in \cite{Agon:2020mvu}, we shall exploit the dependence on the background flow in order to argue the norm bound can be satisfied quite generally, and show this explicitly for variations around vaccuum AdS. In particular, to show $\delta u = \omega(\delta_Y g, \delta g)$ provides a `good' perturbed thread, one needs only to find a \textit{single} background flow $v^\mu$ for which the norm bound $|v_{\eta}| \geq 1$ is satisfied.

Since the background flow $v^\mu$ already obeys $|v| \geq 1$ everywhere, the perturbed vector field $v_{\eta}$ is only at risk of exceeding this bound by an amount of order $\mathcal{O}( \eta)$. For all points which do not saturate the bound, we can take $\eta$ to be small enough so that the perturbed flow still saturates the bound, $|v_{ \eta} | = |v| + \mathcal{O}( \eta) \geq 1$, \emph{i.e.}, there will be some range of $\eta$ for which the perturbative treatment is valid. Within this range of $\eta$, we only have to worry about the neighborhood of points which saturate the bound. Generically, there are many solutions to the convex optimization program, which all yield the same flux along the maximal surface $\Sigma$, but may differ elsewhere. We expect for most space-times, one can always choose a background flow $v^\mu$ such that the norm bound increases away from the maximal surface, with a rapidly increasing norm bound giving a larger range of validity for $\eta$ for which the perturbed thread is valid, although we do not have a general proof of this statement.\footnote{While a general proof is lacking, the intuition is as follows.  In any spacetime, the norm bound is saturated on the maximum volume slice. For any other homologous slice the norm must increase (thus maintaining the bound) in order for the flux to be conserved. This is consistent with the focusing theorem. Consider an infinitesimal tube composed of integral curves to a Lorentzian flow about a point which the curves intersect the maximal volume slice. As one moves to the past or the future of this intersection point, the cross-sectional area of the tube decreases, which forces the integral curves to focus, thereby maintaining the norm bound. Hence, we expect the norm bound to be satisfied for any spacetime obeying appropriate energy conditions.} However, we are able to show this explicitly for perturbations around vacuum AdS using the geodesic construction of the background flow $v^\mu$ in Section \ref{sec:simpleconstructions}.

The norm of the background vector field $v^{\mu}$ for  $\text{AdS}_n$, found using the geodesic method in Section \ref{sec:simpleconstructions}, takes the form
\begin{equation}\label{eq: normBoundMonotonic}
|v| = (1+ \frac{1}{4} \tan^2(t) \sec^2 (\rho_m))^{(n-1)/2}.
\end{equation}
where, recall, we are working with compactified coordinates,  $t \in( -\pi/2, \pi/2)$, $\rho_m, \in( 0, \pi/2)$, and the WDW patch is described by the condition $|\rho_m| + |t| \leq \pi/2$. The norm bound is a monotonically increasing function inside the WDW patch, while outside the WDW patch, from \eqref{eq: functionOutsideWdW} we see we can tune $C$ such that the norm bound is arbitrarily large. Consequently, we need only consider the perturbed norm bound in the vicinity of the maximal slice $\Sigma$ in the interior of the WDW patch.

In particular, moving a distance $\Delta$ in the $t$ direction away from the minimal surface, we see that the norm bound can be written as
\begin{equation}
|v| = 1 + \frac{(n-1)}{8}\Delta^2 \sec^2( \rho_m) + \mathcal{O}( \Delta^4)\;,
\end{equation}
and monotonically increases in $\Delta$ away from this. Since the range of $\sec^2( \rho_m)$ is $\sec^2( \rho_m) \in ( 1, \infty)$, away from the maximal surface the norm bound is strictly satisfied $|v| >1$,  whilst we know from  \eqref{eq: changeInVolumeOnMaximalSlice}  the norm bound is satisfied exactly on the maximal surface. Explicitly, the perturbed norm reads
\begin{equation}\label{eq: etaNormBound}
|v_{\eta}| = |v| -  \eta \left(2 \delta v^{\mu} v_{\mu} - \delta g_{\mu \nu} v^{\mu} v^{\nu} \right)  = 1 + \frac{(n-1)}{8 } \Delta^2 \sec^2( \rho_m)  - \eta \left(2 \delta v^{\mu} v_{\mu} - \delta g_{\mu \nu} v^{\mu} v^{\nu} \right).
\end{equation}
We want to check that for arbitrarily small $\Delta$ we can always find an $\eta$ such that the norm bound is satisfied to $\mathcal{O}( \eta)$. Since the norm bound is satisfied exactly at $t=0$, we know that $\left(2 \delta v^{\mu} v_{\mu} - \delta g_{\mu \nu} v^{\mu} v^{\nu} \right)|_{\Sigma}=0$, while away from the maximal slice we have
\begin{equation}
\left(2 \delta v^{\mu} v_{\mu} - \delta g_{\mu \nu} v^{\mu} v^{\nu} \right)= c( x^{\mu}) \Delta^k + \mathcal{O}( \Delta^{ \bar{k} } ), \ \ \bar{k} > k,
\end{equation}
for some $k> 0$ and $c( x^{\mu})$ which is finite as $t \to 0$. This follows from asking that the perturbative expansion makes sense as we approach the maximal volume slice, such that a power expansion exists in positive powers of $t^k$. From \eqref{eq: etaNormBound} we see the norm bound will be satisfied so long as
\begin{equation}
 \frac{(n-1)}{8 } \Delta^2 \sec^2( \rho_m) \geq \eta \left(2 \delta v^{\mu} v_{\mu} - \delta g_{\mu \nu} v^{\mu} v^{\nu} \right) + \mathcal{O}( \Delta^{ \bar{k} > k} )\;.
\end{equation}
Since the smallest value of $\sec^2( \rho_m)$ is when $\rho_m=0$, the norm bound is satisfied provided
\begin{equation}\label{eq: normBoundSaturated}
 \Delta^{2-k} \geq  \frac{8 c(x^{\mu})}{n-1} \eta   +  \mathcal{O}( \eta \Delta^{ \bar{k} -k } ) .
\end{equation}
If $2-k \leq 0$, then taking $\Delta \to 0$, the left hand side of  \eqref{eq: normBoundSaturated} is either constant or diverges, in which case one can always pick $\eta$ small enough such that the inequality is saturated. Conversely, if $2-k>0$, the inequality can be violated by picking $\Delta$ to be arbitrarily small for fixed $\eta$. However, whenever the inequality \eqref{eq: normBoundSaturated} is violated, $ \eta \left(2 \delta v^{\mu} v_{\mu} - \delta g_{\mu \nu} v^{\mu} v^{\nu} \right) = \mathcal{O}(\eta \Delta^k) \lesssim \mathcal{O}(\eta^{ 1+ \frac{k}{2-k}})  < \mathcal{O}( \eta)$, implying the violations to the norm bound enter at higher than linear order. Thus, we deduce we can always find an $\eta$ such that the norm bound is satisfied to $\mathcal{O}( \eta)$ for any $\Delta$.

Collectively then, the bulk symplectic density $\omega_{\text{bulk}}^{\text{L}}(\delta_{Y}\phi,\delta\phi)$, to leading order, satisfies all of the properties of a perturbative thread form $\delta u$,
\beq \omega_{\text{bulk}}^{\text{L}}(\delta_{Y}\phi,\delta\phi)=\delta u\;,\eeq
and is thus a solution to min flow-max cut program. Since, moreover, $d\omega_{\text{bulk}}^{\text{L}}(\delta_{Y}\phi,\delta\phi)=0$ is equivalent to the linearized Einstein's equations being satisfied, we have that, equivalently,
\beq d(\delta u)=0\Rightarrow \delta E^{\mu\nu}=0\;.\eeq
That is, the  linearized Einstein's equations are captured by the closedness condition of the canonical thread configuration. This is analogous to the case of holographic bit threads, where the canonical candidate for the perturbed thread form is the Iyer-Wald form $\tilde{\chi}$.

Recall from our discussion in Section \ref{sec:interpretation} that Lorentzian threads can be viewed as preparing a CFT state on the maximal volume slice. The time evolution of the CFT state is then characterized by the future of flow $v$. More traditionally the time evolution is characterized by solving bulk gravitational equations subject to specific boundary conditions. Here, we have uncovered a particular thread configuration -- our perturbative thread form $\delta u=\omega_{L}(\delta_{Y}\phi,\delta\phi)$ --  encodes these bulk equations. From this viewpoint, Lorentzian threads genuinely provide a notion of `emergent time': evolution of the boundary CFT, and, correspondingly the dynamics of the bulk, are encapsulated by Lorentzian flows.

We point out, of course, our above arguments only considered perturbative excited states, \emph{i.e.}, for bulk perturbations such that $\Sigma\sim \Sigma_{\eta}$, and, moreover, we only recovered the \emph{linearized} Einstein's equations. The second of these issues we leave for the discussion. The first of these criticisms, that our Lorentzian flows hold perturbatively, is really a limitation of CV duality such that holographic complexity may need to be generalized. In the next section we offer a more general proposal and discuss how it relates to tensor network constructions of spacetime.


\section{Tensor networks and Lorentzian threads} \label{sec:genprop}

Thus far the bulk of this article has concerned itself with optimal flows, \emph{i.e.}, those which solve the min flow problem. We have largely ignored sub-optimal flows, despite the fact they will be present in general. Between this and given the aforementioned limitations of CV duality in regards to highly excited states, here we discuss prospects towards a more general definition of holography complexity. To motivate our proposal we take a cue from tensor network representations of AdS which, as we will discuss below, will be augmented due to our gateline interpretation of Lorentzian threads. We will also take motivation from the covariant formulation for holographic entanglement entropy. In particular, the maximin prescription for computing HRT surfaces will be informing us that sub-optimal tensor networks, and hence flows, have a special role to play.

\subsection{Motivation and physical intuition of tensor networks} \label{subsec:physint}

Tensor networks can be thought of as spatial discretizations of quantum states \cite{Evenbly_2011}, and are known to be a good description in situations of high symmetry, such as the ground state \cite{Evenbly_2015}. They were first introduced in the context of condensed matter quantum many-body simulations as an efficient way of dealing with systems with exponentially-large Hilbert spaces. In this context, different phases/states are often classified in terms of their computational complexity. The strategy of tensor networks is to target a given class of computational complexities and then find, within the given class, an optimal quantum state which most faithfully resembles the original state. In this way, tensor networks avoid the exponential blow-up of the Hilbert space dimension as a function of the system size. This offers a key advantage over a continuum description, overcoming some of the challenges that arise in the definition of complexity in quantum field theory --- see however \cite{Chapman:2017rqy,Jefferson:2017sdb,Caputa:2017urj,Caputa:2017yrh,Chapman:2018hou,Hackl:2018ptj,Camargo:2019isp,Flory:2020eot,Flory:2020dja,Chagnet:2021uvi} for recent progress on this problem.

It behooves us to provide an elementary review of tensor networks and its relation to entanglement and complexity. In terms of tensor networks, a quantum state $|\Psi\rangle$ is described in terms of a set of tensors. These tensors describe a discretized version of the system, where each degree of freedom  correspond to an $M$-level  system, \emph{e.g.}, if $M=2$ we can represent them as spins. In such a description, the wavefunction can be written as
\be\label{state:TN}
|\Psi\rangle=\sum_{k_i=1}^{M}T_{k_1k_2\ldots k_n}|k_1,k_2,\ldots, k_n\rangle\,,
\ee
where each basis state may be expressed as a direct product of local state vectors
\be
|k_1,k_2,\ldots, k_n\rangle=|k_1\rangle\otimes|k_2\rangle\otimes\cdots\otimes|k_n\rangle\,.
\ee
The  state (\ref{state:TN}) is  thus  expressed  by $M^n$ amplitudes $T_{k_1k_2\ldots k_n}$, which implies  that  the  number  of  parameters  that  describe  the  wavefunction is exponentially large in the system size.  The dimension of each index, often called the bond dimension, is given by $M$. One then needs a suitable ansatz for the amplitudes, which is provided by a tensor network.  This ansatz gives the $T_{k_1k_2\ldots k_n}$ in terms of a contraction of complex-valued tensors.  For example, for $n=3$ one has
\be
T_{k_1k_2k_3}=U_{k_1j_3j_1}V_{k_2j_1j_2}W_{k_3j_2j_3}\,
\ee
where $U$, $V$ and $W$ are rank-$3$ tensors and repeated indices imply summation. This operation, and the structure it generates while considering the full  state (\ref{state:TN}), is often represented as a graph, with nodes and edges representing tensors and their indices, respectively.  For a more detailed exposition on the basics of tensor networks, see \emph{e.g.}, \cite{Orus:2013kga,Bridgeman:2016dhh,Jahn:2021uqr}.

\begin{figure}[t]
\centering
\includegraphics[width=2.7in,trim=0 -2.5cm -1cm 0]{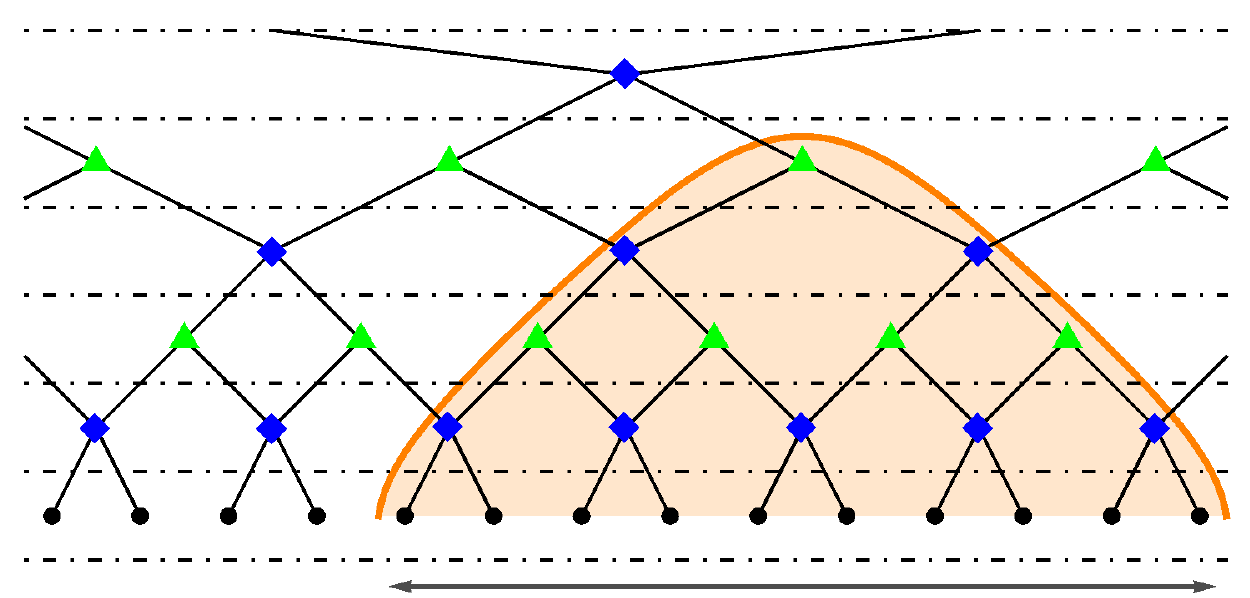}$\qquad\qquad$\includegraphics[width=2.5in,trim=-1cm 0 0 0]{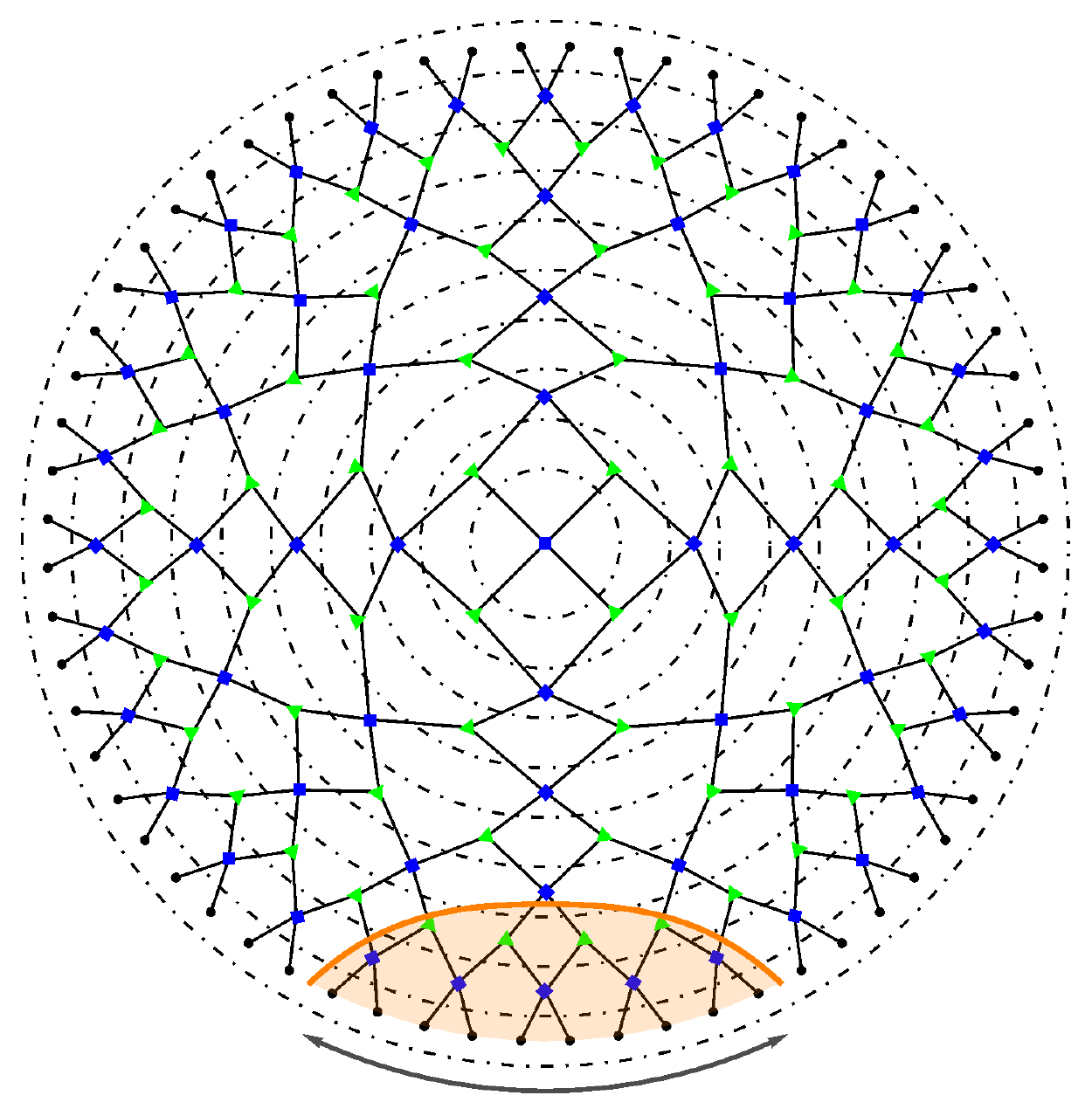}
 \begin{picture}(0,0)
\put(-310.5,29){\scriptsize$A$}
\put(-91,-5){\scriptsize$A$}
\put(-242,44){\scriptsize Layer 0}
\put(-242,56.8){\scriptsize Layer 1}
\put(-242,69.6){\scriptsize Layer 2}
\put(-242,82.4){\scriptsize Layer 3}
\put(-242,95.2){\scriptsize Layer 4}
\put(-242,108){\scriptsize Layer 5}
\put(-231,120.8){\scriptsize $\vdots$}
\end{picture}
\caption{\small Pictorial representation of a binary MERA network on the line (left diagram) and on the circle (right diagram). The black dots represent the original sites in a quantum 1D critical system. Diamonds and triangles are the disentangler tensors and the isometries (coarse-grainings), respectively. The vertical/radial direction represents the renormalization flow and exhibits the property of self-similarity. The orange line denotes a path $\gamma_A$ that splits the network into two smaller networks associated with the subsystem $A$ and its complement. The number of links cut along this trajectory bounds the entanglement entropy of the subsystem, according to the formula (\ref{SA:TNineq}). For some special cases, \emph{e.g.}, the vacuum state, this inequality is saturated so that the bound becomes an equality (\ref{SA:TNeq}).
\label{fig:MERA}}
\end{figure}
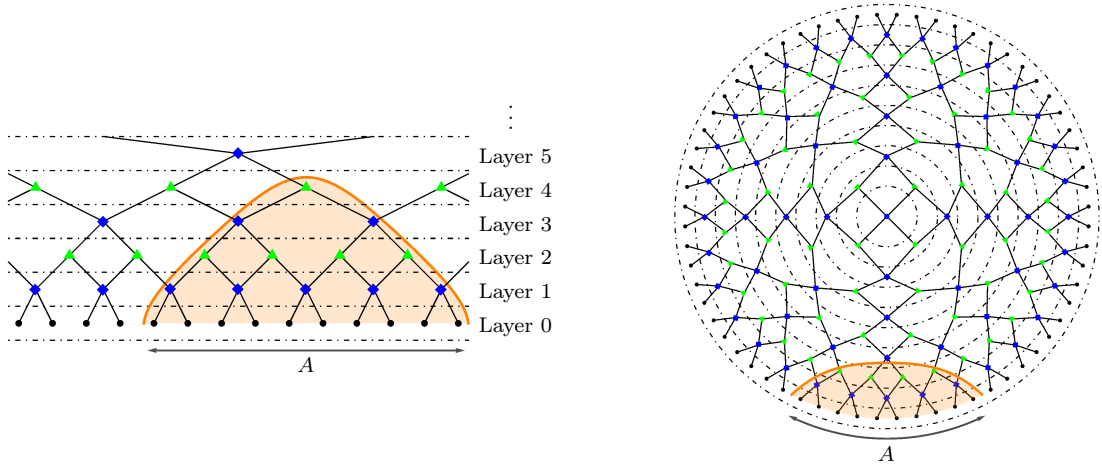

One of the main motivations of using tensor networks to describe holographic spacetimes is that they can be engineered to correctly reproduce the entanglement entropies of subsystems in a given state, in agreement with the RT prescription. This observation was first pointed out in \cite{Swingle:2009bg,Swingle:2012wq}.
A particular class of tensor networks that is relevant to describe systems with conformal symmetry, and hence often discussed in the context of holography, is the Multi-scale Entanglement Renormalization Ansatz (MERA) \cite{Vidal:2008zz}. These tensor networks live naturally on discretizations of hyperbolic space, and, as a result, many statements from holography have a natural realization in such networks. The MERA network has a layered structure that is motivated by the real space renormalization group idea, and is pictorially represented in Figure \ref{fig:MERA} (where we consider $M=2$ for the ease of visualization).  The layers are labeled by an integer $u$ according to their depth, \emph{i.e.}, $u= 0,1,2,...$.  The original system is located on the 0-th layer $(u= 0)$, which is at the bottom of the network. Each layer is then composed by one of two types of tensors, that are referred to as \emph{isometries} or \emph{disentanglers}, respectively.  The isometries combine a set of degrees of freedom by a linear map, which is regarded as a coarse graining procedure or a scale transformation. These tensors represent physical degrees of freedom at any given depth in the network. The isometries on their own are not enough to correctly reproduce features of the ground state of the system.\footnote{They lead to a tree-like structure, called tree tensor network (TTN), that does not capture the dependence of correlations between different degrees of freedom with their separation.} To remedy this, one generally needs to add extra bonds in between the layers that are referred to as disentanglers. These extra tensors do not represent physical degrees of freedom. Their role is to perform an appropriate unitary transformation on the Hilbert space of a subset of the degrees of freedom. This transforms the state $|\Psi_u\rangle$ into a less entangled state $|\Psi_{u+2}\rangle$. One then repeatedly applies this two-step procedure as many times as one wishes.\footnote{If the system is defined on a circle, the number of layers is cut off due to the finiteness of the volume. Locally, tensor networks defined on a circle look the same those defined on a plane, but they are topologically inequivalent \cite{Milsted:2018san,Hung:2021tsu}. See Figure \ref{fig:MERA} (right) for an illustration.} We point out that a continuous version of the MERA network, relevant to describe quantum field theories, was proposed in \cite{Haegeman:2011uy}. This is referred to as continuousMERA (or cMERA, for simplicity). Intuitively, this framework is used to describe networks where the original lattice spacing is sent to zero, and hence, is particularly useful in trying make the connection with AdS/CFT more precise -- see \emph{e.g.}, \cite{Nozaki:2012zj,Miyaji:2015fia,Mollabashi:2013lya,Miyaji:2016mxg}.

To better understand the connection with the RT prescription, we can consider the entanglement entropy for a subsystem $A$, composed of $L$ of the original degrees of freedom, with its complement $B$. The relevant quantity to compute here is the reduced density matrix $\rho_A$, which is found by tracing over the degrees of freedom of the complement. If these two parts are not entangled, the total state can be written as a tensor product $|\Psi\rangle=|\Psi\rangle_A\otimes|\Psi\rangle_B$, such that $\rho_A=\text{tr}_B\,\rho=|\Psi\rangle_B\langle\Psi|$ and $S_A=-\text{tr}\,\rho_A\log \rho_A=0$. More generally,
the global state can be written using the Schmidt decomposition
\be\label{Schmidt}
|\Psi\rangle=\sum_k\lambda_k|\Psi_k\rangle_A\otimes|\Psi_k\rangle_B\,,
\ee
where the $|\Psi_k\rangle_A$ and $|\Psi_k\rangle_B$ each form an orthogonal set of state vectors on $A$ and $B$, respectively. For a properly normalized state, \emph{i.e.}, for $\sum_k |\lambda_k^2|=1$, one then finds that
\be
S_A=-\sum_k|\lambda_k|^2\log|\lambda_k|^2\,.
\ee
In a tensor network, any path $\gamma_A$ cutting through the network ending on the boundary $\partial A$ can be associated with a decomposition like (\ref{Schmidt}). Any of these cuts effectively splits the tensor network in two smaller networks which are associated with the subsystems $A$ and $B$. The contraction between them over the chosen cut entangles both boundary regions. Each of the contractions, \emph{i.e.}, those connecting a pair of legs, contain at most $M^2$ terms, where $M$ is the bond dimension. This limits the number of terms that can appear in the decomposition (\ref{Schmidt}). Conversely, this sets a bound on the contribution to $S_A$ by the contraction of a single pair of legs, which is maximized if all Schmidt values are identical, namely, $|\lambda_k^2|= 1/M^2$. This bound implies that for a general path $\gamma_A$
\be
S_A\leq (\# \text{ cuts}) \times \log M\,.
\ee
Now, the number of cuts will depend on the chosen path $\gamma_A$. In particular, this bound is tightest for the path $\gamma_A$ with fewest number of cuts, thus leading to the well-known bound
\be\label{SA:TNineq}
S_A\leq \min[\# \text{ cuts}] \times \log M\,.
\ee
Remarkably, for a MERA network this bound can be strictly saturated for particular states \emph{given any bipartition of the Hilbert space}. In these cases (\ref{SA:TNineq}) becomes an equality which is reminiscent of the RT prescription,
\be\label{SA:TNeq}
S_A=\min[\# \text{ cuts}] \times \log M\,,
\ee
For instance, in the vacuum state one can estimate the minimal number of cuts to be proportional to the logarithm of the number of degrees of freedom included in the subsystem $A$ \cite{Vidal:2008zz}, $\min[\# \text{ cuts}]\sim \log L$, hence (\ref{SA:TNeq}) agrees with the standard result in CFT\footnote{In this formula $\epsilon$ is a UV cutoff which can be identified as the discretized lattice spacing and $\ell$ is the length of the subsystem.}
\be
S_A=\frac{c}{3}\log\left(\frac{\ell}{\epsilon}\right)\,,
\ee
upon identification $c=3 \log M$. For more general states, the bound can be saturated by particular bipartitions of the Hilbert space but not general ones.

\begin{figure}[t]
\centering
 \includegraphics[width=3.1in]{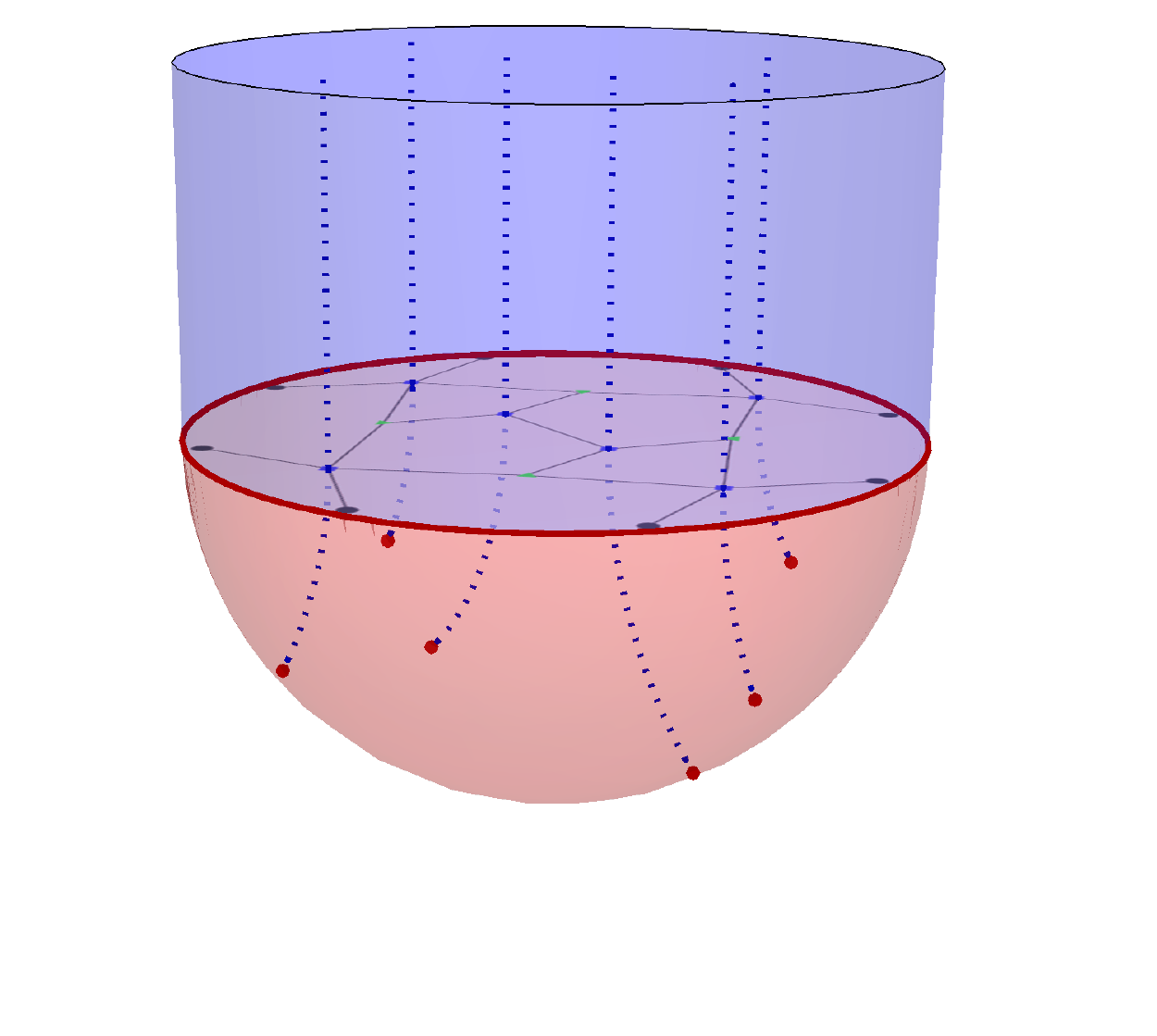}\includegraphics[width=3in]{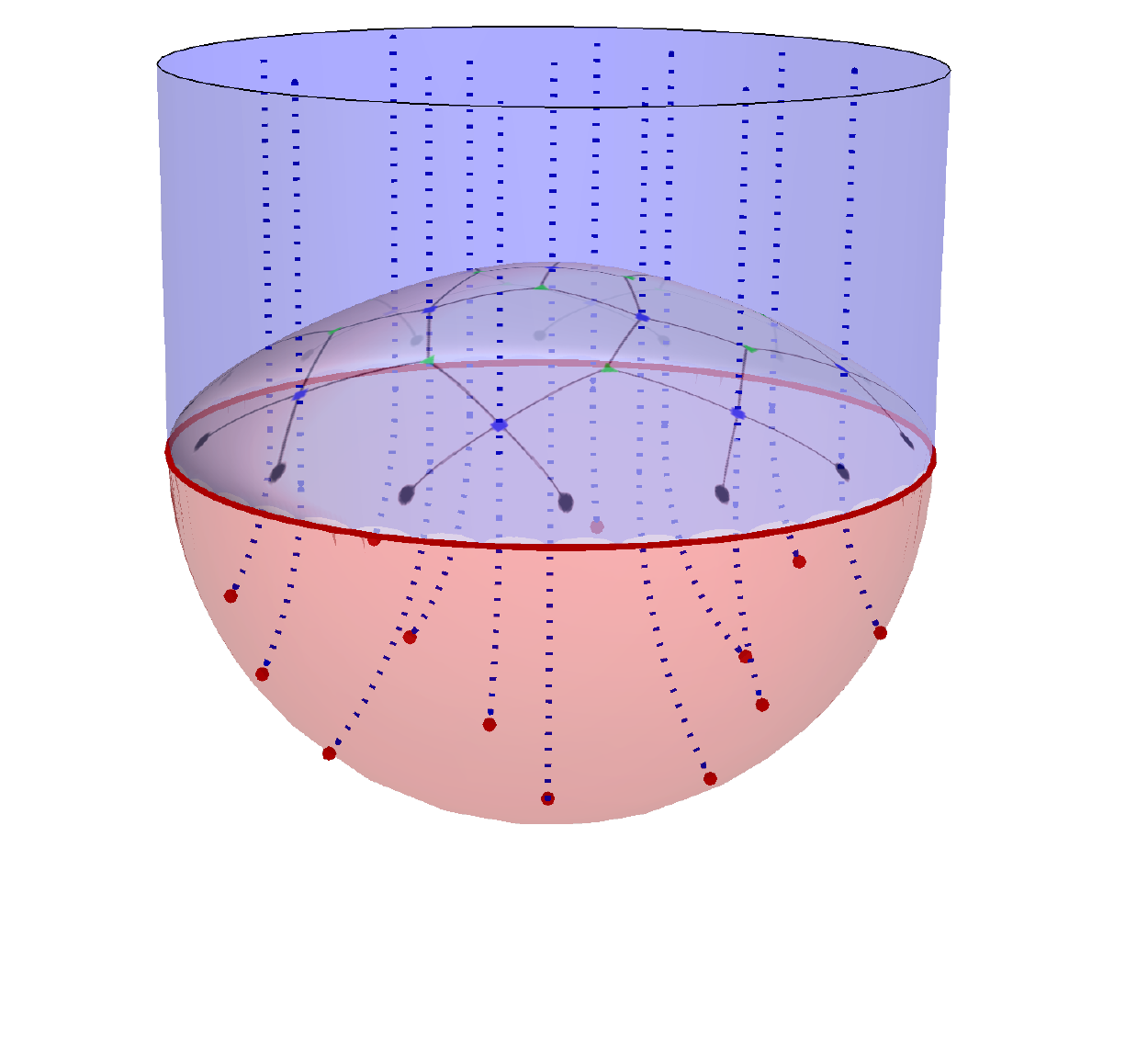}
 \begin{picture}(0,0)
\put(-62,127){$\Sigma$}
\put(-36,127){$A$}
\put(167,127){$\Sigma'$}
\put(186,127){$A$}
\put(116,72){$\mathcal{M}_{-}$}
\put(-112,72){$\mathcal{M}_{-}$}
\put(-98,175){$v$}
\put(134,175){$v'$}
\put(-60,171){$\tilde{\mathcal{M}}$}
\put(169,171){$\tilde{\mathcal{M}}$}
\end{picture}
\vspace{-2cm}
\caption{Via convex optimization, we can find a flow program that is dual to the CV conjecture. The discrete version of the dual program implies that complexity is can be interpreted as the minimum number of threads, or gatelines, preparing a state on the maximal volume slice $\Sigma$ (left). Suboptimal flows prepare more complex TNs, which can be defined over non-maximal volume slices $\Sigma'$ (right).
\label{fig:TNs}}
\end{figure}
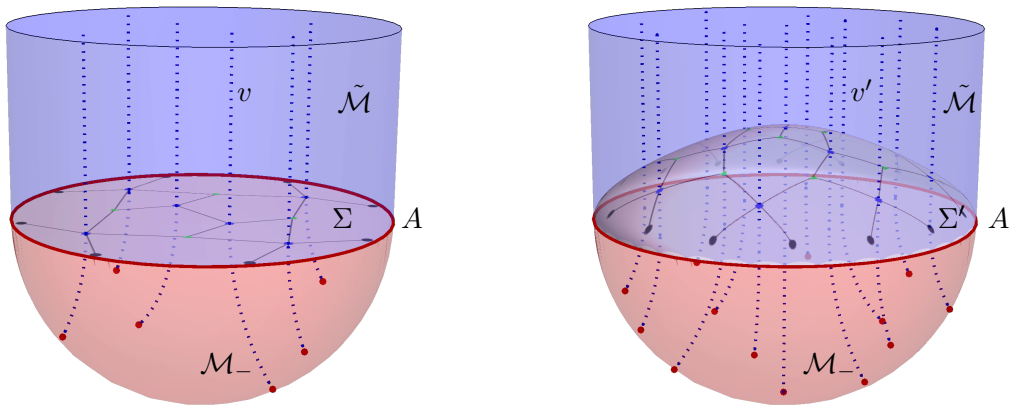

As pointed out in \cite{Stanford:2014jda}, the `complexity=volume' conjecture is naturally realized in tensor network constructions by associating a fixed spatial volume to each (physical) tensor.  On a given tensor network, then, a notion of state complexity can be given in terms of the number of (physical) tensors required to describe the network:
\be\label{Comp-TN}
\mathcal{C}\sim \# \text{ of tensors.}
\ee
It is also true that the same state can be represented by different tensor networks, all with different numbers of tensors (\emph{i.e.}, depending on the structure, bond dimension, etc.), and hence, different complexities. Thus one can always aim to describe the state in the most optimal way (computationally speaking), such that
\be\label{Comp-TN}
\mathcal{C}\sim \text{min}[\# \text{ of tensors}]\,.
\ee
This has resemblance to the thread-based formula for complexity derived in section \ref{sec:interpretation}. There, we argued that a solution to the min flow problem prepares the state on the maximal volume slice. We have also discussed that, upon discretization, we can think of the integral lines of the flows, or threads, as `gatelines' such that
\be\label{Comp-BT}
\mathcal{C}\sim \text{min}[\# \text{ threads}] \sim\text{min}[\# \text{ of gates to prepare the state}]\,.
\ee
The discrete threads also occupy a fixed spatial volume, as the physical tensors. Comparing (\ref{Comp-TN}) and (\ref{Comp-BT}), it is then natural to combine the two prescriptions and conjecture that the optimal thread configuration prepares the tensor network on the maximal volume slice $\Sigma$. That is, we can imagine attaching a unitary (gate) to each thread, which will connect physical tensors of the network from the reference state, defined on $\Sigma_{-}$, to physical tensors of the network on $\Sigma$, so that
\be
\text{min}[\# \text{ threads}] \sim\text{min}[\# \text{ of gates to prepare the state}] \sim \text{min}[\# \text{ of tensors}]\,.
\ee
 One may view these unitaries as playing a similar role as the disentanglers in a MERA network. They do not represent physical degrees of freedom but, instead, implement an appropriate unitary transformation on the Hilbert space that connect degrees of freedom on $\Sigma_{-}$ to those in $\Sigma$, which can now be viewed as two `layers' on a network with an extra dimension $\tau$. In other words, they transform the state $|\Psi_{-}\rangle$ into the state $|\Psi\rangle$. In a Lorentazian setting, these unitaries implement time evolution, in the same sense that the disentanglers implement evolution along the radial (holographic) direction. Thus, time evolution emerges intrinsically from complexity. A pictorial diagram illustrating this interpretation is shown in Figure \ref{fig:TNs}.

\subsection{A lesson from the maximin prescription: suboptimal flows are relevant} \label{subsec:lessonsmaximin}

As mentioned above, equation (\ref{SA:TNeq}) resembles the well-known prescription to compute entanglement entropy in the context of AdS/CFT.
In this context, entanglement entropy can be calculated via the RT formula \cite{Ryu:2006bv}
\be
S_A=\underset{\gamma_A\sim A}{\text{min}}\left[\frac{A(\gamma_A)}{4 G_N}\right]\, .
\ee
This prescription holds for static states, or for time dependent scenarios provided it is used in a moment of time reflection symmetry. More generally, in a covariant setting the formula is upgraded to the HRT prescription \cite{Hubeny:2007xt},
\be
S_A=\underset{\gamma_A}{\text{ext}}\left[\frac{A(\gamma_A)}{4 G_N}\right]\, ,
\ee
where the minimization condition is replaced by extremization. A useful observation that is not often appreciated is that
even for static cases, the HRT formula is more powerful and hence supersedes the RT prescription. The reason is that it does not assume the choice of a particular Cauchy slice $\Sigma$, while RT only works provided one picks the standard constant-$t$ slice.  For other choices $\Sigma'$ the minimization procedure simply does not work. As an example, consider a slice $\Sigma'$ that is arbitrarily close to the upper edge of the WDW patch. In this case the minimization would yield a fake RT surface $\gamma_A'$ with vanishingly small area. This means that even in static cases, the correct way to compute entanglement entropy is via the HRT prescription.

On the other hand, the HRT formula can be rewritten as a maximin prescription \cite{Wall:2012uf}
\be
S_A=\underset{\Sigma'}{\text{max}}\,\underset{\gamma_A'}{\text{min}}\left[\frac{A(\gamma_A')}{4 G_N}\right]\, .
\ee
This formula involves two steps: (i) pick an arbitrary slice $\Sigma'$ and find the minimal area surface $\gamma_A'$ within this slice,  and (ii) vary over all possible slices $\Sigma'$ and find one that maximizes the area of $\gamma_A'$. This two-step algorithm gets rid of all spurious solutions $\gamma_A'$ and gives the true $\gamma_A$, whose area computes the entanglement entropy. We note that in general the maximin prescription does not select a unique slice $\Sigma$, but it imposes a series of constraints that it should satisfy.

Extrapolating the maximin prescription to the realm of tensor networks, this implies that for a generic network defined over an arbitrary $\Sigma'$, we generally expect that \cite{May:2016dgv}:
\be
\text{min}\left[\#\text{ cuts}\right] \times \log M\leq S_A\, .
\ee
However, one can still optimize over these tensor networks defined on arbitrary $\Sigma'$s and arrive to a formula akin to the maximin prescription for holographic entanglement entropy, which should now   be valid for arbitrary, possibly time-dependent states \cite{May:2016dgv}
\be
S_A=\underset{\Sigma'}{\text{max}}\,\text{min}\left[\#\text{ cuts}\right] \times \log M\, .
\ee
An important lesson can be drawn from the above formula: at least for the computation of entanglement entropies, not only the tensor network on the maximal volume slice plays a role but also other non-optimal tensor networks defined over other slices. However, we have argued that these tensor networks have in general higher complexity, as one generally needs more tensors to prepare the state. This highlights the need of a more refined measure of complexity, one that captures the notion of state independence and considers all these tensor networks as part of its definition.

We can also rephrase this discussion by thinking about the Lorentzian flow interpretation of the complexity=volume conjecture. Previously, we have argued that an optimal flow $v$ which solves the min flux problem effectively prepares the state on the maximal volume slice $\Sigma$. We conjectured that the discrete version of this statement still makes sense, so that the thread configuration that can be extracted from one of these flows could be thought of as preparing a tensor network over the maximal volume slice. Similarly, in this context, non-optimal flows $v'$, \emph{i.e.} those with higher flux than $v$, and their corresponding thread configurations can be said to prepare tensor networks over different slices $\Sigma'$. See Figure \ref{fig:TNs} (right) for an illustration. Thus, if we are to construct an averaged measure of complexity that account for all these multiple tensor networks, we could alternatively average over non-minimal thread/flow configurations. We will come back to this point in next section.

\subsection{A more general proposal for holographic complexity} \label{subsec:genprop}

Before discussing a concrete proposal, let us offer a couple of comments to further clarify the idea that we want to put forward. First and foremost, there might be particular states in which the optimal tensor network is enough to compute the full set of entanglement entropies. The prototypical example is the vacuum state, but this also applies to any other static configuration, as well as linear (but possibly time dependent) perturbations of these states.\footnote{Notice that RT surfaces are stationary under arbitrary linear perturbations.} In these cases we have the property that the RT surfaces that compute entanglement entropy of different subsystems all lie on a constant-$t$ slice $\Sigma$, which is a maximal volume slice. Then, the standard notion of complexity as the volume of the maximal slice should be accurate (to a certain extent). However, we could argue that a rigorous measure of complexity should be state independent, in which case this notion of complexity could lead to some challenges. The problem is that for generic out-of-equilibrium configurations, the above observation is not true. Namely, the maximal volume slice cannot be foliated by HRT surfaces in general. This leads us to the next point: for these out-of-equilibrium states it is true that we need to consider tensor networks defined over all possible slices $\Sigma'$. Since these networks have different number of tensors and therefore different complexities, we must then consider an appropriate average over these to fully characterize the state. Likewise, appealing to the idea of state independence, we can argue that a similar average must be considered in general, even for the static cases discussed above.

More concretely, we propose that for a general state we should consider an ensemble of all possible tensor networks defined over all bulk slices $\Sigma'$. Formally, we can formulate this ensemble as a path integral, such that
\be
\mathcal{Z}\sim\int \mathcal{D}[\Sigma']e^{-\frac{1}{\hbar}\mathcal{S}[\Sigma']}\,,\qquad \Sigma'\in \text{WDW patch},
\ee
for a given measure of integration $\mathcal{D}[\Sigma']$ and weight $\mathcal{S}[\Sigma']$ that we have left unspecified. A couple of comments are in order. First, notice that we have chosen to integrate over slices $\Sigma'$ instead of tensor networks. We can formally do the latter provided we compensate with an appropriate change in the measure; however, the first option seems easier to work with so we will stick with it in the following discussion. Second, we have introduced for convenience a control parameter `$\hbar$' such that, for small $\hbar$, it is valid to work in the saddle point approximation. Given our previous discussion, then, we can assume that the maximal volume slice $\Sigma$ should emerge as a `classical' saddle, at least for the case of static spacetimes. For example, a canonical choice could be $\mathcal{S}[\Sigma']\sim\text{Vol}[\Sigma']$,
 however, there is ample room for maneuver to make this work. We also note that we cannot take $\hbar$ to be a universal value. For instance, we cannot take $\hbar\sim G_N$ because we have seen that even in the limit $G_N\to0$ there are situations where the maximal volume slice is not enough. A natural conjecture is that $\hbar$ is a covariantly defined quantity that can take values depending on the background. For instance, it could be associated to a time-scale of the state, so that for static cases $\hbar\to0$, while for dynamical cases one has $\hbar\neq0$. We will come back to this point below.

Assuming that this notion of ensemble average is well defined, we can now ask how an averaged measure of complexity would look. As usual, we could take the expectation value in the path integral language, \emph{i.e.}, (omitting constants of proportionality)
\be
\mathcal{C}\sim\frac{1}{\mathcal{Z}}\int \mathcal{D}[\Sigma']\,\text{Vol}[\Sigma']\,e^{-\frac{1}{\hbar}\mathcal{S}[\Sigma']}\,,
\label{eq:genCprop}\ee
for appropriate optimized choices of $\mathcal{S}$, $\hbar$, and measure of integration. In the saddle point approximation, or when $\hbar\to0$, we recover the complexity=volume formula. For general states, however, this formula gives a weighted average that deviates from CV. In terms of flows, we could imagine defining a weighted average,
\be
v_{\text{avg}}\sim\frac{1}{\mathcal{Z}}\int \mathcal{D}[v']\,v'\,e^{-\frac{1}{\hbar}\mathcal{S}[v']}\,,
\ee
from which we can compute the complexity (\ref{eq:genCprop}) as its flux. In this formula we are summing over arbitrary flow configurations $v'$ such that $\nabla \cdot v'=0$ but do not necessarily correspond to a solution to the min flux problem. The averaged version of the flow will still obey the divergenceless condition $\nabla \cdot v_{\text{avg}}=0$, however, it will generally relax the norm bound. Such a modification is reminiscent of the flow prescription for entanglement entropy for cases where higher curvature corrections are included \cite{Harper:2018sdd}. It will be interesting to use convex optimization techniques to derive a flow-based prescription like this one for other proposals of holographic complexity that do take into account other Cauchy slices, not necessarily the maximal, \emph{e.g.,} the CA or CV 2.0 proposals. We leave this for a future exploration.

It is worthwhile to compare our proposal of holographic complexity to the recent conjecture in \cite{Boruch:2020wax,Boruch:2021hqs}, which is intimately tied to path integral complexity \cite{Caputa:2017urj,Caputa:2017yrh} (briefly reviewed in Section \ref{subsec:prescripts}). The claim is the path integral optimization procedure in holographic CFTs is equivalent to maximizing the Hartle-Hawking (HH) wavefunction in AdS. More precisely, consider the case of the vacuum of a CFT on $\mathbb{R}^{d}$, such that bulk geometry is dual to Euclidean Poincar\'e $\text{AdS}_{d+1}$, $ds^{2}=z^{-2}(dz^{2}+d\tau^{2}+\sum_{i=1}^{d-1}dx_{i}^{2})$. The HH wavefunction of interest is
\beq \Psi_{\text{HH}}^{(T)}[\phi]=\int \mathcal{D}g_{\mu\nu}e^{-I^{E}_{\text{bulk}}[g]-I_{T}[e^{2\phi}]}\delta(g_{ab}|_{Q}-e^{2\phi}\delta_{ab})\;,\label{eq:HHwavefunc}\eeq
where $I^{E}_{\text{bulk}}$ is the bulk gravitational action in Euclidean signature, $Q$ is a bulk codimension-1 surface specified by $z=f(\tau)$ meeting the boundary cutoff surface $\sigma$ located at $z=\epsilon$, and $g_{ab}$ is the induced metric on $Q$ with line element $ds^{2}_{Q}=e^{2\phi(w)}[dw^{2}+\sum_{i=1}^{d-1}dx_{i}^{2}]$, with $w=w(\tau)$. The action $I_{T}$ ascribes a tension $T$ to $Q$, such that $Q$ may be interpreted as an end-of-world-brane with
\beq I_{T}[h]=\frac{T}{8\pi G_{N}}\int_{Q}\sqrt{h}\;.\eeq
Thus, the HH wavefunction $\Psi^{(T)}_{\text{HH}}$ is evaluated over a bulk region $M$ between $\sigma$ and $Q$, where the metric on $Q$ is viewed as the holographic dual of the conformally flat space metric used in the path integral optimization procedure, and $T\neq0$ provides a deformation of the usual HH wavefunction.

Via a saddle-point analysis, one evaluates the HH wavefunction semiclassically for on-shell configurations to leading order, such that $Q$ is a probe brane, \emph{i.e.}, $Q$ doesn't backreact on the bulk AdS geometry. The HH wavefunction is then maximized by choosing an appropriate metric on $Q$. Specifically, as detailed in \cite{Caputa:2017yrh}, the extrinsic curvature $K$ on $Q$ is proportional to the tension,
\beq K|_{Q}=\frac{d}{d-1}T\;,\eeq
such that maximization implies $Q$ provides a CMC slicing of the empty $\text{AdS}_{d+1}$.  The tension provides a measure of the complexity such that when $K=0$ the path integral complexity functional is optimized, where the conformal factor evaluated at the boundary becomes $e^{2\phi}=\epsilon^{-2}$. CMC slices $T\neq0$ correspond to partially optimized or sub-optimal path integral tensor networks.  Specifically, the holographic path integral complexity $\mathcal{C}^{(e)}_{T}$ of the partially optimized tensor network $Q$ at fixed $T$ is estimated by the on-shell bulk gravity action evaluated on $M$ plus a modified Hayward term $I_{H}$ \cite{Hayward:1993my} to deal with a non-smooth corner appearing in $M$ due to $Q$. The minimum value of $\mathcal{C}^{(e)}_{T}$ occurs when $T=0$, where $\mathcal{C}^{(e)}_{T}$ corresponds to complexity=volume-like duality.

Let us now directly compare our general proposal (\ref{eq:genCprop}) to the procedure of maximizing the deformed HH wavefunction (\ref{eq:HHwavefunc}). The HH wavefunction is a path integral over different metrics, which, upon maximization, can be understood as a path integral over different CMC slices $Q$, where the `classical' saddle corresponds to the maximal $K=0$ slice and $\mathcal{C}^{(e)}_{T=0}$ is proportional to the volume on $Q$. Our proposal is more general in that we are not specifying the particular geometry of the slice $\Sigma'$ (just that it lives inside the WDW patch), where generally $\Sigma'$ is an example of a sub-optimal tensor network. However, when $\Sigma'=\Sigma$, the tensor network is optimal such that complexity reduces to the CV relation. Thus, both proposals for holographic complexity (\ref{eq:genCprop}) and (\ref{eq:HHwavefunc}) make use of sub-optimal tensor networks, and we suspect, at least in particular contexts, the two proposals will coincide. In particular, for  Euclidean vacuum AdS and when $M$ is the interior of the Euclideanized WDW patch, slices $Q$ correspond to the CMC slices foliating the WDW patch, such that the two proposals  (\ref{eq:genCprop}) and (\ref{eq:HHwavefunc}) will be equivalent when $\Sigma'$ in (\ref{eq:genCprop}) are restricted to have constant mean curvature and where $\hbar$ corresponds to a scale such that $\hbar\to0$ corresponds to $\Sigma'\to\Sigma$, \emph{e.g.}, $\hbar\sim T$. While the precise form of  $\mathcal{S}[\Sigma']$ is not specified, a natural guess would be the bulk Euclidean action plus the Nambu-Goto like action for $Q$.\footnote{An important caveat worth mentioning: Euclidean path integral optimization described in \cite{Boruch:2020wax,Boruch:2021hqs} is closer to CV duality, while Lorentzian path integral optimization is similar to CA duality. Our proposal above is more like CV duality, however, the average does deviate from CV. In fact, depending on the choice of $\mathcal{S}$, it's possible the ensemble proposal interpolates between CA and CV. We thank Tadashi Takayanagi for a discussion on this point.} It would be interesting to explore the connection between the two proposals further.



\section{Conclusions and outlook} \label{sec:disc}

Let us now very briefly summarize the main ideas of this article before discussing avenues for future work. In this article we provided a Lorentzian flow based reformulation of the complexity=volume proposal, following an application of the min flow-max cut principle from network theory. In the language of flows, boundary computational complexity with respect to boundary regions covering the future boundary of a Lorentzian AdS spacetime is the minimum flux of a Lorentzian flow passing through the maximal Cauchy slice homologous to the boundary region. Crucially, the equivalence between flows and volumes \emph{requires} we attach Euclidean sections in the past and future of the Lorentzian manifold, akin to standard Lorentzian AdS/CFT. Moreover, unlike the usual CV and CA proposals, our interpretation relies on state preparation in AdS/CFT, for which the nature of the reference state (specified in the past Euclidean boundary) plays a crucial role. We used the nesting of bulk Cauchy slices to argue holographic complexity should obey particular inequalities, particularly a bound on the  complexity rate by the `conditional complexity' and, via the generalized PVC relation, we recovered the second time derivative of complexity is equal to the time derivative of the integrated momentum flux. Most of all, we argued that discretized Lorentzian flows should be interpreted in terms of gatelines, such that the complexity is given by the minimal number of gatelines preparing a CFT state on the maximal volume slice, given a choice of reference state. This interpretation, moreover, is aligned with tensor network constructions of AdS spacetimes, where the optimal tensor network is the one with the fewest number tensors, each connected by a gateline.

Following the abstract discussion of Lorentzian flows and their interpretation, we provided multiple geometric realizations of the thread configurations. Using general algorithms, these included radial timelike geodesics foliating the WDW patch in empty AdS and the BTZ black hole, as well as a foliation of the background by a family of constant extrinsic curvature slices. In the case of the BTZ black hole, we also showed how the flow configurations differ at both early and late times, where in the latter case the maximal volume slice wraps around the future singularity. Then, we explained how the second law of quantum complexity can be interpreted in terms of the number of threads crossing the portion of the maximal volume surface in the black hole interior (thus reaching the singularity) such that complexity monotonically increases as time moves forward. The maximal complexity is then interpreted as the complexity associated to the `final state' for which all Lorentzian threads crossing the maximal volume slice necessarily reach the singularity.

We studied perturbations to AdS spacetimes by developing a notion of perturbative Lorentzian threads, which we found most beneficial to describe using differential forms. In particular, we showed a canonical choice for the perturbative Lorentzian thread form is the symplectic current evaluated with respect to the new York deformation. The closedness condition of this form, required by consistency,  was shown to be equivalent to the  linearized Einstein equations holding in the bulk. Moreover, we provided a detailed derivation of linearized Einstein's equations by showing they follow from a first law for holographic complexity, where the complexity is dual to the volume and the boundary and bulk symplectic forms are dual. Our argument does not rely on the flow based reformulation of CV duality, though is certainly augmented by it.

Indeed, the boundary first law of complexity useful for us is given by the variation of the complexity with respect to the boundary sources, which is given by the symplectic form describing the space of sources. Here we have pictured threads as starting from the boundary attached to sources. Varying the sources then corresponds to varying the endpoints of the threads, each of which connect to physical tensors in a tensor network discretization of the bulk Cauchy slice. Consequently, the source variation directly corresponds to a variation of the tensors in the network and hence a changing volume (which is equivalent to the bulk symplectic form), leading the linearized Einstein's equations being satisfied in the bulk. Thus, if we take the viewpoint Lorentzian threads are fundamental objects, we can imagine spacetime being sewn together by thread configurations  attached to the boundary, encoding bulk spacetime dynamics. This picture advocates a notion of `spacetime complexity' where gravity emerges from varying complexity. At any rate, our derivation of Einstein's equations from the first law of complexity is disjoint from Lorentzian threads and can be considered a novel result of independent interest. Moreover, since the equivalence between bulk and boundary symplectic forms is valid for perturbations over general states, not just around vacuum AdS, suggests \emph{any} asymptotically AdS spacetime obeying CV duality and the first law for first order state/metric deformations around the background necessarily satisfies the full non-linear Einstein equations.

Building off of our observations we proposed a generalized notion of holographic complexity based on an ensemble average. Our conjecture is partly motivated by the fact CV duality does not seem suitable when describing highly excited dynamical states. In particular, in these cases not all HRT surfaces lie on the bulk maximal volume slice, such that calculating  holographic entanglement entropy requires one evaluate the entropy over bulk Cauchy slices other than the maximal volume one. From a tensor network perspective, optimal and sub-optimal tensor networks, and their associated thread configurations, are thus expected to play a role in defining a more general and state-independent measure of complexity, where one averages over all such possible tensor networks.

Lastly, at multiple points throughout this manuscript we advocated Lorentzian threads provide a notion of emergent time. Let us clarify what we mean by this.

\section*{Emergence of time}

 We begin with some philosophical remarks. There are several, sometimes competing, notions of time, and each depending on what is treated as `fundamental'. For example, in classical gravity we can distinguish between coordinate time $x^{0}=t$ and proper time $\tau$, where the former can be arbitrarily reparameterized and generally non-observable, while the latter depends on the background metric, determined by the dynamical fields in the theory, and is different with respect to each worldline.  Neither coordinate or proper time are considered to be a `physical time', \emph{i.e.}, a parameter which does not depend on a choice of coordinates or a metric representation of the background geometry. Note, moreover, neither of these notions of time provide the sense of forward progression, unlike, for example, thermodynamical time, which maintains a distinction between past and future by means of an arrow of time granted by the second law of thermodynamics.

In an ADM split (where the `ADM time' $t$ characterizing the split is typically a coordinate time),  the dynamical equations of general relativity describe how the spatial geometry, \emph{i.e.}, metric $h_{ij}$, changes with respect to the ADM time. As a theory of classical mechanics, moreover, the space of solutions are subject to momentum constraints and a Hamiltonian constraint, which, in the case of general relativity, leads to a non-trivial initial value problem. That is, in GR it is difficult to specify a full set of Cauchy data, namely $h_{ij}$ and the conjugate momenta $\pi^{ij}$, on a spatial slice that satisfy all of the constraints such that the evolution is uniquely determined at all future and past times. It is well known solving the initial value problem in GR is highly non-trivial since the Hamiltonian constraint is generally not satisfied uniquely. As described previously, York \cite{York:1972sj} provided an elegant solution to the initial value problem of GR where he showed only the scale free (conformal) part of the metric, and its conjugate momenta provide the necessary Cauchy data, however, only satisfy all of the constraints uniquely when restricted to surfaces of constant mean curvature. York's solution, then, provides a natural foliation of spacetime -- one by hypersurfaces of constant curvature -- where curvature $K$ gives a preferred notion of time, referred to as the `York time'.

 Let us now collect the ways in which the Lorentzian threads and our reformulation of complexity lead to an emergent notion of time. First, Lorentzian threads are naturally thought of as preparing CFT states on the bulk maximal volume slice. Time evolution of the CFT, typically found by solving the dynamical  bulk equations of motion, is captured by the closedness condition of the perturbative thread form $\delta u=\omega_{L}$. This is further exemplified in the tensor network picture, where unitaries attached to the threads act similar to the disentanglers in an Euclidean MERA network, transforming the reference state int the target state (defined over different layers of the network). Upon analytic continuation, these unitaries  naturally implement time evolution, in the same sense that the disentanglers implement evolution along the radial (holographic) direction. Second, the canonical flow solution  $\delta u=\omega_{L}$ makes use of a constant mean curvature slicing of the AdS vacuum, where the bulk is foliated by surfaces of constant extrinsic curvature $K$. In lieu of the above discussion, particular configurations of Lorentzian threads provide a natural foliation of the spacetime by hypersurfaces of constant $K$ (York time), such that data specified on these slices solve the initial value problem (whilst describing boundary time evolution of the CFT state). If we take the threads to be fundamental, while generally highly non-unique, there is a particular configuration that makes bulk locality explicit, $\delta u =\omega_{L}$, and York time emerges as a preferred notion of time.

This latter point is very reminiscent of another sense of emergent time suggested recently by path integral optimization definitions of boundary complexity \cite{Boruch:2020wax,Boruch:2021hqs}. In this set up, the maximization of a bulk Hartle-Hawking wavefunction is equivalent to the path integral optimization procedure in the CFT. The Hartle-Hawking wavefunction (\ref{eq:HHwavefunc}) is given in terms of a Euclidean path integral of the bulk gravitational action plus the probe brane action  of fixed tension $T$, where $T=0$ corresponds to the most optimal tensor network configuration and $T\neq0$ to suboptimal ones. The tension was seen as emergent in that it can be interpreted as the conformal radial coordinate in a bulk $d+1$-dimensional spacetime arising from the boundary optimization procedure. Since the tension $T$ is proportional to the extrinsic curvature $K$ of the $Q$ slices, surfaces $Q$ of tension $T$ foliate the bulk, and, consequently, $T$ may be interpreted as an emergent time. Similarly, though stated without proof, we expect a similar optimization argument would yield York time as an emergent physical time. Deepening the connection between Lorentzian threads and path integral optimization, particularly the related notions of emergent time, is worthy of further investigation.

\section*{Future directions}

Below we list a few potential avenues for research we find promising, in order of increasing speculation.

\noindent \textbf{`Complexity=generalized volume' and higher derivative theories}

In this article we focused on holographic CFTs dual to bulk theories of gravity governed by Einstein's equations. More generally one could consider other bulk theories of gravity, particularly higher derivative theories such as Gauss-Bonnet gravity. The CA conjecture for such holographic set-ups is then easily to generalize: simply work with the bulk action defining the alternative theory. CV duality is not as straightforward to extend as one needs identify a suitable generalization of the `volume'. Such a generalization was proposed in \cite{Bueno:2016gnv}, where the volume $V=\int_{\Sigma}d^{d}x\sqrt{h}$ is replaced with the local functional $W$:
\beq W=\frac{1}{(d-2)E_{0}}\int_{\Sigma}d^{d-1}x\sqrt{h}(E^{abcd}u_{a}u_{d}h_{bc}-E_{0})\;,\label{eq:genvol}\eeq
where $E^{abcd}\equiv\frac{\partial \mathcal{L}_{\text{grav}}}{\partial R_{abcd}}$, $u^{a}$ is a timelike unit normal to $\Sigma$, and $E_{0}$ is some constant dependent on the specific theory of gravity in question; when $\mathcal{L}_{\text{grav}}=E_{0}R$, one recovers the standard expression for volume. More recently, a generalized CV proposal was introduced in \cite{Hernandez:2020nem} to describe holographic complexity for boundary subregions in the `island phase', which modifies (\ref{eq:genvol}) by including additional extrinsic curvature corrections (we can think of volume $W$ (\ref{eq:genvol}) like a Wald functional for volume, analogous to Wald entropy, while the extrinsic curvature corrected volume functional is akin to generalized holographic entanglement entropy \cite{Dong:2013qoa}).

It would be interesting to extend the Lorentzian flow construction such that these CV generalizations could be incorporated. We expect, similar to the case of the bit thread reformulation of holographic entanglement entropy for CFTs dual to higher curvature theories \cite{Harper:2018sdd}, that one would need to replace the norm bound condition, altering the convex optimization program slightly. Separately, we point out our derivation of the linearized Einstein's equations could be extended to derive higher derivative theories of gravity where the volume in the first law of holographic complexity is now replaced by the generalized volume $W$. Indeed, it is straightforward to show
\beq \omega_{L}(\delta_{Y},\delta)\propto \delta W\;,\eeq
for higher derivative theories of gravity. Proof of the linearized equations of motion then follow from the same basic arguments we described above.  It would be interesting to see whether the symplectic current $\omega_{L}$ remains a canonical thread configuration, given the expected modification to the optimization program.

Moreover, the generalized CV proposal presented in \cite{Hernandez:2020nem} makes use of models of `double holography' \cite{Almheiri:2019hni,Chen:2020uac,Chen:2020hmv}. We thus expect one would need to extend the Lorentzian thread prescription to doubly holographic set-ups, for which the recent work on holographic bit threads to these same systems will be of use \cite{Rolph:2021hgz}.

\vspace{2mm}

\noindent \textbf{Comparing to other notions of complexity}

Here we primarily focused on two candidates of holographic complexity: CV and CA conjectures. As already discussed, these conjectures might come from a more general definition of complexity based on ensemble averages. This definition is distinct from but similar in spirit to path integral optimization complexity and its holographic dual \cite{Boruch:2020wax,Boruch:2021hqs}. Indeed, above we argued in specific contexts in which our ensemble proposal for holographic complexity coincides with the maximization of the Hartle-Hawking wavefunction. Our comparison was only for Euclidean AdS in vacuum, however, the maximization procedure was also applied to Lorentzian signature spacetimes, where for Lorentzian spacetimes the complexity functional seems to satisfy complexity=action duality. It would be very interesting to better understand how our ensemble proposal connects to the maximization of the Hartle-Hawking wavefunction,  including its relation to the quantum circuit models \cite{Caputa:2018kdj,Camargo:2019isp,Geng:2019yxo,Chandra:2021kdv}. For example, in \cite{Chandra:2021kdv} complexity of quantum circuits is given by a gravitational action, generalizes CV duality and shows CMC slices play a role in optimizing the circuits. A derivation of the `action' $\mathcal{S}$ from first principles should clarify this question.

We also point out CV duality was recently invoked to provide a CFT interpretation of the first law of causal diamonds $\text{AdS}_{3}$ \cite{Jacobson:2018ahi}, where the volume of a circular disk in $\text{AdS}_{3}$ is dual to the complexity of a cutoff CFT \cite{Sarkar:2020yjs}, such that the bulk law is understood as a first law of differential entropy on the boundary. It would be worthwhile to understand how our derivation of Einstein's equations from the first law of complexity fits in with this picture, though a challenge to overcome is knowing how to incorporate cutoff CFTs into our framework. Perhaps understanding the quantum circuit model analyzed in \cite{Chandra:2021kdv}, which generalizes the path integral optimization to holographic CFTs with finite cutoff, would lend insight.

\vspace{2mm}

\noindent \textbf{Non-linear Einstein's equations from complexity}

Our derivation of the linearized Einstein's equations made use of the first law of complexity, where we only considered linear perturbations about vacuum AdS. More generally one could consider perturbations around excited CFT states, which would carry information about non-linear contributions to bulk gravity equations. Indeed, the bulk-boundary symplectic form equivalence holds for perturbations over general states. This is precisely how non-linear equations were recovered in the gravity from entanglement scenario \cite{Faulkner:2017tkh,Haehl:2017sot}, where one must consider second order variations to the relative entropy (capturing second order contributions in sources $\lambda$ of the Euclidean path integral definition of excited CFT states). In the context of complexity, there are immediately two challenges. The first is identifying the equivalent of the relative entropy for complexity, and understanding its variations. A more difficult challenge, as we have discussed at length, is that for highly excited states CV duality should be generalized, however, a regime where we can still work with CV duality even with these higher order perturbations.

Alternatively, in \cite{Lewkowycz:2018sgn} it was argued that equality between bulk and boundary modular flows for undeformed, but arbitrary states and subregions, the double shape and state deformation of the of the entanglement entropy and modular Hamiltonian reproduces the JLMS formula \cite{Jafferis:2015del}. Then, going in the reverse direction, any asymptotically AdS geometry satisfying JLMS (or equivalently the RT formula) for first order state and metric deformations around the background necessarily satisfies both the linearized and non-linear Einstein equations. Above we applied similar logic to suggest the non-linear equations of motion are satisfied, however, without an explicit calculation. It is worth showing this in detail.

\vspace{2mm}

\noindent \textbf{Bulk reconstruction from complexity}

The Lorentzian thread construction of the canonical thread configuration, \emph{i.e.}, the thread form $u$ and its perturbation $\delta u$, make use of the property of bulk locality. Thus, in principle, it should be possible to invert this problem and recover the bulk spacetime metric from (linear) excitations  of the boundary quantum state. This is the essence of the program of `bulk reconstruction' (see, \emph{e.g.},
\cite{deHaro:2000vlm,Hammersley:2006cp,Czech:2012bh,Balasubramanian:2013rqa,Balasubramanian:2013lsa,Myers:2014jia,Czech:2014wka,Headrick:2014eia,Czech:2014ppa,Czech:2015qta,Engelhardt:2016wgb,Faulkner:2018faa,Roy:2018ehv,Espindola:2017jil,Espindola:2018ozt,Balasubramanian:2018uus,Bao:2019bib,Jokela:2020auu,Cao:2020uvb,Bao:2020abm}), which aims to build the bulk metric purely from boundary data, typically field theory entanglement, by inverting some differential operator encoding information about bulk metric perturbations. Since complexity describes the interior geometry of black holes, reconstruction of the geometry inside of a horizon should be possible starting from complexity, as was recently carried out in \cite{Hashimoto:2021umd}, where input knowledge of the time derivatives of complexity and the Hartman-Maldacena entanglement entropy, and the metric outside of the horizon is required. More generally, from the perspective pursued in this article, we would start with a manifold $M$ with boundary $\partial M$, and a set of forms $\delta u$ which encode the local details of complexity associated with timelike boundary subregions. We will have to assume some knowledge of the bulk, namely, the zeroth order, pure AdS metric $g_{\mu\nu}$. Then, at linear order at least, aside from this initial input,  we would then use the forms $\delta u$ -- which are uniquely specified from CFT data -- capturing the change in the CFT complexity for perturbative excited states, to reconstruct metric perturbations $\delta g_{\mu\nu}$.  A similar treatment of bulk reconstruction in the context of bit threads was carried out in \cite{Agon:2020mvu}, where it was found to reconstruct the metric beyond linear order in perturbations one must invert a higher order differential operator, however, this can be accomplished recursively in the order parameter $\eta$. Lessons from this bit thread approach may shed light onto the method of bulk reconstruction via complexity. In fact, metric reconstruction from complexity may be more powerful as we know the canonical thread satisfies $\delta u=\omega_{L}(\delta_{Y},\delta)$ and the duality between bulk and boundary symplectic structure holds for generic states.

\vspace{2mm}

\noindent \textbf{Quantum corrections to holographic complexity}

Both proposals for holographic complexity are purely classical, \emph{i.e.}, there is not a $1/N$ quantum corrected proposal for either duality, unlike the case for holographic entanglement entropy \cite{Faulkner:2013ana}: entanglement entropy is dual to the RT formula plus bulk entanglement entropy, summarized by the Faulkner-Lewkowycz-Maldacena (FLM) formula. Naively, one might imagine $1/N$-quantum corrected holographic complexity goes as, say, the volume plus `bulk complexity'.  While such a proposal has not been explicitly offered, lessons from quantum corrected entanglement entropy may lend insight. In particular, the recent `quantum bit thread' description of the FLM formula and quantum extremal surfaces \cite{Agon:2021tia,Rolph:2021hgz}, which, due to the similarities between Riemannian and Lorentzian flow principles, may shed light onto how one might engineer a quantum corrected version of holographic complexity and its Lorentzian thread interpretation.

Understanding the role of $1/N$ corrections may lend insight into an important conceptual question when interpreting Lorentzian flows as gatelines: do the gatelines `commute'? This is relevant since in real space entanglement buildup often requires gates which do not commute. The Lorentzian reformulation of CV, via the divergenceless condition, suggests  individual gates do commute.\footnote{We thank Michal Heller for bringing this important point to our attention.} To understand this better, let's first consider the Riemannian flow reformulation of holographic entanglement entropy \cite{Freedman:2016zud}, where the equivalent question is whether any CFT microstate dual to a classical background can be built with only bipartite entanglement, \emph{i.e.}, EPR pairs, or if more than this  is needed, multipartite generalizations. Based on the bit-thread reformulation of the RT formula, it was  conjectured in \cite{Cui:2018dyq} that EPR pairs are enough, however, as argued in \cite{Akers:2019gcv}, this is true only for simply connected regions, while for regions that are not simply connected \emph{e.g.} disjoint subsystems, one generally needs multipartite entanglement.\footnote{Though it should be noted \cite{Harper:2020wad},  even in non-simply connected cases the entanglement building blocks may be interpreted in terms of bipartite degrees of freedom, however, at the expense of changing the topology of the boundary manifold via purifications involving replicated manifolds.}

 Returning to the Lorentzian flow picture of CV, we believe the fact gates do not commute is not an inconsistency, at least not at leading order in $1/N$ in the large $N$ approximation. In particular, we can have a more entangled state as the target state since we are not only evolving the reference state, but also turning on non-trivial sources. Thus, since the threads attach to physical tensors (representing physical degrees of freedom) on a TN defined on $\Sigma$ and, likewise, we can imagine that they attach both to physical tensors and sources on $\Sigma_{-}$, the bulk Cauchy slice capping the southern hemisphere of Euclidean AdS. This situation may change when one considers $1/N$ corrections. Intuitively, in terms of Lorentzian threads, as proposed for the Riemannian case in \cite{Agon:2021tia}, the $1/N$ corrections may allow threads to split such that they connect multiple degrees of freedom on one slice or multiple degrees of freedom on another slice.

\vspace{2mm}

\noindent \textbf{Threads in flat and de Sitter space and complexity}

The continuous version of the Lorentzian min flow-max cut theorem is sufficiently general that it applies to bulk backgrounds beyond asymptotically AdS. In particular, the principle holds in asymptotically de Sitter and Minkowski spacetimes with a suitable cutoff, as first described in \cite{Headrick:2017ucz}. For de Sitter space, one important change is the boundary region $R$ used to define relative homology must be empty. Consequently, the boundary subregion $A$ homologous to bulk slice $\Sigma_{A}$ is one of three options: a proper subset of a future cutoff boundary, where $\Sigma_{A}$ is anchored along $\partial A$; the entire future cutoff such that $\Sigma_{A}$ is any Cauchy slice, or $A$ is the union of the future cutoff and a proper subset of the past cutoff, where $\Sigma_{A}$ is anchored at $\partial A$ along the past cutoff (for an illustration, see Figure 7 of \cite{Headrick:2017ucz}). Despite this change, and since holographic complexity conjectures have not been systematically derived from fundamental principles of AdS/CFT,  it is natural to propose holographic complexity, specifically CV duality, holds in these other spacetimes. The complexity=action conjecture was previously considered to hold for flat spacetimes in \cite{Fareghbal:2018ngr}, while both CV and CA proposals have been analyzed in de Sitter space  \cite{Reynolds:2017lwq}; de Sitter space even has a tensor network description \cite{Bao:2017qmt}. It would be interesting to further investigate Lorentzian flows in these spacetimes and extend the gateline picture developed here accordingly.

\vspace{2mm}

\noindent \textbf{Lorentzian multiflows and de Sitter entanglement}

An interesting open question is whether Lorentzian multiflows exist and are useful. To briefly recap (and further analyzed in the Appendix below), fluxes through the disconnected boundary subregions are not representative of complexity. This is because these regions are not homologous to bulk Cauchy slices, nor can they be replaced by Euclidean caps as in the state preparation picture. Moreover, a naive definition of Lorentzian flows (following the Riemannian multiflow construction \cite{Cui:2018dyq} with the norm bound flipped to the opposite direction) doesn't appear to work due to the future directed condition, typically leading to uncorrelated flows. One can imagine more exotic scenarios in which the future (and past) boundaries of a Lorentzian spacetime are split into multiple subregions such that each subregion is homologous to a Cauchy slice.\footnote{An even more exotic spacetime would be one with a causal structure allowing for multiple futures.} In such a set-up it seems an inequality like the monogamy of mutual information may hold as the usual cut-and-paste proof with surfaces doesn't appear to breakdown. A potentially simple and illuminating setting to explore this question is to allow for the future boundary be two dimensional such that the future can be split into three intervals which touch.\footnote{For example, consider a $1+1$-dimensional FLRW closed spacetime with a big crunch like singularity with space is compactified into a circle. We thank Matthew Headrick for pointing out this example and further clarification on the conceptual hurdles of Lorentzian multiflows.} The maximal cuts for each of the boundary subregions are then spacelike geodesics connecting the boundary points. Via a cut-and-paste argument the pairwise `mutual complexities' are each negative such that the `tripartite complexity' is positive, suggesting a non-monogamy relation. It would be interesting to study this problem in detail from the flow perspective. Understanding the case when both past and future boundaries are segmented may be applicable to the study of de Sitter entanglement (\emph{e.g.}, \cite{Narayan:2017xca,Narayan:2019pjl,Arias:2019pzy,Narayan:2020nsc}) , where rotated Hartmann-Maldacena surfaces, timelike codimensnion-2 extremal surfaces stretching between the past and future boundaries in de Sitter space, describe de Sitter entropy as entanglement via dS/CFT.


\section*{Acknowledgments}

It is a pleasure to thank Cesar Ag\'on, Jos\'e Barb\'on, Elena C\'aceres, Willy Fischler, Matthew Headrick, Michal Heller, Ted Jacobson, Javier Mart\'\i{}n-Garc\'\i{}a, Rob Myers, Martin Sasieta, Leonard Susskind, Tadashi Takayanagi, and Manus Visser for useful discussions and comments on the manuscript. This material is based upon work supported by the Simons Foundation through \emph{It from Qubit: Simons Collaboration on Quantum Fields, Gravity, and Information}. AR and ZWD acknowledge additional financial support from EPSRC.

\appendix

\section{Riemannian flows and holographic entanglement entropy}\label{app:examples}

To keep this article self-contained, here we briefly review the Riemannian flow reformulation of the Ryu-Takayanagi formula as well as its interpretation. Following \cite{Headrick:2017ucz}, we first outline the statement of the Riemannian max flow-min cut theorem and show that it naturally leads to an alternative prescription for holographic entanglement entropy in terms of bit threads. We then  describe some general properties of Riemannian flows and discuss various relevant explicit constructions built in \cite{Agon:2018lwq,Agon:2020mvu}. These constructions are the Riemannian analogs of the Lorentzian flows discussed in the body of this paper.

\subsection{Max flow-min cut theorem and bit threads}

Let $M$ be a compact, oriented manifold with boundary, endowed with a Riemannian metric $g_{\mu\nu}$. A bulk codimension-1 surface $m$ is said to be homologous to a boundary region $A$, $m\sim A$, if there exists a bulk region $r$ whose boundary is $\partial r=A-m$, such that
\beq
\partial r\setminus \partial M=-(m\setminus\partial M)\,.
\label{eq:relhomoRie}
\eeq
A \emph{Riemannian} flow is defined to be a vector field $v^{\mu}$ on $M$ that is divergenceless, whose norm is bounded above by some real positive constant $\alpha$,
\beq
\nabla_{\mu}v^{\mu}=0\,,\qquad |v|\leq \alpha\,.
\label{eq:Rieflowdef}
\eeq
Using these properties, Stokes' theorem then implies the following inequality for any bulk surface (a cut) $m\sim A$ and any flow $v^{\mu}$
\beq\label{MFMCRIE}
\int_{A}v=\int_{m}v\leq\alpha\,\text{area}(m)\,.
\eeq
The max flow-min cut theorem is the statement that follows from the saturation of (\ref{MFMCRIE}):
\beq
\underset{v}{\text{max}}\int_{A}v=\alpha\,\underset{m\sim A}{\text{min}}\,\text{area}(m)\,.\label{eq:maxflowmcthm}
\eeq
On the left hand side, `maximum' is to mean the supremum, `sup', while on the right hand side `minimum' means infimum, `inf'. The max flow-min cut theorem was proven in \cite{Headrick:2017ucz} using convex optimization techniques. We will provide an outline of the proof for the (Lorentzian) min flow-max cut theorem in Appendix \ref{append:Lorentzmfmc}, which follows analogously to the Riemannian case.

As first shown in \cite{Freedman:2016zud}, by identifying the normalization constant to be $\alpha\equiv\frac{1}{4G_{N}}$, the max flow-min cut theorem (\ref{eq:maxflowmcthm}) provides a reformulation of the Ryu-Takayanagi formula for holographic entanglement entropy $S(A)$ in terms of Riemannian flows
\beq
S(A)=\underset{v}{\text{max}}\int_{A}v=\frac{1}{4G_{N}}\underset{m\sim A}{\text{min}}\,\text{area}(m)\,,\label{eq:RTreformflow}
\eeq
where $v$ is a flow with maximal flux through $A$. Written in this way, the RT formula can be given a sharper information theoretic meaning by interpreting the integral curves of a flow $v$ as a set of Planck-thickness `bit threads'. A thread configuration is more precisely defined as a set (possibly unoriented) integral curves of a flow $v$ on $M$ satisfying two properties:
\begin{enumerate}
  \item They start and end on $\partial M$.
  \item Their density is nowhere larger than $|v|$.
\end{enumerate}
Heuristically, each thread can be visualized as a channel that communicates a single bit of quantum information between spatially separated boundary regions. The total amount of information being communicated between two boundary regions, say $A$ and its complement $\bar{A}$, is given by the number of channels connecting the two regions allowed by the bulk geometry. That is, the entanglement entropy $S(A)$ is equal to the maximum number of allowed threads connecting $A$ and $\bar{A}$,
\beq
S(A)=\text{max}\,N_{A\bar{A}}\,,
\eeq
where $N_{A\bar{A}}$ is the number of threads connecting region $A$ to $\bar{A}$.

\subsection{General properties of Riemannian flows}

An important lemma of the max flow-min cut theorem (\ref{eq:maxflowmcthm}) is the nesting property. More specifically, for two disjoint but not necessarily separate boundary regions $A$ and $B$, it follows that the bulk region homologous to $A$, $r(A)$, is a proper subset of the bulk region homologous to the union $AB$, \emph{i.e.}, $r(A)\subset r(AB)$. Correspondingly, the nesting property implies that there exists a flow $v(A,B)$ that simultaneously maximizes the flux through $A$ and $AB$,
\beq
S(A)=\int_{A}v(A,B)\,,\qquad S(AB)=\int_{AB}v(A,B)\,.\label{eq:nestRie}
\eeq
The nesting property can be invoked to elegantly prove the subadditivity and strong subadditivity inequalities of entanglement entropy,
\be
S(A)+S(B)\geq S(AB)\,,
\ee
\be
S(AB)+S(BC)\geq S(B)+S(ABC)\,,
\ee
respectively, which must be satisfied for any quantum system. A proof for the nesting property using convex optimization methods is provided in \cite{Headrick:2017ucz}.

The nesting property of Riemannian flows can be generalized to the case of multiple nested surfaces or, equivalently, to the case of multiple non-intersecting boundary regions. Such a flow is called a multicommodity flow or max multiflow, and their existence was proven in  \cite{Cui:2018dyq} using convex optimization techniques. More precisely, let $\{A_{i}\}$ with $i=1,...,n$ be a set of non-overlapping boundary regions that cover the entire boundary $\partial M$, \emph{i.e.}, $\cup_{i}A_{i}=\partial M$. A multiflow is a set of vector fields $v_{ij}$ on $M$ obeying
\beq
v_{ij}=-v_{ji}\,,\qquad \nabla\cdot v_{ij}=0\,,\qquad \sum_{i<j}^{n}|v_{ij}|\leq\alpha\,,\label{eq:multiflowdef}
\eeq
such that $\hat{n}\cdot v_{ij}=0$ on $A_{k}$ for $k\neq i,j$. Given a multiflow $\{v_{ij}\}$, it can be shown that the set of $n$ vector fields $v_{i}\equiv\sum_{j=1}^{n}v_{ij}$ also satisfy the defining properties of flows, i.e., $\nabla\cdot v_i=0$ and $|v_i|\leq \alpha$, and therefore satisfy the bound
\beq
\int_{A_{i}}v_{i}\leq S(A_{i})\,.
\eeq
A max multiflow in then a set of flows $\{v_{ij}\}$ such that for each $i$, the flow $v_{i}$ defined above yields the maximum flux through $A_{i}$, \emph{i.e.},
\beq
\int_{A_{i}}v_{i}=S(A_{i})\,.
\label{eq:maxmultiflowrie}\eeq
When $n=3$, the existence of a max multiflow implies the nesting property (\ref{eq:nestRie}). Another interesting corollary is when $n=4$, leading to the monogamy of mutual information,
\beq\label{eq:monogamy}
S(AB)+S(AC)+S(BC)\geq S(A)+S(B)+S(C)+S(D)\,.
\eeq
This inequality does not hold for general quantum systems, but it is satisfied by those with holographic duals.


\subsection{Explicit constructions and applications}

There is one crucial difference between computing entanglement entropy via max flows or via minimal surfaces:  while the surfaces are in most cases unique, the solution to the max flow problem is highly degenerate. Intuitively, this degeneracy could be associated to a choice of microstate (or a particular class of microstates) that give rise to the same amount of entanglement between the region $A$ and its complement. This non-uniqueness might seem problematic at first sight, however, it turns out to be very useful in many instances. Indeed, one can exploit this non-uniqueness in clever ways to get information that would be hidden otherwise, for example, by constructing specific solutions that are adapted to the problem at hand or that realize certain property that makes a physical interpretation manifest. Some of these construction were introduced in \cite{Agon:2018lwq,Agon:2020mvu}, and are the Riemannian analog to the Lorentzian flows constructed in Section \ref{sec:simpleconstructions}. Here we will very briefly summarize some of these methods, their properties and applications. In particular, we will discuss the methods of construction based on: (i) \emph{integral lines}, (ii) \emph{level sets}, and (iii) \emph{the Iyer-Wald formalism}.

\subsection*{Construction via integral lines}

The first method of construction starts with a set of integral lines satisfying certain properties:
\begin{itemize}
  \item They must be orthogonal to the minimal surface $m$.
  \item They must be continuous and not self-intersecting.
  \item They must start and end at the boundary of the manifold $\partial M$.
\end{itemize}
Given a family of curves satisfying these properties one then constructs a vector flow $v=|v|\hat{\tau}$, such that $\hat{\tau}$ coincides with the tangent vector associated to
the curves. The magnitude $|v|$ is obtained by applying Gauss's law in its integral form. In order to do so, one parametrizes the integral curves collectively as $X(\vec{x}_{m},\lambda)$, where $\vec{x}_{m}$ is the point at which they intersect the minimal surface and $\lambda$ is an affine parameter running along each of the curves. The magnitude is then obtained by integrating the flux across a Gaussian surface, taken to be an infinitesimal cylinder enclosing one of the curves. This leads to the following formula for the magnitude:
\beq
|v(\vec{x}_{m},\lambda)|=\sqrt{\frac{h(\vec{x}_{m},\lambda_{m})}{h(\vec{x}_{m},\lambda)}}\,,
\eeq
where $h$ is the determinant of the (transverse) metric and $\lambda_m$ is the value of the affine parameter at the location of the minimal surface. In this construction the norm bound is automatically saturated at the location of the minimal surface, however, it must be checked everywhere else \emph{a posteriori}.

A particular class of flows constructed with the above algorithm are the so-called \emph{geodesic flows}. These have the particular property that their integral lines are taken to be space-like geodesics, hence they satisfy a number of nice geometric properties which can be connected with corresponding bulk energy conditions. This method was used to explicitly construct holographic bit threads associated with highly symmetric boundary regions, namely spheres and strips, in empty AdS and planar black brane geometries. Studying more generic bulk metrics, one can make use of standard tools of Riemannian geometry to arrive to some generic results. For instance, the Raychauduri equation on a Riemannian manifold implies that nearby space-like geodesics have positive expansions provided a sufficient set of conditions on the geometry and matter fields are satisfied. Assuming they are, this in turns implies that the norm of the vector field decreases monotonically away from the minimal surface, thus, ensuring the norm bound is satisfied everywhere. Interestingly, for non-vacuum solutions in a negatively curved space, the energy density of the matter fields turns out to be bounded above by the background cosmological constant.

\subsection*{Construction via level sets}

The second method starts with a set of level set surfaces satisfying these properties:
\begin{itemize}
  \item They must contain the minimal surface as one of its members.
  \item They must be continuous and not self-intersecting.
  \item They must not include closed bulk surfaces.
\end{itemize}
Given a family satisfying the above one can proceed as follows: first, one can generate the integral lines by requiring them to be orthogonal to the level set surfaces. With the integral lines at hand, the problem then reduces to the method explained above. Another equivalent way of proceeding is by constructing an appropriate scalar function $\varphi(x^i)$ such that the $\varphi=$ constant surfaces give us the desired level sets. We can then write the vector flow as:
\be\label{v:levelsets}
v=\Upsilon(x^i)\nabla \varphi(x^i)\,,
\ee
and determine $\Upsilon$ by applying Gauss's law in its differential form. This amounts to solve a first order differential equation
\be\label{eq:PsiPDE}
(\nabla\varphi)\cdot(\nabla\Upsilon)+(\nabla^2\varphi) \Upsilon=0\,,
\ee
subject to the boundary condition
\bea
\Upsilon\big|_{m} =\frac{1}{|\nabla \varphi|}\,.
\eea
In any of the two methods we would need to check if norm bound is satisfied \emph{a posteriori}. However, we can bypass this obstacle in a clever way; this is, we can ensure that the norm bound is satisfied by construction if we impose an extra condition on the level set surfaces:
\begin{enumerate}
\setcounter{enumi}{3}
\item They must be homologous to the region $A$.
\end{enumerate}
This condition is not a strict requirement, but a useful one in many situations. More generally, we do not have to assume 4 but in such cases we do need to care about the norm bound.

A particular class of flows constructed from the level set method are the so-called \emph{maximally packed flows}. These flows are constructed by taking the family of level set surfaces as a set of continuously nested minimal surfaces. In this case it can be shown that $v=\hat{n}$ is a solution to the max flow problem, where $\hat{n}$ is the unit normal associated with the level sets. As a result, these particular flows have the peculiar property of saturating the norm bound, $|v|=1$, everywhere in the bulk (or, at least, in the region that is foliated by the family of minimal surfaces) hence their name \emph{maximally packed}. It is clear these flows represent valid bit thread configurations since the extrinsic curvature of the chosen level set surfaces is zero for every member of the family, thus, the divergenceless condition is trivially satisfied, $K=\nabla\cdot\hat{n}=0$. Importantly, maximally packed flows were used in \cite{Agon:2018lwq} in combination with geodesic flows to define \emph{mixed flows}. Among various applications, these flows were used to compute the entanglement wedge cross section of various configurations and to give it a neat information-theoretic interpretation in terms of \emph{entanglement distillation}. Moreover, they were also used as basic ingredients in the construction of explicit examples of \emph{max multiflows}, which were in turn used to illustrate the monogamy of mutual information (\ref{eq:monogamy})

\subsection*{Construction via Iyer-Wald}

The Iyer-Wald formalism is an application of Noether's theorem for a symmetry generated by a Killing vector $\xi$ \cite{Iyer:1994ys}. For on-shell linear perturbations around AdS, this formalism was used to interpret the entanglement entropy associated with spherical regions in the boundary theory as the Noether's charge associated with a specific symmetry \cite{Faulkner:2013ica}. Conversely, starting from the laws of entanglement in the dual theory, it was found that Iyer-Wald leads to the linearized Einstein's equation around AdS. More recently in \cite{Agon:2020mvu}, it was shown that the Iyer-Wald formalism can be used to specify a canonical thread configuration for an arbitrary perturbative state in the bulk. Such a proposal makes the property of \emph{bulk locality} manifest. In particular, this construction was shown to elegantly encode the linearized Einstein's equations at every point in space, through the divergenceless condition.

To understand the details of this construction it is useful to express the max flow problem in terms of differential forms. Given a $d-$dimensional manifold $M$ endowed with a Riemannian metric, $g_{ab}$, there is an explicit map between divergenceless vector fields $v$, \emph{i.e.}, flows, and closed $(d-1)$-forms $\omega$, such that the divergenceless condition $\nabla\cdot v=0$ corresponds to the closedness condition $d\omega=0$, i.e.,
\beq
d\omega=(\nabla\cdot v)\epsilon=0\,,\label{eq:closedformflow}
\eeq
where $\epsilon$ is the natural volume form.\footnote{An independent reformulation of the Ryu-Takayanagi prescription in terms of calibrations (closed forms) was worked out in \cite{Bakhmatov:2017ihw}.} Here the vector field $v^{a}$ is related to the $(d-1)$-form via $v^{a}=g^{ab}(\star\omega)_{b}$, where $\star\omega$ represents the Hodge dual of $\omega$, defined as $(\star\omega)_{b}=\frac{1}{(d-1)!}\sqrt{g}\omega^{a_{1}...a_{d-1}}\varepsilon_{a_{1}...a_{d-1}b}$, with $\varepsilon_{a_{1}...a_{d}}$ being the total antisymmetric Levi-Civita symbol. The inverse map $\omega=\frac{1}{(d-1)!}\epsilon_{a_{1}...a_{d-1}}v^{b}dx^{a_{1}}\wedge...\wedge dx^{a_{d-1}}$, relates the form $\omega$ directly to the vector field in terms of the natural volume form $\epsilon=\frac{1}{d!}\epsilon_{a_{1}...a_{d}}dx^{a_{1}}\wedge...\wedge dx^{a_{d}}$.

As shown in \cite{Agon:2020mvu} (see also \cite{Freedman:2016zud,Headrick:2017ucz} for further discussion) the relationship (\ref{eq:closedformflow}) allows for a reformulation of the bit thread description of the Ryu-Takayanagi formula in terms of closed differential forms. More precisely, using that the form $\omega$ evaluated on an arbitrary codimension-1 bulk surface $\Gamma$ can be cast as
\beq
\omega|_{\Gamma}=(n_{a}v^{a})\tilde{\epsilon}\,,
\eeq
where $n$ is the local unit normal and $\tilde{\epsilon}$ is the $(d-1)$-dimensional volume form induced on $\Gamma$, the max flow-min cut theorem is reexpressed as
\beq
\underset{\omega\in W}{\text{max}}\int_{A}\omega=\underset{m\sim A}{\text{min}}\int_{M}\tilde{\epsilon}\,.\label{eq:maxflowmincutforms}
\eeq
Here $W$ is the set of closed forms obeying the norm bound (in the presence of a metric).
Consequently, the RT formula for entanglement entropy is given by
\beq
S_{A}=\frac{1}{4G_{N}}\underset{\omega\in W}{\text{max}}\int_{A}\omega\,.\label{eq:diffformHEE}
\eeq

The differential form reformulation of the RT formula (\ref{eq:diffformHEE}) is particularly useful when one does not know the explicit form of the background metric. For example, in \cite{Agon:2020mvu} this alternative description was useful in characterizing \emph{perturbative bit threads} --- Riemannian bit threads corresponding to linear perturbations around AdS. Along the way, the use of the Iyer-Wald formalism turned out particularly useful, as it provides a canonical choice for the pertubed thread configuration, $\omega=\omega_0+\delta\omega$, where $\omega_0$ is a vacuum solution to the max flow problem and $\delta\omega\equiv 4G_{N}\tilde{\chi}$. Here $\chi$ is a $(d-1)$-form that arises from Noether's theorem for a symmetry generated by a Killing vector $\xi$ that generates the RT surface; $\tilde{\chi}$ is the form evaluated on the Cauchy hypersurface $\Sigma$ containing the (spherical) boundary region $A$ and its associated RT surface $m$. Moreover, the linearized Einstein's equations $\delta E_{\mu\nu}=0$ arise from $\tilde{\chi}$ being closed, $d\tilde{\chi}=-2\xi^{\mu}\delta E_{\mu\nu}\epsilon^{\mu}=0$. This property was used to recast the problem of metric reconstruction in terms of the inversion of a specific differential operator.


\section{Some proofs involving Lorentzian flows} \label{append:Lorentzmfmc}

\subsection{Min flow-max cut}

As discussed in Section \ref{sec:prelims}, the max flow-min cut theorem for flows on Riemannian manifolds becomes the min flow-max cut theorem for flows on Lorentzian spacetimes \cite{Headrick:2017ucz}
\beq \underset{v}{\text{min}}\int_{A}v=\underset{\Sigma\sim A}{\text{max}}\;\text{vol}(\Sigma)\;,\label{eq:minflowmaxcutapp}\eeq
where for now we set $|v|\geq \alpha=1$. By `minimum' we mean the infimum over flows $v$, and by `maximum' we mean supremum. The proof for the theorem relies on convex optimization techniques and takes a similar form as the proof in the Riemannian case. For completeness, let us present a slightly more detailed proof than the one presented in \cite{Headrick:2017ucz}.

First let us very quickly state the necessary elements of convex optimization to be used in the proof. For more details see \cite{Headrick:2017ucz} and the references therein.


\subsection*{Convex optimization basics}

 A convex program, or optimization problem $P$ can be generically written as follows:
$$P\;: \text{minimize} \;f_{0}(y)\;\text{over}\; y\in\mathcal{D},\;\text{subject to}\; f_{i}(y)\leq 0\;\text{and}\; h_{j}(y)=0$$
for all $i,j$. Here $y$ are vectors living in some vector space $Y$ for which the domain $\mathcal{D}\subset Y$ is a non-empty convex subset. The convex function $f_{0}:\mathcal{D}\to\mathbb{R}$ is the objective function; $f_{i}:\mathcal{D}\to\mathbb{R}$ denotes a set of (convex) inequality constraint functions, and $h_{j}:\mathcal{D}\to\mathbb{R}$ is an (affine) equality constraint function. The solution $p^{\ast}$ of $P$ is the infimum of objective $f_{0}$ on the feasible set $\mathcal{F}$, \emph{i.e.}, the set of all points in $\mathcal{D}$ for which $f_{i}(y)\leq0$ and $h_{j}(y)=0$ for all $i,j$. Succinctly, $p^{\ast}=\text{inf}f_{0}(\mathcal{F})$.

As is the case for the proof of the max flow-min cut and min flow-max cut theorems, it is often useful to write the original (primal) optimization problem in terms of an equivalent dual problem using Lagrangian duality. Lagrangian duality has one introduce Lagrange multipliers to enforce the constraints, solving for the original variables to rewrite the original (primal) problem as an equivalent dual optimization problem. More precisely, the Lagrangian $L(y,\lambda,\nu)$ is defined as
$$L(y,\lambda,\nu)\equiv f_{0}(y)+\sum_{i=1}^{m}\lambda_{i}f_{i}(y)+\sum_{j=1}^{n}\nu_{j}h_{j}(y)$$
where $\lambda\in(\mathbb{R}^{+})^{m}$  and $\nu_{j}\in\mathbb{R}^{n}$ are Lagrange multipliers enforcing the inequality constraints $f_{i}(y)\leq0$, and equality constraints $h_{j}(y)=0$, respectively. The solution $p^{\ast}=\text{inf}f_{0}(y\in\mathcal{F})$ to the primal problem is then the solution to the dual program when the supremum of $L(y,\lambda,\nu)$ is equal to $f_{0}(y\in\mathcal{F})$. Precisely, $\underset{y\in\mathcal{D}}{\text{inf}}\underset{\lambda,\nu}{\text{sup}}L(y,\lambda,\nu)=\text{inf}\,f_{0}(y\in\mathcal{F})$. The dual program $P'$ is therefore a concave program, where one maximizes $g_{0}(\lambda,\nu)\equiv\underset{y\in\mathcal{D}}{\text{inf}}L(y,\lambda,\nu)$ over $\lambda,\nu$.

The primal and dual programs are said to exhibit strong duality when the solution of the dual program $d^{\ast}$ is equal to the solution of the primal program, $d^{\ast}=p^{\ast}$. Strong duality is implied by Slater's condition, which formally states the convex problem admits a feasible point $y_{0}$ in the interior of $\mathcal{D}$ such that $f_{i}(y_{0})<0$ for all $i$.

These definitions will be enough for us to prove the min flow-max cut theorem (\ref{eq:minflowmaxcutapp}), as shown in \cite{Headrick:2017ucz}.


\subsection*{Proof}

The proof of the min flow-max cut theorem largely involves three steps: (i) Express the left hand side of (\ref{eq:minflowmaxcutapp}) as a convex optimization problem and show Slater's condition is satisfied; (ii) Use Lagrangian duality to rewrite the program as an equivalent optimization problem, and (iii) verify the solution to the dual program is $\text{sup}_{\Sigma\sim A}\text{vol}(\Sigma)$.

\textbf{(i)} It is straightforward to state the left hand side of the min flow-max cut theorem (\ref{eq:minflowmaxcutapp}) as a convex program. Using the definition of Lorentzian flows (\ref{eq:Lorflowdef}), one has
\beq
\begin{split}
 \text{min flow}\;:\;\text{minimize}\;\int_{A}\sqrt{h}n_{\mu}v^{\mu}&\;\;\text{over}\; v^{\mu} \;\text{with}\;\sqrt{h}n_{\mu}v^{\mu}|_{R}=0\,,\\
&\text{subject to}\;\nabla_{\mu}v^{\mu}=0\;,\;1-|v^{\mu}|\leq0\;,
\end{split}
\label{eq:minflowprog}\eeq
where the future directed causal vector fields $v^{\mu}$ living in $M$ form a convex set. We now want to show Slater's condition holds. The argument is as follows. Given boundary region $R^{c}=\partial M\setminus R$ satisfies $J^{+}(R^{c})=J^{-}(R^{c})=M$, every point in the interior of $M$ lies on a timelike curve that begins an ends on $R^{c}$, \emph{e.g.}, from past to future timelike infinity. Consequently, $M$ may be covered with overlapping timelike tubes of some proper constant thickness, where inside each tube a Lorentzian flow $v$ may be placed. In the language of Slater's condition, there exists a feasible vector field $v^{\mu}$ with $1-|v|<0$ everywhere.

\textbf{(ii)} Next dualize the min flow program (\ref{eq:minflowprog}) using Lagrangian duality. This is accomplished by introducing Lagrange multiplier scalar fields $\psi,\phi$ for the explicit constraints $\nabla\cdot v=0$ and $1-|v|<0$, respectively. We take $\phi\geq0$. The Lagrangian function is then
\beq
\begin{split}
L(v^{\mu},\psi,\phi)&=\int_{A}\sqrt{h}n_{\mu}v^{\mu}+\int_{M}\sqrt{g}(-\psi\nabla_{\mu}v^{\mu}+\phi(1-|v^{\mu}|))\\
&=\int_{\partial M}\sqrt{h}n_{\mu}v^{\mu}(\chi_{A}-\psi)+\int_{M}\sqrt{g}(v^{\mu}\partial_{\mu}\psi-|v^{\mu}|\phi+\phi)\;,
\end{split}
\label{eq:Lagdualmfmc}\eeq
where to get to the second line integration by parts was used and the characteristic function $\chi_{A}$ for $A$ on $\partial M$ was introduced, such that $\chi_{A}=1$ on $A$ and $\chi_{A}=0$ on $A^{c}$. The task now is to minimize $L$ with respect to $v^{\mu}$, where we are interested in when $L$ is bounded from below.  We do this by analyzing the two terms on the right hand side of (\ref{eq:Lagdualmfmc}) separately.

 Begin with the first term, $\int_{\partial M}\sqrt{h}n_{\mu}v^{\mu}(\chi_{A}-\psi)$.  Keeping in mind the implicit constraint $\sqrt{h}n_{\mu}v^{\mu}|_{R}=0$, this term is non-zero only on $R^{c}$. The boundary region $R^{c}$ is comprised of three components:  $R_{0}^{c}, R_{-}^{c}$, and $R_{+}^{c}$ are the timelike, past spacelike, and future spacelike parts of $R^{c}$, respectively. By past (future) spacelike, we mean $n_{\mu}$ is past (future) directed timelike covector fields. Each of these components lead to three different scenarios when the first term in (\ref{eq:Lagdualmfmc}) is bounded from below, \emph{i.e.}, has a finite infimum: (1) On $R^{c}_{0}$, $n\cdot v$ can take either sign since $n_{\mu}$ is spacelike, such that the infimum is finite if and only if $\psi|_{R^{c}_{0}}=\chi_{A}$; (2) on $R^{c}_{-}$ $n\cdot v$ is negative and therefore the infimum is finite if and only if $\psi|_{R_{-}^{c}}\geq\chi_{A}$, and (3) on $R^{c}_{+}$ $n\cdot v$ is positive such that the infimum is finite if and only if $\psi|_{R_{-}^{c}}\leq\chi_{A}$.

Now consider the second term, $\int_{M}\sqrt{g}(v^{\mu}\partial_{\mu}\psi-|v^{\mu}|\phi+\phi)$. This term will have a finite infimum if and only if $v^{\mu}\partial_{\mu}\psi\geq0$, and thus $\partial_{\mu}\psi$ is a future directed causal covector field (such that $\partial^{\mu}\psi$ is a past directed causal vector field). Moreover, given that $v^{\mu}$ is timelike, by the reverse Cauchy-Schwarz inequality we have $v^{\mu}\partial_{\mu}\psi=|v^{\mu}||\partial_{\mu}\psi|\cosh(\eta)$, for $\eta$ an angle between $v^{\mu}$ and $\partial_{\mu}\psi$; consequently, $0\leq |v^{\mu}|(|\partial_{\mu}\psi|\cosh(\eta)-\phi)$. Therefore, the second term will have a finite infimum if and only if $0\leq \phi\leq|\partial_{\mu}\psi|$.

Imposing each of these conditions above, $\int_{M}\sqrt{g}\phi$ is the dual objective to the corresponding dual program
\beq
\begin{split}
&\text{max cut 1}\;:\;\text{maximize}\;\int_{M}\sqrt{g}\phi \;\;\text{over}\;\psi,\phi\\
&\text{with}\; \psi|_{R_{0}^{c}}=\chi_{A}\;, \psi|_{R_{-}^{c}}\geq\chi_{A}\;,\;\psi|_{R_{+}^{c}}\leq\chi_{A}\;,0\leq\phi\leq|\partial_{\mu}\psi|
\end{split}
\label{eq:dualprogLorentzv1}\eeq
with $\partial_{\mu}\psi$ being a future directed causal vector field.

\textbf{(iii)} To complete the proof of the min flow-max cut theorem, the task now is to show the solution to the dual program is $\text{sup}_{\Sigma\sim A}\text{vol}(\Sigma)$. We first do this by eliminating $\phi$ from the max-cut 1 program by replacing the objective $\phi$ for $|\partial_{\mu}\psi|$, for which the maximum of the objective is clearly achieved.

We can also simplify the form of the boundary conditions, such that $\psi=\chi_{A}$ on all of $R^{c}$. This is accomplished in the following way. First note that the future directed causal condition on $\psi$ implies it is non-decreasing along any causal curve.\footnote{To see this, let $m^{\mu}$ be any future directed causal vector field. Then $m^{\mu}\partial_{\mu}\psi\geq0$.} Collectively the boundary conditions imply $0\leq\psi\leq 1$ holds everywhere, since every point in $M$ lies on a causal curve starting and ending on $R^{c}$. Let's now consider when $\psi$ doesn't saturate the inequalities, \emph{e.g.}, $\psi<\chi_{A}$ on some subset $q$ of $R^{c}_{+}$. Since $0\leq\psi\leq 1$, $q$ must be a subset of boundary region $A$. Then define a new function $\tilde{\psi}$ such that it equals $\psi$ everywhere except within a small neighborhood of $q$, where $\tilde{\psi}=1$. Since $q$ is spacelike, this can be done such that $\partial_{\mu}\tilde{\psi}$ is a future directed causal covector field satisfying $|\partial_{\mu}\tilde{\psi}|>|\partial_{\mu}\psi|$. In other words, the objective for $\tilde{\psi}$ is larger than for $\psi$, and, consequently, without changing the supremum we may have $\psi=\chi_{A}$ on $R^{c}_{+}$. A very similar argument yields $\psi=\chi_{A}$ on $R^{c}_{-}$.

With these simplifications, we may recast the dual program (\ref{eq:dualprogLorentzv1}) more succinctly
\beq
\begin{split}
&\text{max cut 2}\;:\;\text{maximize}\;\int_{M}\sqrt{g}|\partial_{\mu}\psi|\;\;\text{over}\;\psi\;\;\text{with}\; \psi|_{R^{c}}=\chi_{A}\;,\partial_{\mu}\psi \;\text{future directed causal}\;.
\end{split}
\label{eq:dualprogLorentzv2}\eeq
Our last task is to now show the solution $d^{\ast}$ to this dual program is equal to $\text{sup}_{\Sigma\sim A}\text{vol}(\Sigma)$.\footnote{We will show an example of \emph{convex relaxation}, which replaces a non-convex optimization problem, namely, finding the maximal volume slice in a given homology class, with an equivalent convex program (max cut 2).} Following the logic outlined in \cite{Headrick:2017ucz}, we will first work to rewrite the objective of the dual program (\ref{eq:dualprogLorentzv2}) such that it is related to the volume of a bulk slice $\Sigma\sim A$, such that $\int_{M}\sqrt{g}|\partial_{\mu}\psi|=\text{vol}(\Sigma)$. Consequently, taking the supremum of both sides of this relation demonstrates the solution to (\ref{eq:dualprogLorentzv2}) is equal to right hand side of the min flow-max cut theorem (\ref{eq:minflowmaxcutapp}).

\begin{figure}[t]
\includegraphics[width=5.5cm]{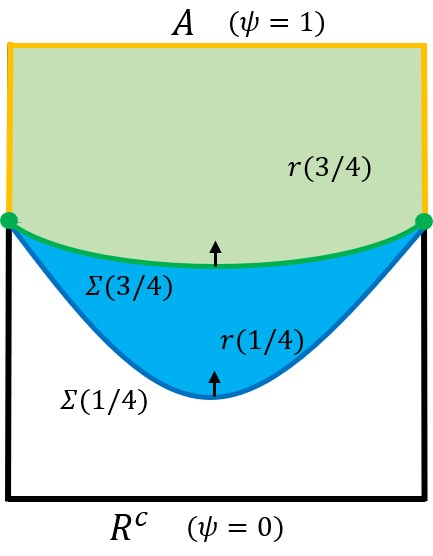}
\centering
\caption{A depiction of the level sets of $\Sigma(p)$ for $0<p<1$. The boundary region $A$ is the thick upper gold region, while the lower black boundary represents $R^{c}$. The boundary condition $\psi|_{R^{c}}=\chi_{A}$ sets $\psi=1$ on $A$ and $0$ on $A^{c}$. The bulk region $r(3/4)$ associated with $\Sigma(3/4)$ is shaded in green, while the bulk region $r(1/4)$ is shaded in blue. Note that unlike the Riemannian case, since $\partial_{\mu}\psi$ is future directed and causal, $0\leq\psi\leq1$, disallowing for any level sets for $p>1$ or $p<0$. }
\label{fig:mfmcpfnest}\end{figure}

Let us therefore focus on the objective $\int_{M}\sqrt{g}|\partial_{\mu}\psi|$. Let us assume for simplicity that $\psi$ living on $M$ is differentiable.\footnote{This is not strictly true as $\psi$ is not differentiable on $\partial A$ where $\psi|_{R^{c}}=\chi_{A}$ leads to a discontinuous jump.} We will consider the level sets of $\psi$, \emph{i.e.}, $\{\psi(x)=p\}$ for $p\in\mathbb{R}$. Given that $\psi|_{R^{c}}=\chi_{A}$, we can consider a one-parameter family of bulk regions $r(p)$ associated with the level sets of $\psi$ such that
\beq r(p)\equiv\{x\in M\;:\;\psi(x)\geq p\}\;,\label{eq:bulklevelregs}\eeq
for which we emphasize the special values $r(0)$ and $r(1)$. By continuity of $\psi$, we have $\psi=p$ on $\partial r(p)\setminus\partial M$. The bulk slice $\Sigma(p)$ (the level set) is defined as the closure of $\partial r(p)\setminus\partial M$, with an orientation covector $n_{\mu}$ parallel to $\partial_{\mu}\psi$, such that it points into bulk region $r(p)$ (see Figure \ref{fig:mfmcpfnest}). If we consider $\psi$ as a coordinate on the part of $M$ where $\partial_{\mu}\psi\neq0$, we may rewrite the objective as an integral over $p$ of the volume of the level sets $\Sigma(p)$:
\beq \int_{M}\sqrt{g}|\partial_{\mu}\psi|=\int_{-\infty}^{\infty}dp\;\text{vol}(\Sigma(p))\;.\label{eq:objfuncv1}\eeq

Now use the boundary condition on $\psi$, $\psi|_{R^{c}}=\chi_{A}$ (recall $\chi_{A}=1$ on $A$ and $\chi_{A}=0$ on $A^{c}$). Then, using (\ref{eq:bulklevelregs}), for $0<p<1$, we have $r(p)\cap R^{c}=A$, such that $\Sigma(p)$ is homologous to $A$.\footnote{Unlike the Riemannian case, since $0\leq\psi\leq 1$, there are no level sets with $p<0$ or $p>1$.} Using that $\Sigma(p)\sim A$ and taking into account that the volume is a non-negative quantity, the integrated volume of the level sets is bounded from above by the volume of the maximal slice homologous to $A$. That is,
\beq \int_{-\infty}^{\infty}\;dp\;\text{vol}(\Sigma(p))=\int_{0}^{1}dp\;\text{vol}(\Sigma(p))\leq \underset{\Sigma\sim A}{\text{sup}}\;\text{vol}(\Sigma)\;.\eeq
By (\ref{eq:objfuncv1}), we have then
\beq \underset{\underset{\psi|_{R^{c}}=\chi_{A}}{\psi}}{\text{sup}}\int_{M}\sqrt{g}|\partial_{\mu}\psi|\leq \underset{\Sigma\sim A}{\text{sup}}\;\text{vol}(\Sigma)\;.\eeq
All that is left to show is this bound is saturated. Consider any bulk slice $\Sigma\sim A$ with an associated bulk region $r$.  Let $\chi_{r}$ be its characteristic function (such that $\chi_{r}=1$ on $r$ and $\chi_{r}=0$ on $M\setminus r$). While $\chi_{r}$ is not generally differentiable due to this jump on $\Sigma$, it can be approximated arbitrarily well by a differentiable function $\psi$, such that the level sets of $\psi$ lie arbitrarily close to $\Sigma$. Consequently, the objective $\int_{M}\sqrt{g}|\partial_{\mu}\psi|$ is arbitrarily close to $\text{vol}(\Sigma)$. In other words, the supremum over all $\Sigma$ is equal to the supremum over all $\psi$, \emph{i.e.},
\beq \underset{\Sigma\sim A}{\text{sup}}\;\text{vol}(\Sigma)=\underset{\underset{\psi|_{R^{c}}=\chi_{A}}{\psi}}{\text{sup}}\int_{M}\sqrt{g}|\partial_{\mu}\psi|\;.\eeq
Thus, the solution $d^{\ast}$ to the dual program is equal to $\underset{\Sigma\sim A}{\text{sup}}\;\text{vol}(\Sigma)$, the right hand side of the min flow-max cut theorem (\ref{eq:minflowmaxcutapp}), and the proof is complete.


\subsection{Nesting property}

Let us now prove the nesting property for bulk slices used in Section \ref{sec:prelims} to derive the conditions of superadditivity and weak superadditivity. For convenience, we simplify the notation above a bit by using the shorthand
\beq \mathcal{C}(\sigma_{A})=\underset{v}{\text{inf}}\int_{A}v=\underset{\Sigma\sim A}{\text{sup}}\;\text{vol}(\Sigma)\;.\eeq
We will also assume the supremum is uniquely achieved, such that there exists a maximal volume slice $\Sigma^{\ast}$, for which we denote $\text{vol}(\Sigma^{\ast})=\mathcal{V}(\Sigma^{\ast})$. We also leave the condition $n_{\mu}v^{\mu}|_{R}=0$ implicit.

Let $A$ and $B$ be two disjoint boundary regions (though possibly sharing a common boundary, see, \emph{e.g.} Figure \ref{fig:maxslicenest}) and denote $AB$ as their union. The nesting property for Lorentzian flows is a lemma of the min flow-max cut theorem which states there exists a flow $v(A,B)$ that simultaneously minimizes the flux through $A$ and $AB$. That is, the nesting property says
\beq \int_{AB}v(A,AB)+\int_{A}v(A,AB)=\mathcal{C}(\sigma_{AB})+\mathcal{C}(\sigma_{A})\;.\label{eq:nestpropapp}\eeq
Generally $v(A,AB)$ will not also minimize the flux through $B$, but in fact maximize it. The nesting property for cuts is simply that $r(A)\subset r(AB)$.

The proof for the (continuum) Riemannian version of the nesting property was given in \cite{Headrick:2017ucz} using convex optimization techniques. A completely analogous proof for the Lorentzian case is given below, which takes the min flow-max cut theorem (\ref{eq:minflowmaxcutapp}) as an input and will make use of strong duality.

On the flow side, the proof follows a similar format as for the min flow max cut theorem, involving three steps in which we: (i) write a convex program for a combined sum of flows through $A$ and $AB$ (see the left hand side of (\ref{eq:nestpropapp})); (ii) find the Lagrange dual optimization problem, and (iii) show the solution to the dual problem is the right hand side of (\ref{eq:nestpropapp}).

\textbf{(i)} The left hand side of (\ref{eq:nestpropapp}) is straightforward to write as a convex optimization program we refer to as the min flow program:
\beq \text{min flow}\;:\;\text{minimize}\;\left(\int_{A}v+\int_{AB}v\right)\;\text{over}\;v\,,\;\text{subject to}\; \nabla\cdot v^{\mu}=0\,,\;1-|v^{\mu}|\leq0\;,\label{eq:minflowprognest}\eeq
where $v=v(A,B)$ is a future directed causal vector field. The solution $p^{\ast}$ to the min flow program is clearly bounded below by the sum of the individual minima of the two, \emph{i.e.},
\beq \underset{v}{\text{inf}}\left(\int_{A}v+\int_{AB}v\right)\geq\mathcal{C}(\sigma_{A})+\mathcal{C}(\sigma_{AB})\;.\eeq
When this bound is saturated we have the nesting for flows (\ref{eq:minflowmaxcutapp}). Moreover, since $v$ is a feasible vector field with $|v|>1$ everywhere, Slater's condition is satisfied.

\text{(ii)} We now dualize the program in nearly identically the same way as for the min flow-max cut theorem, introducing to Lagrange multiplier scalar fields $\psi,\phi$ for the respective explicit constraints. We then likewise replace the objective $\int_{M}\sqrt{g}\phi$ with $\int_{M}\sqrt{g}|\partial_{\mu}\psi|$ where the $\psi$ will obey a particular boundary condition and $\partial_{\mu}\psi$ is a future directed causal covector field. Glossing over some of the algebra, we have the dual program
\beq \text{combined max cut}\;:\; \text{maximize}\;\int_{M}\sqrt{g}|\partial_{\mu}\psi|\;\text{over}\;\psi\;\text{with}\;\psi|_{R^{c}}=\chi_{A}+\chi_{AB}\;,\label{eq:dualmincutnest}\eeq
where $\chi_{A}$ and $\chi_{AB}$ are the characteristic functions for $A$ and $AB$, respectively. Note that since $\chi_{AB}=\chi_{A}+\chi_{B}$, then $\chi_{A}+\chi_{AB}=2\chi_{A}+\chi_{B}$. Since $\partial_{\mu}\psi$ is future directed, then by the boundary condition $\psi|_{R^{c}}=\chi_{A}+\chi_{AB}$, we have $0\leq\psi\leq2$ everywhere.

\textbf{(iii)} We will again consider the level sets of the single function $\psi$, but this time to simultaneously represent the $A$ and $AB$ cuts.  More precisely, we define a bulk region $r(p)$ with $p\in\mathbb{R}$ as $r(p)\equiv \{x\in M\;:\;\psi(x)\geq p\}$ such that the level set $\Sigma(p)$ is the closure of $\partial r(p)\setminus\partial M$. The boundary condition on $\psi$ implies that its level sets $\Sigma(p)$ are homologous to $AB$ for $0<p<1$ and homologous to $A$ for $1<p<2$. Consequently,
\beq \int_{M}\sqrt{g}|\partial_{\mu}\psi|=\int_{-\infty}^{\infty}dp\;\text{vol}(\Sigma(p))\leq\int_{0}^{1}dp\;\text{vol}(\Sigma(p))+\int_{1}^{2}dp\;\text{vol}(\Sigma(p))\leq +\mathcal{C}(\sigma_{A})\;,\label{eq:intstep1}\eeq
where, for example, by $\mathcal{C}(\sigma_{AB})$ is shorthand for $\underset{\Sigma(AB)\sim AB}{\text{sup}}\;\text{vol}(\Sigma(AB))$. The only way for saturation to occur is if all of the $\Sigma(p)$ for $0<p<1$ fall on $\Sigma(AB)$, while $\Sigma(p)$ for $1<p<2$ fall on $\Sigma(A)$. Given $0\leq\psi\leq 2$, there are no level sets for $p<0$ or $p>2$.  By definition, $r(p)\subset r(p')$ for $p'<p$, \emph{i.e.}, $r(A)\subset r(AB)$.

Now take the supremum of the left hand side of (\ref{eq:intstep1}), which we recognize as the objective of the dual program (\ref{eq:dualmincutnest}),
\beq \underset{\underset{\psi|_{R^{c}=\chi_{A}+\chi_{AB}}}{\psi}}{\text{sup}}\int_{M}\sqrt{g}|\partial_{\mu}\psi|\leq\mathcal{C}(\sigma_{A})+\mathcal{C}(\sigma_{AB})\;.\eeq
All that remains is to show this bound is saturated. Following the same steps in the min flow-max cut theorem we have indeed inequality is saturated. Thus, the solution $d^{\ast}$ to the dual program (\ref{eq:dualmincutnest}) is equal to the right hand side of the nesting property (\ref{eq:nestpropapp}). Then, by Slater's condition in step (i), we have strong duality $d^{\ast}=p^{\ast}$, which completes the proof.


\subsection{Lorentzian multiflows: pitfalls and a partial proof}

The existence of Riemannian multiflows was used to prove monogamy of holographic mutual information \cite{Cui:2018dyq}. If  Lorentzian multiflow were to exist, we expect it could be used, in analogy with Riemannian multiflows, to prove inequalities for multiple boundary regions, \emph{e.g.}, generalized versions of the rate of complexity increase found in \cite{Couch:2018phr}, or possible a monogamy type relationship. However, in our attempts in proving the existence of Lorentzian multiflows, we found a number of conceptual and technical obstructions towards finding a general satisfactory definition of such flows. For the interested reader, here we outline some of the struggles in proving Lorentzian multiflows exist in general, with the hope that, if such a flow can be found, the present calculations can provide some use. In particular, here we describe possible obstructions in defining multiflows where the minimal flux out of a collection of boundary regions $A_i$ takes the form $v_i = \sum_{j} \xi_{ij} v_{ij}$, for some $\xi_{ij}$, a relaxation of the condition $v_{i}=\sum_{j}v_{ij}$ used for Riemannian flows. We will also present a partial proof of the existence of Lorentzian multiflows that are uncorrelated, and therefore in our opinion are uninteresting.

To set the stage, recall the definition of Riemannian multiflows $v_{ij}$ (\ref{eq:multiflowdef}), for which one has the max multiflow theorem (\ref{eq:maxmultiflowrie}), where $v_{i}=\sum_{j=1}v_{ij}$ is a max flow for spatial boundary subregion $A_{i}$, with $\int_{A_{i}}v_{i}=S(A_{i})$.  A key point is that the $v_i$ contain correlated components $v_{ij}$. Namely, the vector field maximizing the flux out of region $A_3$ can be written in terms of a linear sum of vector fields, which includes for example $v_{32}$. But the vector field $v_{32}= -v_{23}$ also appears when considering the maximal flux out of the region $A_2$. It is by considering various linear combinations of the $v_{i}$ that one is able to prove MMI \cite{Cui:2018dyq}.

The goal in the Lorentzian case is then to find a collection of flows $\mathcal{V}= \{v \}$, such that some linear combination of the $v \in \mathcal{V}$, $v_i = \sum_{v \in \mathcal{V}} \xi_{i, v} v$,  minimizes the flux out of $n$ boundary regions $A_i$ covering $\partial M$. That is, there exists a Lorentzian multiflow $\{v_{ij}\}$ such that for each $i$, the $n$ vector fields $v_{i}=\sum_{j=1}^{n}\xi_{ij}v_{ij}$, for some $\xi_{ij}$, is a min flow for $A_{i}$, \emph{i.e.},
\beq \Phi(A_{i})=\int_{A_{i}}v_{i}\;.\label{eq:minmultiflow}\eeq
If such a flow exists, where the $v_{i}$ contain correlated components, then we might hope to find new, generalized inequalities. We already know in certain spacetimes such flows exist; indeed in \cite{Couch:2018phr} it was shown when the strong energy condition and Einstein's equations hold in the bulk, a foliation of the boundary induces a maximal foliation of the bulk where the unit vector normal to the bulk Cauchy slices is identified with a min multiflow, minimizing over all boundary regions.

More generally, in analogy with Riemannian multiflows, one might guess a natural definition of Lorentzian multiflows is as given in (\ref{eq:lorentzmultiflowv1}). Then the outline of the proof of existence of a min multiflow is as follows. By the reverse triangle inequality, $v_{i}$ is a Lorentzian flow such that
\beq \sum_{i=1}^{n}\left(\int_{A_{i}}v_{i}\right)\geq \sum_{i=1}^{n}\Phi(A_{i})\;.\eeq
Therefore, to prove (\ref{eq:minmultiflow}) we must show
\beq \sum_{i=1}^{n}\left(\int_{A_{i}}v_{i}\right)\leq\sum_{i=1}^{n}\Phi(A_{i})\;,\label{eq:LHSminmultflowpf}\eeq
as then $\int_{A_{i}}v_{i}\leq\sum_{i=1}^{n}\Phi(A_{i})$. The strategy to verify this is similar to the proof of min flow-max cut theorem where one develops a Lagrangian dual program and invoking strong duality. This again broadly involves three steps: (i) express the left hand side (\ref{eq:LHSminmultflowpf}) as a convex optimization problem, showing Slater's condition is satisfied; (ii) use Lagrangian duality to rewrite the program as an equivalent optimization problem, and (iii) verify the optimal point $d^{\ast}$ of the dual program is bounded from above by
\beq d^{\ast}\leq\sum_{i=1}^{n}\Phi(A_{i})\;.\eeq
Then, following Slater's condition and strong duality, where the optimal solution of the original program $p^{\ast}=d^{\ast}$, one would have $p^{\ast}\leq\sum_{i=1}^{n}\Phi(A_{i})$, completing the proof.

Unfortunately, the definition (\ref{eq:lorentzmultiflowv1}) has a number of misgivings, as we list below, such that even (\ref{eq:LHSminmultflowpf}) is ill-defined. In the obstructions we encounter, we will offer potential resolutions, however, ultimately find the existence of min multiflows on general Lorentzian spacetimes remains elusive.

\noindent \textbf{Obstruction (1): Flows with flux unbounded from below}

In the Riemannian definition of multiflows, the spacetime is covered by non-overlapping boundary regions not necessarily nested, though a corollary to the max multiflow theorem is the nesting property. In the Lorentzian context, the homology condition demands that for each $A_i$, $J^+(A_i) \cap \partial M = A_i$. Regions that do not satisfy this condition, such as region $B$ in Figure \ref{fig:nestedABCv1}, can  have flux $\Phi_{i}$ unbounded from below. By contrast, flows through nested regions $A$ or $AB$ have flux $\Phi(A_{i})=\mathcal{C}(\sigma_{A_{i}})$, unbounded from above, but bounded from below. From the perspective of the min flow-max cut theorem unbounded from above is not a problem since we are interested in the flow with minimal flux, and $A$ and $AB$ obey the relative homology condition. However, we immediately run into an obstacle when $\Phi(A_{i})$ is unbounded from below, namely, we find a contradiction with (\ref{eq:LHSminmultflowpf}), and therefore the outlined strategy above fails.

This suggests we should restrict ourselves to nested boundary regions let us attempt to find a min multiflow in this context. Let the boundary of the Lorentzian manifold $M$ be covered by (overalpping) nested regions $A_{i}$ and their complement $\overline{\cup_{i}A_{i}}=A_{0}$. Denote the region between the intersection of $A_i, A_{i+1}$ as $B_i$, a collection of disjoint boundary regions $B_i$ similar to $B$ in Figure \ref{fig:nestedABCv1}.\footnote{As another example, one can consider three nested boundary regions $A,B,C$, and their complement $D$. Then, correspondingly, we would have $A_{1}=ABC$, $A_{2}=AB$, and $A_{3}=A$, where the $B_{i}$'s are given by $B_{1}=B$ and $B_{2}=C$.} Thus, $\partial M=\cup_{i}A_{i}\cup A_{0}$.  The goal is to find a collection of vector fields $v_{ij}$ such that some linear combination of the $v_{ij}$, $v_i = \sum_{j} \xi_{ij} v_{ij}$ are themselves flows minimizing the flux through each $A_i$. We proceed below, noting further obstacles.

\noindent \textbf{Obstruction (2): Antisymmetry of multiflow}

In the proof of MMI, it is important that each $v_{ij}$ is itself a flow. Analogously, we want to demand  $v_{ij}$ satisfy
\beq \nabla \cdot v_{ij} =0\;,\quad  |v_{ij}| \geq 1\;,\quad v_{ij}\;\;\text{causal future directed}\;,\quad n_{\mu} v^{\mu}_{ij}|_{A_{k}} = 0 \;\forall \;i, j \neq k\;.\eeq
However, the causal future directed (FDC) condition gives rise to an immediate disruption to a common additional requirement of multicommodity flows: $v_{ij} = -v_{ji}$. This condition ensures $v_{ij}$, combined with the condition has the interpretation of a flow between regions $A_i$ and $A_j$, whilst $v_i$ is the net flow out of $A_i$. Keeping to this interpretation, imposing $v_{ij}= -v_{ij}$ where $v_{ij}$ is FDC for $i<j$ and past directed and causal for $j>i$. In doing so, however, we run into a direct contradiction: since each $v_i$ must be a flow, then by the reverse Cauchy-Schwarz inequality, the linear combination $v = \sum_i v_i$ should also be a flow. However, $v= \sum_{ij} v_{ij} =0$ due to the antisymmetry of $v_{ij}$. Thus, $v=0$ and is not a flow since it doesn't satisfy the norm bound $|v|\geq1$. Note that while we don't care to work with the linear combination $v$, for consistency $v$ should still be a flow, but it is not. Moreover, note the antisymmetry is not an obstruction in the Riemannian case, since $v=0$ is a valid flow due to the norm bound being an upper bound. Due to this inconsistency, we drop the condition $v_{ij}=-v_{ji}$.

\noindent \textbf{Obstruction (3): Flux through $A_{i}$}

With overlapping nested regions $A_{i}$ on the Lorentzian boundary, each $A_{k}$ will contain a spacelike boundary region, for which its normal $n_{\mu}$ is timelike and $n_{\mu} v^{\mu}_{ij}|_{A_{k}} \neq 0$ for any $k$, where $v_{ij}$ is FDC (or PDC). Indeed, when considering relative homology with respect to $R$, this is one reason why we must impose the condition that $J^+(R^c) = J^-(R^c) =M$, which precludes the case that $R$ is a purely spacelike boundary region.\footnote{If $R$ is purely spacelike then it cannot be reached via a causal curve from the future or the past, depending on whether $n_{\mu}$ is past or future directed.} This arises in the proof of Slater's condition for the Lorentzian MFMC.  Since $J^+(R^c) = J^-(R^c) =M$ any point in $M$ lies on a causual curve starting/ending on $R_c$ and so we can find a feasible vector field for which $n \cdot v|_R=0$.

In the case of nested surfaces $A_i$, each satisfying the homology condition $J^+(A_i) \cap \partial M = A_i$, there always exists a spacelike surface $\Sigma_i$ homologous to the $A_i$ \cite{Headrick:2017ucz}. Hence the flux through $A_i$ is strictly greater than zero; we cannot impose a flux-like bound through the $A_i$. Put succinctly, since we are working with nested regions, we no longer have $n\cdot v_{ij}|_{A_{k}}=0$ for $k\neq i,j$, and so we drop this from the definition as well.  Also note we cannot impose flux conditions on the $B_i$ in a natural way, since they will not all satisfy $J^+(B_i^c) = J^-(B_i^c) =M$ which is an obstruction to satisfying Slaters condition.

\noindent \textbf{A proof for uncorrelated flows through nested regions}

Despite each of these last two obstructions, let us attempt anyway to prove the existence for a min multiflow when $A_{i}$ form a collection of nested regions. Let $\{v_{ij}\}$ be a collection of flows such that $|v_{ij}| \geq 1$, $v_{ij}$ FDC for $i<j$ and $\nabla \cdot v_{ij} =0$. We try to find $\xi_{ij}$ such that each $v_i = \sum_{j} \xi_{ij} v_{ij}$ is a minimal flow for each of the $A_i$, where the $v_i$ contains correlated components. Since we are considering nested regions, all we need to show is
\beq \sum_{i=1}^{n}\left(\int_{A_{i}}v_{i}\right)\leq\sum_{i=1}^{n}\mathcal{C}(\sigma_{A_{i}})\;,\label{eq:LHSminmultflowpf2}\eeq

We find it insightful to first consider uncorrelated flows, where $v_i = \sum_{j<i} v_{ji}$, \emph{i.e.}, for $\xi_{ij}=\delta_{ij}$ for $j<i$. In this case, since the components which make up each $v_i$ are independent we know the program is \textit{equivalent} to minimizing the flow through each $A_i$ individually, and, consequently, we will not find any new interesting inequalities relating flows connecting to multiple regions. Moreover, we will see arranging $\xi_{ij}$ such that $v_{ij}$ are correlated is unable to properly dualize the program.

The program we aim to dualize is
\beq
\begin{split}
\text{min multiflow}\;\;:\;\;&\text{minimize}\;\sum_{i=1}^{n}\int_{A_{i}}\sqrt{h}n_{\mu}v_{i}^{\mu}\;\;\text{over}\;v_{ij}^{\mu}\; \text{with} \; i<j \;,\\
&\text{subject to} \;\nabla_{\mu}v^{\mu}_{ij}=0\;,\;1-|v_{ij}^{\mu}|\leq0\;
\end{split}
\label{eq:minmultiflowprog}\eeq
Following the same argument as in step (i) of the proof of the min flow-max cut theorem (\ref{eq:minflowprog}), Slater's condition holds, \emph{i.e.}, for each $i,j$ there exists a feasible vector field $v^{\mu}_{ij}$ with $1-|v_{ij}^{\mu}|<0$ everywhere in $M$.

We now dualize the program, similar to what was done for the min flow-max cut theorem. As described above, this is accomplished by introducing a Lagrange multiplier for each constraint, and then integrating out the primal variables, leaving only an equivalent program in terms of Lagrange multipliers. Thus, introduce a set of Lagrange multiplier scalar fields $\psi_{ij}$ with $i<j$ associated to the divergenceless constraint $\nabla_{\mu}v_{ij}^{\mu}=0$, and Lagrange multiplier scalar fields $\phi_{ij}
$  associated to the norm bound constraint $1-|v^{\mu}_{ij}|\leq0$, for which $\phi_{ij} \geq 0$. The Lagrangian associated to the program dual to (\ref{eq:minmultiflowprog}) is then
\begin{equation}
L\left(\{v_{i j}^{\mu} \},\left\{\psi_{i j}\right\}, \{\phi_{ij}\} \right)=\sum_{i<j}^n\int_{A_{j}} \sqrt{h} n_{\mu} v_{i j}^{\mu}+\int_{M} \sqrt{g}\left[-\sum_{i <j}^{n} \psi_{i j} \nabla_{\mu} v_{i j}^{\mu}+\sum_{i <j}^{n} \phi_{ij}\left(1-\left|v_{i j}^{\mu}\right|\right)\right] .
\end{equation}
By induction, the first term can be recast as
\begin{equation}
\sum_{i<j}^n \int_{A_j} \sqrt{h} n_{\mu} v_{i j}^{\mu} = \sum_{i<j}^n \sum_{k=j}^n \int_{B_k} \sqrt{h} n_{\mu} v_{i j}^{\mu}= \sum_{k=1}^n \int_{B_k} \sum_{j=1}^{k} \sum_{i <j} v_{ij}\;.
\end{equation}

Performing integration by parts to move the derivative off of $v_{ij}^{\mu}$ in the second term we find
\beq -\int_{M}\sqrt{g}\sum_{i<j}^{n}\psi_{ij}\nabla_{\mu}v^{\mu}_{ij}=\int_{M}\sqrt{g}\sum_{i<j}^{n}v^{\mu}_{ij}\nabla_{\mu}\psi_{ij}-\sum_{i<j}^{n}\int_{\partial M}\sqrt{h}n_{\mu}v^{\mu}_{ij}\psi_{ij}\;.\eeq
Using $\partial M=B_{i}\cup B_{j}\cup_{k\neq i,j}B_{k}$, for the non-overlapping regions $B_{i}$, which includes $B_0$, we can replace the integral over $\partial M$ as the sum of integrals over $B_{i}$, such that the Lagrangian of the dual program becomes
\beq
\begin{split}
L\left(\{v_{i j}^{\mu} \},\left\{\psi_{i j}\right\}, \{\phi_{ij}\} \right)=& \sum_{k=0}^n \int_{ B_k} \left(\sum_{j =1 }^k \sum_{i <j}^n   \sqrt{h} n_{\mu}  v_{ij}(1- \psi_{ij}) - \sum_{j=k+1}^n \sum_{i<j}^n \psi_{ij}  \right)\\
&+\sum_{i<j}^{n}\int_{M}\sqrt{g}\left(\phi_{ij}+(v^{\mu}_{ij}\nabla_{\mu}\psi_{ij}-|v_{ij}^{\mu}|\phi_{ij}))\right)\;.
\end{split}
\eeq

Following the discussion in (\ref{eq:Lagdualmfmc}), the boundary term will have a finite infimum when $\psi_{ij}|_{A_{k}} = 1$ for all $i<j, j \leq k $ and $\psi_{ij}|A_k =0$ for all $i<j$, $j \geq k$. Meanwhile, the bulk term will only have a finite infimum if and only if we have $v^{\mu}_{ij}\partial_{\mu}\psi_{ij}\geq0$, such that $\partial_{\mu}\psi_{ij}$ is a future directed causal covector field, and $0\leq\phi_{ij}\leq|\partial_{\mu}\psi_{ij}|$ (the latter inequality follows from the reverse Cauchy-Schwarz inequality). We can replace $\phi_{ij}$ by its maximum value, $|\partial_{\mu} \psi_{ij}|$ in which case we arrive at the following dual program
\beq
\begin{split}
\text{dual}\;:\;\;&\text{maximize}\;\sum_{i<j}^n \int_{M}\sqrt{g}  |\partial_{\mu} \psi_{ij}|\;\;\;\text{over}\;\{\psi_{ij}\}\\
&\text{with}\; \psi_{ij}|_{A_{k}}=1\; \forall i<j, j \leq k, \; \psi_{ij}|_{A_{k}}=0 \; \forall i<j, j \geq k \;,
\end{split}
\label{eq:dualprogminmultif}\eeq
with $\partial_{\mu}\psi_{ij}$ being future directed. Defining $\psi\equiv\sum_{i<j}^n \psi_{ij}$, we can consider the level sets of $\psi$. It is easy to see that $\psi|_{A_k} = \sum_{i<k} = k$ so that $0 \leq \psi \leq k$. Furthermore, by the same arguments as in the proof of the min flow-max cut theorem, level sets of $\psi$ will be homologous to $A_k$ for $ k-1<p<k$, $k\geq1$. In particular, sparing the reader some of the details, we have,
 \begin{equation}
\int_{-\infty}^{\infty} |\partial_{\mu} \psi| = \int_0^{k} |\partial_{\mu} \psi| = \int_0^{1}|\partial_{\mu} \psi| + \int_{1}^{2}|\partial_{\mu} \psi| + \dots \leq \sum_i \sup(\text{Vol}(A_i)) = \sum_i \mathcal{C}(\sigma_{A_i}).
\end{equation}
and so the program dualizes as expected. Thus, there exists a min multiflow that is uncorrelated in general Lorentzian spacetimes.

Notice, however, when changing around the components $\xi_{ij}$, so that the flows $v_i$ contain correlated components, the proof seems to invariably break down. Notably, if we took a collection of flows $v_i = \sum_{j}  v_{ij}$ ($\xi_{ij}=1$ for all $j$) where the $v_{ij}$ are symmetric, then the above dualization proceeds in a similar manner, but we end up with an upper bound of $2 \sum_i \mathcal{C}(\sigma_{A_i})$.


\section{Verifying linearized Einstein's equations in $d= 3+1$} \label{app:verifyEineqs}

In four spacetime dimensions, rotations about the $(\tau,x,y)$ axis ( keeping $z$ fixed) are parametrized by
\begin{equation}
R=\left[\begin{array}{ccc}
\cos \theta+u_{\tau}^{2}(1-\cos \theta) & u_{\tau} u_{x}(1-\cos \theta)-u_{z} \sin \theta & u_{\tau} u_{y}(1-\cos \theta)+u_{x} \sin \theta \\
u_{x} u_{\tau}(1-\cos \theta)+u_{y} \sin \theta & \cos \theta+u_{x}^{2}(1-\cos \theta) & u_{x} u_{y}(1-\cos \theta)-u_{\tau} \sin \theta \\
u_{y} u_{\tau}(1-\cos \theta)-u_{x} \sin \theta & u_{y} u_{x}(1-\cos \theta)+u_{\tau} \sin \theta & \cos \theta+u_{y}^{2}(1-\cos \theta)
\end{array}\right]\;,
\end{equation}
which describes a proper rotation by an angle $\theta$ around the axis with unit vector $u = ( u_\tau,u_x,u_y)$. We shall exploit the rotations which mix the spatial and $\tau$ components of Euclidean Ads, which can be seen to be the Wick rotation of the Lorentz boosts.

First consider the rotation about the axis $(0,1,0)$ by an angle $\theta$. Substituting into \eqref{eq: moreGeneral} we are led to the expression
\begin{align}  \label{eq: xboost} \nonumber
& 2 \tau {\delta E}^{\tau\tau} \left(\tau^2 \sin ^2 \theta +\tau^2 \cos ^2 \theta +z^2 \sin ^2 \theta  \right)+
 2 \tau {\delta E}^{\tau y} \left(\tau^2 \sin 2 \theta -2 \tau^2 \sin \theta  \cos \theta -2 z^2 \sin \theta  \cos \theta \right) \\ \nonumber
 & +4 \tau^2 z \cos \theta  {\delta E}^{\tau z}+2 \tau \left(\tau^2+z^2\right) {\delta E}^{xx}+2 \tau {\delta E}^{yy} \left(\tau^2 \sin ^2 \theta +\tau^2 \cos ^2 \theta +z^2 \cos ^2 \theta \right)+\\
 & 4 \tau^2 z \sin \theta  {\delta E}^{zy}+2 \tau z^2 {\delta E}^{zz} =0\;.
\end{align}
The  left hand side of $\eqref{eq: xboost}$ is a real analytic function in $\theta$ in an open domain around $\theta = 0$. Hence, in order for it to vanish its Taylor series about $\theta =0$ must also vanish. Performing an expansion in powers of $\theta^n$ we are led to the four equations
\begin{align} \label{eq: xboostthetaexpansion}
& n=1: \  4 \tau^2 z {\delta E}^{zy}-4 \tau z^2 {\delta E}^{ty}\;, \\
 & n=2: \  -2 \tau^2 z {\delta E}^{\tau z}+2 \tau z^2 {\delta E}^{\tau\tau}-2 \tau z^2 {\delta E}^{yy}\;, \\
 & n=3:\  \frac{8}{3} \tau z^2 {\delta E}^{\tau y}-\frac{2}{3} \tau^2 z {\delta E}^{zy}\;, \\
 & n=4: \ \frac{1}{6} \tau^2 z {\delta E}^{\tau z}-\frac{2}{3} \tau z^2 {\delta E}^{\tau\tau}+\frac{2}{3} \tau z^2 {\delta E}^{yy}\;.
\end{align}
We could include higher powers of $\theta$, but one can verify that the order equations do not lead to new constraints. With a bit of algebra \eqref{eq: xboostthetaexpansion} the constraints are solved by
\begin{equation} \label{eq: solnXboost}
\begin{split}
&\delta E^{yy}= \delta E^{\tau\tau}\;, \quad \delta E^{\tau z}= 0\;,\quad  \delta E^{\tau y}= 0\;, \quad \delta E^{zy}=0\;,\\
&\delta E^{zz}=-\frac{\left(2 \tau^2+z^2\right) {\delta E}^{\tau\tau}}{z^2}-\frac{\left(\tau^2+z^2\right) {\delta E}^{xx}}{z^2}\;.
\end{split}
\end{equation}

By symmetry, rotation about the y axis, $(0,0,1)$ will give us a similar expression to (\ref{eq: solnXboost}) but with $x \leftrightarrow y$, which gives us further
\begin{equation}\label{eq: solnYboost}
{\delta E}^{xx}= {\delta E}^{\tau\tau},  \ {\delta E}^{zz}= -\frac{\left(2 \tau^2+z^2\right) {\delta E}^{\tau\tau}}{z^2}-\frac{\left(\tau^2+z^2\right) {\delta E}^{yy}}{z^2}, \  {\delta E}^{\tau x}= 0, \ {\delta E}^{zx}= 0\;.
\end{equation}
Comparing the two expressions for $\delta E^{zz}$ (or $\delta E^{\tau\tau}$) leads to
\begin{equation}
\delta E^{xx} = \delta E^{yy}\;.
\end{equation}

Finally we can study the mixed rotations, for example along the axis $(0,\frac{1}{\sqrt{2}},\frac{1}{\sqrt{2}})$. Under this boost  \eqref{eq: moreGeneral} becomes
\begin{equation}
4 \tau z^2 \sin ^2 \frac{\theta }{2} (\cos \theta -1) {\delta E}^{\tau\tau}+4 \tau z^2 \sin ^2  \frac{\theta }{2} (1-\cos \theta ) {\delta E}^{xx}+4 \tau z^2 \sin ^2 \frac{\theta }{2} (\cos \theta +1) {\delta E}^{xy} =0 .
\end{equation}
Substituting in for the previous expressions we have from Equations \eqref{eq: solnXboost} and \eqref{eq: solnYboost} gives us
\begin{equation}
{\delta E}^{xy}= 0 .
\end{equation}

This is enough to write all of the expressions in terms of $\delta E^{\tau\tau}$. Explicitly we have seen all of the components vanish expect for
\begin{equation}
\delta E^{\tau\tau} , \ {\delta E}^{xx}= \delta E^{\tau\tau}, \ {\delta E}^{yy}= {\delta E}^{\tau\tau}, \ {\delta E}^{zz}= \frac{\left(-3 \tau^2-2 z^2\right) {\delta E}^{\tau\tau}}{z^2}
\end{equation}
It remains to show that $\delta E^{\tau\tau}=0$. To do this, we shall use the Bianchi identities $\nabla_{ \mu} \delta E^{\mu \nu}=0$, which follow directly from the covariance of the action \cite{Faulkner:2013ica}. It is simple to solve the four Bianchi identities for $\delta E^{\tau\tau}$ to arrive at the following four expressions for $\delta E^{tt}$, which must be consistent\footnote{Here the $T,Z,X,Y$ describe which component of the Bianchi identity was solved.}
\begin{equation}
\delta E^{\tau\tau} = T(z,x,y), \delta E^{\tau\tau} = \frac{z^7 Z(\tau,x,y)}{( 3\tau^2 + 2z^2)^{1/4}}, \delta E^{\tau\tau} = X( \tau,z,y) , \delta E^{\tau\tau} = Y(\tau,z,x)\;.
\end{equation}
To see that the only consistent solution is that $\delta E^{\tau\tau} =0$, we first note the last two equations tell us $\delta E^{\tau\tau}$ has no $x$ or $y$ dependence, which along with the first equation tells us it depends only on $z$, but then the second equation clearly has a $t$ dependence and so the only consistent solution is $T(z,x,y) = Z(\tau,x,y) = X(\tau,z,y) = Y(\tau,z,x)=0$. Thus, $\delta E^{\tau\tau}=0$, which completes the proof.

\bibliographystyle{JHEP}
\bibliography{refs-CBT}

\end{document}